\documentclass[apj,numberedappendix,iop,twocolappendix]{emulateapj}

\usepackage%
[
pagebackref=true, 
colorlinks=true,
linkcolor=red,
filecolor=red,
citecolor=blue,
urlcolor=cyan,
pdftitle={},
pdfauthor={},
pdfsubject={},
pdfcreator={KAZEditor},
pdfproducer={Kazzimiei},
pdfkeywords={Planetary systems, Planets and satellites: formation, 
             Planets and satellites: individual (Kepler 11), 
             Planets and satellites: interiors, Protoplanetary disks, 
             Planet-disk interactions}
bookmarks=true,
bookmarksopen=false,
breaklinks=true,
hyperfootnotes=true,
hypertexnames=false,
pdftoolbar=true,
pdfmenubar=true,
pdfstartview=FitH,
pdffitwindow=false,
pdfnewwindow=true, 
pdfpagemode=None]{hyperref}
\usepackage{breakurl}
\newcommand{\Mearth}{\mbox{$M_{\oplus}$}}            
\newcommand{\Rearth}{\mbox{$R_{\oplus}$}}            
\newcommand{\Mjup}{\mbox{$M_{\mathrm{J}}$}}          
\newcommand{\ts}{\mbox{$t_{i}$}}                     
\newcommand{\Mp}{\mbox{$M_{p}$}}                     
\newcommand{\Ms}{\mbox{$M_{\star}$}}                 
\newcommand{\Msun}{\mbox{$M_{\odot}$}}               
\newcommand{\Rhill}{\mbox{$R_\mathrm{H}$}}           
\newcommand{\Rbondi}{\mbox{$R_\mathrm{B}$}}          
\newcommand{\tiso}{\mbox{$t_{\mathrm{iso}}$}}        
\newcommand{\Rcapt}{\mbox{$R_{\mathrm{A}}$}}         
\newcommand{\densu}{\mbox{$\mathrm{g\,cm^{-2}}$}}    
\newcommand{\K}{\mbox{$\mathrm{K}$}}                 
\newcommand{\AU}{\mbox{au}}                          
\newcommand{\ice}{\mbox{H$_{2}$O}}                   
\newcommand{\hhe}{\mbox{H+He}}                       
\newcommand{\kep}{\mbox{Kepler~11}}                  
\newcommand{\yr}{\mbox{yr}}                          
\newcommand{\Myr}{\mbox{Myr}}                        
\newcommand{\Gyr}{\mbox{Gyr}}                        
\newcommand{\cisec}[1]{Section~\ref{#1}}             
\newcommand{\cifig}[1]{Figure~\ref{#1}}              
\newcommand{\cieq}[1]{Equation~(\ref{#1})}           
\newcommand{\pz}{\phantom{0}}                        
\newcommand{\pp}{\phantom{+}}                        
 
\newcommand{\B}{\rule[-1.2ex]{0pt}{0pt}}             

\slugcomment{The Astrophysical Journal, 828:33, 2016 September 1}

\shorttitle{\footnotesize The Astrophysical Journal, 828:33, 2016 September 1\hfill D'Angelo \& Bodenheimer}
\shortauthors{\footnotesize The Astrophysical Journal, 828:33, 2016 September 1\hfill D'Angelo \& Bodenheimer}

\begin{document}

\title{In Situ and Ex Situ Formation Models of Kepler~11 Planets}

\author{Gennaro D'Angelo\altaffilmark{1,2} and Peter Bodenheimer\altaffilmark{3}}
\altaffiltext{1}{NASA Ames Research Center, MS 245-3, Moffett Field, CA 94035, USA
(\href{mailto:gennaro.dangelo@nasa.gov}{gennaro.dangelo@nasa.gov})}
\altaffiltext{2}{SETI Institute, 189 Bernardo Avenue, Mountain View, CA 94043, USA}
\altaffiltext{3}{UCO/Lick Observatory, University of California, Santa Cruz, CA 95064, USA
(\href{mailto:peter@ucolick.org}{peter@ucolick.org})}

\begin{abstract} 
We present formation simulations of the six \kep\ planets. 
Models assume either \textit{in situ} or \textit{ex situ} assembly,
the latter with migration, and are evolved to the estimated age 
of the system, $\approx 8\,\Gyr$.
Models combine detailed calculations of both the gaseous envelope 
and the condensed core structures, including accretion of gas and solids, 
of the disk's viscous and thermal evolution, including photo-evaporation 
and disk-planet interactions, and of the planets' evaporative mass loss 
after disk dispersal. 
Planet-planet interactions are neglected.
Both sets of simulations successfully reproduce measured radii, masses, 
and orbital distances of the planets, except for the radius of \kep b, 
which loses its entire gaseous envelope shortly after formation.
Gaseous (\hhe) envelopes account for $\lesssim18$\% of the planet 
masses, and between $\approx 35$ and $\approx 60$\% of the planet 
radii.
In situ models predict a very massive inner disk, whose solids' surface 
density ($\sigma_{Z}$) varies from over $10^{4}$ to $\approx 10^{3}\,\densu$
at stellocentric distances $0.1\lesssim r\lesssim 0.5\,\AU$. 
Initial gas densities would be in excess of $10^{5}\,\densu$ if solids 
formed locally.
Given the high disk temperatures ($\gtrsim 1000\,\K$), planetary 
interiors can only be composed of metals and highly refractory materials. 
Sequestration of hydrogen by the core and subsequent outgassing 
is required to account for the observed radius of \kep b.
Ex situ models predict a relatively low-mass disk, whose initial $\sigma_{Z}$
varies from $\approx 10$ to $\approx 5\,\densu$ at $0.5\lesssim r\lesssim 7\,\AU$
and whose initial gas density ranges from $\approx 10^{3}$ to $\approx 100\,\densu$.
All planetary interiors are expected to be rich in \ice, as core assembly 
mostly occurs exterior to the ice condensation front. 
\kep b is expected to have a steam atmosphere, and \ice\ is likely 
mixed with \hhe\ in the envelopes of the other planets.
Results indicate that \kep g may not be more massive than \kep e.
\end{abstract}


\section{Introduction}
\label{sec:Intro}

\defcitealias{lissauer2013}{L13}
\defcitealias{gennaro2010}{DL10}

Numerous planetary systems have been discovered that consist
of two or more planets with masses of a few Earth masses (\Mearth),
orbiting in the same plane within $0.5\,\AU$ of the central star
\citep[e.g.][]{figueira2012,mullally2015}.
Some properties of such systems are reviewed by \citet{winn2015}.  
A particularly well-studied example is the \kep\ system, with
a central solar-type star (age $8.5^{+1.1}_{-1.4}\,\mathrm{Gyr}$)
and six orbiting planets with semi-major axes ranging from $0.091$
to $0.466\,\AU$. Their radii, measured through transit observations
from the \textit{Kepler} spacecraft, are $1.80$--$4.19$ Earth radii (\Rearth),
placing them in the super-Earth/sub-Neptune size range.
Estimates of their masses have been obtained from transit
timing measurements 
\citep[hereafter \citetalias{lissauer2013}]{lissauer2013}; 
for the five inner planets, the values range from $1.9$ to 
$8.0\,\Mearth$ \citep[see also][]{borsato2014,hadden2014}. 
All of these inner planets have densities that are substantially less than 
that of a rocky planet, implying that they could each be composed of
a heavy-element core surrounded by a gaseous envelope. 
In the cases of planets c, d, e, and f, these envelopes are most
likely composed of hydrogen and helium, in roughly solar
proportions, while \kep b's envelope could be composed
either of \hhe\ or \ice\ steam 
(\citeauthor{lopez2012}\ \citeyear{lopez2012}; \citetalias{lissauer2013}). 
The estimated mass fractions of these envelopes range from 
$0.5$\% in the case of \kep b to $15.7$\% for \kep e 
\citepalias[see also \citeauthor{lopez2012}\ \citeyear{lopez2012}]{lissauer2013}. 
However, their volumes are substantial and play an important role 
in determining the observed planet radii.

One of the main issues pertinent to the understanding of this
system is the formation history of the planets. While it is
generally assumed that these planets formed by core-nucleated
accretion \citep{safronov1969}, the formation location is not well
established. Several studies have proposed that planets in systems 
of this type formed in situ 
\citep[e.g.,][]{hansen2012,hansen2013,ikoma2012,chiang2013,tan2015}
and analytic estimates of the envelope-to-core mass ratio in 
the relevant mass range have also been made \citep{lee2015,ginzburg2016}.
Other simulations support the alternative \textit{ex situ} assumption,
that is, that the planets formed farther out in the disk and, during or after 
formation, migrated inward  to their present positions through interactions 
with the protoplanetary disks 
\citep[e.g.,][]{mcneil2010,rogers2011,lopez2012,mordasini2012,bodenheimer2014,%
hands2014,chatterjee2015}.  
The physical mechanisms involved in the process of orbital migration
via disk-planet tidal interactions are reviewed by \citet{kley2012} and 
\citet[pp.\ 667-689]{baruteau2014}.

While the in situ model has been favored because the terrestrial
planets in the solar system presumably formed in this way,
the migration (or ex situ) model has been favored because of 
the difficulties in forming planets in situ in the very inner regions of disks
\citep{bodenheimer2000b}, well inside the orbit of Mercury 
in the solar system.
However, a significant argument against the ex situ picture is that
in a multiple system, if the planets had undergone convergent migration, 
they would be expected to have been captured into mean-motion 
resonances \citep[e.g.,][]{lissauer2011a}, while in fact most of the systems 
observed by Kepler do not appear to be in resonance. 
\citet{goldreich2014} discuss a mechanism, involving an instability in 
resonances, that would allow the planets to move through resonances; 
therefore, migration is not ruled out.
\citet{deck2015} revisit this problem and conclude
that although the instability in resonances is indeed possible,
the time spent by a planet pair in resonance exceeds by a considerable
amount the time during which the planets are out of resonance. 
Thus, they argue that the \citeauthor{goldreich2014} process does not 
solve the problem that most Kepler planet pairs are not in resonance.
However, there are other mechanisms that could move planets out of 
resonance, including dissipative effects \citep{delisle2012,lithwick2012,batygin2013},  
stochastic effects during migration \citep{rein2012}, tidal effects caused 
by planet-wake interactions \citep{baruteau2013} and the effects of small 
eccentricities in the planetary orbits \citep{batygin2015}.
Moreover, the highly complex orbital architecture of compact multiple systems, 
as in the case of \kep, may indeed require a migration-based formation 
scenario \citep{migaszewski2012}.
Nevertheless, there are still many other uncertainties in the theory of 
planet-disk interactions, as discussed in more detail by \citet{chiang2013}
and \citet{kley2012}, leading to legitimate questioning of whether 
the theoretical processes of planet formation with migration
can explain the mass distribution as well as the orbital period
distribution of super-Earth/sub-Neptune planets \citep{howard2010}.

On the other hand, there are several difficulties with the picture
of in situ formation of super-Earths/sub-Neptunes. 
First, even if the planets did form by this process, they still would be 
subject to orbital decay.
If migration is included in simulations of in situ super-Earth formation
\citep{ogihara2015}, the semi-major axis distribution of planets does 
not agree with the distribution observed by Kepler.
Agreement is possible only if orbital migration is suppressed.
Second, for the specific case of the \kep\ system, a very high surface 
density of solid material in the inner disk is required, well above that of
the minimum-mass solar nebula and even above that of minimum-mass 
extrasolar nebulae \citep{chiang2013,schlichting2014}. 
This problem may be alleviated if it is assumed that solid material migrates 
inwards from the outer disk, in the form of either protoplanetary cores 
\citep{ward1997}, planetesimals \citep{hansen2012}, or small pebbles 
\citep{tan2015}, and collects at the appropriate locations. 
Third, Kepler data show an excess of planet pairs just exterior 
of the 2:1 and 3:2 mean-motion resonances, compared to the pairs just 
interior to these resonances \citep{lissauer2011b,fabrycky2014,chatterjee2015}, 
which is not easily accountable by in situ formation.
Fourth, it has often been assumed that in the inner disk the temperatures 
are too high for an appreciable concentration of solid material to exist. 
In fact, the simple assumption that the ratio of sound speed to orbital 
speed, i.e., the ratio of disk scale height to radial distance, $H/r$, is 
$0.03$ at $0.15\,\AU$ gives a temperature of about $1500\,\K$ 
for a solar-mass central star. However, this objection is not 
necessarily significant. Many disk models give cooler 
temperatures at $0.1$--$0.4\,\AU$. The evolving two-dimensional
models of \citet{dodson2009} give mid-plane temperatures
in the inner disk of about $1000\,\K$ at an age of $10^5$ years, 
with cooling at later times. The models of \citet{chiang1999} give even
cooler temperatures in the disk interior at these distances.

This paper considers all the observed planets in the \kep\
system and asks whether they formed in situ or ex situ, i.e.,
including orbital migration. In spite of the difficulties mentioned above,
both of these possibilities remain viable options.
Detailed formation and evolution models are numerically simulated 
for both scenarios with the same or very similar physical assumptions 
used in the construction of the planet models. 
In both cases, the simulations are advanced to an age of $8\,\mathrm{Gyr}$, 
where the computed  masses, radii, and semi-major axes are compared 
with observations. 
The general physical and numerical aspects of the calculations are
reported in \cisec{sec:np}.
The in situ and ex situ models are presented, respectively, in 
\cisec{sec:insitu} and \ref{sec:exsitu} while results are discussed 
in \cisec{sec:dac}. The conclusions are drawn in \cisec{sec:fine}.

\section{Numerical Procedures}
\label{sec:np}

\begin{deluxetable}{cl} 
\tablecolumns{2} 
\tablewidth{0pc} 
\tablecaption{List of Symbols\label{table:qq}} 
\tablehead{ 
\colhead{Symbol}&\colhead{Definition}
} 
\startdata 
$M_{c}$                     & Planet's condensed core mass; \cisec{sec:esc}\\
$R_{c}$                     & Planet's condensed core radius; \cisec{sec:esc}\\
$\Rcapt$                    & Accretion radius; \cieq{eq:RA}\\
$\Rbondi$                   & Planet's Bondi radius; \cieq{eq:RA}\\
$\Rhill$                    & Planet's Hill radius; \cieq{eq:RA}\\
$L_{p}$                     & Planet total luminosity; \cieq{eq:LBC}\\
$T_{\mathrm{eff}}$          & Planet effective temperature; \cieq{eq:LBC}\\
$R_{p}$                     & Planet radius; \cieq{eq:LBC}\\
$\Mp$                       & Planet mass; \cieq{eq:PBC}\\
$\kappa_{\mathrm{R}}$       & Rosseland mean opacity; \cieq{eq:PBC}\\
$P$                         & Pressure; \cieq{eq:PBC}\\
$T_{\mathrm{eq}}$           & Irradiation equilibrium temperature; \cieq{eq:T4eff}\\
$T_{\star}$                 & Stellar effective temperature; \cieq{eq:Teq}\\
$R_{\star}$                 & Stellar radius; \cieq{eq:Teq}\\
$a$                         & Planet orbital radius; \cieq{eq:Teq}\\
$M_{e}$                     & Planet's gaseous envelope mass; \cisec{sec:esc}\\
$\dot{M}_{e}$               & Planet's mass accretion rate of gas; \cisec{sec:esc}\\
$\dot{M}_{c}$               & Planet's mass accretion rate of solids; \cieq{eq:dotmc}\\
$\sigma_{Z}$                & Disk's surface density of solids; \cieq{eq:dotmc}\\
$\Omega$                    & Orbital frequency; \cieq{eq:dotmc}\\
$F_{g}$                     & Gravitational enhancement factor; \cieq{eq:dotmc}\\
$M_{\star}$                 & Stellar mass; \cieq{eq:mciso}\\
$L_{\star}$                 & Stellar luminosity; \cisec{sec:gl}\\
$\dot{M}^{\mathrm{iso}}_{e}$  & Gas mass-loss rate during isolation; \cieq{eq:mexuv}\\
$F_{\mathrm{XUV}}$          & XUV radiation flux during isolation; \cieq{eq:mexuv}\\
$R_{\mathrm{XUV}}$          & XUV radiation absorption radius; \cieq{eq:mexuv}\\
$\varepsilon$               & XUV absorption efficiency; \cieq{eq:mexuv}\\
$r$                         & Stellocentric distance; \cieq{eq:devol0}\\
$\Sigma$                    & Disk's surface density of gas; \cieq{eq:devol0}\\
$\mathcal{T}$               & Gravitational torque; \cieq{eq:devol0}\\
$\mathcal{T}_{\nu}$         & Viscous torque; \cieq{eq:devol0}\\
$\nu$                       & Gas kinematic viscosity; \cieq{eq:devol1}\\
$T$                         & Temperature; \cieq{eq:Qcool}\\
$T_{\mathrm{irr}}$          & Irradiation temperature; \cieq{eq:Qirr}\\
$H$                         & Disk scale height; \cieq{eq:Tirr}\\
$\dot{M}_{\mathrm{scat}}$   & Scattering rate of solids; \cieq{eq:core_scat}\\
$\dot{\Sigma}_{\mathrm{pe}}$& Disk photo-evaporation rate; \cieq{eq:dsig_pe}\\
$r_{\mathrm{crt}}$          & Critical radius for photo-evaporation; \cisec{sec:DPH}
 \enddata
\end{deluxetable}

In this section, we outline the numerical methods applied in the models
presented herein, highlighting major differences between in situ and 
ex situ calculations. As a reference, Table~\ref{table:qq} contains a list 
of some of the symbols used in the paper and the equation in which 
they appear or the section in which they are first mentioned
(physical constants are omitted). 
In order to simplify labels, some quantities apply to both 
the disk and the planet, e.g., $T$ for gas temperature or 
$\kappa_{\mathrm{R}}$ for the Rosseland mean opacity, 
and shall be distinguished by the context in which they are used.

\subsection{Envelope Structure Calculation}
\label{sec:esc}

The calculation of the structure and evolution of the planetary gaseous 
envelope is based on the assumption that the envelope is spherically 
symmetric around its center and evolves through states of hydrostatic 
equilibrium \citep[e.g.,][]{kippenhahn2013}. 
The envelope lies on a core of condensed matter, whose total mass 
$M_{c}$ and radius $R_{c}$ are both functions of time.
The core radius is determined as explained below.
The envelope structure is calculated by solving the equations for mass
conservation, hydrostatic equilibrium, energy conservation, and 
radiation diffusion \citep[see][]{bodenheimer1986}.
The energy equation includes heating produced by in-falling 
planetesimals, the work done by gravity, cooling from the release 
of internal heat, and heating by stellar radiation (when applicable).
In convective unstable shells, where the radiative temperature gradient
exceeds in magnitude the adiabatic gradient \citep{kippenhahn2013}, 
the actual gradient of temperature is set equal to the adiabatic gradient.

The chemical composition of the envelope gas is assumed to be uniform,
with hydrogen and helium mass fractions  $X = 0.74$ and $Y = 0.24$,
respectively. The equation of state for this gas mixture is that computed
by \citet{saumon1995}, which accounts for the partial degeneracy of 
electrons and for non-ideal effects in the gas. 
(Strictly speaking, this equation of state neglects heavier elements and 
uses $Y=1-X$.)

\begin{figure}
\centering%
\resizebox{\linewidth}{!}{%
\includegraphics[clip]{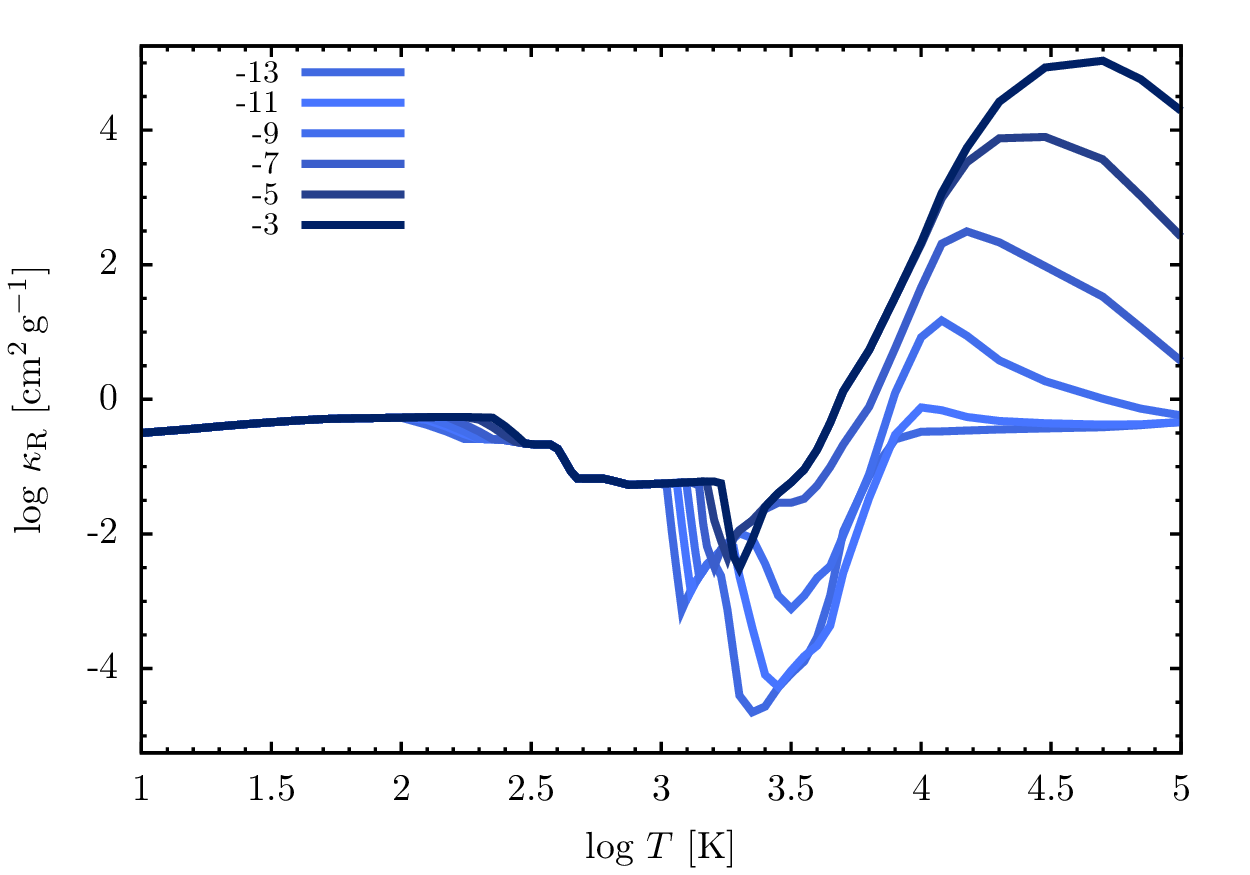}}\\
\resizebox{\linewidth}{!}{%
\includegraphics[clip]{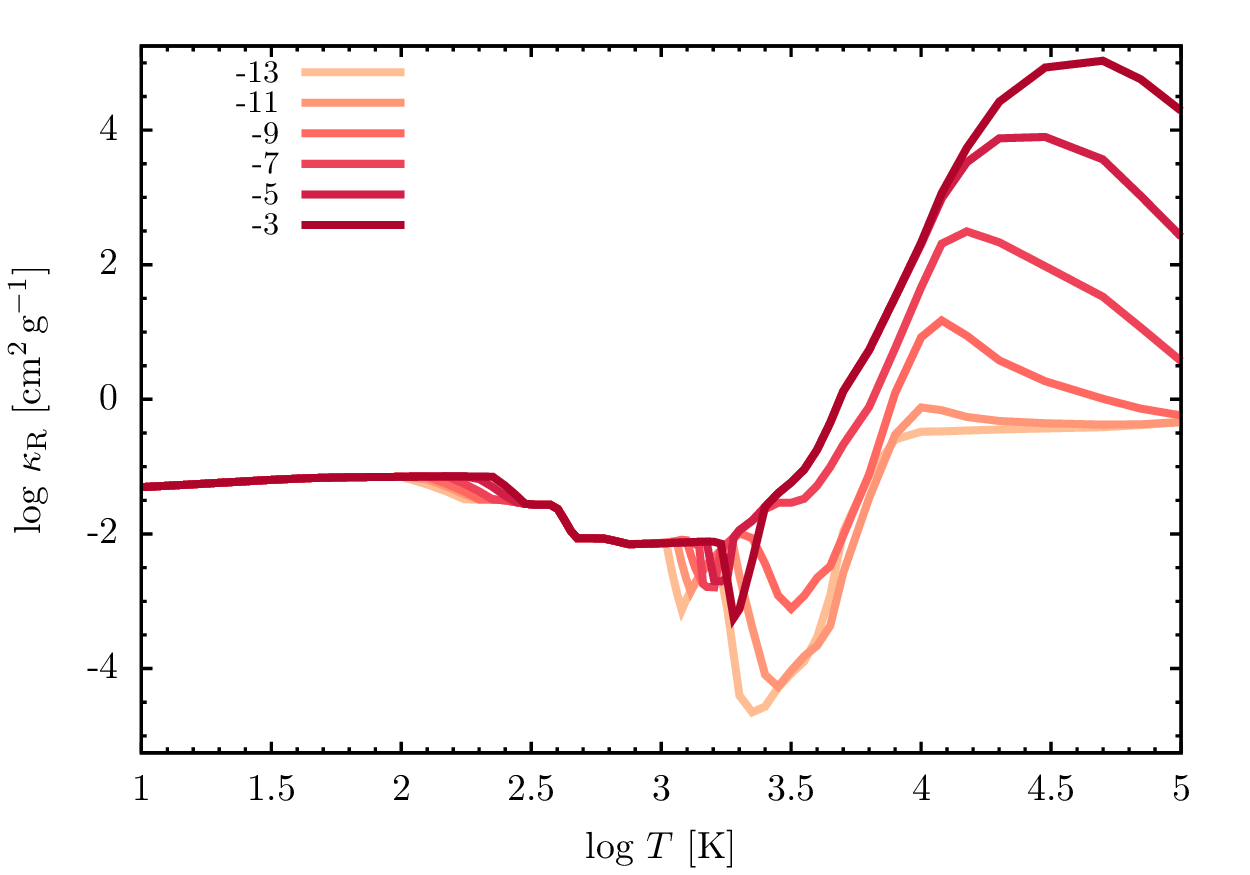}}
\caption{%
             Rosseland mean opacity versus temperature applied in the envelope,
             when there is a supply of dust grains (via accretion of gas and/or solids),
             for various gas densities (the legends refer to the logarithm of $\rho$
             in $\mathrm{g\,cm^{-3}}$).
             The distribution of grains has a minimum radius of $0.005\,\mu\mathrm{m}$
             and a maximum radius of $1$ (top) and $10\,\mathrm{mm}$ (bottom).
             The gas-to-dust mass ratio is $71.5$.
             }
\label{fig:opa}
\end{figure}
The envelope opacity arises from the combined contributions of dust, 
atoms, and molecules.  Dust opacities are calculated as described
in \citet{gennaro2013}, assuming the presence of a number of different
grain species, and grain size distributions with a minimum radius of 
$0.005\,\mu\mathrm{m}$ and a maximum radius of $1$ or 
$10\,\mathrm{mm}$. The number density of dust grains is proportional 
to the $-3$ power of the grain radius. 
Low-temperature gas opacities are taken from \citet{ferguson2005},
whereas high-temperature gas opacities are taken from the OPAL
tables \citep{iglesias1996}. 
\cifig{fig:opa} illustrates the Rosseland mean opacity for the
two grain size distributions considered here, along with gas opacity.
Dust grains are supposed to be present
in the envelope only if solids are supplied via gas and/or planetesimal
accretion. In the absence of a steady supply, dust grains quickly sediment 
to deeper layers and evaporate. When this happens, i.e., during phases 
of zero gas and solids' accretion, low-temperature opacities are replaced
with the molecular opacities of \citet{freedman2008}.

The planetary evolution code is largely the same as that used by
\citet{pollack1996}, \citet{bodenheimer2000b}, \citet{hubickyj2005},
and \citet{lissauer2009}.
The structure equations are solved by means of the Henyey method
\citep[e.g.,][]{bodenheimer2006}, 
supplemented with boundary conditions at the core-envelope interface, 
$R_{c}$, and at the top of the envelope, $R_{p}$. 
At $R_{c}$, the mass is set equal to $M_{c}$ and the luminosity is
set to zero \citep[i.e., there is no energy flux through the core-envelope
interface, see the discussion in][]{bodenheimer2014}.
The planet radius, $R_{p}$, is assumed to have an upper bound 
$\Rcapt$, the accretion radius, defined by
\begin{equation}
 \frac{1}{\Rcapt}\equiv \frac{4}{\Rhill}+\frac{1}{\Rbondi},
 \label{eq:RA}
\end{equation}
where $\Rhill$ and $\Rbondi$ are the Hill and Bondi radius, respectively. 
The limiting envelope radii, $\Rcapt\approx \Rhill/4$ and
$\Rcapt\approx \Rbondi$, were estimated from first principles
by means of three-dimensional (3D) calculations of the flow dynamics 
around planets in disks \citep{lissauer2009,gennaro2013}.

At early stages of evolution, the planet is ``in contact'' with the disk 
(the so-called \textit{nebular} stage), and the density and temperature 
at the top of the envelope are taken as the local disk values.
During the nebular stage, gas is added to the envelope in order to restore 
the condition $R_{p}\approx \Rcapt$ \citep[see][]{pollack1996}. 
Once this condition can no longer be maintained, the envelope contracts 
inside $\Rcapt$, entering a \textit{transition} stage, which coincides 
with the run-away gas accretion phase if the gas accretion rate is sufficiently 
large.
In fact, the gas accretion rate of the envelope, $\dot{M}_{e}$, is limited 
by the maximum rate at which the disk can deliver gas to the planet's vicinity.
If tidal perturbations by the planet are negligible, the disk-limited accretion 
rate can be described in terms of simple analytical arguments \citep{gennaro2008}.
If tidal perturbations are not negligible, the problem becomes highly
complex and depends on the interplay among viscous torques, 
tidal torques, and close-range flow dynamics around the planet. 
Disk-limited accretion rates were derived from 3D hydrodynamics 
high-resolution calculations, as described in \citet{bodenheimer2013}, 
extending the parameter space covered by the fitting functions 
reported therein to include a larger disk viscosity range.
It is important to stress that $\dot{M}_{e}$ cannot
exceed the disk-limited accretion rate, and there are instances 
in which this limit sets  in during the nebular stage of evolution.
The boundary conditions applied at $R_{p}$ during the transition stage 
are discussed in \citet{bodenheimer2000b}.

Eventually, the disk's gas around the planet's orbit is dispersed, typically
via photo-evaporation by stellar radiation, gas supply ceases, and 
the planet enters the \textit{isolation} stage.
Note that the process of gap formation in the disk by tidal torques alone, 
under typical disk conditions of viscosity and temperature and for 
planetary masses up to several times Jupiter's mass ($\Mjup$), 
does not lead to isolation 
\citep[e.g.,][and references therein]{gennaro2008,bodenheimer2013}.
During the isolation stage, standard photospheric boundary conditions 
are applied at $R_{p}$ \citep[e.g.,][]{cox1968}
\begin{eqnarray}
                             L_{p} &=& 4\pi \sigma_{\mathrm{SB}} R^{2}_{p} T^{4}_{\mathrm{eff}}\label{eq:LBC}\\
 \kappa_{\mathrm{R}} P &=& \frac{2}{3} \frac{G\Mp}{R^{2}_{p}}, \label{eq:PBC}
\end{eqnarray}
where $\sigma_{\mathrm{SB}}$ is the Stefan-Boltzmann constant,
$G$ is the gravitational constant, and $\kappa_{\mathrm{R}}$ and $P$ are,
respectively, the photospheric values of the Rosseland mean opacity and 
pressure.
In \cieq{eq:LBC}, the luminosity on the left-hand side comprises both the internal
power generated by the planet and the re-radiated power that arises from the
absorption of stellar radiation. 
Hence, the planet's effective temperature is given by \citep{bodenheimer2014}
\begin{equation}
 T^{4}_{\mathrm{eff}}=T^{4}_{\mathrm{int}}+T^{4}_{\mathrm{eq}},
 \label{eq:T4eff}
\end{equation}
where $T_{\mathrm{int}}$ depends on $R_{p}$ and on the luminosity 
internally generated by the planet. 
The equilibrium temperature is such that \citep[e.g.,][]{guillot2010}
\begin{equation}
 T^{4}_{\mathrm{eq}}=\frac{1}{4}T^{4}_{\star}\left(\frac{R_{\star}}{a}\right)^{2}(1-A_{b}),
 \label{eq:Teq}
\end{equation}
which assumes full redistribution of the incident radiation. 
The albedo $A_{b}$ is taken as a constant equal to $0.1$.
In \cieq{eq:Teq}, $T_{\star}$ and $R_{\star}$ are the effective
temperature and the photospheric radius of the star, respectively.

In calculations that allow for orbital migration, a distinction can be made 
between isolation from the disk's gas and from the disk's solids. Isolation 
from the planetesimals' disk occurs when the migration speed 
$da/dt\approx 0$ and $\dot{M}_{p}=\dot{M}_{c}+\dot{M}_{e}\lesssim 0$, 
or when the solids' surface density at the planet's location becomes very small.
Isolation from the solid disk occurs prior to isolation from the disk's gas or
shortly thereafter. However, the consequences of this delay are negligible.
Thus, when a planet enters the isolation phase, it is basically isolated 
from both the disk's gas and solids.

During all stages, if $R_{p} > \Rcapt$, mass is gradually removed from 
the envelope. In a more realistic context, during the nebular and transition
stages, an inflated planet would lose unbound mass hydrodynamically, 
carried away by the surrounding disk flow \citep{gennaro2013}. 
This mechanism of mass loss is different from those operating during the 
isolation stage (see \cisec{sec:gl}). As a result of gas loss, while $M_{c}$ 
is a monotonic function of time, $M_{e}$ and $\Mp$ may not be.

Accretion of solids is treated as in \citet{pollack1996}. All solids accreted
by the planet are assumed to sink to the core, in a condensed form, and 
increment the core mass. As originally derived by \citet{safronov1969}, 
the accretion rate of solids can be written as
\begin{equation}
 \dot{M}_{c}=\mathcal{S}_{\mathrm{eff}}\,\sigma_{Z}(a)\Omega(a) F_{g},
 \label{eq:dotmc}
\end{equation}
where $\mathcal{S}_{\mathrm{eff}}$ is the effective cross section for
planetesimal capture of the planet, $\sigma_{Z}(a)$ is the solids' surface
density at the planet's orbital radius, $a$, $\Omega(a)$ is the planet's orbital 
frequency, and $F_{g}$ is the ratio of the gravitational to the geometric 
cross section \citep{greenzweig1990,greenzweig1992}, known as 
the gravitational enhancement factor.
Further details can be found in \citet[and references therein]{pollack1996}.
The accretion rate in \cieq{eq:dotmc} neglects the contribution of the dust 
entrained in the accreted gas ($\lesssim 1$\% by mass).
The planetesimal radius is assumed to be $100\,\mathrm{km}$, although
smaller planetesimals were also tested.

A planetary embryo accreting planetesimals at a fixed orbital
radius will deplete an annular region around its orbit of full width about
equal to $8\,\Rhill$, at which point the condensed core becomes detached
from the planetesimals' disk. In these calculations, the secular evolution 
of planetesimals is neglected and therefore, once detached, the core 
reaches its final mass
\begin{equation}
 M^{f}_{c}\approx\sqrt{\frac{(16\pi a^{2} \sigma_{Z})^{3}}{3\Ms}},
 \label{eq:mciso}
\end{equation}
which neglects the contribution of the envelope mass to $\Rhill$ and 
is hence appropriate when $\Mp\approx M_{c}$.
The situation is more complex for a migrating planet, since the depletion
rate of solids in the disk tends to be initially slower than the migration 
rate through the disk. Hence, the planet cuts a swathe through the solids' 
disk, 
which deepens as $|da/dt|$ reduces, eventually detaching itself.
The final mass of the core in this case is more difficult to predict, 
as it depends on both the accretion and migration history.
During the long isolation stage, a planet may be subjected to stochastic 
impacts, which may alter the core and envelope mass and the planet's orbit
if the impactors are sufficiently massive. This possibility is not considered 
here.

As mentioned above, solids sink to the top of the core. 
This process releases energy in the envelope, affecting the local 
energy budget, but not the local chemical composition of the envelope. 
These calculations consider accretion of hydrated, partly hydrated, and 
anhydrous planetesimals, depending on the local disk temperature.
Rocky planetesimals may reach the core nearly intact if they are large 
enough (they may be held together by their own gravity) or if the ram 
pressure does not exceed their compressive strength 
\citep[and references therein]{gennaro2015}.
Ice-rich planetesimals are more easily disrupted or entirely ablated in 
the envelope because the critical temperature of \ice\ is only $\approx 650\,\K$, 
a value reached in relatively shallow layers of the envelope
(their mass is nevertheless assumed to sink to the core).

An important part of the calculation is represented by the capture 
of planetesimals, which determines self-consistently the cross section 
$\mathcal{S}_{\mathrm{eff}}$ in \cieq{eq:dotmc}, and by their interaction 
with the planet's envelope, which determines depth-dependent mass and 
energy deposition rates. This part is based on the protocols described 
in \citet{pollack1996}, enhanced with an improved integration 
algorithm of the planetesimals' trajectories \citep{gennaro2014}.
In brief, a number of trajectories with a varying impact parameter 
($\le R_{p}$) are integrated through the envelope. 
The largest impact parameter for which the body hits the core surface, 
breaks up, or is entirely ablated provides the radius for planetesimal capture
and hence the cross section $\mathcal{S}_{\mathrm{eff}}$.
At this point, an additional series of trajectory integrations is performed, 
with an impact parameter up to the capture radius, to record the ablation history 
and the fate of the body as a function of the impact parameter. This collective 
information provides the mean energy and mass deposition rates in each 
envelope layer.

\begin{figure*}
\centering%
\resizebox{\linewidth}{!}{%
\includegraphics[clip]{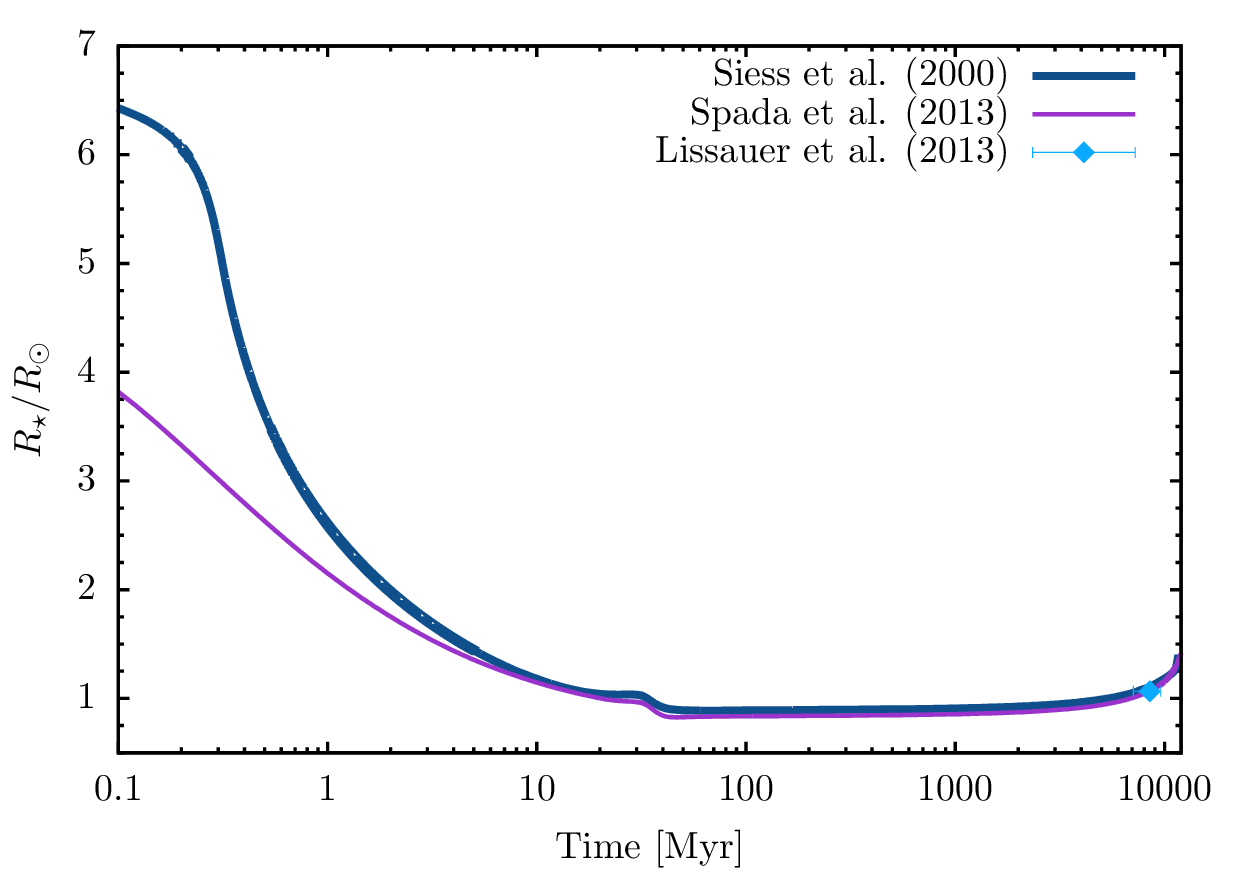}%
\includegraphics[clip]{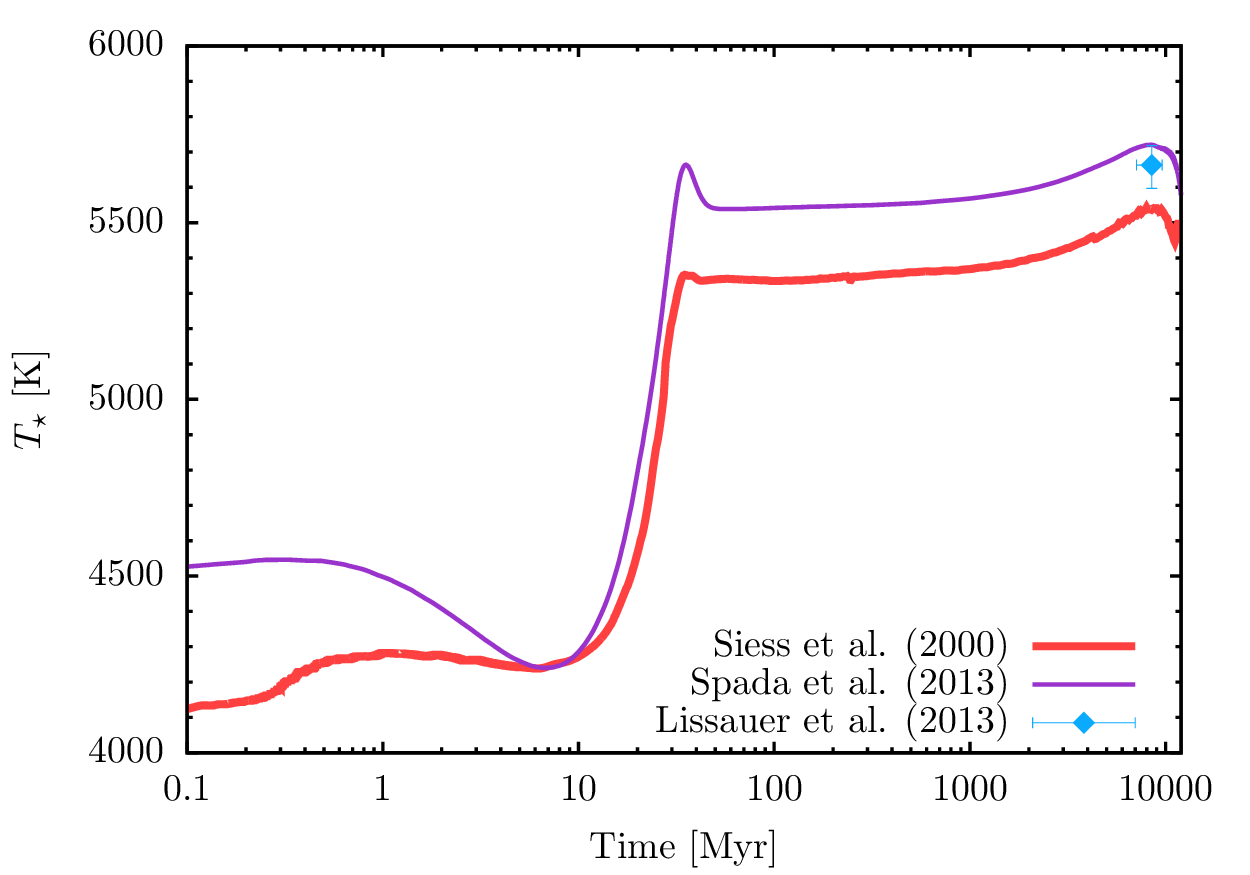}%
\includegraphics[clip]{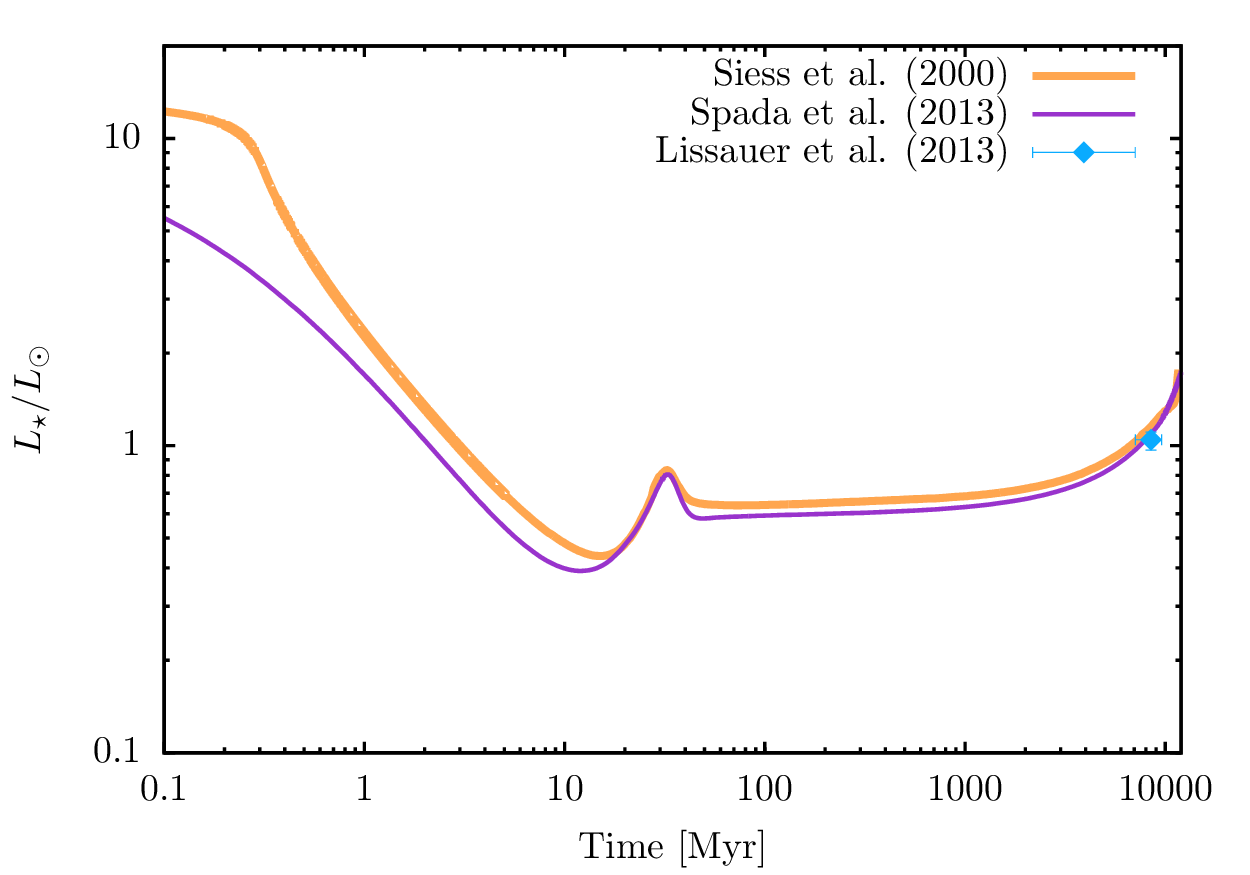}}
\caption{%
             Radius (left), effective temperature (center), and luminosity (right)
             evolution of the star: thick curves are for a protosolar metallicity
             ([Fe/H]$=0.0$), $1\Msun$ stellar model from \citet[][]{siess2000}; 
             thin curves are for a [Fe/H]$=0.0$,
             $0.95\Msun$ Yonsei-Yale model from \citet[][]{spada2013}.
             Estimated age, radius, effective temperature, and luminosity of the
             host star \kep, from \citetalias{lissauer2013}, are also shown as points with 
             error bars. \citet[][]{lissauer2011a} estimated a radius of
             $R_{\star}=1.1\pm 0.1\,R_{\odot}$, and effective temperature of
             $T_{\star}=5680\pm 100\,\K$, and an age between $6$ and $10\,\mathrm{Gyr}$.
             }
\label{fig:star}
\end{figure*}
The methods outlined above apply to both in situ and ex situ calculations.
The basic difference is that the boundary conditions at $R_{p}$ change 
over time in ex situ models, whereas in situ calculations allow only for
a linear decline with time of the gas surface density at $R_{p}$. 
In ex situ models, the surface density $\sigma_{Z}$ in \cieq{eq:dotmc} 
also varies as a function of the distance from the star and of the disk 
temperature. Additionally, ex situ models use stellar properties from 
stellar structure models of solar-type stars to compute the equilibrium
temperature (\cieq{eq:Teq}), the mass-loss rate during isolation 
(\cieq{eq:mexuv}), and the irradiation temperature of the disk (\cieq{eq:Tirr}). 
The calculations apply a stellar model of a solar-mass and protosolar 
metallicity
\citep[\mbox{[Fe/H]}$=0.0$,][]{asplund2009,lodders2010} star from 
\citet{siess2000}, whose radius, effective temperature, and luminosity 
are plotted in \cifig{fig:star}. For comparison, some calculations are
repeated by adopting a Yonsei-Yale model \citep[][]{spada2013} 
for a $0.95\,\Msun$, [Fe/H]$=0.0$ star, also represented in \cifig{fig:star}.
In contrast, in situ models are based on fixed, solar-type values
for radius, effective temperature, and luminosity of the star.

\subsection{Core Structure Calculation}
\label{sec:csc}

In situ formation calculations determine the core radius, $R_{c}$,
from its current mass, $M_{c}$, using tables of results from \citet{rogers2011}.
The core is composed of iron and silicates, with Earth-like mass 
fractions of $30$\% and $70$\%, respectively. 
Applied to an Earth-mass planet, the results predict a radius within
$2.9$\% of the Earth radius, $\Rearth$.
The radius $R_{c}$ is then used as an inner boundary condition for
the \hhe\ envelope.

Ex situ formation calculations allow for accretion of planetesimals whose
composition varies as a function of time and distance from the star. 
At disk temperatures below $150\,\K$, the planetesimals are ice-rich, 
$50$\% by mass ($45$\% silicates and $5$\% iron).
They become progressively ice-poor (and rich in silicates and iron) at 
higher disk temperatures and are anhydrous at temperatures above 
$250\,\K$ \citep{gennaro2015}, where a terrestrial-type composition 
is adopted ($70$\% silicates and $30$\% iron by mass).
The mass fractions of iron, silicates, and ice are linearly interpolated
in temperature between $150$ and $250\,\K$.

Since the composition of the condensed core may vary during 
the evolution of ex situ models, as the composition of accreted 
solids changes, detailed calculations of the core structure are 
performed. The core is assumed to be spherically symmetric 
about its center and described by the equation of mass conservation
and hydrostatic equilibrium (see Appendix~\ref{sec:ics}). 
The core is taken to be fully differentiated into an iron nucleus, 
a silicate mantle, and an outer shell of condensed \ice.
Two-layer (with any combination of the three materials) or one-layer 
structures are also possible.
Each material is characterized by a ``cold'' equation of state (EoS) 
relating density and pressure.
We experimented with a combination of Birch-Murnaghan, Vinet, 
and Generalized Rydberg EoS \citep[for a review, see][]{stacey2008}, 
and extend them to very high pressures by means of 
the Thomas-Fermi-Dirac theory. 
Details are given in Appendix~\ref{sec:ics}.
Temperature effects on the EoS are neglected since thermal pressure
is expected to provide only a minor correction to $R_{c}$ for the core 
masses considered here
\citep[see, e.g.,][]{valencia2006,seager2007,sotin2007,sohl2014a}. 
However, temperature effects are included to account for phase
transitions within the layers (see Appendix~\ref{sec:ics} for details).
Applying this module to an Earth-mass planet with a composition
of $70$\% silicates and $30$\% iron by mass, the calculated radius 
is within $0.1$\% of $\Rearth$. For a cold-Earth analog 
\citep[$67.3$\% silicates and $32.7$\% iron by mass,][pp.\ 27-68]{sohl2007}, the
resulting radius is within $0.8$\% of $\Rearth$.
More detailed structure calculations, including heat transfer and additional
material phases, are also presented in Appendix~\ref{sec:ics}. They are
compared to isothermal core structures in Appendix~\ref{sec:cot}.

The integration of the equations proceeds from the center outward, varying 
the central pressure in an iterative fashion, until the pressure at the core radius,
$R_{c}$, matches (within $1$\%) the pressure at the bottom of the envelope, 
which is provided by the envelope structure module (\cisec{sec:esc}).  
The integration of the core structure equations is performed whenever 
$M_{c}$ increases by $\ge 1$\% or if the pressure or temperature at $R_{c}$ 
changes by $\ge 10$\%.

\subsection{Radiation-Driven Gas Loss During Isolation}
\label{sec:gl}

Removal of envelope gas after the planet becomes isolated, when
stellar photons can directly impinge on the planet dayside, is based 
on the energy-limited hydrodynamics escape driven by stellar X-ray 
and EUV (XUV) radiation 
\citep{watson1981,erkaev2007,murray-clay2009,lopez2012}. In this
limit, the mass-loss rate of the envelope can be approximated as
\begin{equation}
 \dot{M}^{\mathrm{iso}}_{e}=%
                      \frac{\varepsilon \pi R^{2}_{\mathrm{XUV}} F_{\mathrm{XUV}}}{K(\xi)}%
                      \left(\frac{R_{p}}{G \Mp}\right),
 \label{eq:mexuv}
\end{equation}
where $R_{\mathrm{XUV}}\approx 1.1\,R_{p}$ is the envelope radius 
at which the atmosphere becomes optically thick to the incoming 
stellar XUV radiation and most of the flux $F_{\mathrm{XUV}}$ is 
absorbed \citep{erkaev2007,murray-clay2009},
$K(\xi)=1-3/(2\xi)+1/(2\xi^{3})$ is a reduction factor of the planet's 
potential energy caused by tidal forces of the star, and 
$\xi=\Rhill/R_{p}>1$ \citep{erkaev2007}. 
Escape is assumed to take place at the equipotential surface passing 
through the collinear Lagrange point $L_{1}$. Note that \cieq{eq:mexuv} 
diverges for $\xi\rightarrow 1$. During the evolution in isolation,
it is assumed that $\dot{M}_{e}=-\dot{M}^{\mathrm{iso}}_{e}$.

Observations of the histories of EUV and X-ray fluxes of solar-type stars 
suggest that mass loss is most vigorous at ages of $t<0.1\,\mathrm{Gyr}$
\citep{ribas2005}. 
Here, we set $F_{\mathrm{XUV}}=3\times 10^{-4} L_{\star}/(4\pi a^{2})$ for
$t\lesssim 0.1\,\mathrm{Gyr}$ and
$F_{\mathrm{XUV}}=3\times 10^{-6} (5\,\mathrm{Gyr}/t)^{1.23} L_{\star}/(4\pi a^{2})$
at later times \citep{ribas2005}, 
where $L_{\star}$ is the stellar bolometric luminosity.
The quantity $\varepsilon$ is an efficiency factor intended to roughly 
account for radiative losses from the envelope \citep{erkaev2007}, 
so that only a fraction of the incident flux $F_{\mathrm{XUV}}$ can 
effectively drive mass loss. This factor is quite uncertain.
These calculations are based on an efficiency of 
$\varepsilon=0.1$ \citep[e.g.,][]{murray-clay2009,lopez2012}. 
However, other values of $\varepsilon$ are considered for a sensitivity 
study. 

In situ and ex situ calculations handle radiation-induced mass loss 
of the envelope in similar ways. 
The only basic difference is that, in \cieq{eq:mexuv} and in the function
$K(\xi)$, $R_{\mathrm{XUV}}$ replaces $R_{p}$ in the in situ models.
Additionally, the stellar luminosity of ex situ models varies in time, 
according to the applied stellar evolution model (see \cifig{fig:star}),
whereas in situ simulations use $L_{\star}=L_{\sun}$. 
Since the XUV stellar output is assumed to be proportional to $L_{\star}$, 
the mass-loss history of the envelope during isolation may differ from that 
occurring at a constant value of $L_{\star}$, as used by in situ calculations.

Other mass-loss mechanisms during the beginning of the isolation 
phase have been considered. \citet{ikoma2012} and \citet{ginzburg2016} 
studied the effect of loss of pressure support at the planet's outer 
boundary once the disk's gas disperses; the energy for the mass loss 
is supplied by the planet's cooling luminosity. 
\citet{owen2016} considered a similar mechanism, initiated by the loss 
of pressure from the disk, in which a ``Parker'' wind is driven by 
a combination of the stellar continuum radiation and the gravitational 
energy released as the planet contracts. 
The assumptions made in these works are considerably different from 
those made here, and it is not clear if these processes would be significant 
in our models.  
First, the planet radius in the above papers is implicitly assumed to be 
close to \Rbondi, while in the models discussed here, just before disk 
dispersal, $R_{p}$ is roughly a factor of $\approx 3$ to $\approx 6$ smaller 
than \Rbondi\ (see also \cisec{sec:insitu}). 
Second, as the disk disperses, the possible up-lifting of the outer envelope 
layers due to loss of disk pressure during the nebular stage is implicitly 
included in the structure calculation through the boundary conditions at 
$R_{p}$.  After disk dispersal, the boundary conditions transition to those 
of an isolated photosphere (Equations~(\ref{eq:LBC}) and (\ref{eq:PBC})). 
The transition occurs on the cooling timescale of the outer envelope layers, 
which is shorter than the disk dispersal timescale. As a result, the 
photospheric pressure increases and the planet further contracts inside 
\Rbondi. The \citeauthor{owen2016} mechanism, for example, is not 
significant for $R_{p}\lesssim 0.1\,\Rbondi$. 
Clearly, whether these wind mechanisms are in fact unimportant in 
calculations like ours needs to be tested in more detail.

\subsection{Disk Evolution}
\label{sec:de}

These models consider the evolution of an axisymmetric gaseous disk, 
of surface density $\Sigma$, driven by turbulence viscosity $\nu$ 
(of some nature), stellar-induced photo-evaporation, tidal torques due 
to gravitational interactions with an embedded planet, and accretion 
of gas on the planet. 
Indicating with $\mathcal{T}_{\nu}(r)$ the viscous torque acting between
two adjacent disk's annuli at a distance $r$ from the star and with $-\mathcal{T}(r)$ 
the tidal torque exerted by the planet on the disk's gas at radius $r$, 
conservation of mass and momentum within the disk leads to 
the following disk's evolution equation \citep[e.g.,][]{lin1986b}
\begin{equation}
 \pi r\frac{\partial}{\partial t}(\Sigma+\Sigma_{\mathrm{pe}}+\Sigma_{\mathrm{ac}})=%
 \frac{\partial}{\partial r}\!\left\{%
 \frac{1}{r\,\Omega}\frac{\partial}{\partial r}\left[%
 \mathcal{T}_{\nu}(r)-\mathcal{T}(r)\right]\right\},
 \label{eq:devol0}
\end{equation}
where $\Omega$ is the disk's rotation rate, $\dot{\Sigma}_{\mathrm{pe}}$
is gas mass removed by photo-evaporation per unit disk surface and 
unit time (see \cisec{sec:DPH}), 
and $\dot{\Sigma}_{\mathrm{ac}}$ is gas mass removed by accretion
on the planet per unit disk surface and unit time. 
Recalling that 
$\mathcal{T}_{\nu}=-2\pi r^{3}\nu\Sigma\,\partial\Omega/\partial r$
\citep{lynden-bell1974} and approximating $\Omega$ to the Keplerian 
rotation rate $\sqrt{G\Ms/r^3}$, the viscous torque becomes 
$\mathcal{T}_{\nu}=3\pi r^{2}\nu\Sigma\Omega$ and 
\cieq{eq:devol0} assumes the more familiar form
\begin{eqnarray}
 &r&\frac{\partial}{\partial t}(\Sigma+\Sigma_{\mathrm{pe}}+\Sigma_{\mathrm{ac}})=\nonumber\\%
 &\pz&
 \frac{\partial}{\partial r}\!\left[%
 3\sqrt{r}\frac{\partial}{\partial r}\!\left(\nu\Sigma\sqrt{r}\right)%
-\frac{2\,\Sigma}{\Omega}\frac{\partial\mathcal{T}}{\partial m}\right].
 \label{eq:devol1}
\end{eqnarray}
The quantity $\partial\mathcal{T}/\partial m$ is the torque density distribution, 
i.e., the gravitational torque per unit disk mass ($dm=2\pi \Sigma r dr$)
arising from tidal interactions with the planet. This function
is discussed in \cisec{sec:om}. If multiple planets orbit in the disk,
$\partial\mathcal{T}/\partial m$ is the sum of all partial torque density
distributions. The presence of the tidal torque term in \cieq{eq:devol1}
naturally accounts for (planet-induced) gap formation in the density 
distribution.
Notice that, if $\nu\Sigma$ is constant in radius, a viscously evolving 
planet-less disk is in a steady state.

The mass removed from the disk, via accretion on the planet, per 
unit surface and unit time, is written as 
$\dot{\Sigma}_{\mathrm{ac}}=\delta(r-a)\dot{M}_{e}/(2\pi a)$
so that $\int 2\pi \dot{\Sigma}_{\mathrm{ac}} r dr=\dot{M}_{e}$,
the planet's gas accretion rate, which ensures conservation of
the mass transferred between the planet and the disk.
Contrary to $\dot{\Sigma}_{\mathrm{pe}}$
(which is zero or positive), $\dot{\Sigma}_{\mathrm{ac}}$ can be positive, 
null, or negative. In the latter case, mass is transferred from the planet 
to the disk. The planet's envelope can lose mass ($\dot{M}_{e}<0$) 
if its radius exceeds the accretion radius $\Rcapt$
(see \cisec{sec:esc}).
Numerically, to avoid discontinuities, the mass added to (removed from) 
the planet is removed from (added to) a disk region around the planet's 
orbit of radial width a few times $\Rhill$.

The presence of an accreting planet can change the mass transfer
through the disk and thus alter the disk's surface density \citep{lubow2006}. 
This effect cannot be described by \cieq{eq:devol1}.
To account for it, the approach of \citet{lubow2006} is applied. 
The accretion rate through the disk is
$dm/dt=2/(r\Omega)\partial\left(\mathcal{T}_{\nu}-\mathcal{T}\right)/\partial r$
(adopting the convention that $dm/dt>0$ for an inward transfer 
of mass). By indicating with $\langle dm/dt\rangle_{\mathrm{ext}}$ and
$\langle dm/dt\rangle_{\mathrm{int}}$ accretion rates averaged
over narrow rings, respectively, exterior and interior to the planet's 
orbit (and sufficiently apart from it), the condition is imposed that
\begin{equation}
 \left\langle\frac{dm}{dt}\right\rangle_{\!\mathrm{ext}}=%
 \left\langle\frac{dm}{dt}\right\rangle_{\!\mathrm{int}}+\dot{M}_{e}.
 \label{eq:brigap}
\end{equation}
\cieq{eq:brigap} is used to adjust $dm/dt$ (i.e., $\Sigma$) inside
and outside of the planet's orbit in a mass-conservative manner. 
This correction is applied only if 
$\langle dm/dt\rangle_{\mathrm{ext}}-\dot{M}_{e}>%
\langle dm/dt\rangle_{\mathrm{int}}>0$. 
For stability reasons, mass adjustments are spread over 
several grid zones.

The thermal structure of the disk is determined by imposing
a simple energy balance involving viscous heating, radiative 
cooling from the surface of the disk, and irradiation heating 
by the star
\begin{equation}
 Q_{\nu}-Q_{\mathrm{cool}}+Q_{\mathrm{irr}}=0.
 \label{eq:EEq}
\end{equation}
In the case of Keplerian rotation, the energy flux produced by 
viscous dissipation is \citep{m&m}
\begin{equation}
 Q_{\nu}=\frac{9}{4}\nu\Sigma\Omega^{2}.
 \label{eq:Qnu}
\end{equation}
The energy flux escaping from the disk's surface and the heating
flux generated by stellar photons are \citep{hubeny1990}
\begin{equation}
 Q_{\mathrm{cool}} =  2\sigma_{\mathrm{SB}}\,T^{4}\left(%
         \frac{3}{16}\kappa_{\mathrm{R}}\Sigma+\frac{1}{2}+\frac{1}{2\kappa_{\mathrm{P}}\Sigma}%
         \right)^{-1}
 \label{eq:Qcool}
\end{equation}
and
\begin{equation}
 Q_{\mathrm{irr}}  =  2\sigma_{\mathrm{SB}}\,T^{4}_{\mathrm{irr}}\left(%
         \frac{3}{16}\kappa_{\mathrm{R}}\Sigma+\frac{1}{2}+\frac{1}{2\kappa_{\mathrm{P}}\Sigma}%
         \right)^{-1}, 
 \label{eq:Qirr}
\end{equation}
respectively.
In the equations above, $T$ is the mid-plane temperature of the disk, 
$T_{\mathrm{irr}}$ is the stellar irradiation temperature, and 
$\kappa_{\mathrm{R}}$ and $\kappa_{\mathrm{P}}$ are the Rosseland
and Planck mean opacities. These opacity coefficients are calculated 
following the method of \citet{gennaro2013}, for grain size distributions
of up to $1\,\mathrm{mm}$ in radius, and connected to the gas
opacities of \citet{ferguson2005}.
The advantage of using the fluxes in Equations~(\ref{eq:Qcool}) and 
(\ref{eq:Qirr}) is that, in the approximation of vertically integrated quantities,
they also describe optically thin disks. 
The irradiation temperature is written as \citep[e.g.,][]{menou2004}
\begin{equation}
 T^{4}_{\mathrm{irr}}=\frac{1}{2}%
                                   \left(T^{4}_{\star}+T^{4}_{\mathrm{acc}}\right)
                                   \left(\frac{R_{\star}}{r}\right)^{2}%
                                   \left[\frac{2}{5}\left(\frac{R_{\star}}{r}\right)+%
                                          \frac{2}{7}\frac{H}{r}\right].
 \label{eq:Tirr}
\end{equation}
The disk scale height, assuming vertical hydrostatic equilibrium, 
is given by
$H=\sqrt{\gamma\,k_{\mathrm{B}}T/(\mu m_{\mathrm{H}})}/\Omega$. 
The adiabatic index $\gamma$ is set between $1.4$ and $1.6$
and the mean molecular weight is $\mu=2.39$ 
($k_{\mathrm{B}}$ is the Boltzmann constant and 
$m_{\mathrm{H}}$ the hydrogen mass).
\cieq{eq:Tirr} includes the contribution from luminosity released by accretion 
on the star \citep{pringle1981}
\begin{equation}
 T^{4}_{\mathrm{acc}}=\frac{1}{8\pi}%
 \left(\frac{G\Ms\dot{M}_{\star}}{\sigma_{\mathrm{SB}}R^{3}_{\star}}\right).
\label{eq:Tacc}
\end{equation}
The stellar accretion rate is computed as
$\dot{M}_{\star}=2/(r \Omega) \partial \mathcal{T}_{\nu}/\partial r$,
from the solution of \cieq{eq:devol1} at the inner boundary of the disk
(where $\mathcal{T}=0$).

The disk also contains a solid component, assumed to be formed of
$100\,\mathrm{km}$-radius planetesimals, whose surface density is 
$\sigma_{Z}$.
In principle, this planetesimal disk would evolve through gravitational
encounters, including collisions, and interactions with any embedded planet. 
Gas drag would also affect the evolution of this solid component, but over
rather long timescales, given the size of the bodies considered here. 
However, for the sake of simplicity and tractability,
it is assumed that $\sigma_{Z}$ only varies because of depletion 
by accretion of solids on the planet ($\dot{M}_{c}$) and 
because of scattering by the planet's gravity \citep{ida2004}
\begin{equation}
 \dot{M}_{\mathrm{scat}}=\left(\frac{G M_{c}}{R_{c}}\right)^{2}\!%
                                         \left(\frac{a}{2 G \Ms}\right)^{2}\dot{M}_{c}.
 \label{eq:core_scat}
\end{equation} 
The equation above is a very simple approximation, based on energy 
arguments, and assumes that $\Mp\approx M_{c}$, which is appropriate 
for the planets modeled here.
The surface density $\sigma_{Z}$ also changes in response to  temperature 
variations in the disk, allowing for the vaporization of ice (see \cisec{sec:csc}).

\cieq{eq:devol1} is solved by means of a hybrid implicit/explicit numerical 
scheme, based on the fourth/fifth-order Dormand-Prince method 
(embedding backward differentiation) with an adaptive step-size control 
for the global accuracy of the solution \citep{hairer1993}.
Additional details and tests can be found in \citet{gennaro2012}.
Toward the end of the disk's life, when the evolution is entirely driven 
by photo-evaporation and the disk is quickly dispersed, the algorithm 
transitions from implicit to explicit, with a time step condition that 
constrains the maximum amount of mass removed from any disk 
annulus. 
Details on the solution method of \cieq{eq:EEq} for the energy 
balance are also given in \citet{gennaro2012}.
The disk extends in radius from the larger of $0.01\,\AU$ and 
$R_{\star}=R_{\star}(t)$ to 
$900\,\AU$, and is discretized over $6000$ grid points by imposing
a constant ratio $\Delta r/r$. Beyond $900\,\AU$, to ensure
that the grid boundary does not interfere with viscous spreading,
a buffer zone of $400$ additional grid points (at a degraded
resolution) brings the outer disk edge to $\approx 3.5\times 10^{5}\,\AU$.

\subsection{Tidal Interactions and Orbital Migration}
\label{sec:om}

In order to describe tidal interactions between the disk and the planet, 
we apply the formalism of \citet{gennaro2008,gennaro2010} for local 
isothermal disks. This is based on the torque density distribution, 
which is defined by the integral
\begin{equation}
 \mathcal{T}(a)\equiv 2\pi\!\int_{0}^{\infty}{\frac{\partial \mathcal{T}}{\partial m}%
                                      \,\Sigma(r)\, r\, dr},
 \label{eq:dTdMdef}
\end{equation}
where $\mathcal{T}(a)$ is the total torque applied to the planet.
In actuality, the integral is performed over the disk's radial extent. 
In a disk whose properties vary smoothly with radius, the theory 
of disk resonances  \citep[e.g.,][]{meyervernet1987,ward1997}
suggests that
\begin{equation}
 \frac{\partial \mathcal{T}}{\partial m}=%
 \mathcal{F}\!\left(x, \beta, \zeta \right)%
 \Omega^{2}(a) \, a^{2}\!%
 \left(\frac{\Mp}{\Ms}\right)^{2}\! \left(\frac{a}{\Delta_{p}}\right)^{4}\!,
 \label{eq:dTdM}
\end{equation}
where $\mathcal{F}$ is a dimensionless parametric function of 
$x=(r-a)/\Delta_{p}$ with $\Delta_{p}=\max{[H(a),\Rhill]}$, whose extrema
are at $x\approx \mp 1$.
The parameters $\beta=-d\ln{\Sigma}/d\ln{r}$ and $\zeta=-d\ln{T}/d\ln{r}$ 
are calculated as averages between $x=-4$ and $4$ (the function
$\mathcal{F}$ is practically zero outside of these limits).
\citet[hereafter \citetalias{gennaro2010}]{gennaro2010} 
tested the validity of \cieq{eq:dTdM} and provided 
analytic approximations of the function $\mathcal{F}$, for wide ranges of 
the parameters $\beta$ and $\zeta$, based on 3D hydrodynamics 
calculations of disk-planet interactions.

Gravitational interactions transition from a linear to a nonlinear regime
when $|\mathcal{T}(a)|\gtrsim |\mathcal{T}_{\nu}(a)|$, or
\begin{equation}
\left(\frac{\Mp}{\Ms}\right)^{2}\gtrsim 3\pi\left(\frac{\nu}{a^{2}\Omega}\right)%
       \left(\frac{\Delta_{p}}{a}\right)^{3}.
 \label{eq:gapcon}
\end{equation}
Nonlinear interactions can cause order-of-magnitude variations in 
the surface density (relative to the unperturbed disk), as the planet 
mass grows.
However, under typical disk conditions, the torque density 
$\partial\mathcal{T}/\partial m$ varies smoothly across the transition,
and the maximum and minimum of the function $\mathcal{F}$ change only
by factors of the order of unity. 
The variation of $\partial\mathcal{T}/\partial m$ across the transition
is implemented as explained in \citetalias{gennaro2010}.

The rate of change of the planet's orbital radius is found by imposing
conservation of orbital angular momentum, which yields
$da/dt=2\mathcal{T}(a)/[\Mp\,a\,\Omega(a)]$. 
Since $\mathcal{T}(a)$ is defined through \cieq{eq:dTdMdef},
the migration speed becomes
\begin{eqnarray}
 \frac{da}{dt} & = & 4\pi\,\Omega(a)\!%
 \left(\frac{a}{\Ms}\right)\!%
 \left(\frac{\Mp}{\Ms}\right)\!\left(\frac{a}{\Delta_{p}}\right)^{4} \nonumber\\
                     &    &\times\int{\!\!\mathcal{F}\!\left(x, \beta, \zeta \right)\Sigma(r)\, r\, dr},
 \label{eq:dadt}
\end{eqnarray}
where the integration is performed over the entire disk.
In the linear regime, one can show that the integral on the right-hand
side of \cieq{eq:dadt} is $\propto H^{2}\Sigma(a)$, hence
$da/dt\propto (\Mp/\Ms)(\pi a^{2}\Sigma(a)/\Ms)(a/H)^{2}$, which
is proportional to both the planet mass and the local disk mass
($\pi a^{2}\Sigma$).
A comparison between a direct 3D calculation of planet migration
and \cieq{eq:dadt} is shown in Figure~9 of \citetalias{gennaro2010}.
In the nonlinear regime, the integral depends on the planet mass
through functions $\mathcal{F}$ and $\Sigma$. As the density
gap deepens, the integral has nonzero contributions mostly from 
regions near the gap edges. \citet{gennaro2006} showed that
this formalism provides a good agreement with results from 
hydrodynamics calculations of planet migration also in the 
nonlinear regime (for non-highly eccentric orbits).
It should be noted that there are regimes of fast orbital 
migration in which $\partial\mathcal{T}/\partial m$ can also 
depend on $da/dt$ and which may not be fully captured by 
the formalism applied here \citep{gennaro2008}. 
However, the conditions required by these extreme regimes are 
not met in this study.

The formalism used here for disk-planet tidal interactions relies 
on the local isothermal approximation of disk's gas. 
The resulting torques agree well with analytical estimates 
\citep{tanaka2002}, when the comparison is possible 
\citep[\citetalias{gennaro2010};][]{masset2010}.
Adiabatic disks can produce torques that may behave differently 
\citep[see][pp.\ 667-689, for recent reviews]{kley2012,baruteau2014}. 
However, 
while prescriptions are available for the total torque $\mathcal{T}(a)$ 
acting on a low-mass planet in the adiabatic limit 
\citep{masset2010,paardekooper2011}, there is no formalism for 
the description of the torque density $\partial\mathcal{T}/\partial m$ 
in this limit. 
It is important to stress that the use of the distribution function
$\partial\mathcal{T}/\partial m$, but not of $\mathcal{T}(a)$, fulfills 
the action-reaction principle within the disk-planet system, thus
accounting for disk-planet tidal interactions.
Additionally, a description based on $\partial\mathcal{T}/\partial m$, 
but not on $\mathcal{T}(a)$, allows for a \textit{continuous} transition 
between different regimes of orbital migration, without 
the need of relying on some gap formation criterion and imposing 
different migration rates. 
In fact, as planet mass and disk thermodynamical conditions change, 
the tidal interactions (and hence $da/dt$) adapt consistently to 
the changing conditions.
Finally, inside $\approx 5\,\AU$, outward migration in adiabatic disks 
may occur for planet masses somewhat greater than $\approx 7\,\Mearth$ 
\citep[pp.\ 667-689]{baruteau2014}, possibly affecting the largest simulated planet.
However, by the time this planet attains that mass, the local disk has 
become radiatively efficient.

Disk-planet tidal interactions also affect orbital eccentricity.
In the linear regime, orbits tend to be circularized on timescales 
shorter than the migration timescales \citep[e.g.,][]{pawel1993,tanaka2004}.
In the strong nonlinear regime, the outcome of tidal interactions is 
more complex \citep[e.g.,][pp.\ 347-371]{lubow2011}, though this regime is not 
relevant in these calculations.

Orbital migration of a planet during formation may also be driven
by interactions with planetesimals 
\citep[e.g.,][and references therein]{minton2014}. 
However, since the secular evolution of the planetesimals' disk is 
neglected, so is planetesimal-induced migration.

\subsection{Disk Photo-evaporation}
\label{sec:DPH}

The disk photo-evaporation follows an approach along the lines of 
\citet{alexander2007}, in which the total amount of gas removed
from the disk per unit surface and unit time is
\begin{equation}
 \dot{\Sigma}_{\mathrm{pe}}=\dot{\Sigma}_{\mathrm{dif}}+\dot{\Sigma}_{\mathrm{rim}}.
 \label{eq:dsig_pe}
\end{equation}
By assumption, photo-evaporation is essentially driven by stellar 
EUV radiation. Gas removal by FUV and X-ray radiation is not 
considered \citep[but see the discussion in][]{gorti2009b,gorti2009a}.
The emission rate of EUV ionizing photons by the star is 
$10^{42}\,\mathrm{s}^{-1}$ \citep{alexander2007}.
The component $\dot{\Sigma}_{\mathrm{dif}}$
represents the removal rate due to the ``diffuse'' stellar radiation,
whereas the additional component $\dot{\Sigma}_{\mathrm{rim}}$ is
activated after the disk becomes radially optically thin to stellar photons
inside some radius $r_{\mathrm{rim}}$ (``rim'' photo-evaporation).

Diffuse photo-evaporation depends on the gravitational radius 
$r_{g}=G\Ms/c^{2}_{s}$ \citep{hollenbach1994}, where $c_{s}$ 
is the sound speed of an ionized hydrogen/helium mixture at 
$T\approx 10^{4}\,\K$, the nearly constant temperature of 
the upper layers of a disk heated by EUV radiation 
\citep[e.g.,][]{gorti2009b}. 
It is assumed that $\dot{\Sigma}_{\mathrm{pe}}\approx 0$ inside 
of the critical radius $r_{\mathrm{crt}}=r_{g}/10$
\citep[$\approx 0.7\,\AU$,][]{liffman2003,gorti2009a},
where gas lies too deeply in the gravitational field of the star 
to escape. The maximum of $\dot{\Sigma}_{\mathrm{dif}}$ is
around the radius $r\approx r_{\mathrm{crt}}$.

For most of the disk evolution, $\dot{\Sigma}_{\mathrm{rim}}=0$.
At later times, when the mass supply rate operated by viscous 
stresses cannot keep up with the removal rate caused by
$\dot{\Sigma}_{\mathrm{dif}}$, the disk's gas becomes locally 
depleted (typically, somewhat inward of $1\,\AU$). Inside of this
density gap induced by photo-evaporation, gas viscously drains
toward the star on relatively short timescales, of the order of $10^{4}$ 
years for the kinematic viscosity adopted in this study.
Once the disk develops an inner cavity, becoming optically thin 
interior to $r=r_{\mathrm{rim}}$, $\dot{\Sigma}_{\mathrm{rim}}$
provides an additional contribution to photo-evaporation at and 
around the rim region. As a result of the enhanced
$\dot{\Sigma}_{\mathrm{pe}}$, the rim radius $r_{\mathrm{rim}}$ 
increases as the disk disperses from the inside out.
The presence of a sufficiently massive planet, with a semi-major axis
of $a\approx r_{\mathrm{crt}}$, can aid in the formation of the 
photo-evaporation induced gap through gas depletion by tidal 
torques. Beyond the critical radius, a planet accreting gas at high 
rates can reduce the gas density interior to its orbit (see \cieq{eq:brigap}) 
and hence facilitate gap formation by photo-evaporation.

\section{In Situ Formation Models}
\label{sec:insitu}

The calculations consider two phases: the formation phase during 
which accretion of gas and solids takes place, and the evolutionary
or isolation phase, during which the core mass remains constant 
but the envelope is subject to evaporative mass loss. During this
latter phase, the planet is assumed to be completely isolated. 
These phases, up to an age of $8\,\Gyr$, 
are followed numerically for all six of the \kep\ planets.
Each planet is assumed to form at its present orbital position; 
migration is not considered, either of the planet or of the solid 
material that forms its core. 
The initial core mass is $\approx 1\,\Mearth$ at a time of 
$2\times 10^{5}\,\yr$; the corresponding envelope mass is 
$\approx 10^{-3}\,\Mearth$, consistently calculated with the core
mass and the nebular boundary conditions.
The surface density of solids $\sigma_{Z}$ at each formation
radius is adjusted so that the final model at an age of $8\,\Gyr$ 
matches, as closely as possible, the radius of the planet as measured 
by Kepler. The corresponding total planetary masses are then compared 
to those measured via transit timing variations \citepalias{lissauer2013}.
As pointed out by \citet{bodenheimer2014}, these required surface 
densities are high (see Table~\ref{table:sumin}), a factor of roughly four 
to eight times those given by the minimum-mass extrasolar nebula of 
\citet{chiang2013} and three to nine times the values estimated by 
\citet{schlichting2014}. Compared to the densities extrapolated 
from the minimum-mass solar nebula of \citet{hayashi1981}, 
these factors would be much larger, between $25$ and $60$.

The disk temperature during the formation phase, which serves as 
a boundary condition on the planetary structure, is assumed to be 
$T=1000\,\K$ in all cases -- the same assumption is made by
\citet{chiang2013}. The disk gas density during that phase is derived 
assuming that the gas-to-solid mass ratio is $200$, and that the ratio 
of the disk scale height to the orbital distance is $H/a=0.03$.
The disk gas density is  assumed to decrease linearly with time, 
with an assumed cutoff time for the presence of the gas of $3.5\,\Myr$,
which in these models represents the isolation time, \tiso.
The outer radius of the planet, $R_{p}$, during the formation phase
is given by the accretion radius $\Rcapt$ in \cieq{eq:RA}.
As mentioned in \cisec{sec:esc},  the factor four approximately 
describes the results of hydrodynamics simulations of a planet 
embedded in a disk \citep{lissauer2009}, which show that only 
the gas within $\approx \Rhill/4$ remains bound to the planet. 
In these in situ models, the value of $\Rhill/4$ is always smaller than 
the Bondi radius, \Rbondi, by a factor of $\approx 2$ to $\approx5$. 
Thus, \cieq{eq:RA} implies that $R_{p}$ is only weakly dependent 
on disk temperature. Furthermore, as the disk cools with time \Rbondi\ 
gets larger. \citet{stevenson1982} showed that the planet structure is 
only marginally dependent on $R_{p}$ for radiative envelopes (which 
is the case for the outer part of the envelope). Hence, the assumption 
that the disk temperature is constant with disk radius is not expected 
to significantly affect the results.

At the cutoff time, the model makes a transition from disk boundary 
conditions (i.e., the nebular stage, see \cisec{sec:esc}) 
to isolated conditions, basically stellar photospheric boundary 
conditions with the inclusion of the radiation input from the central star, 
as given by Equations~(\ref{eq:LBC})--(\ref{eq:Teq}).
The surface temperature of the planet during the evolutionary phase 
is normally close to the equilibrium temperature, $T_{\mathrm{eq}}$ 
in the stellar radiation field; the approximation is made that this 
temperature is constant with time. The outer layers rapidly thermally
adjust to this new temperature, which is between $500$ and $1000\,\K$.
In all cases, the mass of the gaseous envelope is considerably less
than that of the heavy-element core at the time of this transition. 
The phase of rapid gas accretion (see \cisec{sec:esc}) is never reached. 
When accretion stops, the radius of the planet decreases considerably 
on a short timescale, then declines slowly as the planet contracts and cools.

During the isolation phase, mass loss from the planet's atmosphere
as a result of energy input from stellar X-ray and EUV photons can be 
important. This process is included, starting immediately after disk dispersal
($t=3.5\,\Myr$), according to the energy-limited approximation outlined
in \cisec{sec:gl}. 
These calculations apply a standard value for the efficiency parameter, 
$\varepsilon=0.1$. 
The mass loss turns out to be quite important for the inner planet \kep b, 
but not significant for the outer planet \kep g.

The present calculations differ from those published in earlier papers
\citep[e.g.,][]{hubickyj2005,lissauer2009} because of the dust and 
molecular opacities applied during the formation phase. 
The opacity table includes grain sizes in the range from 
$0.005\,\mu\mathrm{m}$ to $1\,\mathrm{mm}$, with a power-law size 
distribution proportional to the grain radius to the power of $-3$ 
(see \cifig{fig:opa}, top). 
This grain size distribution matches the observations of T~Tauri disks 
better than do opacities based on an interstellar size distribution, 
reduced by a constant factor of about $50$, as used in our earlier 
papers. In contrast to the calculations of \citet{bodenheimer2014},
where grain settling and coagulation were included according 
to the method of \citet{naor2010}, the present calculations use 
pre-computed tables of opacity as a function of temperature and density. 
These simulations require numerous trials based on adjustment of
the main parameter, which is the solid surface density, and inclusion 
of the detailed opacity simulations would have been too time-consuming.
During the isolation phase, since there is no input of solid material,
the grains are assumed to have settled into the envelope's interior and 
evaporated; 
the molecular opacities of \citet{freedman2008} are used during this phase.

\subsection{In Situ Model Results}
\label{sec:insitu_res}

\begin{deluxetable*}{cccccccccc}
\tablecolumns{10}
\tablewidth{0pc}
\tablecaption{Summary of Results for In Situ Formation of \kep\ Planets\tablenotemark{a}\label{table:sumin}}
\tablehead{
\colhead{Planet}&\colhead{$M_{c}/\Mearth$}&\colhead{$M_{e}/\Mearth$}&\colhead{$R_{c}/\Rearth$}&
\colhead{$R_{p}/\Rearth$}&\colhead{$L_{p}/L_{\odot}$}&
\colhead{(Fe,Si)\%\tablenotemark{b}}&\colhead{(Fe,Si,\hhe)\%\tablenotemark{c}}&\colhead{$a$ [\AU]}&
\colhead{$\sigma_{Z}$ [$\mathrm{g\,cm^{-2}}$]}
}
\startdata
b & $1.96$ & $0.00$ & $1.19$ & $1.19$ &                                 & $(30,70)$ &$(30,70,0.0)$ & $0.091$ & 
$10000$\\
c & $5.76$ & $0.26$ & $1.60$ & $2.91$ & $3.9\times 10^{-7}$ & $(30,70)$ &$(28.7,67.0,4.3)$ & $0.107$ & $14500$\\
d & $5.01$ & $0.49$ & $1.53 $ & $3.24$ & $2.3\times 10^{-7}$ & $(30,70)$ &$(27.7,64.5,7.8)$ & $0.155$ &
$\pz6200$\\
e & $6.66$ & $1.45$ & $1.67$ & $4.24$ & $2.1\times 10^{-7}$ & $(30,70)$ &$(24.6,57.5,17.9)$ & $0.195$ &
$\pz4600$\\
f & $2.84$ & $0.11$ & $1.33$ & $2.42$ & $4.1\times 10^{-8}$ & $(30,70)$ &$(28.9,67.4,3.7)$ & $0.250$ &
$\pz1680$\\
g & $5.01$ & $0.74$ & $1.53$ & $3.34$ & $2.4\times 10^{-8}$ & $(30,70)$ &$(26.1,61.0,12.9)$ & $0.466$ &
$\pz\pz685$
\enddata
\tablenotetext{a}{Values at time $t=8\,\mathrm{Gyr}$. 
The last column refers to the local density of solids at $t=0$.}
\tablenotetext{b}{Percentage of the core mass.
`Fe' and `Si' indicate the core's iron nucleus and the silicate mantle, respectively.}
\tablenotetext{c}{Percentage of the planet mass. `\hhe' represents the envelope gas.}
\end{deluxetable*}

\begin{figure*}[]
\centering%
\resizebox{0.75\linewidth}{!}{%
\includegraphics[clip]{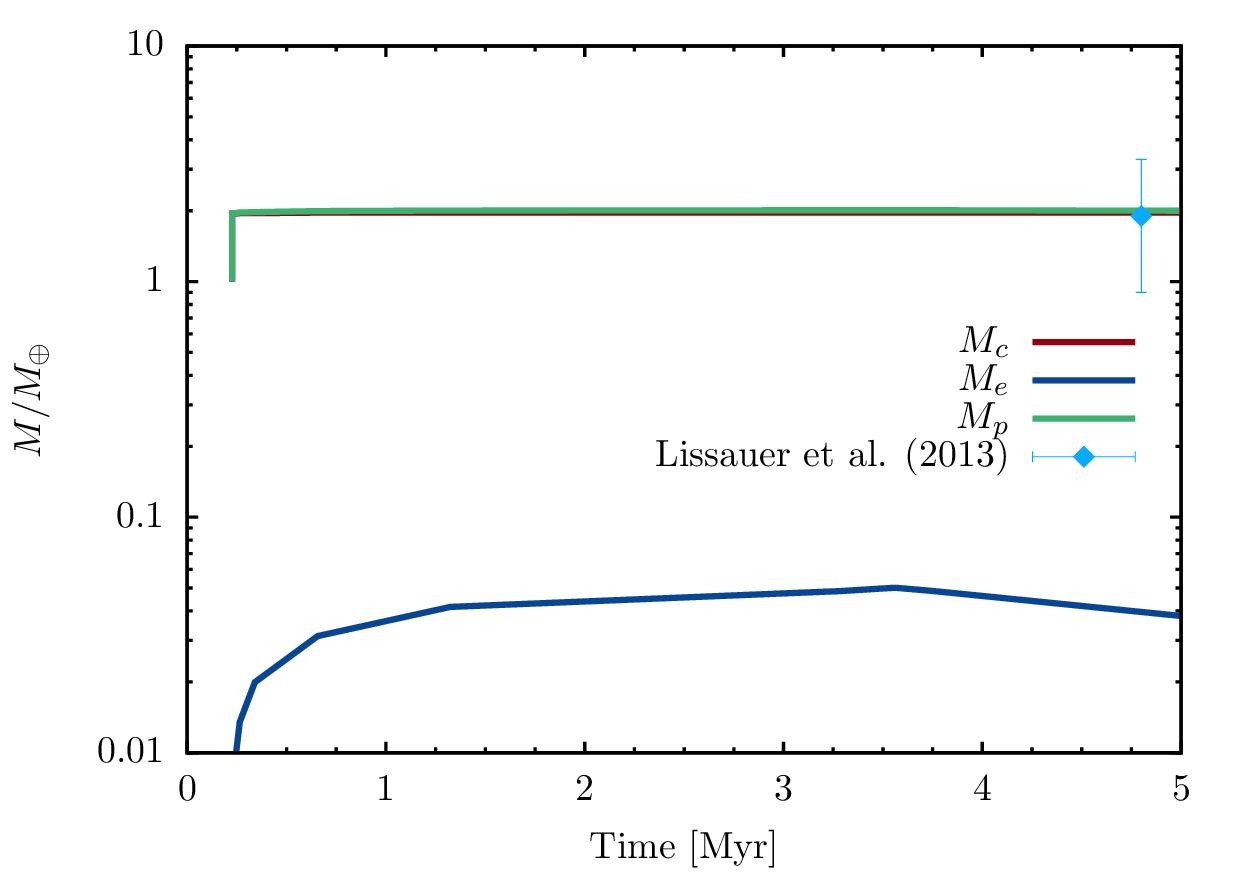}%
\includegraphics[clip]{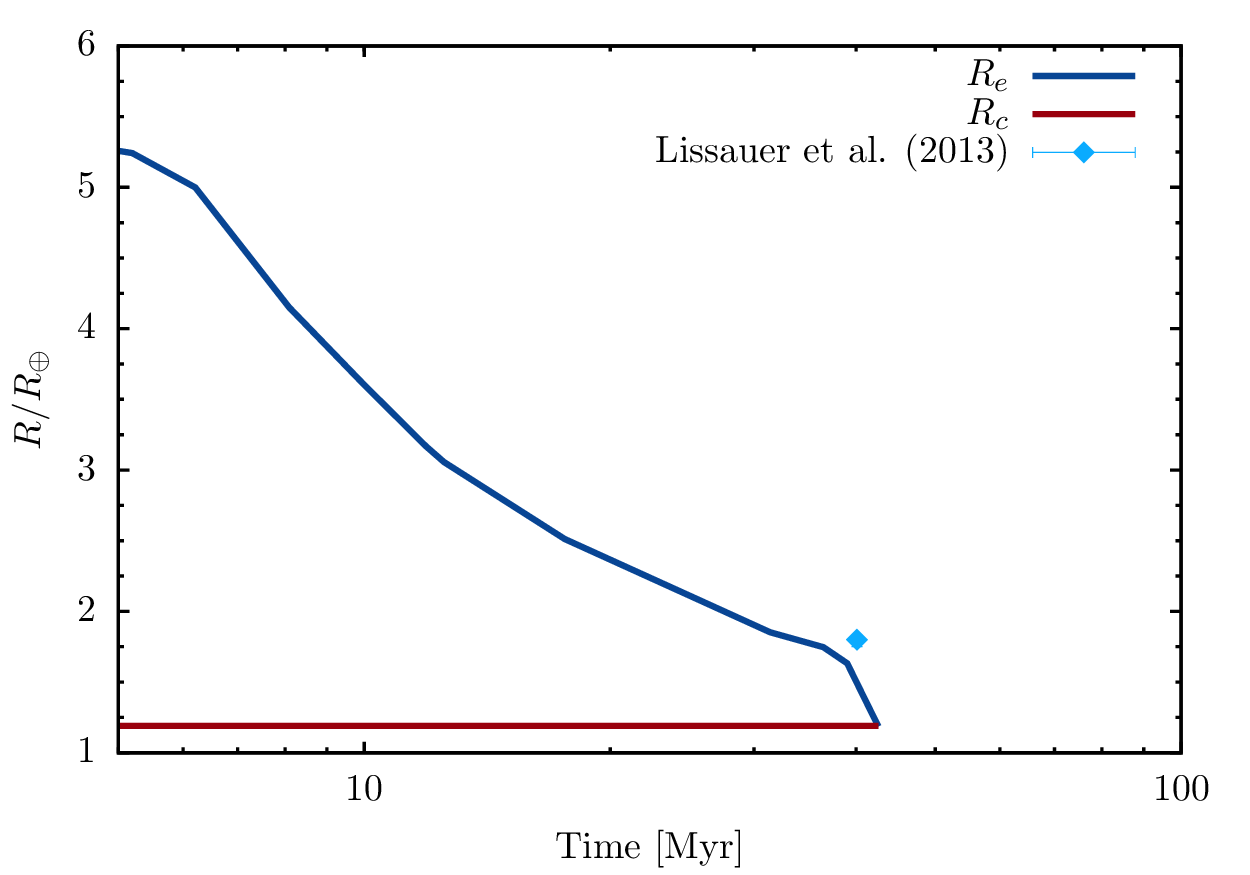}}
\resizebox{0.75\linewidth}{!}{%
\includegraphics[clip]{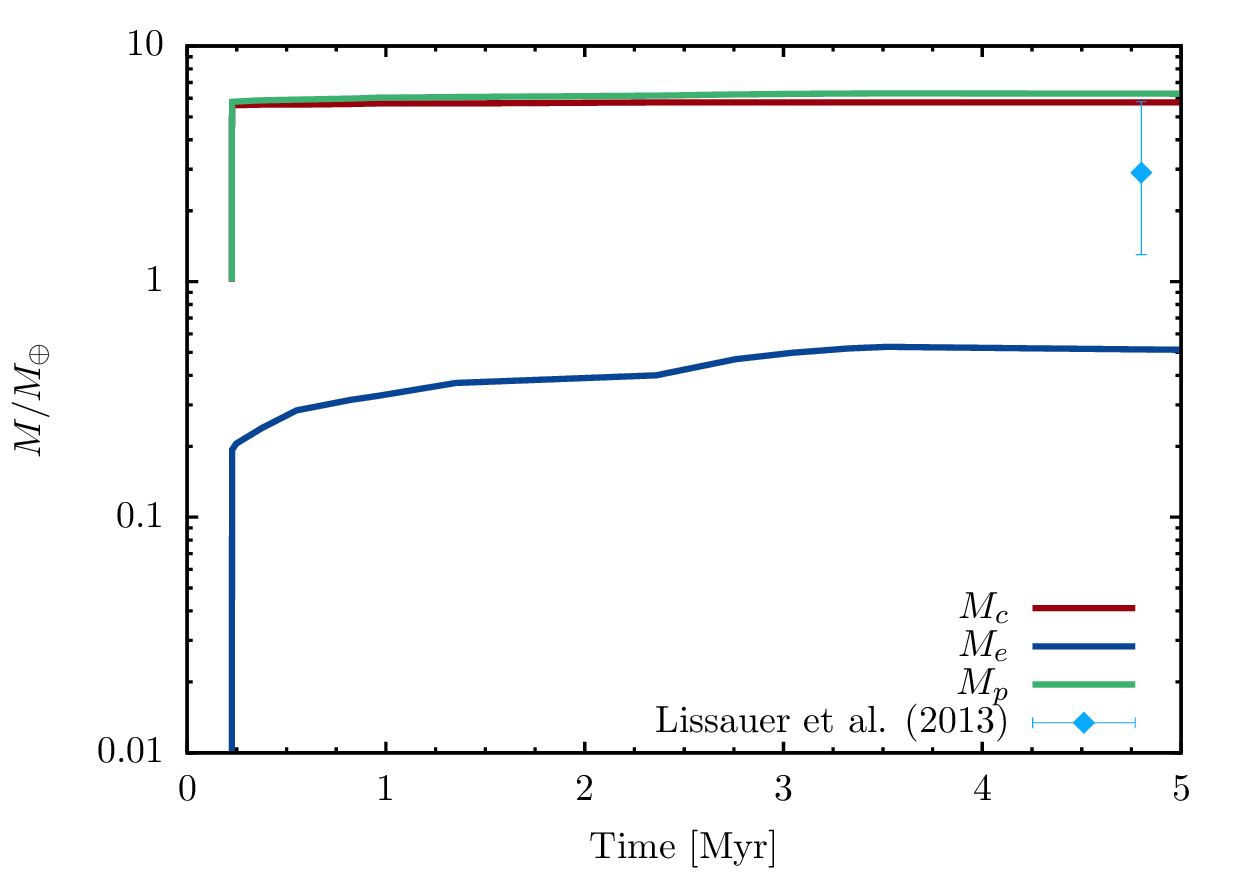}%
\includegraphics[clip]{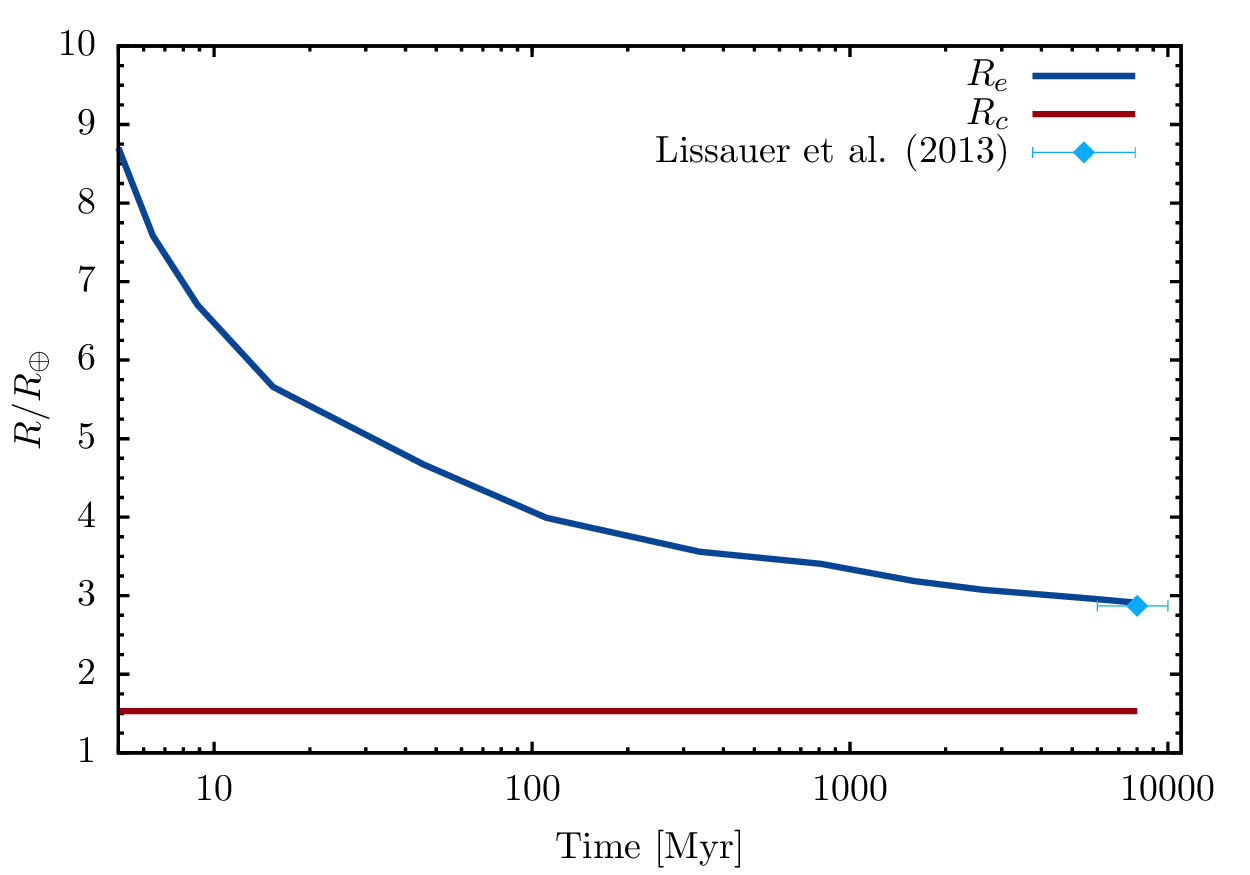}}
\resizebox{0.75\linewidth}{!}{%
\includegraphics[clip]{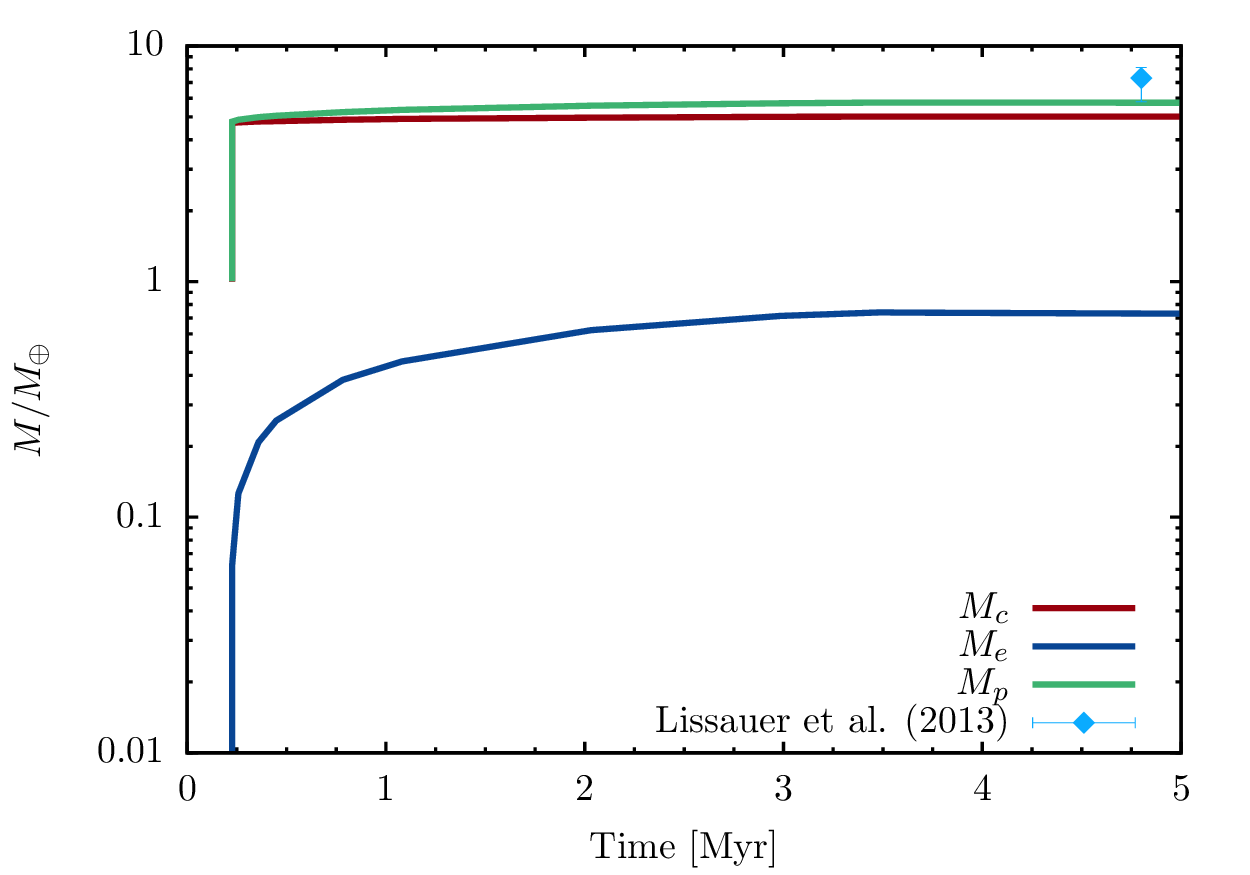}%
\includegraphics[clip]{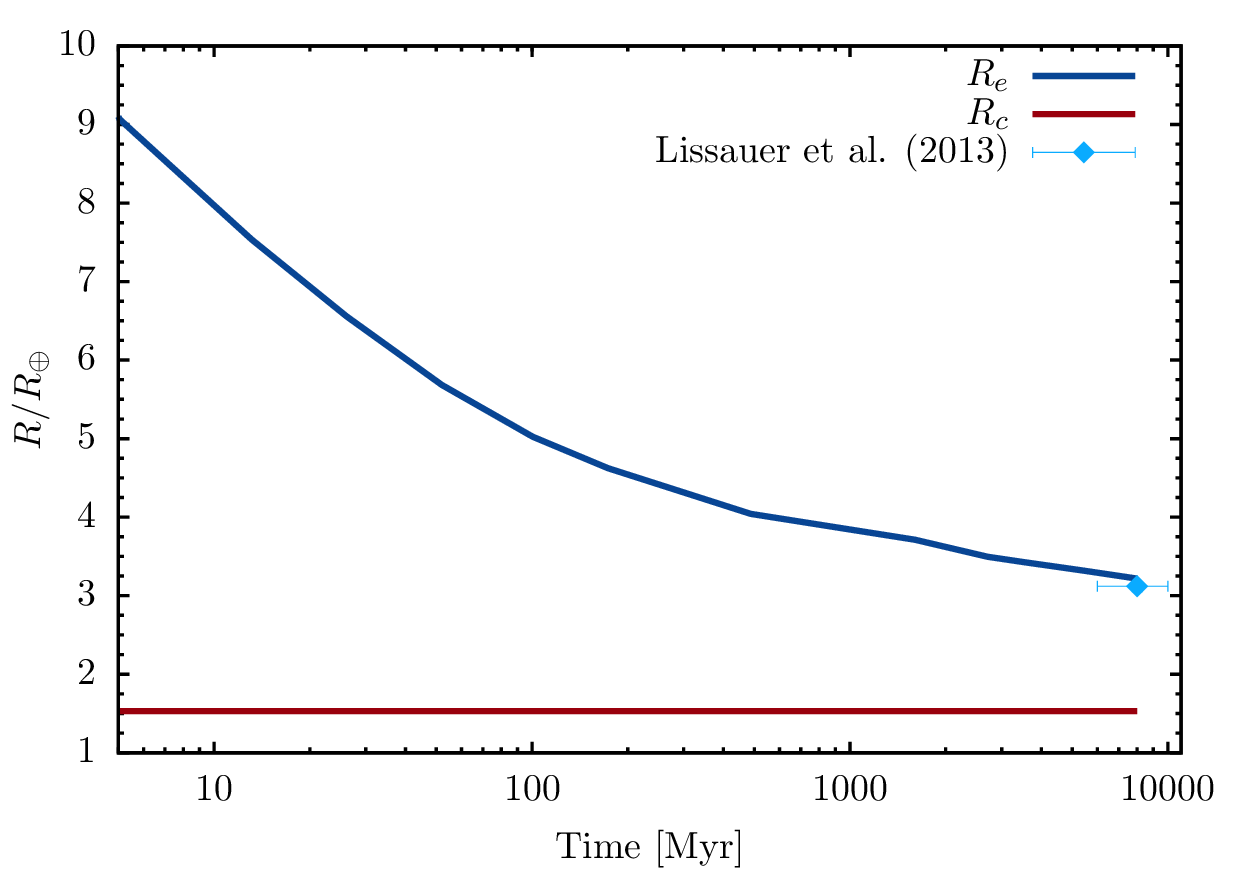}}
\caption{%
             Formation and evolutionary (i.e., isolation) phases of in situ models 
             for planets \kep b (top), \kep c, and \kep d (bottom). 
             Core, envelope, and total mass (left panels, as indicated), and core 
             and envelope radius (right panels as indicated; $R_{e}=R_{p}$) are 
             illustrated as functions of time. 
             The disk's gas around the planet's orbit is assumed to disperse in 
             $\tiso=3.5\,\Myr$. 
             The models start from a $1\,\Mearth$ planetary core at a time of 
             $2\times 10^{5}\,\yr$.
             The vertical error bars indicate the results from \citetalias{lissauer2013}.
             To account for uncertainty, the age of the star is assumed to be 
             $8\pm2\,\mathrm{Gyr}$ 
             \citep[\href{http://exoplanetarchive.ipac.caltech.edu/index.html}{Nasa Exoplanet Archive}, 
             and 
             \href{http://exoplanets.eu/}{The Extrasolar Planets Encyclopaedia}]{lissauer2011a}.
             }
\label{fig:pw1}
\end{figure*}

\begin{deluxetable*}{cccccccccc}
\tablecolumns{9}
\tablewidth{0pc}
\tablecaption{Properties of In Situ Formation Models of \kep\ Planets at Isolation\tablenotemark{a}\label{table:sumin_iso}}
\tablehead{
\colhead{Planet}&\colhead{\tiso [\Myr]}&\colhead{$M_{c}/\Mearth$}&\colhead{$M_{e}/\Mearth$}&\colhead{$R_{c}/\Rearth$}&
\colhead{$R_{p}/\Rearth$}&\colhead{(Fe,Si,\hhe)\%\tablenotemark{b}}&\colhead{$T_{\mathrm{eq}}$\tablenotemark{c} [\K]}&\colhead{$\langle\dot{M}_{e}\rangle$\tablenotemark{d} [$\Mearth\,\yr^{-1}$]}
}
\startdata
b & $3.5$ &$1.96$ & $0.05$ & $1.19$ & $\pz5.3$ &$(29.2,68.3,2.5)$ & $927$ & $-4.0\times 10^{-9\pz}$\\
c & $3.5$ &$5.76$ & $0.53$ & $1.60$ & $\pz9.5$ &$(27.5,64.1,8.4)$ & $880$ & $-1.3\times 10^{-9\pz}$\\
d & $3.5$ &$5.01$ & $0.74$ & $1.53$ & $11.7$ &$(26.1,61.0,12.9)$ & $731$ & $-1.2\times 10^{-9\pz}$\\
e & $3.5$ &$6.66$ & $1.65$ & $1.67$ & $14.3$ &$(24.0,56.1,19.9)$ & $623$ & $-1.0\times 10^{-9\pz}$\\
f & $3.5$ &$2.84$ & $0.16$ & $1.33$ & $\pz8.4$ &$(28.4,66.3,5.3)$ & $550$ & $-2.6\times 10^{-10}$\\
g & $3.5$ &$5.01$ & $0.75$ & $1.53$ & $\pz9.4$ &$(25.7,59.9,14.4)$ & $409$ & $-1.0\times 10^{-10}$
\enddata
\tablenotetext{a}{The isolation time, \tiso, is the time at which disk's gas is assumed to disperse.}
\tablenotetext{b}{Percentage of the planet mass at time $t=\tiso$.}
\tablenotetext{c}{Constant equilibrium temperature, $T_{\mathrm{eq}}$, during the isolation phase.}
\tablenotetext{d}{Rate of change of the envelope mass averaged over the first $100\,\Myr$ 
                           of evolution in isolation. For \kep b, $\langle\dot{M}_{e}\rangle$ is an average 
                           over $10\,\Myr$.}
\end{deluxetable*}

\begin{figure*}[]
\centering%
\resizebox{0.75\linewidth}{!}{%
\includegraphics[clip]{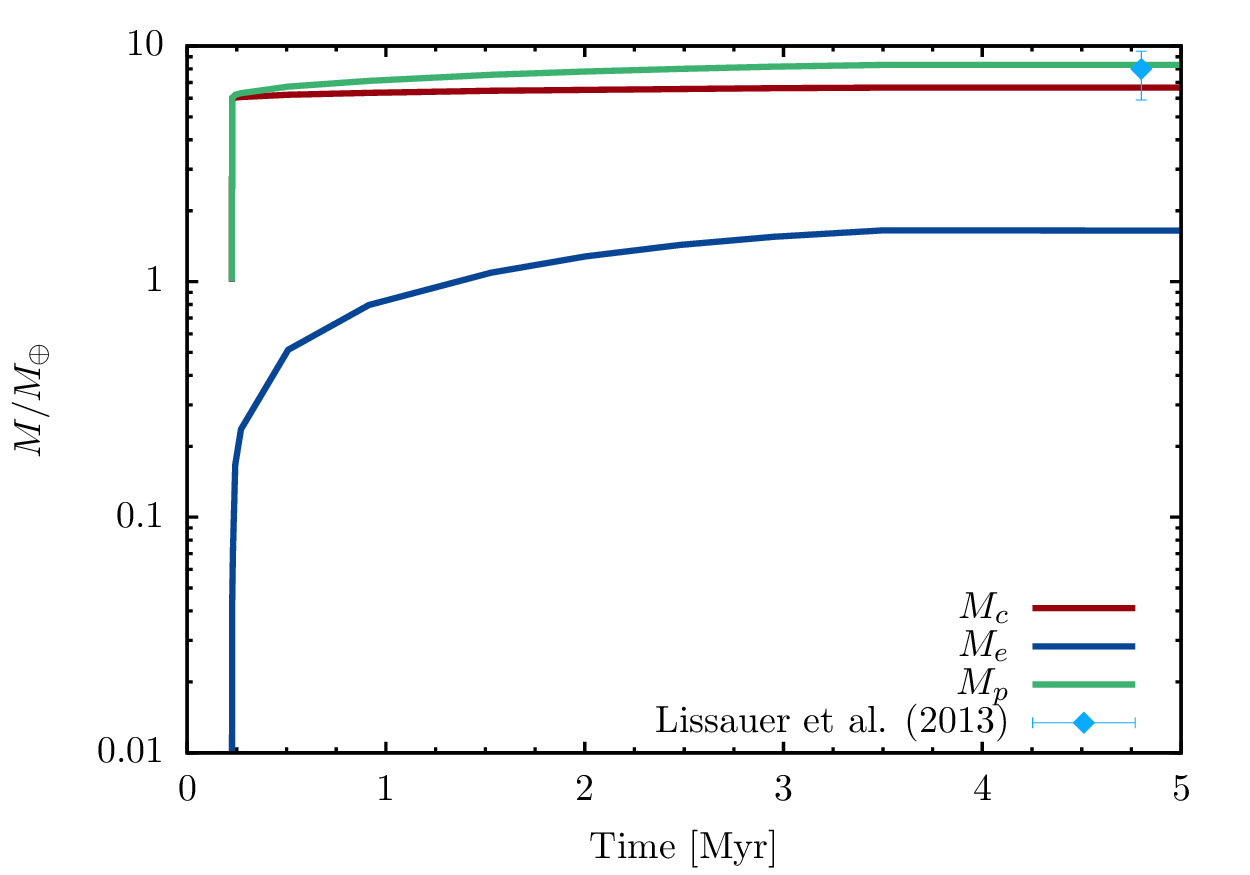}%
\includegraphics[clip]{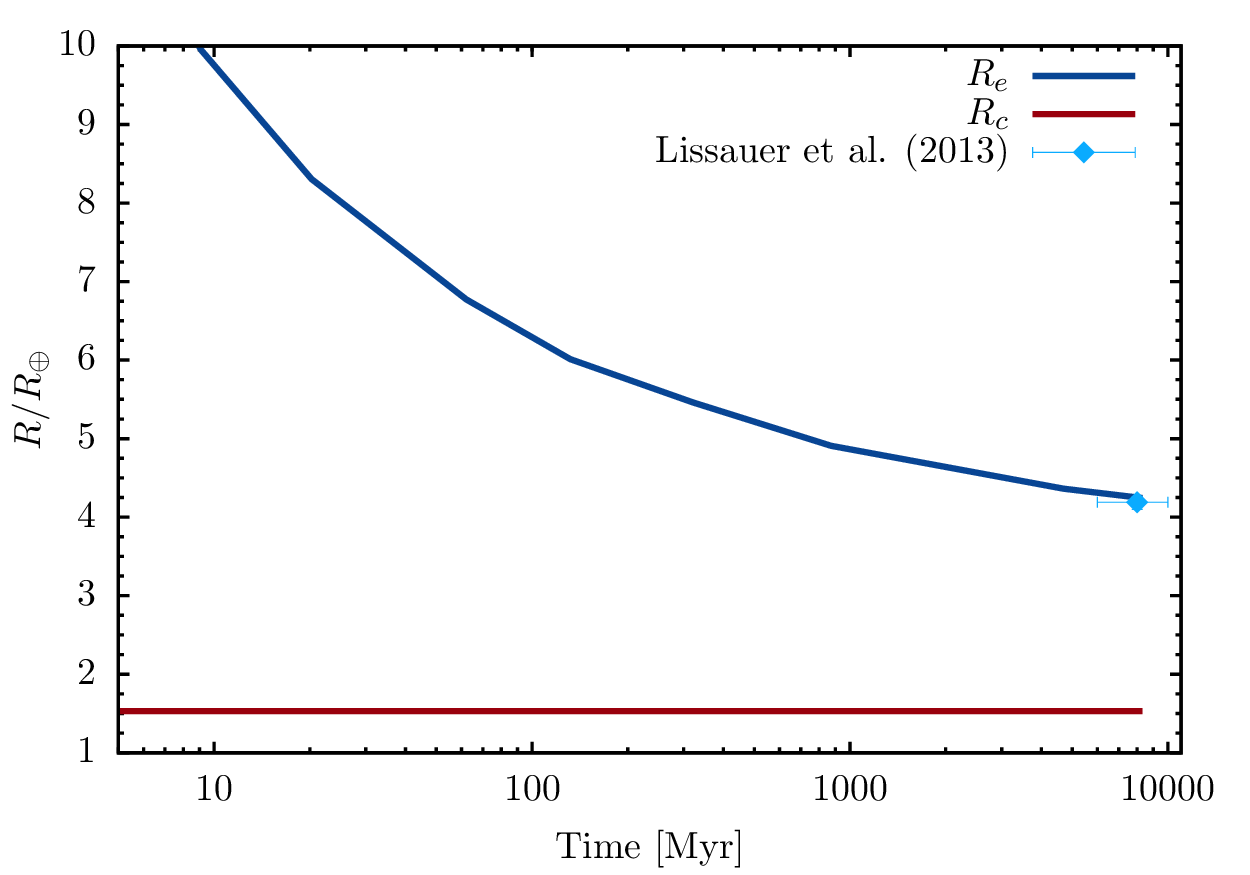}}
\resizebox{0.75\linewidth}{!}{%
\includegraphics[clip]{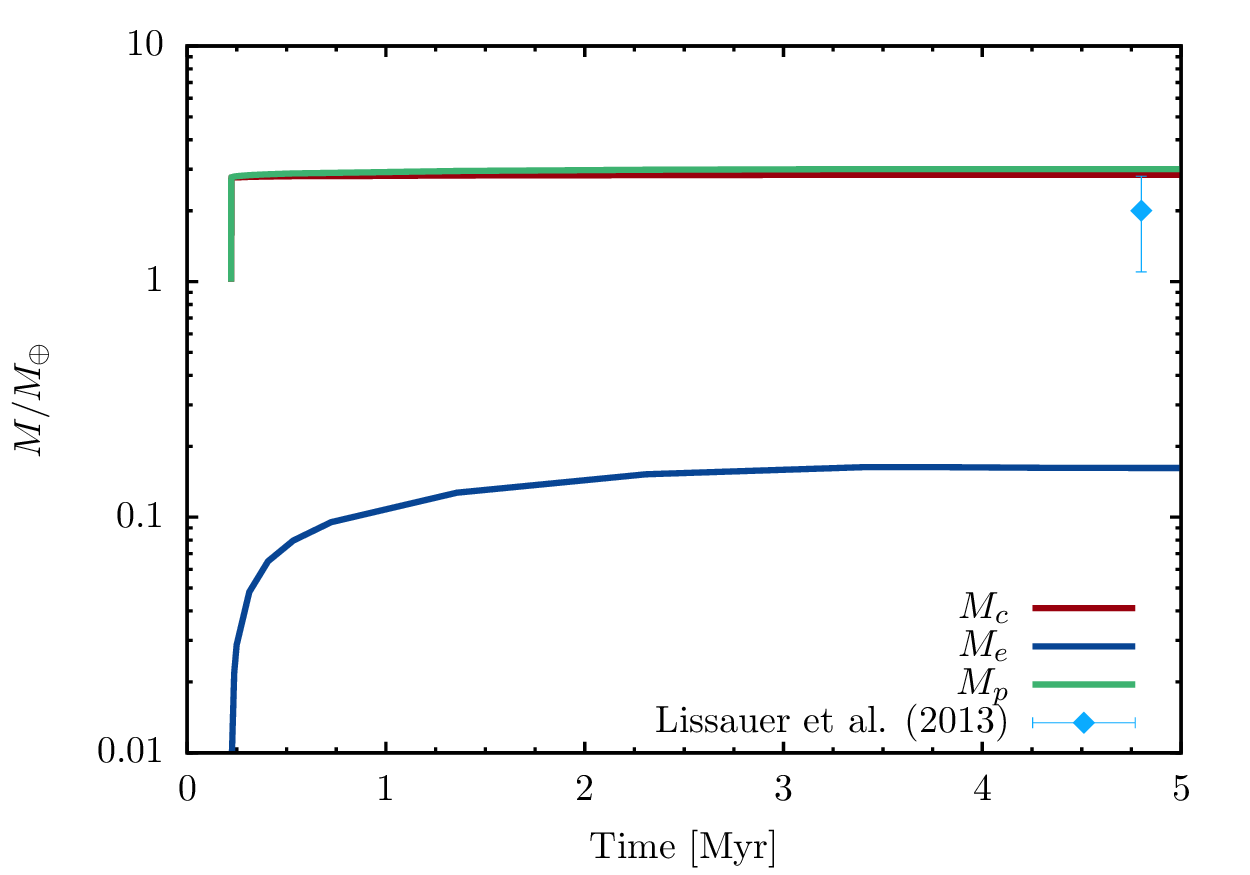}%
\includegraphics[clip]{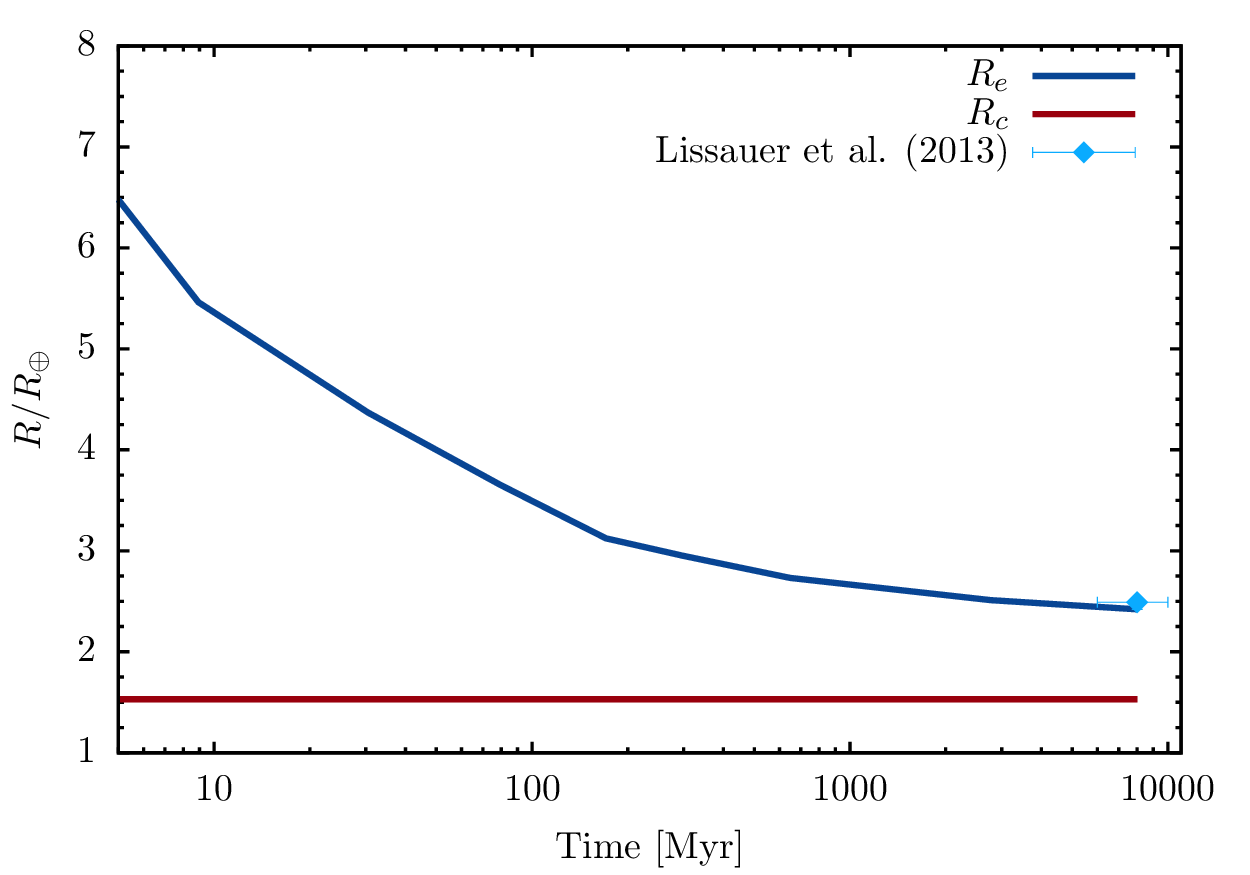}}
\resizebox{0.75\linewidth}{!}{%
\includegraphics[clip]{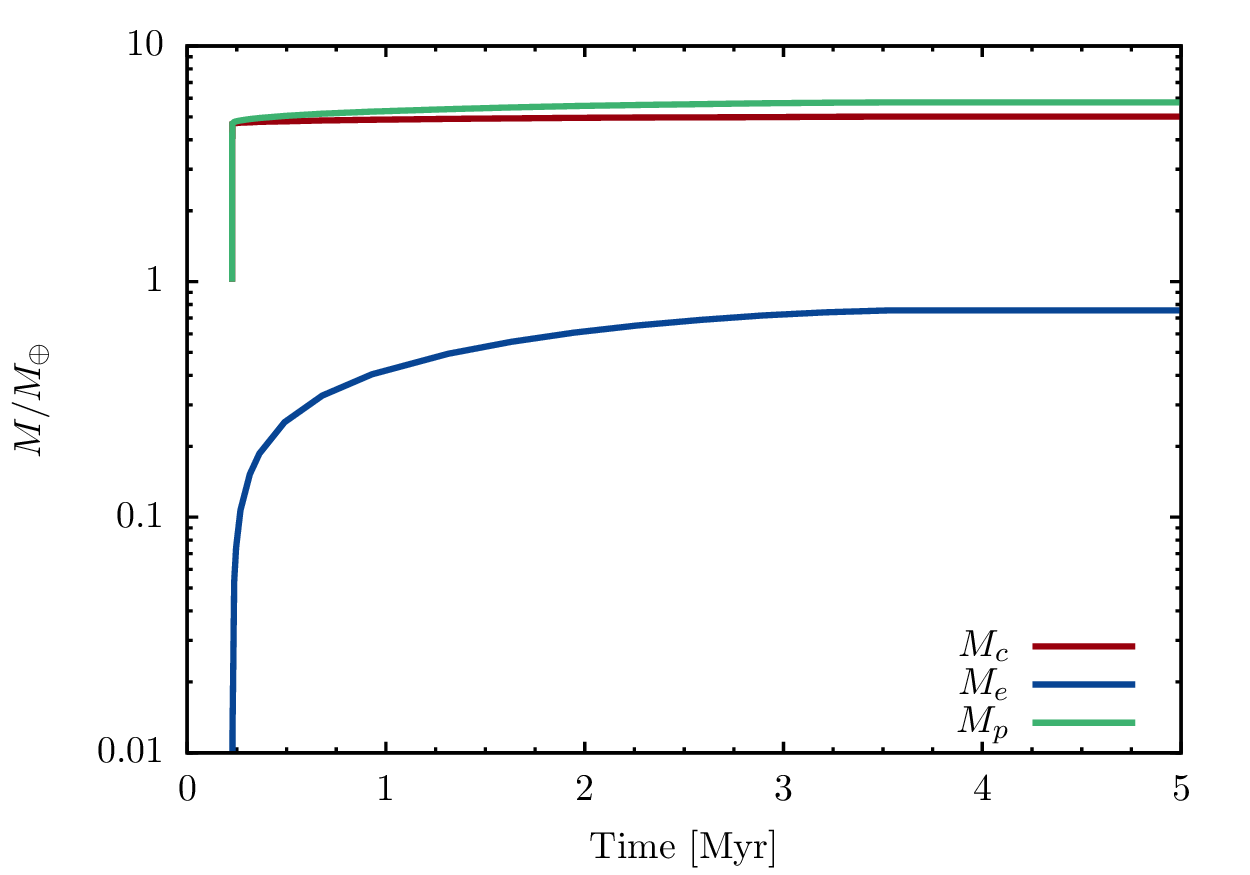}%
\includegraphics[clip]{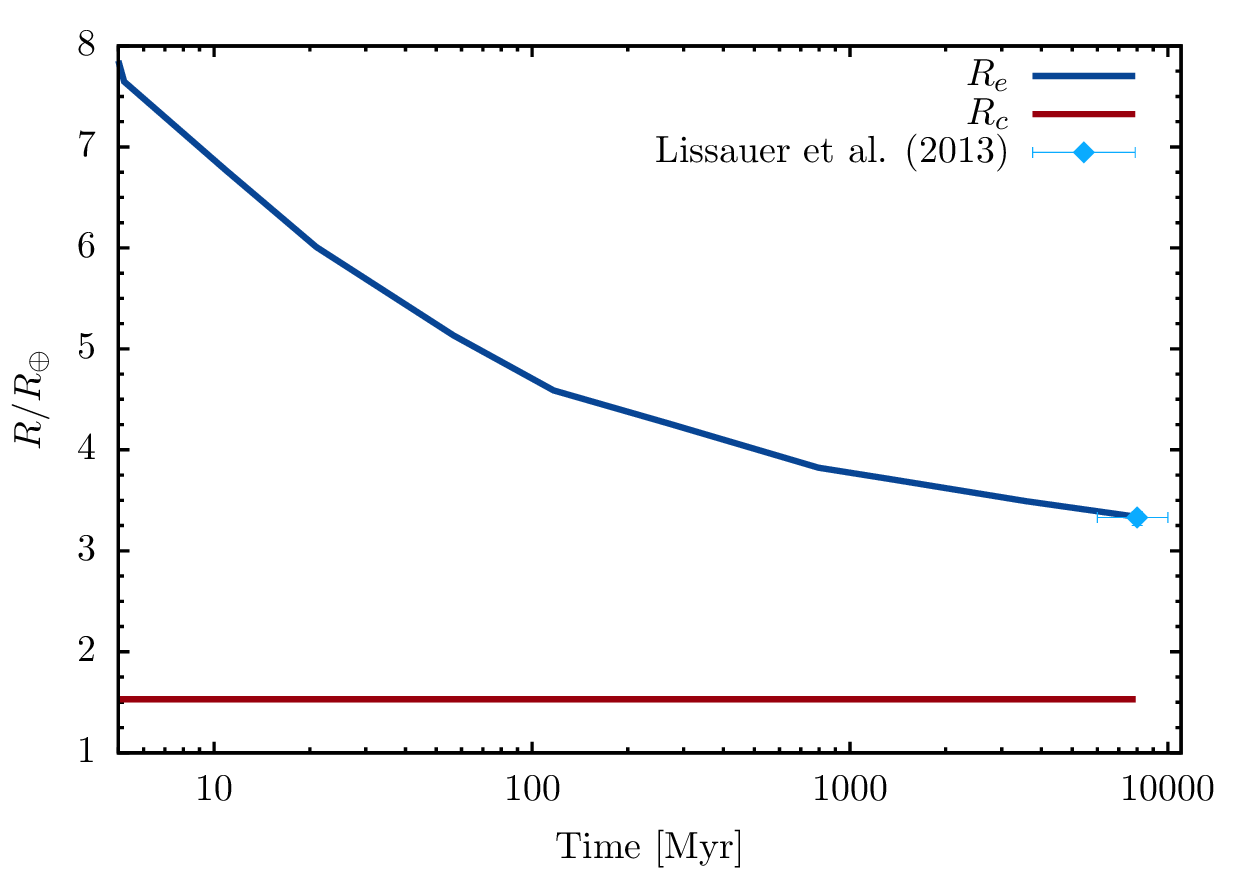}}
\caption{%
             Same as in Figure~\ref{fig:pw1}, but
             for in situ models of \kep e (top), \kep f, and \kep g (bottom).
             }
\label{fig:pw2}
\end{figure*}

Table~\ref{table:sumin} gives a summary of the final properties of 
the six simulated \kep\ planets, along with the deduced value
of $\sigma_{Z}$.
The final core mass, the final envelope mass, the final core radius, 
the final planet radius, the final planet luminosity, the composition, 
and the orbital position are presented.

Figures~\ref{fig:pw1} and \ref{fig:pw2} show, for the various planets,
the evolution of the core mass, envelope mass, and outer
radius. In general, because of the high solid surface densities
required, the core mass increases rapidly, on a timescale $< 10^5$ years. 
In fact, by using \cieq{eq:dotmc}, the accretion timescale of the core is
\begin{equation}
 \frac{M_{c}}{\dot{M}_{c}}=\frac{4}{3}\frac{\rho_{c} R_{c}}{\sigma_{Z} F_{g}}%
                                          \sqrt{\frac{a^{3}}{G\Ms}},
 \label{eq:tcg}
\end{equation}
where the cross section for planetesimals' capture is
$\mathcal{S}_{\mathrm{eff}}\approx \pi R^{2}_{c}$ when the gas bound 
to the core is very tenuous. If $F_{g}\approx 1$, \cieq{eq:tcg} gives 
initial accretion timescales (see $a$ and $\sigma_{Z}$ in 
Table~\ref{table:sumin}) $\lesssim 10^{5}\,\yr$.
Nonetheless, at an initial envelope mass of $\approx 10^{-3}\,\Mearth$,
even $100\,\mathrm{km}$-size planetesimals are affected by gas drag
in the envelope and $\mathcal{S}_{\mathrm{eff}}$ becomes 
$\gg \pi R^{2}_{c}$ \citep{gennaro2014}. In the calculations, the actual 
timescales are all of the order of $10^{4}\,\yr$ (see Figures~\ref{fig:pw1} 
and \ref{fig:pw2}).
Thus, any uncertainties in Equation (\ref{eq:dotmc}) or in the 
choice of the initial core mass have practically no effect on the 
final result. The core mass
levels off at the isolation mass, as given by \cieq{eq:mciso}.
The envelope mass increases more slowly, on a timescale of 
$10^{6}\,\yr$. The outer radius during the formation 
phases shows an initial rapid rise corresponding to the rapid 
core growth, and then a nearly flat section since the outer 
boundary condition is essentially determined by the nearly 
constant core mass. 
Once the transition at $\tiso=3.5\,\Myr$ is reached, 
the radius decreases rapidly as a result of the transition to 
isolated boundary conditions. Beyond that time, the radius 
decreases slowly as a result of contraction and cooling.
As a further effect, the envelope mass and outer radius decline
as a result of mass loss induced by stellar XUV radiation.

Some properties of the \kep\ planets at the time of disk dispersal
($t=\tiso$), according to our in situ models, are reported in 
Table~\ref{table:sumin_iso}.
Planet \kep b would lose its entire \hhe\ envelope in 
$4 \times 10^7$ years. For planets \kep f through c, proceeding inwards, 
mass-loss rates are $1$ to $5\times 10^{-9}\,\Mearth\,\yr^{-1}$
at $10\,\Myr$, decreasing to $1$ to $5\times 10^{-10}\,\Mearth\,\yr^{-1}$ 
at $100\,\Myr$. 
At the final age of $8\,\Gyr$, these rates are down to
$2.3\times 10^{-13}$ to $1.3\times 10^{-12}\,\Mearth\,\yr^{-1}$.

Specifically, the main results from our in situ models can be summarized 
as follows.

\textbf{\kep b}.
a low-mass \hhe\ envelope forms around the core mass of 
$1.96\,\Mearth$, but during the isolation phase this 
envelope is entirely lost. The final mass is consistent with the measured
mass $1.9^{+1.4}_{-1.0}\,\Mearth$ \citepalias{lissauer2013}, but the final radius, 
the core radius of $1.19\,\Rearth$, is far below the measured value of 
$1.80^ {+0.03}_{-0.05}\,\Rearth$, as shown in the top-right panel of 
\cifig{fig:pw1}. Even by taking the upper limit of the measured mass,
a $100$\% silicate core would still have too small a radius, only $1.5\,\Rearth$.
A steam envelope (not modeled here) is probably required to achieve
consistency \citep{lopez2012}. 
This possibility, however, is inconsistent with in situ formation inside 
$0.1\,\AU$ of a solar-type star because of the lack of ice in the core. 
Another possibility is the release of gas sequestered by the core during 
formation. 
The result that the entire envelope mass is lost remains valid even if 
the assumed core mass is increased to $3\,\Mearth$.

\textbf{\kep c}.
about half of the accreted envelope mass is lost during the isolation
phase, but the final radius agrees with the measured radius of
$2.87^{+0.05}_{-0.06}\,\Rearth$ (see \cifig{fig:pw1}, 
center-right). 
The final total mass of $6.02\,\Mearth$ falls just above the 
one-standard-deviation upper limit for the measured mass 
of $2.9^{+2.9}_{-1.6}\,\Mearth$, but is certainly within the 
uncertainties in the theoretical models.

\textbf{\kep d}.
about one-third of the accreted envelope mass is lost during 
the isolation phase. The final computed radius of $3.24\,\Rearth$ 
is within $4$\% of the measured value of $3.12^{+0.06}_{-0.07}\,\Rearth$
(see \cifig{fig:pw1}, bottom-right). 
The final computed  total mass of $5.5\,\Mearth$ is just below the 
one-standard-deviation lower limit for the measured mass of 
$7.3^{+0.8}_{-1.5}\,\Mearth$. These small discrepancies are well 
within the uncertainties of the models.
To reduce the radius to agree with the measured value would require 
reducing the mass, increasing the discrepancy with the measured value.

\textbf{\kep e}.
at a separation from the star of about $0.2\,\AU$, the accreted \hhe\ 
envelope of this object loses only about $12$\% of its mass during 
the isolation phase. The final computed radius of $4.24\,\Rearth$ 
agrees to within $1.5$\% with the measured value of 
$4.19^{+0.07}_{-0.09}\,\Rearth$, 
as indicated the in top-right panel of \cifig{fig:pw2}. 
The final computed total mass of
$8.11\,\Mearth$ agrees with the measured mass of $8.0^{+1.5}_{-2.1}\,\Mearth$.
A slight reduction in the assumed mass ($\approx 5$\%) would bring 
the radius into agreement with the observed value and the planet mass 
would still agree with the observed mass, well within one-standard-deviation 
uncertainties.

\textbf{\kep f}.
even though the planet lies farther from the star ($0.25\,\AU$) 
than \kep e, its lower core mass ($2.8$ vs $6.7\,\Mearth$)
results in $31$\% of the \hhe\ envelope being lost during the
isolation phase. The final computed radius of $2.42\,\Rearth$ 
agrees with the measured value of $2.49^{+0.04}_{-0.07}\,\Rearth$ 
to within $2.5$\% (see \cifig{fig:pw2}, center-right).
The final total mass of $2.95\,\Mearth$ is slightly above the
one-standard-deviation upper limit for the measured value of 
$2.0^{+0.8}_{-0.9}\,\Mearth$.
To improve the agreement with the observed radius, the assumed 
mass would have to increase by about $0.1\,\Mearth$, increasing 
the (small) discrepancy with the observed value. In any case, 
the agreement is within the uncertainties of the theoretical model.

\textbf{\kep g}.
at a distance of $0.466\,\AU$, the planet loses only about $2$\% 
of its \hhe\ envelope mass during the isolation phase.
The computed final radius of $3.34\,\Rearth$ agrees with the measured
value of $3.33^ {+0.06}_{-0.08}\,\Rearth$. The corresponding computed
total mass is $5.75\,\Mearth$. The observed mass in this case is not well 
constrained; it is less than $25\,\Mearth$.

\section{Ex Situ Formation Models}
\label{sec:exsitu}

\subsection{General Results}
\label{sec:gen_res}

\begin{figure}
\centering%
\resizebox{\linewidth}{!}{%
\includegraphics[clip]{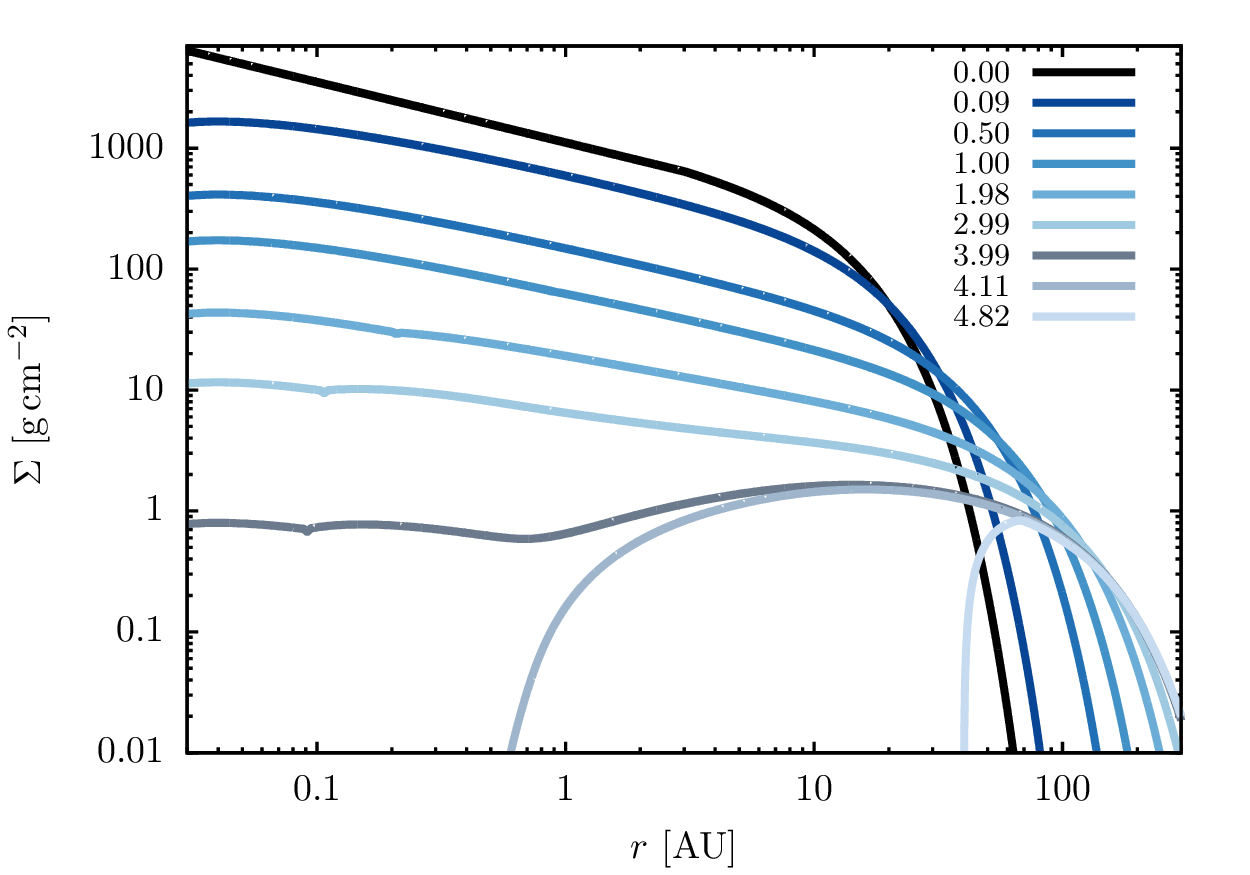}}\\
\resizebox{\linewidth}{!}{%
\includegraphics[clip]{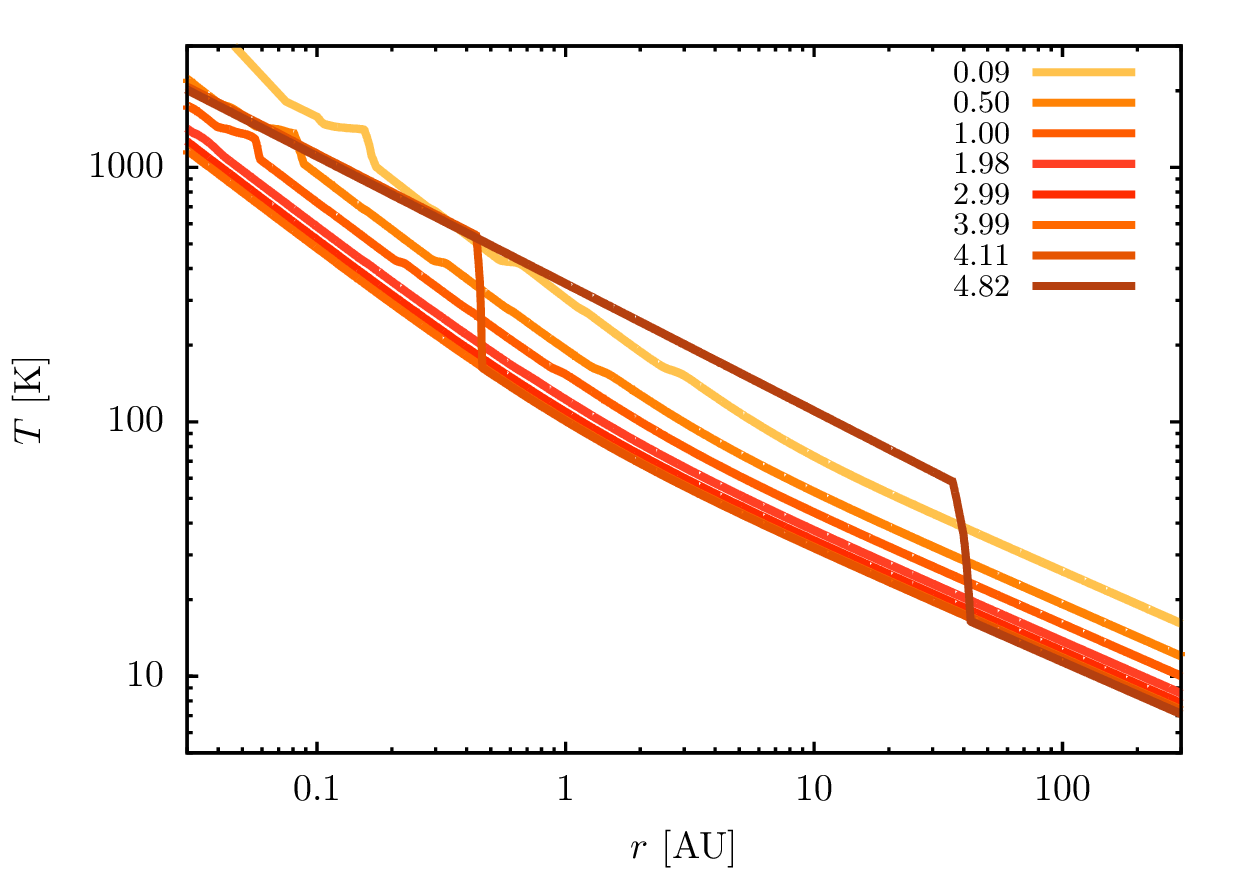}}
\caption{%
             Surface density (top) and temperature distribution (bottom) of the disk's gas,
             from the model of \kep b, as a function of time (as indicated in the legends in
             units of megayears). A cavity forms in the inner disk at $t\approx 4.1\,\Myr$.
             Photo-evaporation of gas caused by stellar irradiation first induces
             the formation of a density gap, which turns into a cavity because gas 
             interior to the gap is rapidly removed by viscous diffusion toward the star. 
             The disk rim, i.e., the external edge of the gap, now exposed to direct stellar 
             irradiation, quickly recedes away (see \cisec{sec:DPH} for details). 
             The temperature transitions in the profiles at $4.1$ and 
             $4.8\,\Myr$ represent transitions between the local irradiation temperature
             (in \cieq{eq:Tirr}) and the gas temperature (from \cieq{eq:EEq}).
             }
\label{fig:devo_b}
\end{figure}

The two phases of planet evolution identified in \cisec{sec:insitu} 
can also be defined in ex situ models. The time $t=0$ coincides 
with the time at which the disk evolution starts from the imposed
initial conditions. The gaseous disk evolution depends on several
quantities (see \cisec{sec:de}--\ref{sec:DPH}).
The gas surface density at $t=0$ is 
$\Sigma=1110\sqrt{1\,\AU/r}\,\densu$ with an exponential 
cut-off beyond some radius, so that a total disk mass of 
$0.03\,\Msun$ is initially confined within $\approx 60\,\AU$ of 
the star \citep[see, e.g.,][]{williams2011}.
This surface density distribution, $\Sigma(t=0)$, was determined 
after a number of attempts aimed at reproducing the observed 
physical properties of planet \kep b and at tentatively matching
those of \kep f (the planet farthest from the star for which both 
$R_{p}$ and $\Mp$ are constrained by transit observations). 
Only later it was realized that, fortuitously, $\Sigma(t=0)$ inside 
$\approx 10\,\AU$ matches quite closely the minimum-mass 
solar nebula density of \citet{davis2005}.
Although no other slope $d\ln{\Sigma}/d\ln{r}$ was tested for 
the initial $\Sigma$, it is unlikely that the adopted initial surface 
density provides a unique solution to the problem.
In any case, the choice of the initial density values is likely 
to influence the outcomes of the models more than does 
the choice of the initial slope $d\ln{\Sigma}/d\ln{r}$.

These calculations apply a time-constant kinematic viscosity 
$\nu=\nu_{1}\sqrt{r/1\,\AU}$, where 
$\nu_{1}\approx 4\times 10^{14}\,\mathrm{cm^{2}\,s^{-1}}$.
In terms of the viscosity prescription of \citet{S&S1973}, the 
$\alpha$-parameter quantifying turbulence varies with time 
and distance from the star.
The value around the starting orbital radius of the simulated 
planets is between $\approx 10^{-3}$ and $\approx 10^{-2}$.
The disk provides an initial accretion rate toward the star of 
order $10^{-7}\,\Msun\,\yr^{-1}$.

The lifetime of the gaseous disk is determined by $\Sigma(t=0)$,
$\nu$, the initial disk mass, and the photo-evaporation rate.
All of these quantities are the same for all planet models. 
As mentioned
in \cisec{sec:DPH}, the presence of a planet may affect disk
dispersal, to a smaller or larger extent depending on the planet
mass, its gas accretion rate, and orbital radius. In fact, although 
the planets end up inside the critical radius $r_{\mathrm{crt}}$,
they do spend most of their disk-embedded evolution at larger 
radii and, therefore, they can potentially influence 
$\dot{\Sigma}_{\mathrm{pe}}$. 
However, in the models presented here, this effect appears 
to be marginal and the gas inside $1\,\AU$ is dispersed in 
$\approx 4\,\Myr$, with a time-spread among models of 
about $2$\%.

The evolution of $\Sigma$ and $T$ is illustrated in 
\cifig{fig:devo_b} (see the figure's caption for details).
Dust opacity transitions are visible in the temperature profiles,
the most prominent of which are represented by 
the evaporation of the silicate species above $1000\,\K$.
The fainter opacity transitions associated with the evaporation
of icy grains are also visible (located around $3\,\AU$ at
$t\approx 10^{5}\,\yr$ and $1\,\AU$ at $t\approx 10^{6}\,\yr$).
In the forming region of \kep\ planets ($r>0.09\,\AU$),
the gas temperature is $\lesssim 1600\,\K$ at 
$t\approx 10^{5}\,\yr$ and becomes $\lesssim 1000\,\K$ at 
times $t\gtrsim 5\times 10^{5}\,\yr$. By the time the simulated
planets have settled on their final orbits, the local gas temperature
varies between $\approx 160$ and $\approx 500\,\K$.

Gas photo-evaporation by stellar irradiation produces a density gap
somewhat inward of $1\,\AU$, around the radius
$r_{\mathrm{crt}}\approx 0.7\,\AU$, where the ratio between the 
photo-evaporation timescale and the accretion timescale through 
the disk is smallest. 
Viscous diffusion quickly removes gas inward of $r_{\mathrm{crt}}$
on a timescale of $\sim r^{2}_{\mathrm{crt}}/\nu$
(see \cisec{sec:DPH}), generating a cavity at $t\approx 4.1\,\Myr$.
Afterwards, rim photo-evaporation dissipates gas inside-out,
pushing the cavity edge outward to $r\approx 40\,\AU$ by 
$t\approx 4.8\,\Myr$ (see \cifig{fig:devo_b}).
Inside the disk cavity, which is virtually devoid of gas, the temperature
$T$ is set equal to the irradiation temperature, so that 
$T^{4}=L_{\star}/(4\pi \sigma_ {\mathrm{SB}} r^{2})$.
Beyond the cavity edge, the temperature is set by the gas thermal 
balance, \cieq{eq:EEq} (hence the large temperature transitions in 
the bottom panel of \cifig{fig:devo_b} for $t>4\,\Myr$).
The evolution of $\Sigma$ and $T$ in \cifig{fig:devo_b} is the same 
for all models, except for variations induced by disk-planet tidal 
interactions and gas accretion on the planet.

\begin{deluxetable*}{cccccccccc}
\tablecolumns{10}
\tablewidth{0pc}
\tablecaption{Summary of Results for Ex Situ Formation of \kep\ Planets\tablenotemark{a}\label{table:sumex}}
\tablehead{
\colhead{Planet}&\colhead{$M_{c}/\Mearth$}&\colhead{$M_{e}/\Mearth$}&\colhead{$R_{c}/\Rearth$}&
\colhead{$R_{p}/\Rearth$}&\colhead{$L_{p}/L_{\odot}$}&
\colhead{(Fe,Si,\ice)\%\tablenotemark{b}}&\colhead{(Fe,Si,\ice,\hhe)\%\tablenotemark{c}}&
\colhead{$a_{i}$ [\AU]}&\colhead{$a_{f}$ [\AU]}
}
\startdata
b & $2.10$  & $0.00$  & $1.47$  & $1.47$  & & $(10.6,50.6,38.8)$ &
$(10.6,50.6,38.8,0.0)$ & $2.14$ & $0.091$ \\
c & $4.56$  & $0.10$  & $1.84$  & $2.84$  & $2.1\times 10^{-7}$ & $(6.6,46.5,46.9)$ &
$(6.5,45.5,45.9,2.1)$ & $3.94$ & $0.109$ \\
d & $5.58$  & $0.31$  & $1.93$  & $3.14$  & $1.4\times 10^{-7}$ & $(6.0,46.0,48.0)$ &
$(5.7,43.6,45.5,5.2)$ & $4.68$ & $0.156$ \\
e & $6.90$ & $1.29$  & $2.00$  & $4.14$  & $1.6\times 10^{-7}$ & $(5.7,45.7,48.6)$ &
$(4.8,38.5,41.0,15.7)$ & $5.35$ & $0.194$ \\
f & $2.74$  & $0.07$ & $1.62$ & $2.49$ & $3.4\times 10^{-8}$ & $(6.1,46.1,47.8)$ & 
$(6.0,44.9,46.6,2.5)$ & $2.10$ & $0.248$ \\
g & $5.57$ & $0.69$ & $1.92$ & $3.40$ & $1.9\times 10^{-8}$ & $(5.0,45.0,50.0)$ & 
$(4.5,40.1,44.4,11.0)$ & $4.41$ & $0.469$ 
\enddata
\tablenotetext{a}{Values at time $t= 8\,\mathrm{Gyr}$.}
\tablenotetext{b}{Percentage of the core mass. 
`Fe', `Si', and `\ice' indicate, respectively, the iron nucleus, the silicate mantle, and 
the \ice\ outer shell of the core (see Appendix~\ref{sec:ics} for details).}
\tablenotetext{c}{Percentage of the planet mass.}
\end{deluxetable*}

\begin{figure*}
\centering%
\resizebox{\linewidth}{!}{%
\includegraphics[clip]{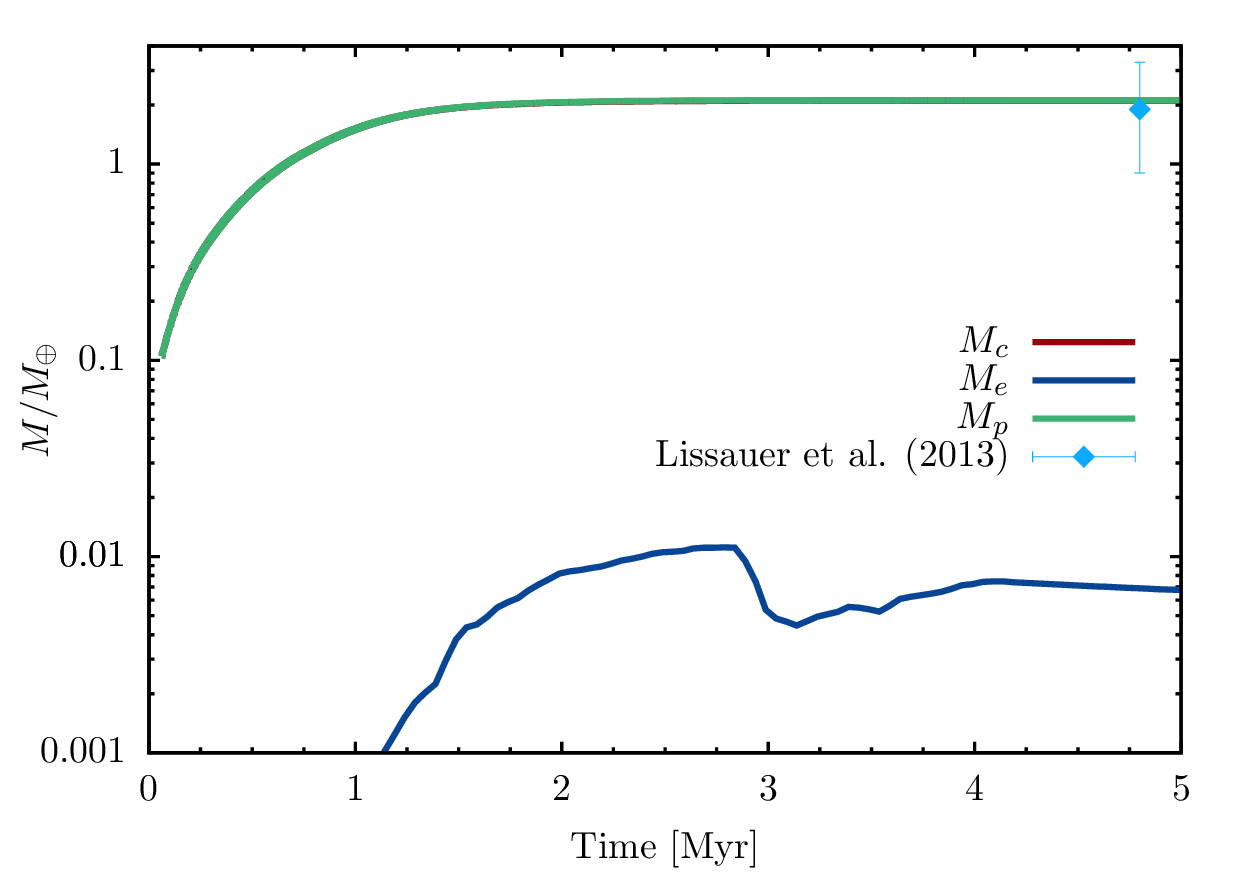}%
\includegraphics[clip]{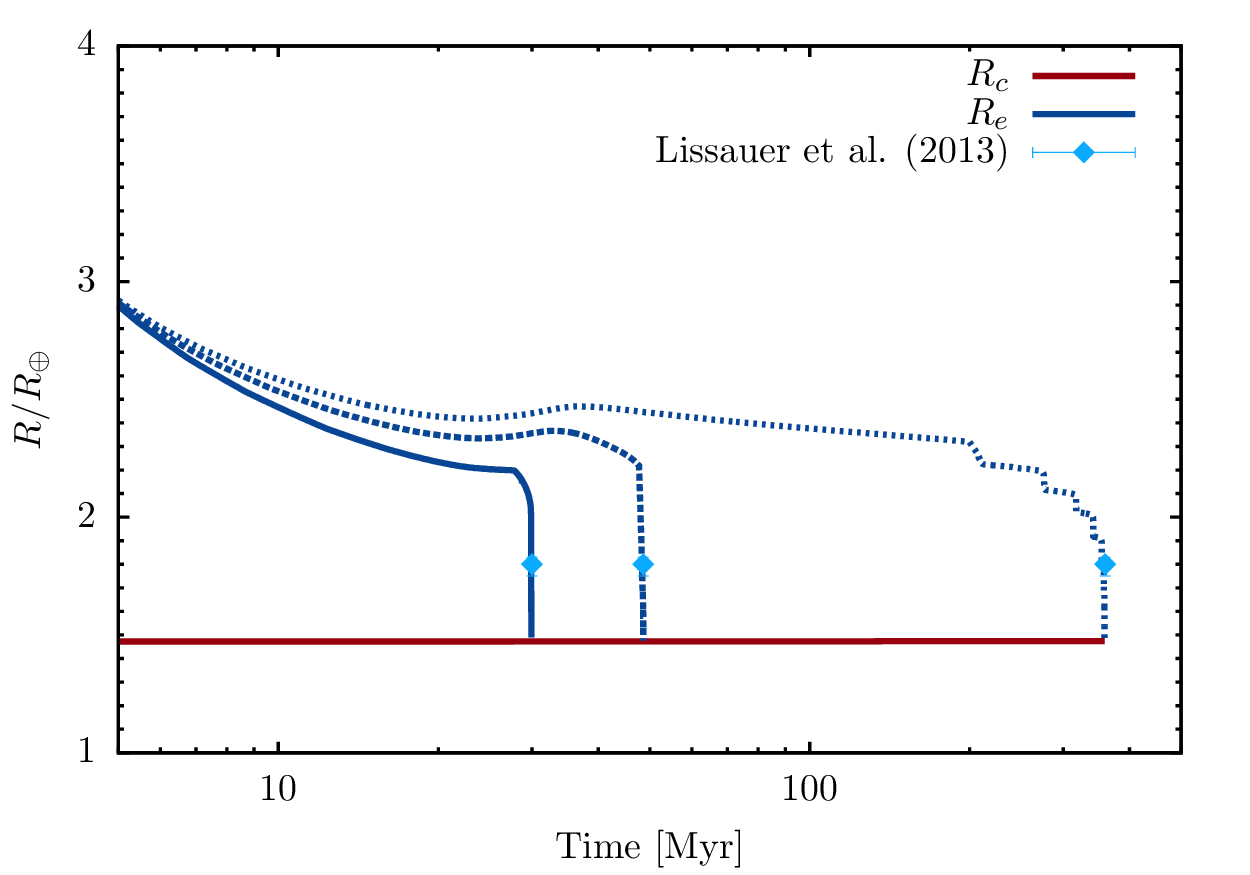}%
\includegraphics[clip]{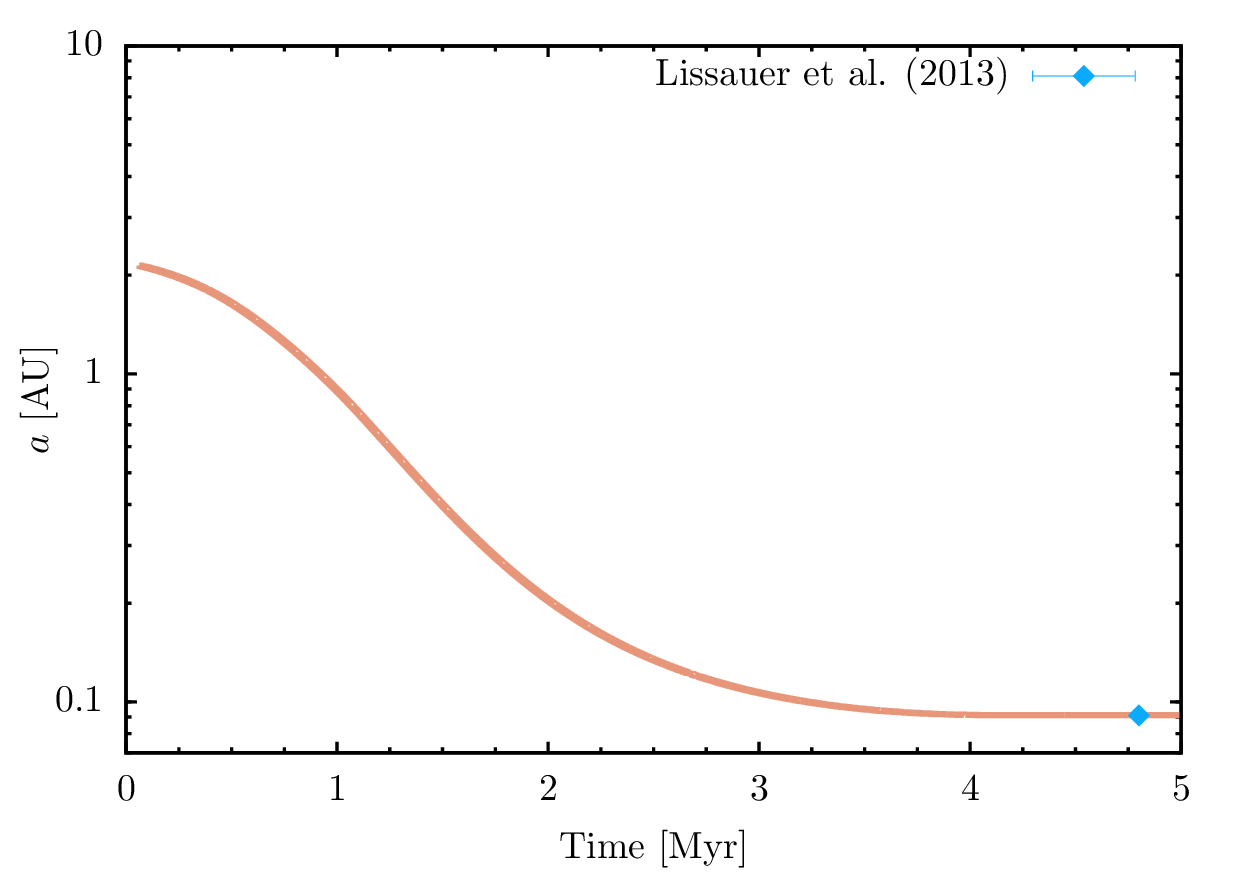}}
\resizebox{\linewidth}{!}{%
\includegraphics[clip]{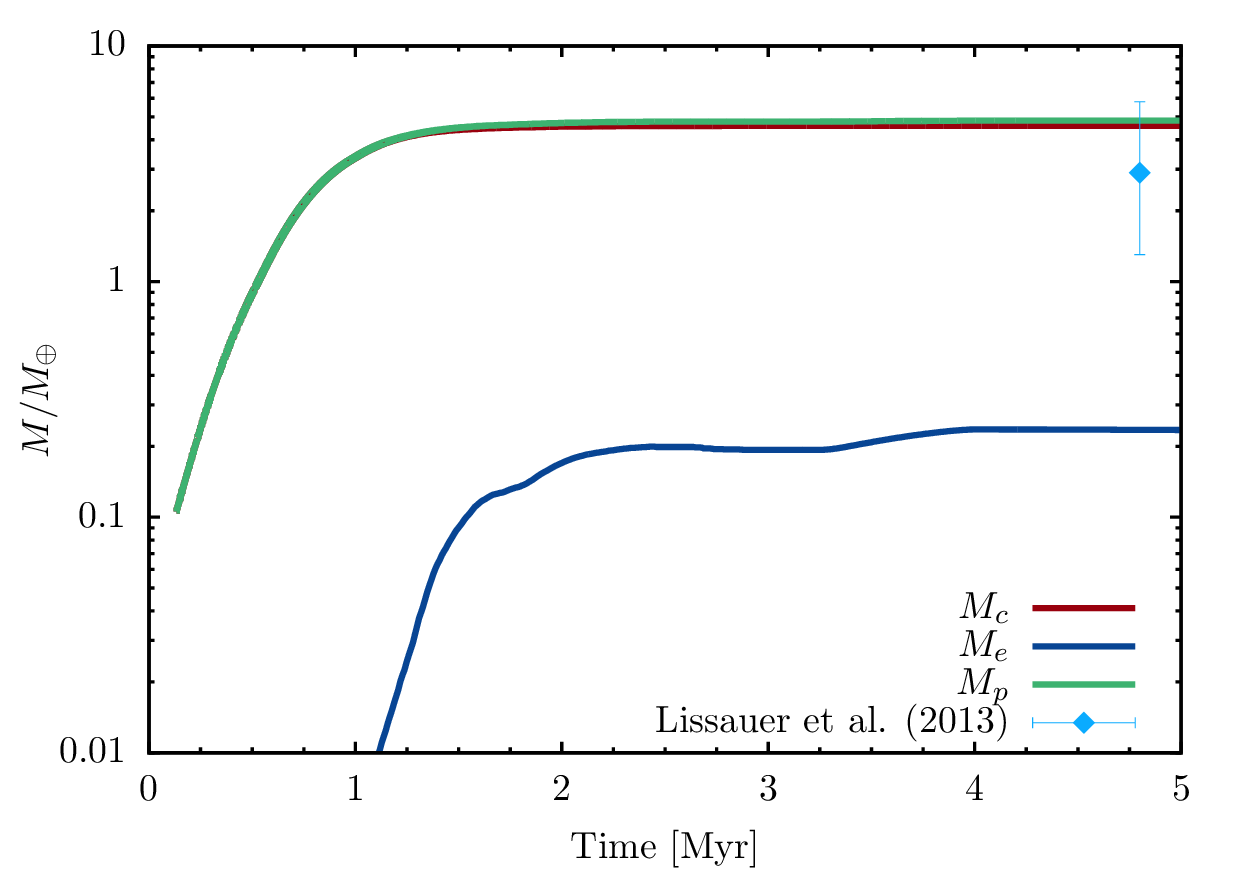}%
\includegraphics[clip]{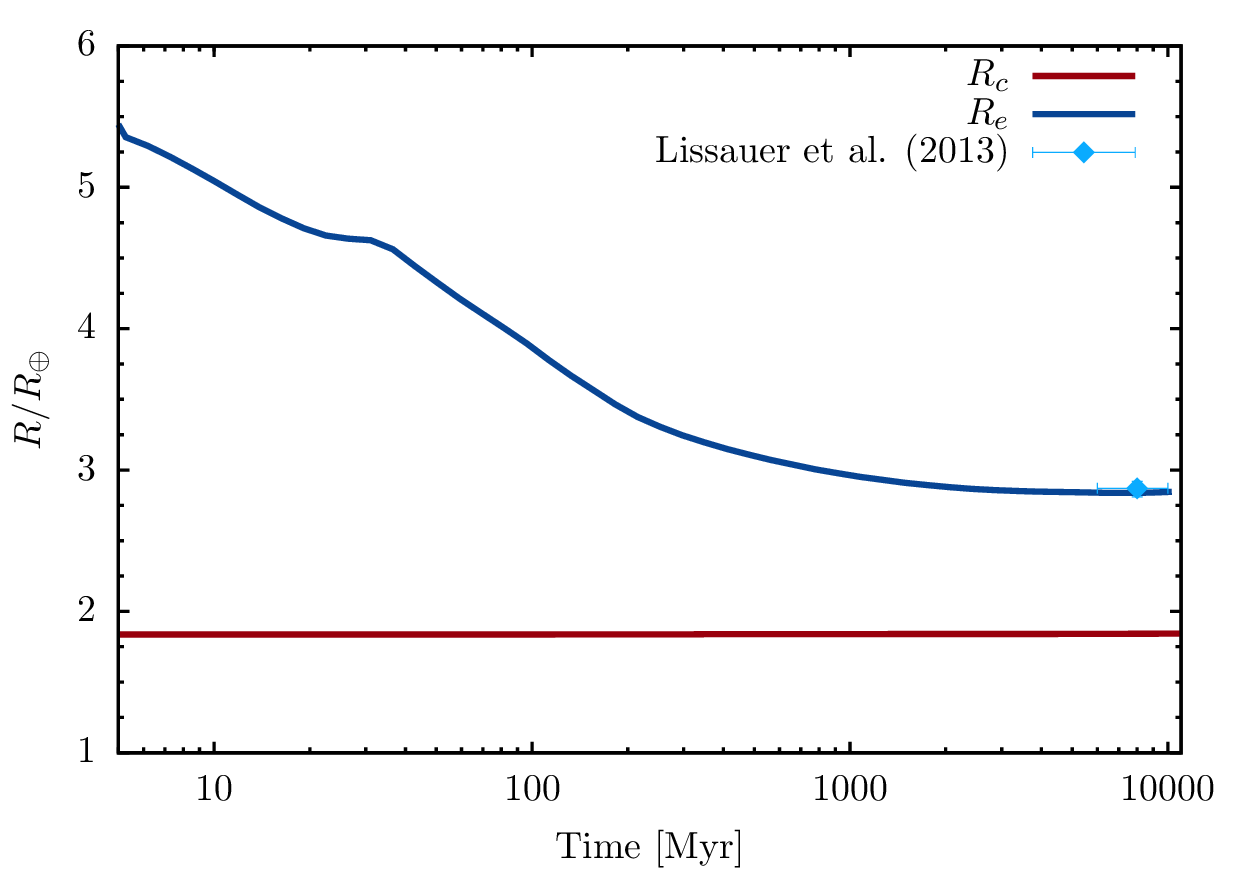}%
\includegraphics[clip]{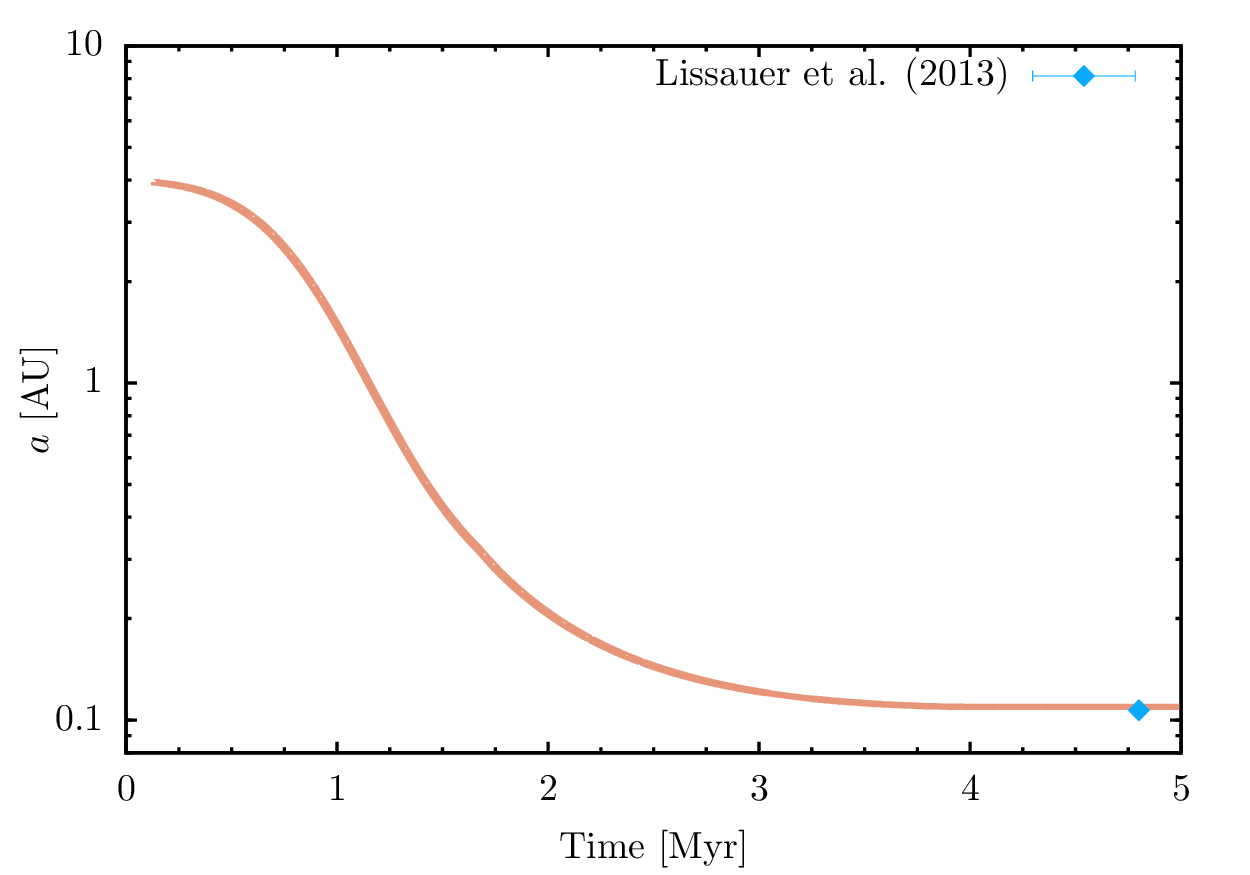}}
\resizebox{\linewidth}{!}{%
\includegraphics[clip]{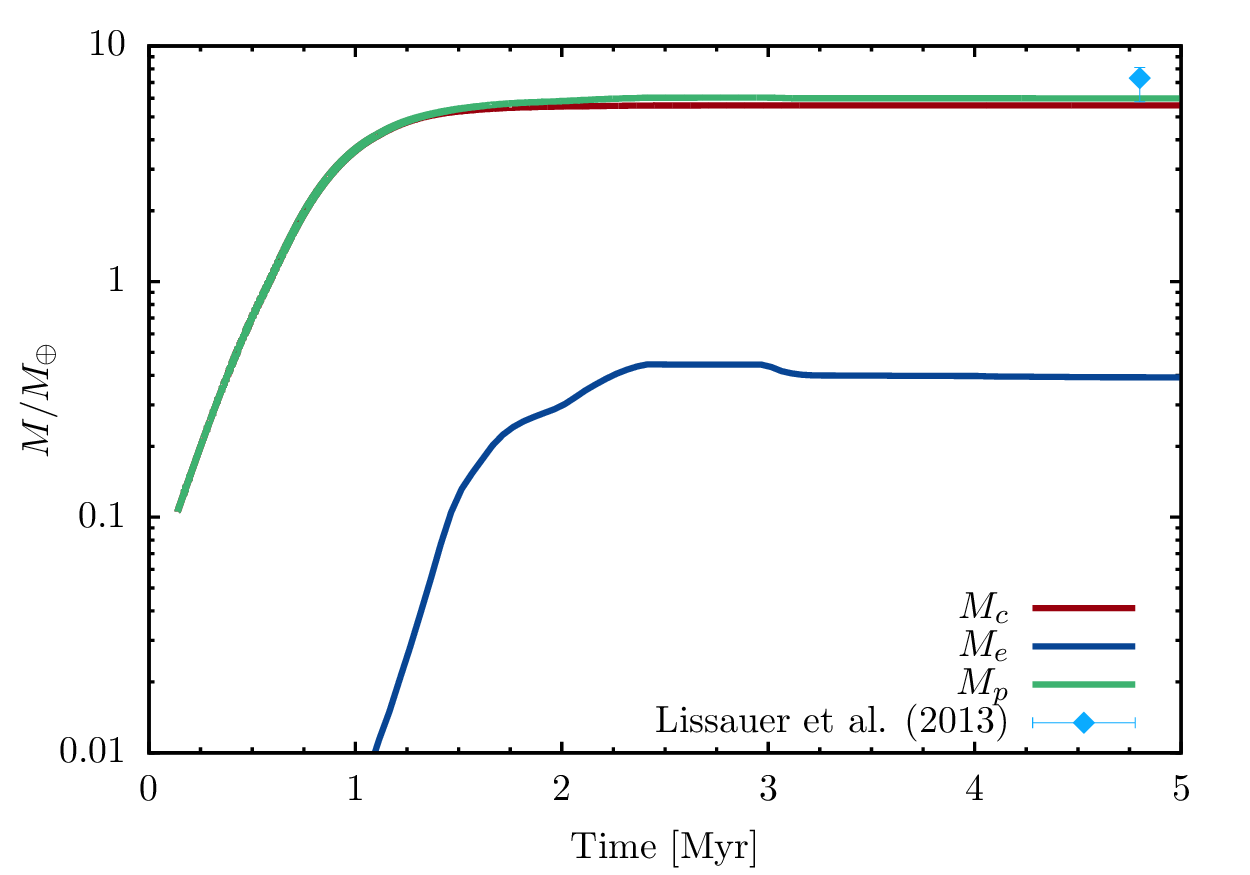}%
\includegraphics[clip]{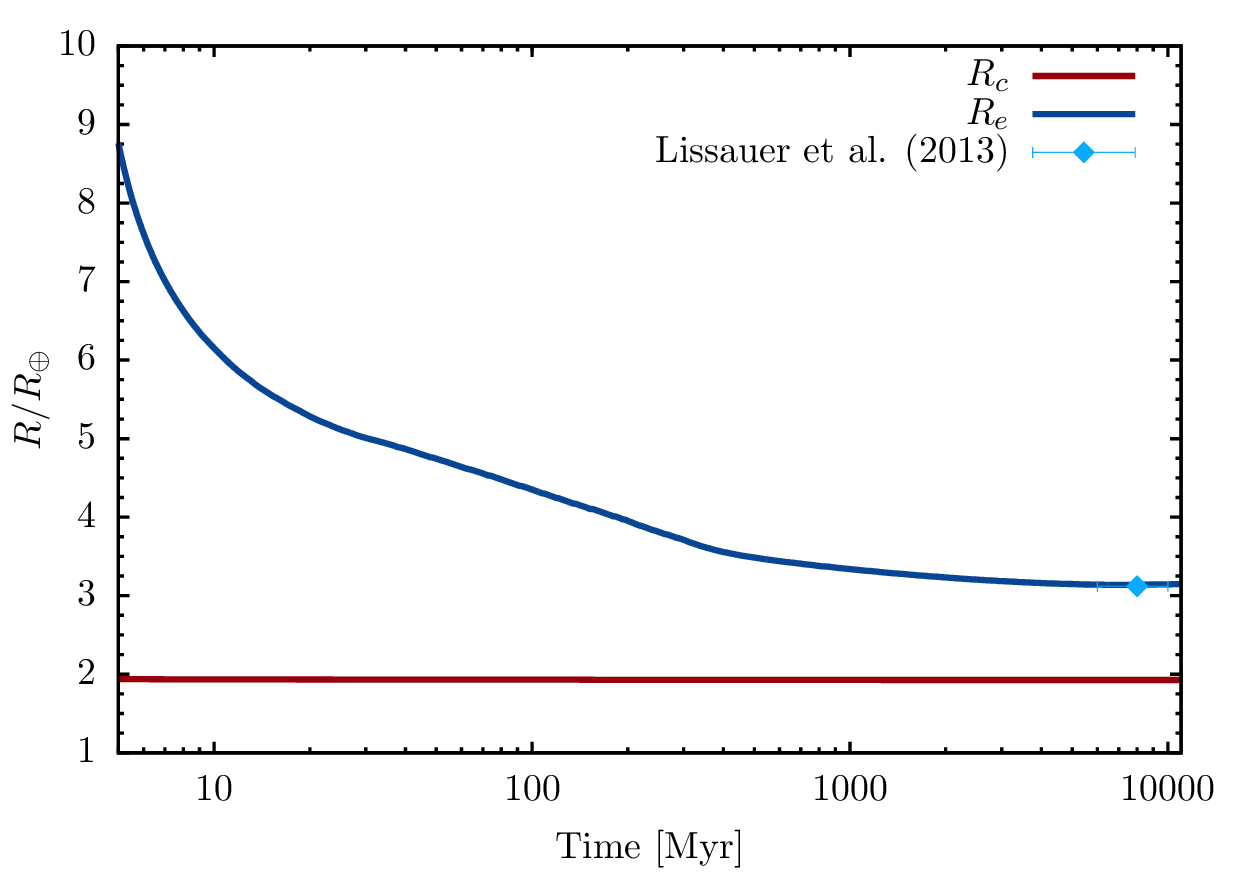}%
\includegraphics[clip]{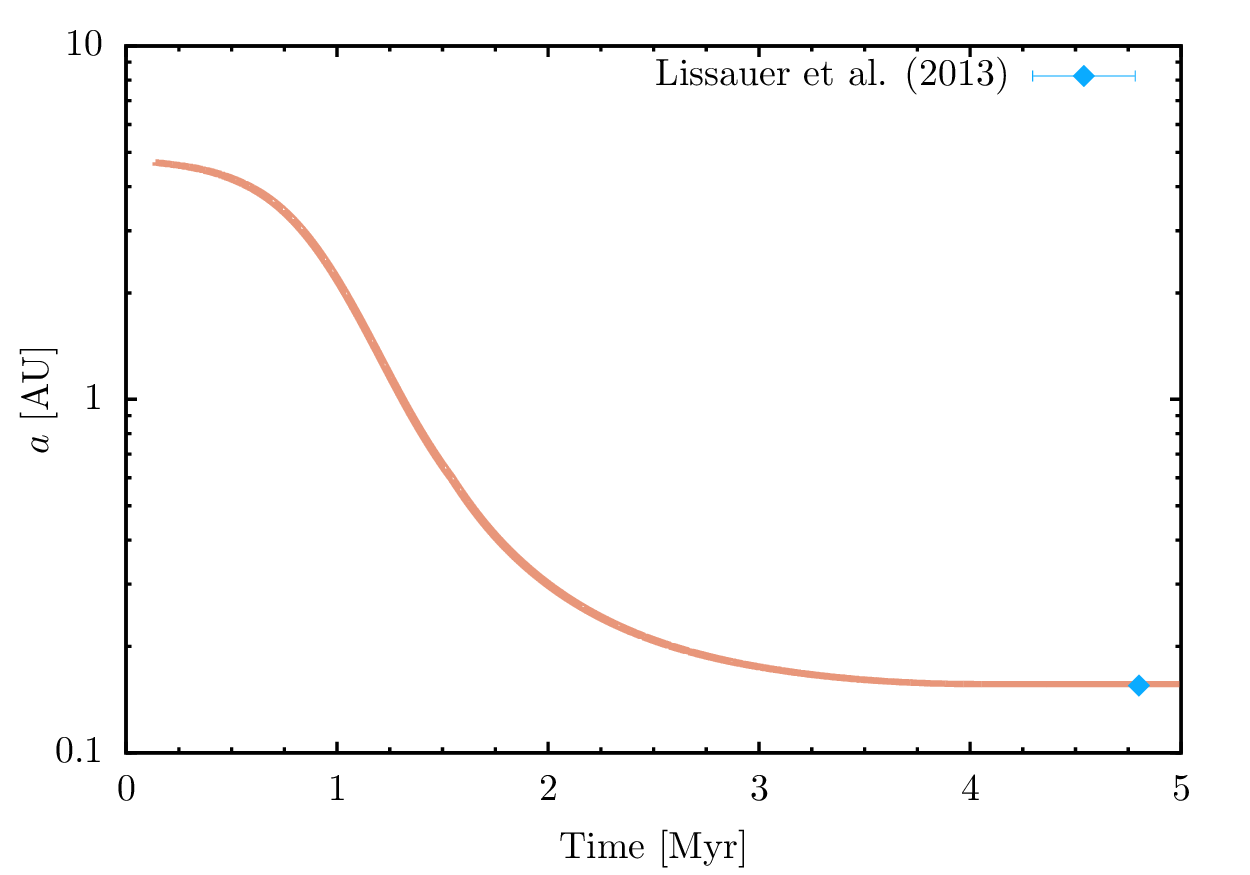}}
\caption{%
             Results from ex situ models of \kep b (top), \kep c, and \kep d (bottom). 
             Plots show, as a function of time, core, envelope, and total mass (left), 
             core and envelope radius (center; $R_{e}=R_{p}$), and orbital radius 
             (right). Disk's gas inside $\approx 1\,\AU$ dissipates within a little over 
             $4\,\Myr$ (see Table~\ref{table:sumex_iso}). 
              For \kep b, the plot of the radius evolution includes the reference case
             with an efficiency for mass loss due to photo-evaporation of
             $\varepsilon=0.1$ (solid line), and the cases with 
             $\varepsilon=0.05$ (dashed line) and $0.01$ (dotted line).
             All models start from a $0.1\,\Mearth$ planetary embryo, at time \ts.
             The vertical error bars indicate the results from \citetalias{lissauer2013}.
             To account for uncertainty, the age of the star is assumed to be 
             $8\pm2\,\mathrm{Gyr}$ \citep[\href{http://exoplanetarchive.ipac.caltech.edu/index.html}{Nasa    Exoplanet Archive}, and \href{http://exoplanets.eu/}{The Extrasolar Planets Encyclopaedia}]{lissauer2011a}.
             }
\label{fig:ow1_sd}
\end{figure*}

\begin{deluxetable*}{cccccccccc}
\tablecolumns{9}
\tablewidth{0pc}
\tablecaption{Properties of Ex Situ Formation Models of \kep\ Planets at Isolation\tablenotemark{a}\label{table:sumex_iso}}
\tablehead{
\colhead{Planet}&\colhead{\tiso [\Myr]}&\colhead{$M_{c}/\Mearth$}&\colhead{$M_{e}/\Mearth$}&\colhead{$R_{c}/\Rearth$}&
\colhead{$R_{p}/\Rearth$}&\colhead{(Fe,Si,\ice,\hhe)\%\tablenotemark{b}}&\colhead{$T_{\mathrm{eq}}$\tablenotemark{c} [\K]}&\colhead{$\langle\dot{M}_{e}\rangle$\tablenotemark{d} [$\Mearth\,\yr^{-1}$]}
}
\startdata
b & $4.11$ &$2.10$ & $0.007$ & $1.47$ & $\pz3.2$ &$(10.6,50.4,38.6,0.4)$ & $819$ & $-3.8\times 10^{-10}$\\
c & $4.08$ &$4.56$ & $0.235$ & $1.84$ & $\pz8.0$ &$(6.2,44.3,44.6,4.9)$ & $749$ & $-5.8\times 10^{-10}$\\
d & $4.06$ &$5.58$ & $0.395$ & $1.94$ & $12.7$ &$(5.6,42.9,44.9,6.6)$ & $627$ & $-4.0\times 10^{-10}$\\
e & $4.04$ &$6.90$ & $1.399$ & $2.02$ & $14.9$ &$(4.8,38.0,40.4,16.8)$ & $563$ & $-4.7\times 10^{-10}$\\
f & $4.09$ &$2.74$ & $0.098$ & $1.62$ & $\pz6.8$ &$(5.9,44.5,46.1,3.5)$ & $492$ & $-1.1\times 10^{-10}$\\
g & $4.04$ &$5.57$ & $0.701$ & $1.94$ & $\pz9.8$ &$(4.5,40.0,44.3,11.2)$ & $361$ & $-1.0\times 10^{-10}$
\enddata
\tablenotetext{a}{The isolation time, \tiso, is the time at which the disk's gas at $r\lesssim a$ disperses.}
\tablenotetext{b}{Percentage of the planet mass at time $t=\tiso$.}
\tablenotetext{c}{Equilibrium temperature of the planet, \cieq{eq:Teq}, at $t=\tiso$.}
\tablenotetext{d}{Rate of change of the planet's envelope mass averaged over the first $100\,\Myr$ 
                           of evolution in isolation. For \kep b, $\langle\dot{M}_{e}\rangle$ is an average over 
                           $10\,\Myr$.}
\end{deluxetable*}

\begin{figure*}[t!]
\centering%
\resizebox{\linewidth}{!}{%
\includegraphics[clip]{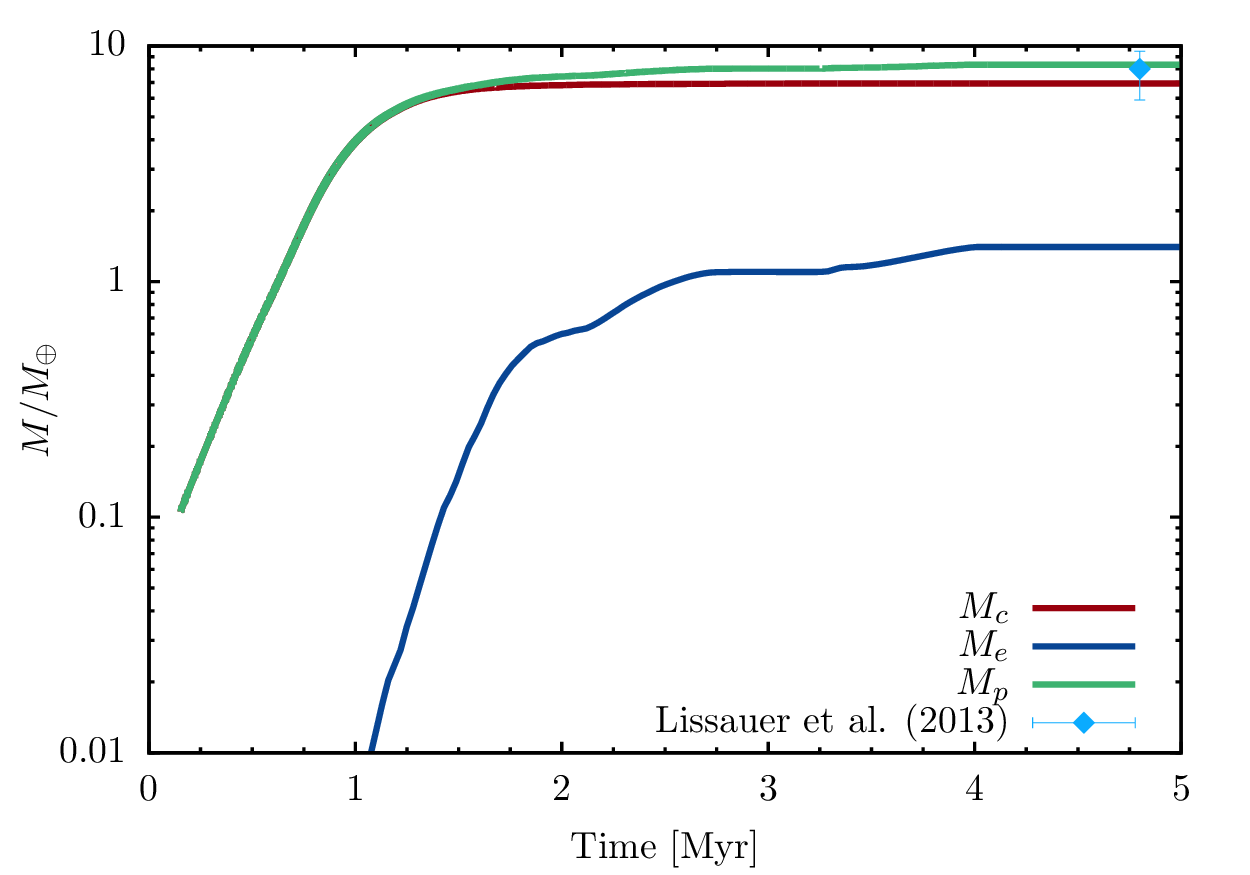}%
\includegraphics[clip]{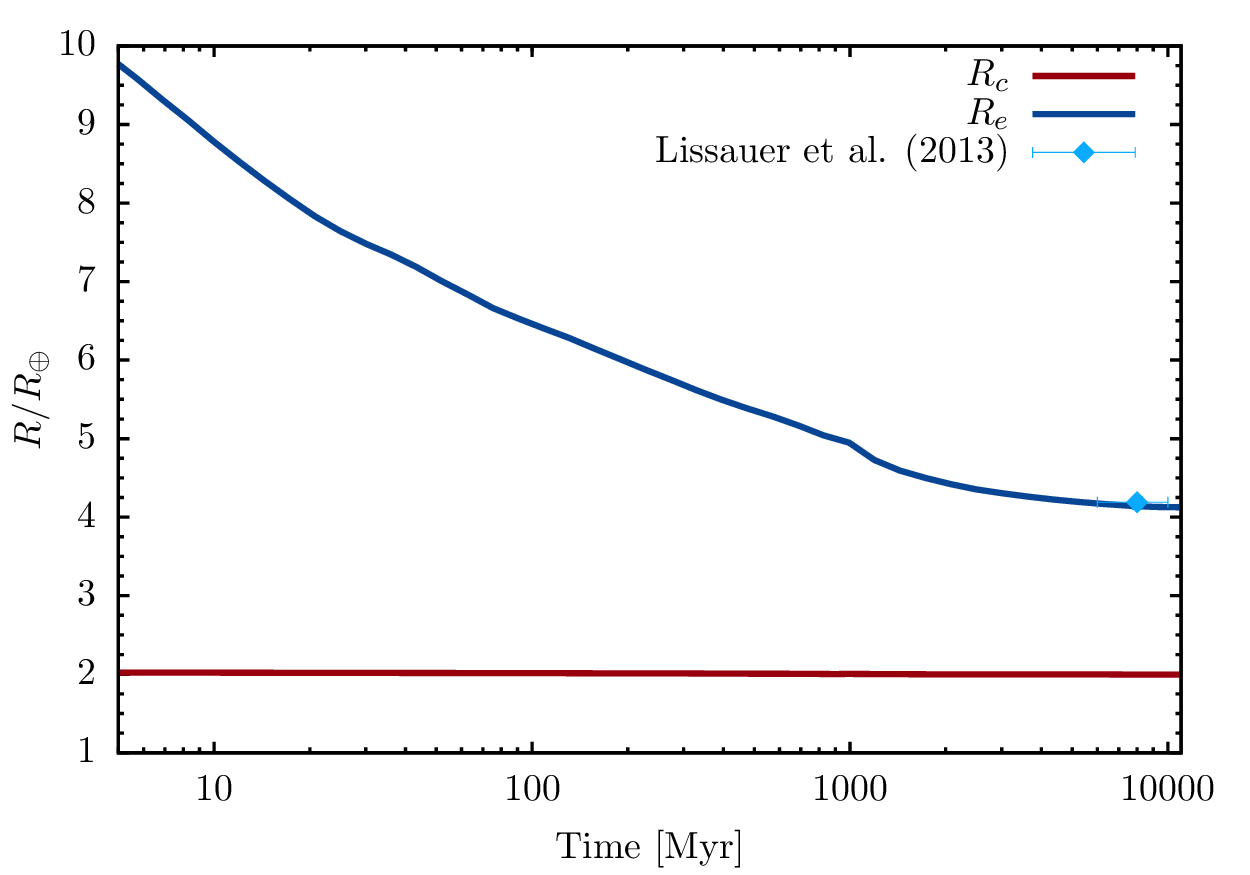}%
\includegraphics[clip]{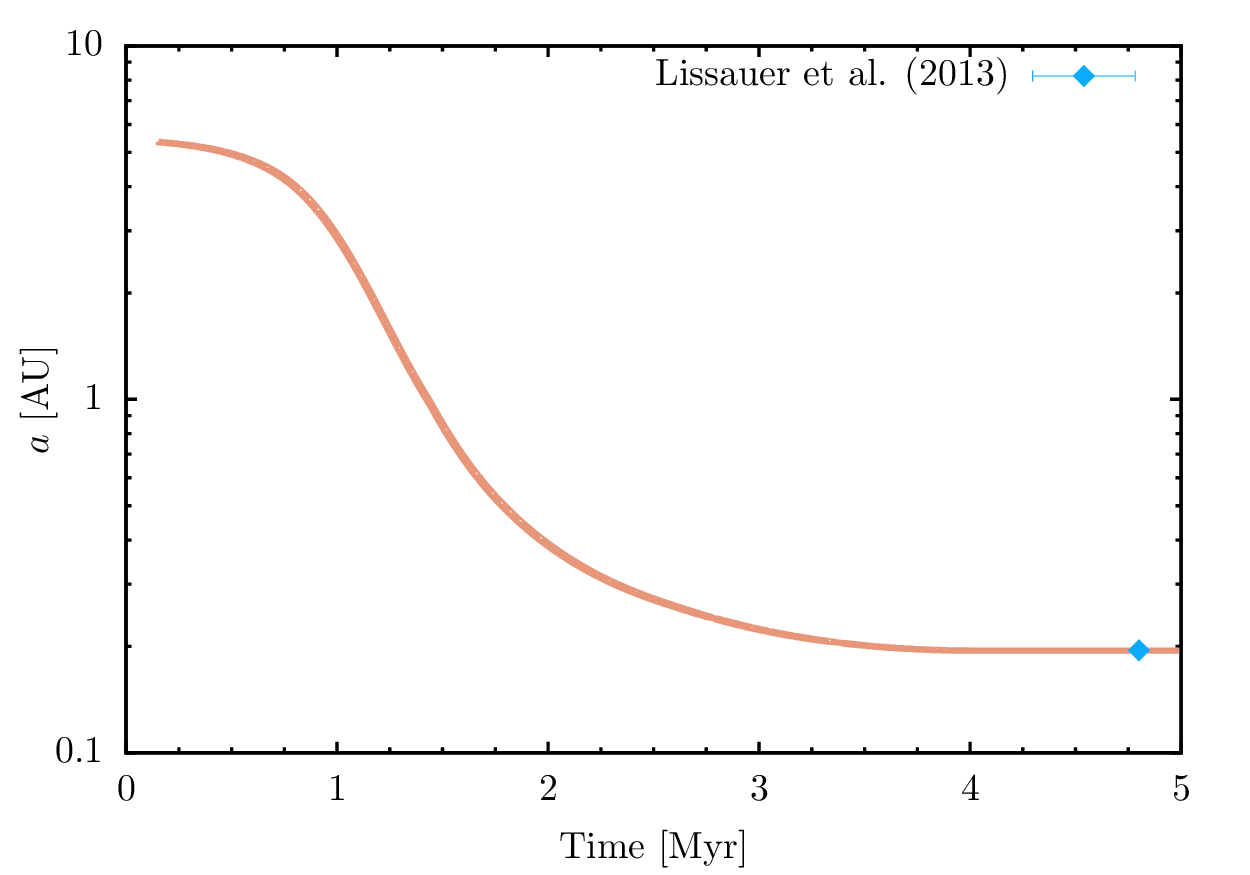}}
\resizebox{\linewidth}{!}{%
\includegraphics[clip]{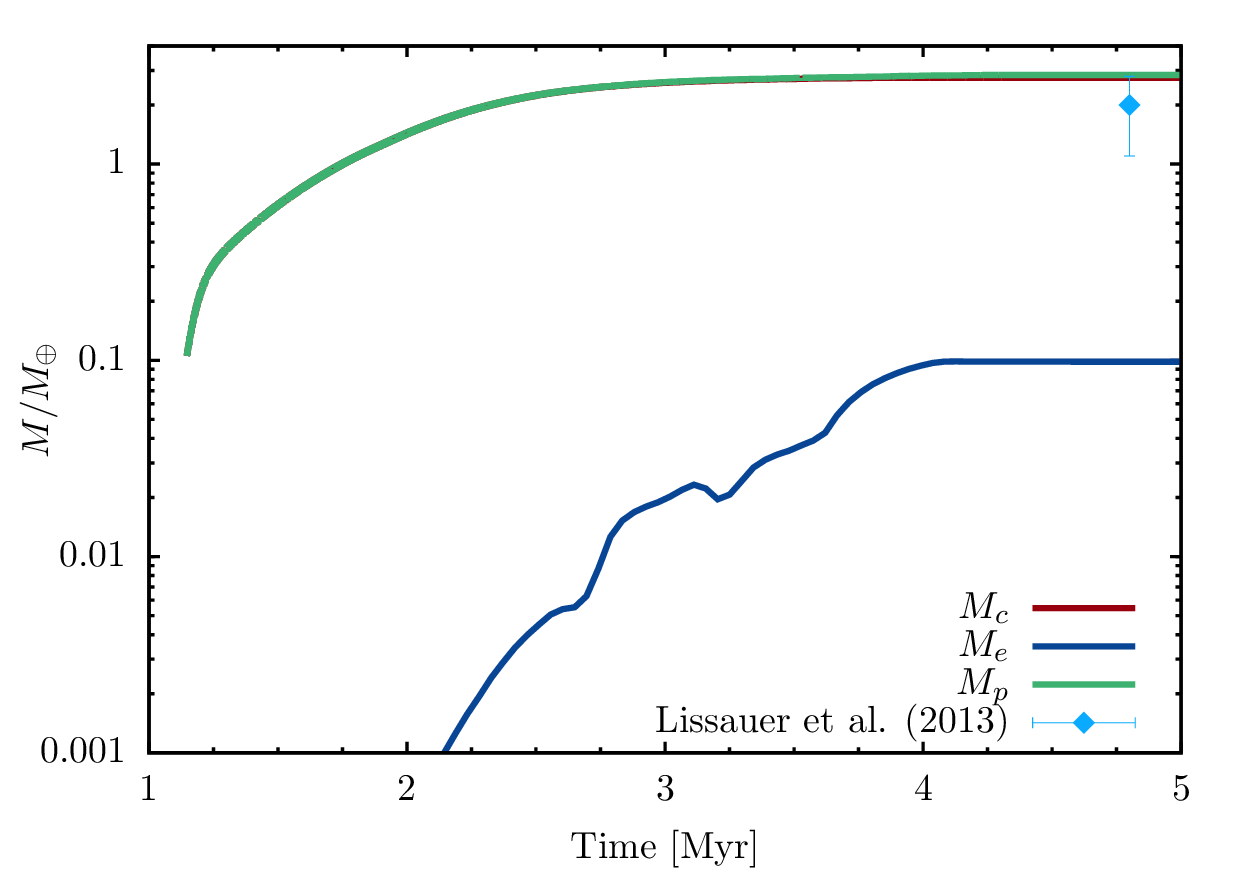}%
\includegraphics[clip]{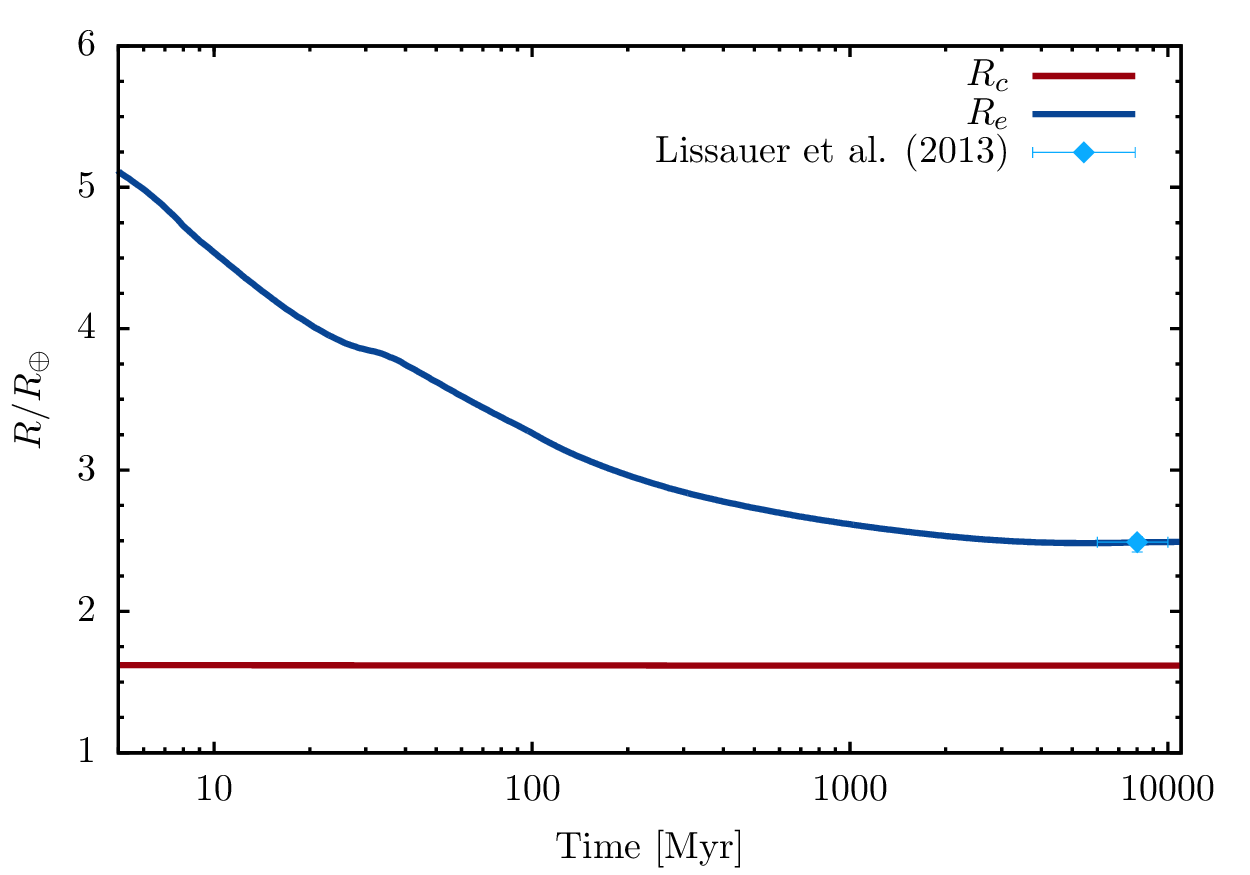}%
\includegraphics[clip]{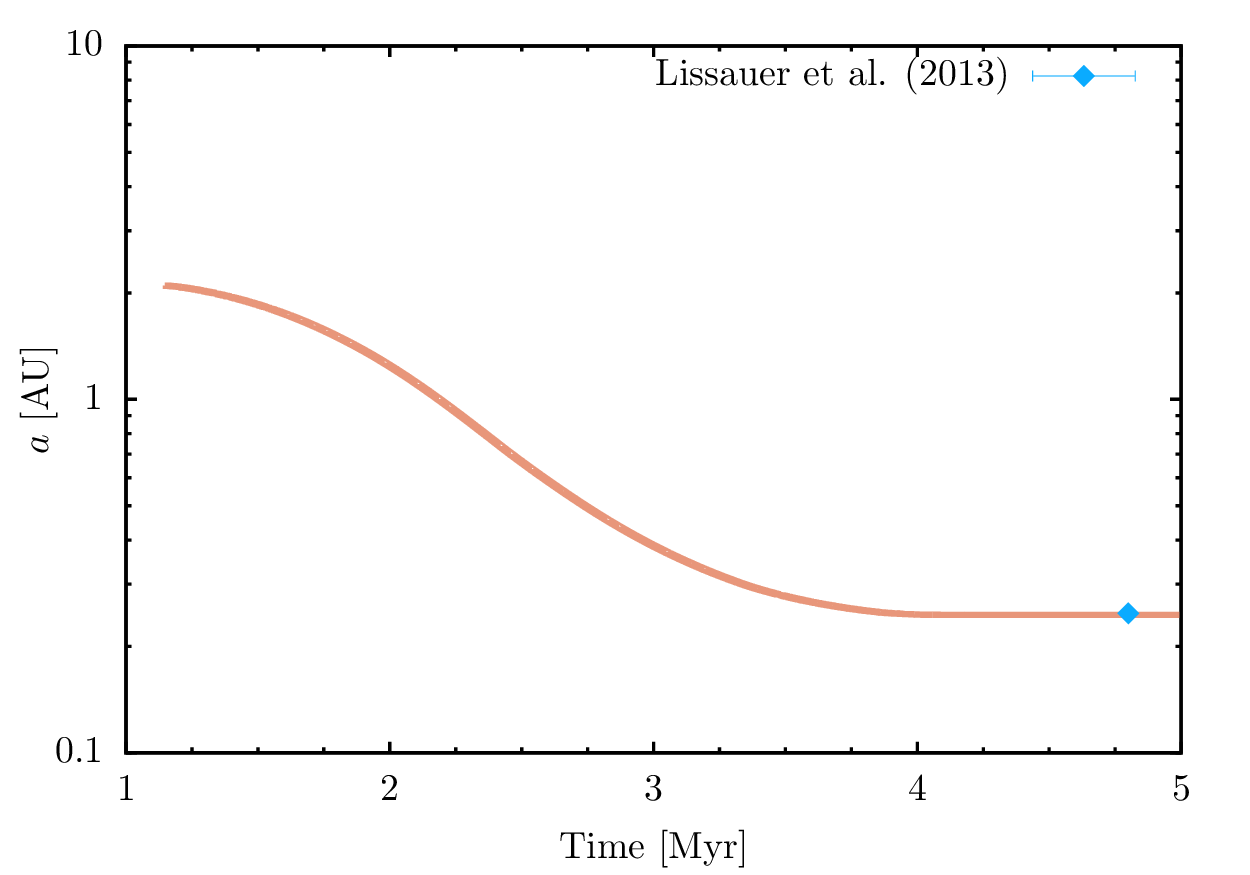}}
\resizebox{\linewidth}{!}{%
\includegraphics[clip]{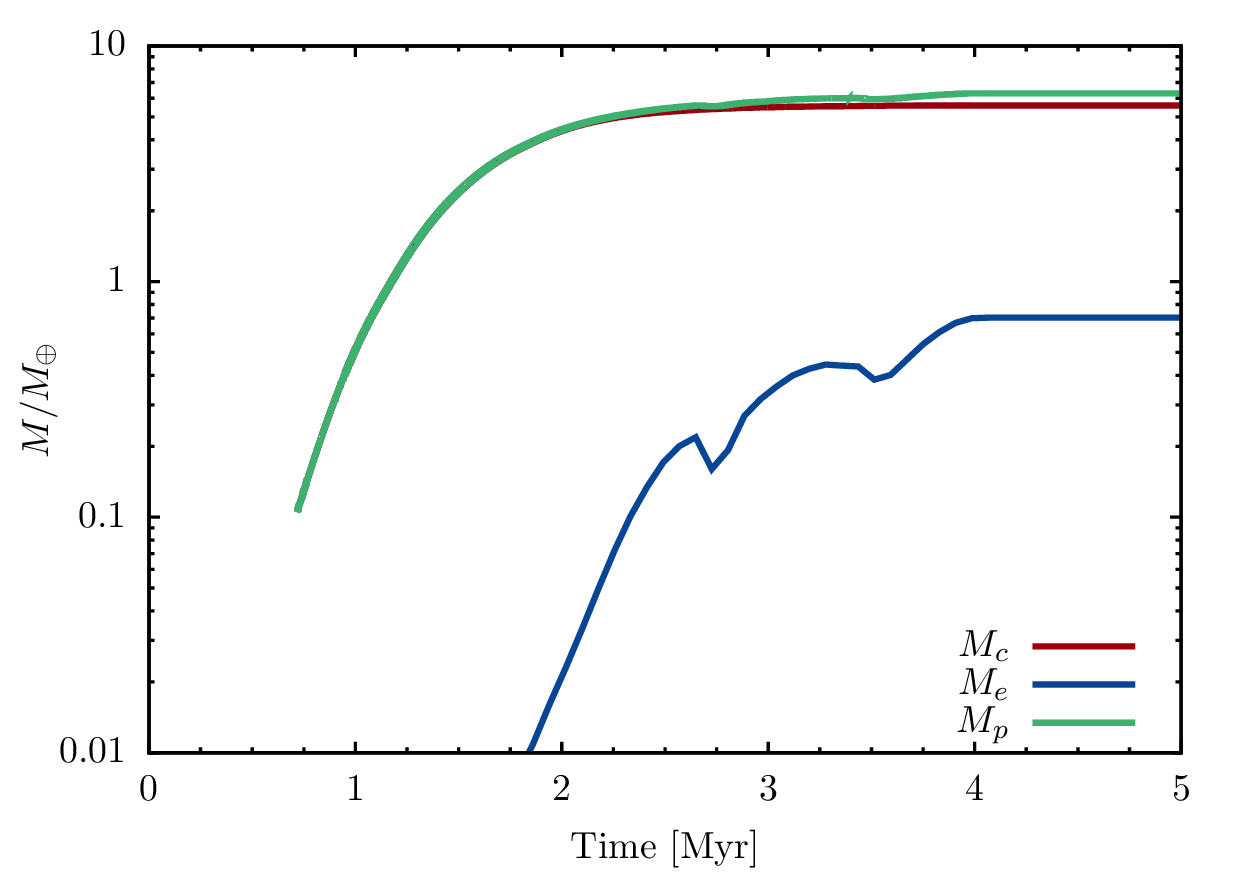}%
\includegraphics[clip]{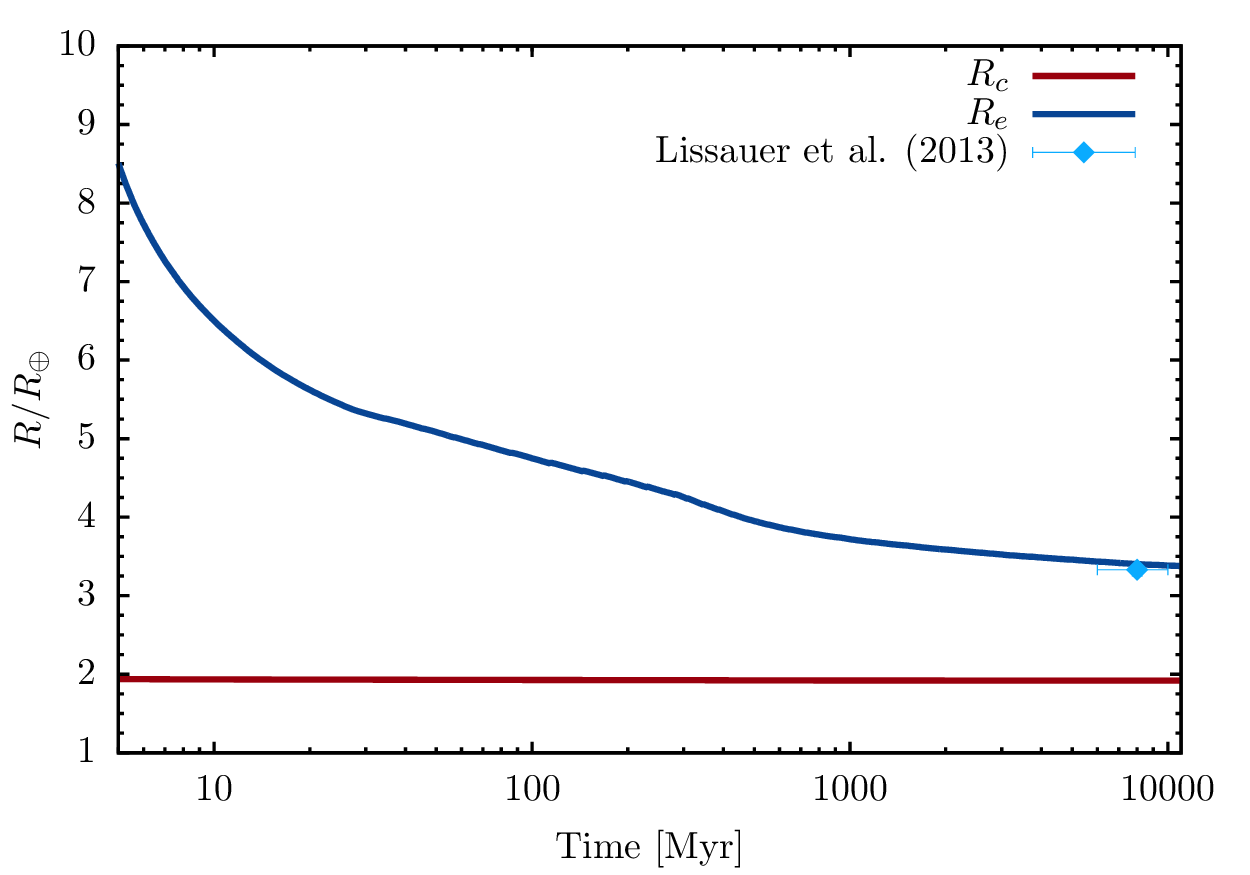}%
\includegraphics[clip]{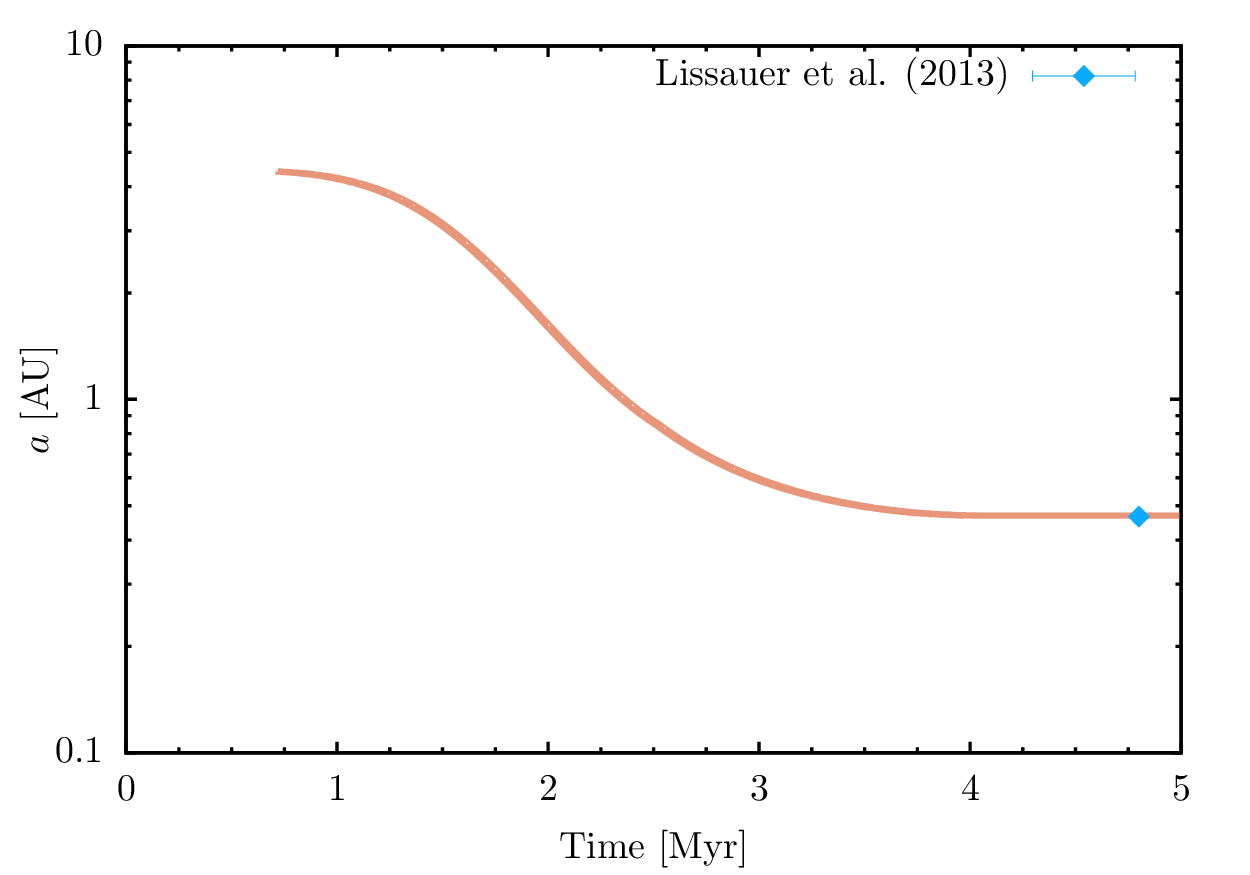}}
\caption{%
             Same as in Figure~\ref{fig:ow1_sd}, but for ex situ models of \kep e (top), 
             \kep f, and \kep g (bottom). The simulated evolution of \kep f begins
             slightly inside \kep b's initial orbit, but at a much later time. \kep g 
             starts in between the initial orbits of \kep c and d, but is delayed 
             until these planets are inside $3.5\,\AU$. 
             }
\label{fig:ow2_sd}
\end{figure*}

The energy output of the central star can impact both
the evolution of the disk and that of the planet. The two stellar
models considered here (see \cisec{sec:esc}) show similar
luminosities for $t\gtrsim 10\,\Myr$ (see \cifig{fig:star}), implying 
similar evolution of the planets in isolation. However, there are
differences at earlier times, specifically in the effective temperature
$T_{\star}$ and radius $R_{\star}$, which enter the irradiation 
temperature in \cieq{eq:Tirr}. Therefore, the disk thermal budget 
may be affected and hence the planet migration history may differ 
somewhat (see \cisec{sec:om}). The temperature at the planet
surface may also change.
These differences are assessed for the cases of \kep b and c
(see \cisec{sec:exsitu_res}).

Among the main simplifications of this study are the neglect of 
planet-planet interactions and the fact that only a single planet
evolves in the disk, though the models account for 
the depletion of the planetesimal disk generated by planets
that have already initiated the formation process. 
It is assumed that by the time $t=\ts$, a solid core of mass 
$M_{c}=0.1\,\Mearth$ (corresponding to 
$M_{e}\approx 10^{-7}\,\Mearth$, calculated consistently with 
$M_{c}$ and the disk boundary conditions) has formed at 
a current orbital radius of $r=a_{i}$. 
There is no speculation about its previous accretion/migration 
history and $a_{i}$ is considered to be the initial orbital radius of 
the simulated planet.
(However, the small initial planet mass implies that orbital migration via 
disk-planet tidal interactions may be negligible at times $t<\ts$).
Both \ts\ and $a_{i}$ are free parameters, constrained by the 
requirement, among others, that no two orbital paths intersect each other. 
Strictly speaking, this is not a physical requirement \citep[e.g.,][]{hands2014}
but rather a necessity dictated by the absence of gravitational 
interactions among planets. 
The time \ts\ ranges from $\approx 6\times 10^{4}\,\yr$ (\kep b) 
to $\approx 10^{6}\,\yr$ (\kep f).
Clearly, the time \ts\ is also determined by the choice of using
the same initial core mass for all planets. The orbital radius
$a_{i}$ ranges from $\approx 2.1\,\AU$ for \kep f to $5.35\,\AU$ 
for \kep e.

Table~\ref{table:sumex} summarizes the final properties (and $a_{i}$) 
of the six simulated \kep\ planets, assumed to have formed ex situ.
These are referred to as reference models.
The initial orbital radius of each planet, $a_{i}$, and the epoch
\ts\ are found by trial and error so that the final model provides 
reasonable matches to $a$, $R_{p}$, and $\Mp$ at an age 
of $8\,\Gyr$, as reported by \citetalias{lissauer2013}. Additionally, 
as mentioned above, orbital paths must not intersect. 
Plots of various quantities versus time from the resulting models
are illustrated in Figures~\ref{fig:ow1_sd} and \ref{fig:ow2_sd}. 
Since all planets start well beyond $1\,\AU$, contrary to in situ 
models, the local density of solids at $a_{i}$ and \ts\ is moderate
to low: $\sigma_{Z}\approx 9\,\densu$ for \kep b, 
$\approx 6\,\densu$ for \kep f, and between $\approx 3$ and 
$\approx 4\,\densu$ for the other planets (in ascending order of $a_{i}$).

Models for each planet were constructed in ascending order of 
final orbital radius, $a_{f}$.
At time $t=0$, the gas-to-solid mass ratio is set to about $70$ 
beyond the ice condensation line at around $3\,\AU$ ($T<150\,\K$). 
Planetesimals are anhydrous interior to $\approx 1\,\AU$ 
($T>250\,\K$), where the gas-to-solid mass ratio becomes 
approximately $140$ (see \cisec{sec:csc} for details).
As the disk evolves, the ice sublimation line moves inward to 
$r\approx 1\,\AU$ by $t\approx 1\,\Myr$ (see \cifig{fig:devo_b}).
The model for \kep b is constructed from this initial surface density
of solids, $\sigma_{Z}(t=0)$.
The distribution $\sigma_{Z}$ is depleted by the passage of \kep b. 
Indicating with $a^{c}_{i}$ the initial and $a^{c}_{f}$ the final
(i.e., observed) orbital radii of \kep c, the distribution $\sigma_{Z}$ 
for modeling this planet is determined by taking the depleted mass 
in solids between $a^{c}_{f}$ and $\approx 1.25\,a^{c}_{i}$, and 
redistributing the mass over the region according to a $1/\sqrt{r}$ 
power law. The same procedure is used to determine
$\sigma_{Z}$ for the construction of models for \kep d and e 
(each based on the depleted reservoir of solids left by 
the preceding planet). 
For the models of \kep f and g, which start inside the orbits of
preceding planets at significantly later times, the depleted mass 
in solids is redistributed interior to $\approx 1.3\,a^{e}_{i}$ 
(down to their observed orbital radii).

Typically, models make a transition from disk to photospheric 
boundary conditions (see \cisec{sec:esc}) at the isolation time, \tiso.
Although in some models gas accretion can be disk-limited during 
late stages of formation, as for in situ models a proper phase of rapid 
gas accretion (i.e., $R_{p}$ significantly smaller than $\Rcapt$) 
is never reached during the formation phase. 
Accretion of solids could in principle (and does on occasion) continue 
beyond the isolation time, until the feeding zone is emptied 
(which requires $da/dt\approx 0$).
However,  since orbital migration becomes very slow much earlier 
than $\tiso$ (see Figures~\ref{fig:ow1_sd} and \ref{fig:ow2_sd}), 
$M_{c}$ plateaus well before isolation is achieved, as can be seen 
in the left panels of Figures~\ref{fig:ow1_sd} and \ref{fig:ow2_sd}.
For all practical purposes, a planet is isolated from both the disk's 
gas and solids at $t>\tiso$.
Some properties of the reference models of \kep\ planets at $t=\tiso$ 
are listed in Table~\ref{table:sumex_iso}.
The small scatter in isolation times is likely caused by the removal 
of gas via accretion on the planet (when $a>r_{\mathrm{crt}}$), 
which tends to lower the accretion rate through the disk for 
$r\lesssim a$ \citep[][see also \cieq{eq:brigap}]{lubow2006} 
and thus operates in concert with disk photo-evaporation 
to augment gas depletion inside $r\approx a$.
In fact, the time $\tiso$ is shorter for planets with larger
envelope masses, i.e., with larger $\langle\dot{M}_{e}\rangle$.
However, the contribution of $\dot{M}_{e}$ to 
$\dot{\Sigma}_{\mathrm{pe}}$ is quite marginal in these calculations.
In the more realistic situation in which all planets migrated in the disk,
the time $\tiso$ would be set by the largest planet, \kep e. 
However,  since the formation phases of all planets are basically complete 
by that time, no significant consequences would be anticipated.

The relative gas content is largest for \kep e, accounting for 
$\approx 17$\% of the total mass at $\tiso\approx 4\,\Myr$, and 
somewhat less at $8\,\mathrm{Gyr}$ (see Table~\ref{table:sumex}).
The light elements (\hhe) in the other planets make $\lesssim 10$\%
of the total mass at the isolation time and, in most cases, only a few
to several percent at $8\,\mathrm{Gyr}$. 
At this age, however, the gaseous envelope always accounts for 
$\approx 35$\% to $\approx 50$\% of the planet radius.
For all planets, the condensible mass fraction of \ice\ is $\gtrsim 39$\%, 
indicating that their cores mostly form behind the ice condensation line.
Although the composition of the initial core is dictated by the local 
disk composition of the solids at $t=\ts$, its mass ($0.1\,\Mearth$) 
is small enough to not affect the final core composition much.
\kep b contains the smallest mass fraction of \ice\ and the largest
mass fractions of silicates and iron, due to its small initial orbital radius
($\approx 2\,\AU$) and early start time, $\ts\approx 6\times 10^{4}\,\yr$.
Nonetheless, also in this case the substantial fraction of \ice\ 
($\approx 39$\% by mass) implies that the planet accumulates 
its condensible inventory mostly behind the ice condensation front.
Despite an equally small starting orbit, the core composition of 
\kep f is instead more similar to that of neighboring planets because 
of its late start time ($\ts\approx 1.1\,\Myr$) and growth in a colder
disk environment.

\begin{figure}
\centering%
\resizebox{\linewidth}{!}{%
\includegraphics[clip]{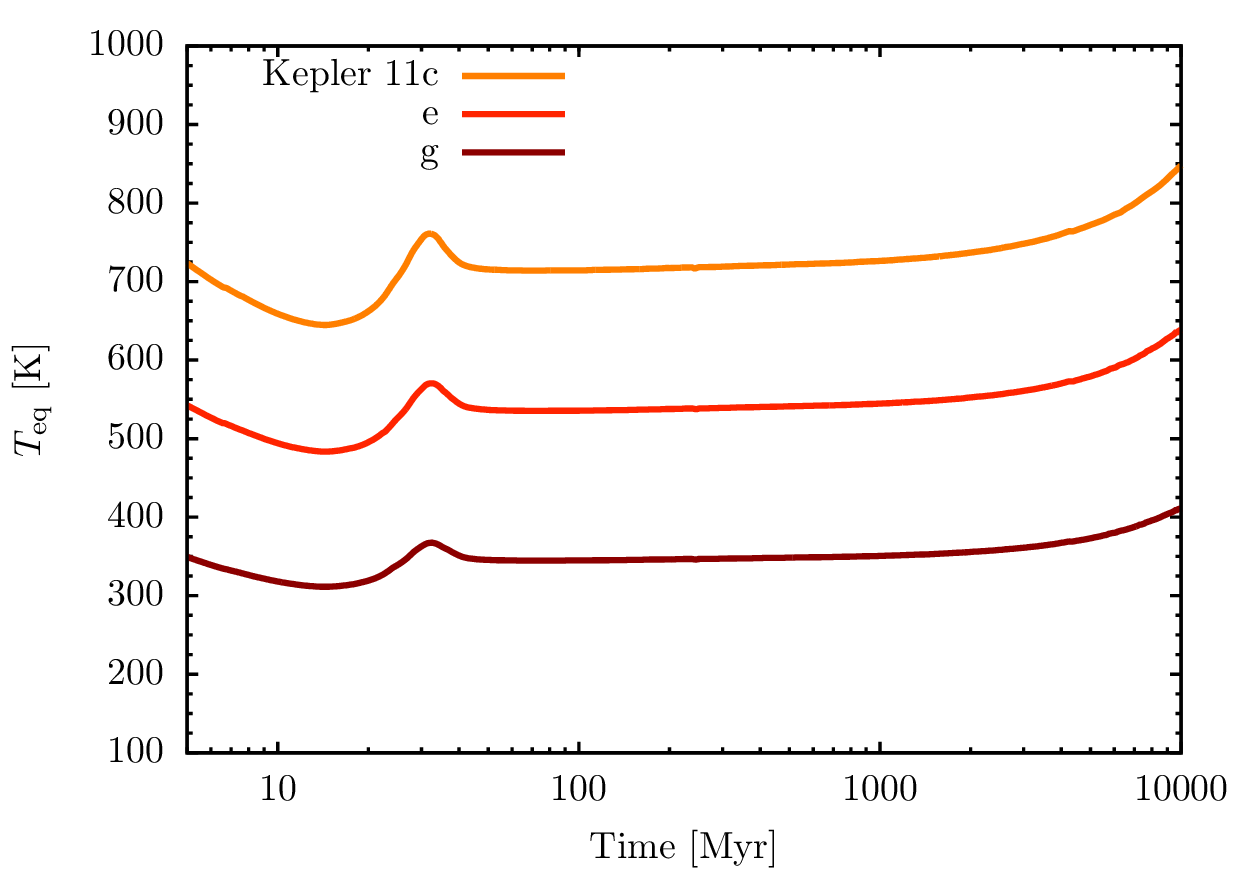}}
\caption{%
             Equilibrium temperature in the radiation field of the star,
             \cieq{eq:Teq}, from the stellar model of
             \citet{siess2000} in \cifig{fig:star}. The temperature
             is plotted at the final orbital radius of planets \kep c, e,
             and g (see Table~\ref{table:sumex}), as indicated.
             }
\label{fig:Teq}
\end{figure}
Except for the varying stellar properties (see \cifig{fig:star}), 
the isolation evolution of ex situ models behaves as that 
of in situ models.
The surface temperature of the planet closely follows the equilibrium 
temperature, $T_{\mathrm{eq}}$, which is shown in \cifig{fig:Teq}
for the isolation evolution of \kep c, e, and g.

As for in situ models, the planet radius $R_{p}$ decreases on 
a short timescale once the planet becomes isolated. Afterwards,
$R_{p}$ steadily declines as the planet cools. The radius of planets 
\kep c and d reaches a minimum at an age between $7$ and 
$7.5\,\mathrm{Gyr}$, after which the envelope begins to slowly expand, 
following the rise of $T_{\mathrm{eq}}$ (see \cifig{fig:Teq}).
However, the expansion is very modest and by an age of 
$12\,\mathrm{Gyr}$, $R_{p}$ increases over its minimum value by 
$< 1$\%. \kep f follows a similar trend, achieving a minimum
radius around the age of $7.5\,\mathrm{Gyr}$ and then inflating
slightly ($R_{p}$ changing by $<1$\%).
Possibly due to their more massive envelopes and hence
larger internal energy, the simulated planets \kep e and g 
still contract at an age of $\approx 11\,\mathrm{Gyr}$.

\begin{figure}
\centering%
\resizebox{\linewidth}{!}{%
\includegraphics[clip]{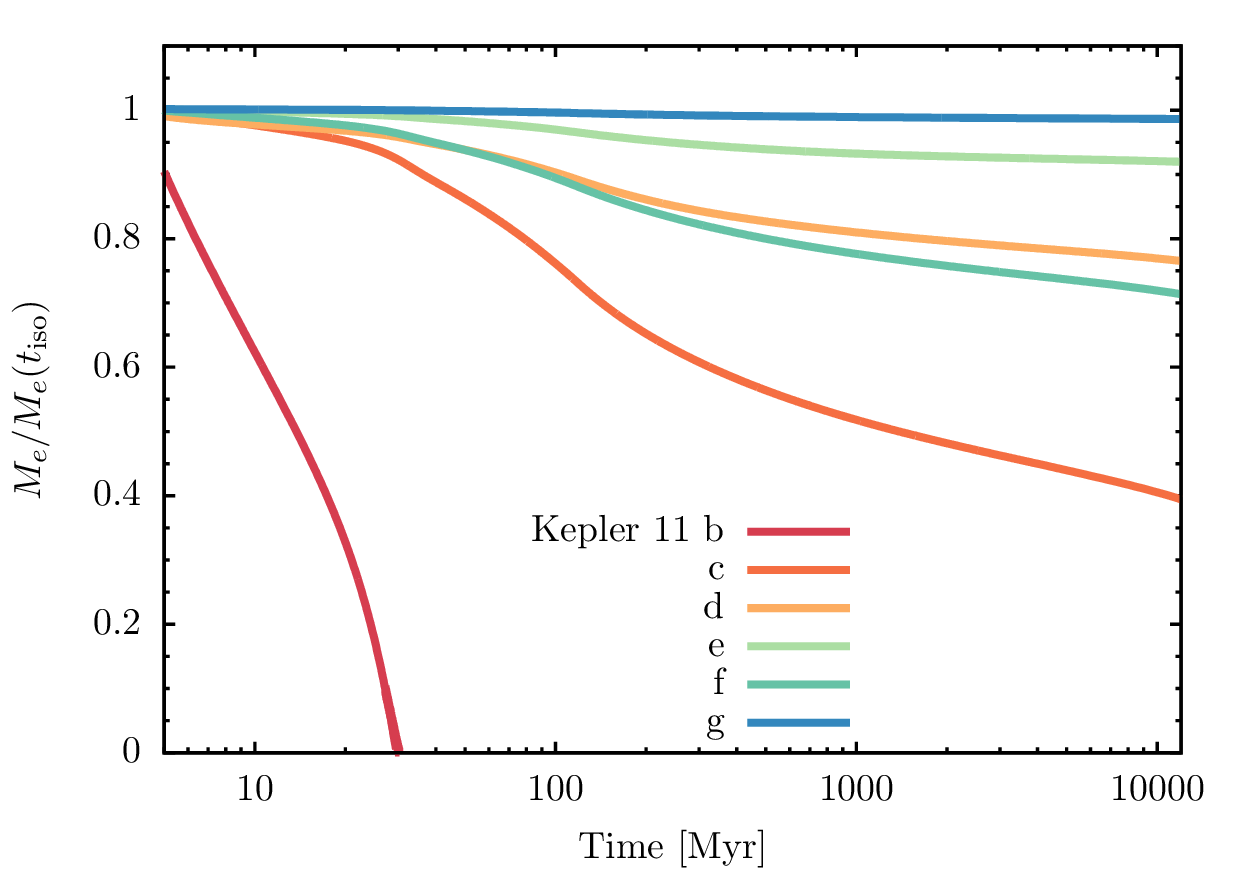}}
\caption{%
             Normalized envelope mass, $M_{e}$, as a function of time during 
             the phase of evaporative mass loss, after the planets become
             isolated from the disk. 
             The envelope (\hhe) mass is normalized to 
             $M_{e}$ at $t=\tiso$ (see Table~\ref{table:sumex_iso}).
             By $t=5\,\Myr$, the simulated \kep b planet has already lost about 
             $10$\% of its \hhe\ mass at $t=\tiso$.
             }
\label{fig:mem5}
\end{figure}

\begin{figure*}[]
\centering%
\resizebox{0.85\linewidth}{!}{%
\includegraphics[clip]{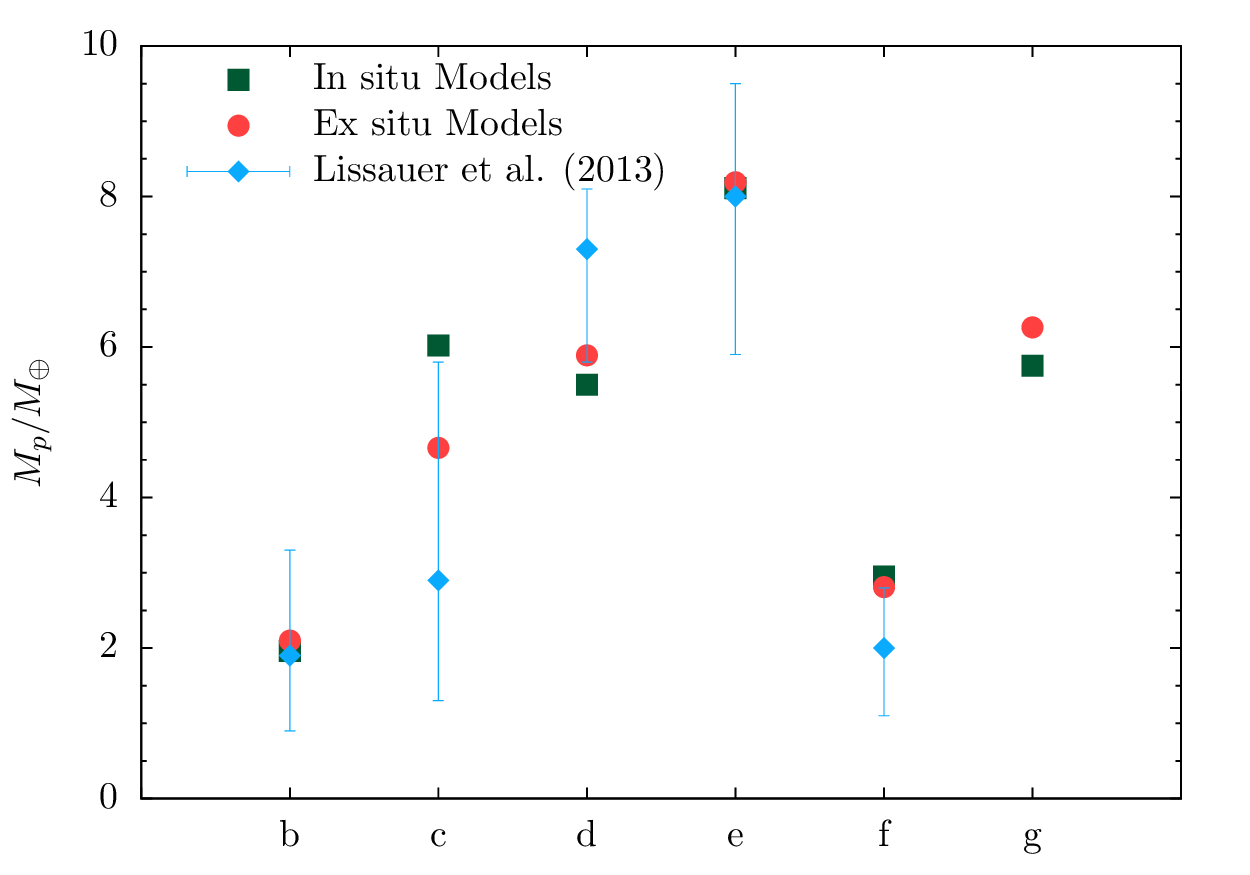}%
\includegraphics[clip]{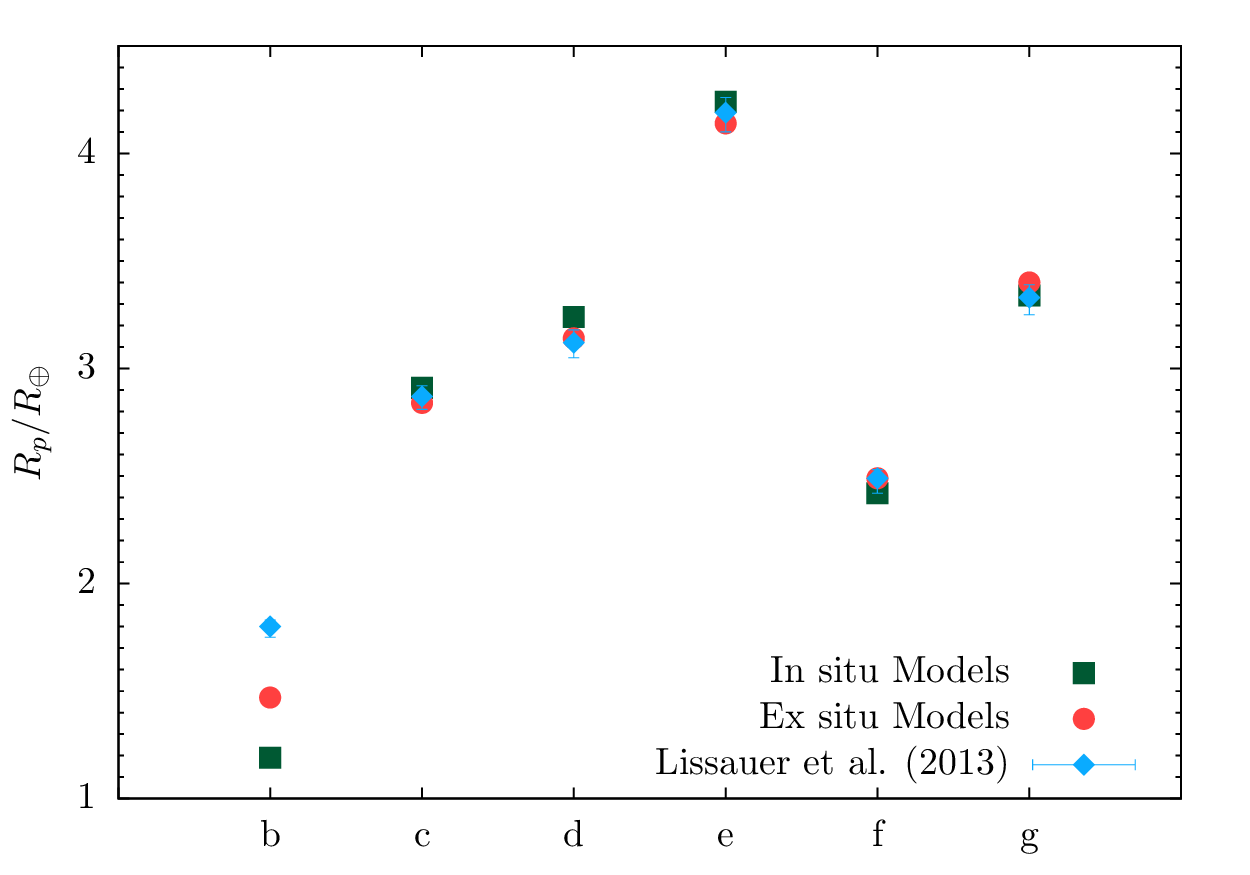}}
\caption{%
             Comparison between results from simulated planets and data from 
             \citetalias{lissauer2013}: planet mass (left) and planet radius (right).
             The mass of \kep g is only loosely constrained by transit observations 
             ($\Mp<25\,\Mearth$).
             The final orbital distances from ex situ models are in very close agreement
             with measurements (compare entries in Tables~\ref{table:sumin} and
             \ref{table:sumex}) and are not plotted.
             }
\label{fig:comp}
\end{figure*}
The evaporative loss of envelope gas, caused by the absorption of stellar 
X-ray and EUV photons and given by \cieq{eq:mexuv}, starts at 
$t=\tiso\approx 4\,\Myr$, i.e., as soon as the gaseous disk interior to 
the planets' orbit is cleared and the isolation phase begins.
As for in situ models, a standard value for the efficiency parameter, 
$\varepsilon=0.1$, is applied (but see also \cisec{sec:exsitu_res}). 
The resulting envelope masses versus time are illustrated in 
\cifig{fig:mem5}. 
The rate $|\dot{M}_{e}|$ is especially large at early ages, 
$t\lesssim 100\,\Myr$, because both $R_{p}$ and the flux 
$F_{\mathrm{XUV}}$ (see \cisec{sec:gl}) are large.                    
Values of the gas loss rates, averaged over the first $100\,\Myr$ 
($10\,\Myr$ for \kep b) of evolution, are listed in 
Table~\ref{table:sumex_iso} and range from $10^{-10}$ to around 
$6\times 10^{-10}\,\Mearth\,\yr^{-1}$, i.e., between
$\approx 10^{10}$ and $\approx 10^{11}\,\mathrm{g\,s}^{-1}$.
Mass loss is significant for the inner planets \kep b, c, and d. 
By $8\,\Gyr$ these planets lose, respectively, $100$\%, $\approx 57$\%, 
and $\approx 22$\% of their $M_{e}$ at $t=\tiso$. Envelope loss is also 
substantial for \kep f, which loses nearly $29$\% of its gaseous mass 
during the isolation phase.
In relative terms, \kep e and g are more immune to evaporative mass 
loss, as $M_{e}$ reduces by about $8$\% and $2$\%, respectively, 
from $\approx 4\,\Myr$ to $8\,\Gyr$.
Of the \hhe\ mass removed during isolation, at least $38$--$40$\% 
evaporates within $\approx 100\,\Myr$.
As planets contract, $R_{p}/\Rhill\ll 1$ hence $K(\xi)\approx 1$ in
\cieq{eq:mexuv}, and \hhe\ gas is stripped from the envelope at 
a rate $\dot{M}^{\mathrm{iso}}_{e}\propto R^{3}_{p}/(\Mp a^{2})$.
After $8\,\Gyr$, mass-loss rates become quite small for all planets,
ranging from around $10^{-13}$ (for \kep f and g) to 
$\approx 10^{-12}\,\Mearth\,\yr^{-1}$ (for \kep c and d).

Comparing the evolution of $R_{p}$ in Figures~\ref{fig:pw1} and 
\ref{fig:ow1_sd} after isolation, one can see that planets formed 
in situ remain somewhat more inflated than do planets formed ex situ. 
The difference is especially large for the case of \kep b, probably due 
to the fact that the planet formed in situ acquires a more massive 
\hhe\ envelope (because of the rapid core growth). 
Differences tend to vanish at later times.
This behavior accounts for the differences in the average rates 
$\langle\dot{M}_{e}\rangle$ in Tables~\ref{table:sumin_iso} and 
\ref{table:sumex_iso}. Although there is a factor of $10$ difference
in the gas loss rates between in situ and ex situ models of \kep b, 
in this case, the difference is immaterial since both simulated 
planets lose their entire envelopes within $30$--$40\,\Myr$. 

\subsection{Results for Individual Planets}
\label{sec:exsitu_res}

The main results from our ex situ reference models can be 
summarized as follows.

\textbf{\kep b}.
at an initial orbital distance of $2.14\,\AU$ ($\ts\approx 6.3\times 10^{4}\,\yr$) 
and a surface density of solids $\sigma_{Z}\approx 9\,\densu$, 
according to \cieq{eq:mciso} a non-migrating core would achieve 
a final mass of $\approx 0.7\,\Mearth$, and smaller if depletion via 
scattering (\cieq{eq:core_scat}) was taken into account. 
About $90$\% of the core mass is accreted from solids orbiting beyond
$0.5\,\AU$ from the star, hence the presence of large amounts of \ice\
in the planet. The envelope mass is maximum around $2.8\,\Myr$, 
before part of it becomes unbound and is released back to the disk.
After the planet becomes isolated, the remaining  \hhe\ gas is
removed by stellar X-ray and EUV radiation within $\approx 30\,\Myr$.
Reducing the efficiency of the evaporative mass-loss rate, $\varepsilon$
in \cieq{eq:mexuv}, allows the planet to retain an atmosphere for
somewhat longer. However, even a value as low as $\varepsilon=0.01$ 
predicts a complete removal of the primordial \hhe\ gas within 
a few times $100\,\Myr$ (see \cifig{fig:ow1_sd}). 
Despite the presence of abundant \ice\ in the core ($\approx 40$\% 
by mass), $R_{c}$ is still significantly smaller than the observed radius.
Although not included, this model naturally accounts for the formation 
of a steam atmosphere that was proposed to reconcile simulated and
observed radii \citep{lopez2012}.
Both final orbital radius and planet mass agree with measured values
(see \cifig{fig:comp}).

\textbf{\kep c}.
the planet starts at roughly $4\,\AU$ ($\ts\approx 1.3\times 10^{5}\,\yr$)
and grows about $90$\% of its condensible final mass at $r\gtrsim 0.8\,\AU$. 
Nearly $46$\% of the planet's final mass is in \ice, the second largest 
(after \kep f) relative fraction of all simulated planets (although \kep d, 
e, and g contain more \ice\ in absolute measure).
When the local disk disperses and the planet becomes isolated, 
the envelope includes about $5$\% of $\Mp$. Roughly
$90$\% of this \hhe\ mass is accreted from the disk's gas at
$170\lesssim T\lesssim 460\,\K$. 
Over the course of the isolation phase, stellar radiation removes 
more than half of the envelope mass, leaving only $\approx 2$\% 
of $\Mp$ in primordial \hhe\ gas at $8\,\Gyr$, the smallest relative
fraction in all simulated planets that retain an envelope.
The evolution of $R_{p}$ in \cifig{fig:ow1_sd} shows a period, between 
$t\approx 20$ and $\approx 40\,\Myr$, in which the planet contraction 
slows down. This feature is likely associated with the rise in 
$T_{\mathrm{eq}}$ at that age (see \cifig{fig:Teq}), following the 
brightening of the star (\cifig{fig:star}). A similar feature
appears in the radius evolution of \kep b and f planets, which 
can more promptly respond to changes in the external incident 
flux having (with \kep c) the least massive envelopes.
The final planet mass and radius agree with measurements, whereas 
the final orbital distance is just above the one-standard-deviation 
upper limit of the measured value ($0.108\,\AU$).

\textbf{\kep d}.
the starting orbital radius is less than $1\,\AU$ larger than
that of \kep c and the start time is comparable 
($\ts\approx 1.4\times 10^{5}\,\yr$). Therefore, 
similarly to its inner neighbor, the planet accumulates 
$\approx 90$\% of $M_{c}$ outside of $\approx 0.9\,\AU$. 
The maximum value of $M_{e}$ is attained shortly prior to $2.5\,\Myr$,
but afterwards some envelope gas becomes unbound and returns 
to the disk. For $\approx 1\,\Myr$ prior to isolation, $R_{p}$ remains 
very close to the accretion radius $\Rcapt$, preventing further  
accretion of gas.
During its evolution in isolation, the envelope loses somewhat
less than $0.1\,\Mearth$, about a fifth of $M_{e}$ at $t=\tiso$.
Comparing the core radii in Tables~\ref{table:sumex} and 
\ref{table:sumex_iso}, a small difference can be noticed.
The core masses at $t=\tiso$ and at $8\,\Gyr$ are virtually
identical, yet the pressure applied at the top of the core 
at the later epoch is nearly twice as large 
($\approx 9\,\mathrm{GPa}$ vs.\ $\approx 4.8\,\mathrm{GPa}$)
because of the cooler temperature,
which in this case accounts for the $85\,\mathrm{km}$ reduction 
in $R_{c}$. 
The values of the planet's mass and radius and the orbital 
distance at $8\,\Gyr$ are all within measurement errors.

\textbf{\kep e}.
the most massive of the six, in terms of both core and
envelope mass, this planet also has the farthest initial orbit at 
$5.35\,\AU$ ($\ts\approx 1.5\times 10^{5}\,\yr$) from the star. 
The initial local density of solids is $\approx 3\,\densu$.
Neglecting scattering, a non-migrating planet at that distance would 
achieve a final core mass of around $2\,\Mearth$ before emptying 
its feeding zone.
The planet grows $90$\% of its core mass beyond $\approx 1\,\AU$,
but most of its \hhe\ inventory is accreted inside this radius.
Essentially, the planet's core fully forms behind the ice condensation 
line ($\approx 49$\% of $M_{c}$ is \ice). 
The average mass-loss rate at the beginning of the isolation phase
($-\langle\dot{M}_{e}\rangle$, see Table~\ref{table:sumex_iso})
is higher than that of \kep d despite the larger mass and orbital radius 
of \kep e. The reason is the strong dependence of 
$\dot{M}^{\mathrm{iso}}_{e}$ in \cieq{eq:mexuv} on planet radius. 
In the case of \kep d, $R_{p}$ drops below $6\,\Rearth$ by $\approx 10\,\Myr$, 
whereas the radius of \kep e remains $>6\,\Rearth$ well after $100\,\Myr$
(see \cifig{fig:ow2_sd}). In absolute terms, the planet loses the second
largest amount of primordial \hhe\ during the isolation phase
($0.11\,\Mearth$).
The difference in $R_{c}$ between $t=\tiso$ and $t=8\,\Gyr$ is again 
caused by the pressure difference at the bottom of the envelope.
The final values of $a$, $\Mp$, and $R_{p}$ all agree with measurements.

\textbf{\kep f}.
the observed mass of the planet is significantly smaller than those 
of its neighbors, possibly suggesting a formation at a smaller orbital 
distance. To achieve a correspondingly smaller core mass with our 
$\sigma_{Z}$, the planet's initial orbit is interior to the initial orbits 
of the other planets. 
Consequently, the planet requires a late start, $\ts\approx1.1\,\Myr$, 
to avoid crossing other orbital paths\footnote{
This solution is unlikely unique, and an earlier start may be possible 
together with a wider initial orbit and a lower $\sigma_{Z}$. In this 
case, however, the slower growth of $M_{c}$ would entail larger 
gas densities to account for the required amount of orbital migration.}.
The assembly of the core takes place for the most part beyond 
$\approx 0.6\,\AU$ and at disk temperatures of $T\lesssim 160\,\K$. 
As a result, the core composition is very similar to that of fully 
hydrated planetesimals ($50$\% \ice\ and $45$\% silicates by mass).
Though rich in \ice\ in relative terms, $47$\% of $\Mp$ (the richest, 
in fact), because of its small mass the planet contains an amount of 
\ice\ ($1.3\,\Mearth$) greater only than the \ice\ mass of \kep b.
During the evolution in isolation, stellar radiation strips off about 
$29$\% of the \hhe\ gas accreted during the formation phase.
At $8\,\Gyr$, the planet is left with a gas content of $2.5$\% 
by mass.
The final planet radius and orbital distance agree with observations,
whereas the final total mass, $2.81\,\Mearth$, is close to 
the one-standard-deviation upper limit of the measured 
value  ($2.8\,\Mearth$).

\textbf{\kep g}.
similarly to its inner neighbor, the planet starts its simulated 
evolution at an advanced stage of the disk's life 
($\ts\approx 7\times 10^{5}\,\yr$) in between the initial orbits 
of \kep c and d. 
Roughly $90$\% of its condensible inventory is collected beyond 
$\approx 1\,\AU$, and all of the core is accumulated behind the 
ice condensation front. In fact, its formation occurs at disk 
temperatures below $\approx 170\,\K$ and its core has virtually 
the same composition as that of fully hydrated planetesimals.
Partially hydrated planetesimals account for only $0.03$\% of 
the core mass. 
Mass loss during isolation is negligible, and the planet basically
contracts at a constant mass.
Again, the core shrinks somewhat as the envelope cools down
and the applied pressure at $R_{c}$ increases.
The planet radius at $8\,\Gyr$ is just above the one-standard-deviation 
upper limit of the measured value ($3.39\,\Rearth$), whereas $a_{f}$ 
agrees with measurements.
The planet mass is not really constrained by transit observations, but
the value provided by this ex situ model is consistent with that from 
the in situ calculations in Table~\ref{table:sumin}. Both
simulations point at a planet somewhat less massive than \kep e
and of comparable mass to that of \kep d.

\subsection{Effects of Changes in Model Assumptions}
\label{sec:effects}

\begin{figure}
\centering%
\resizebox{\linewidth}{!}{%
\includegraphics[clip]{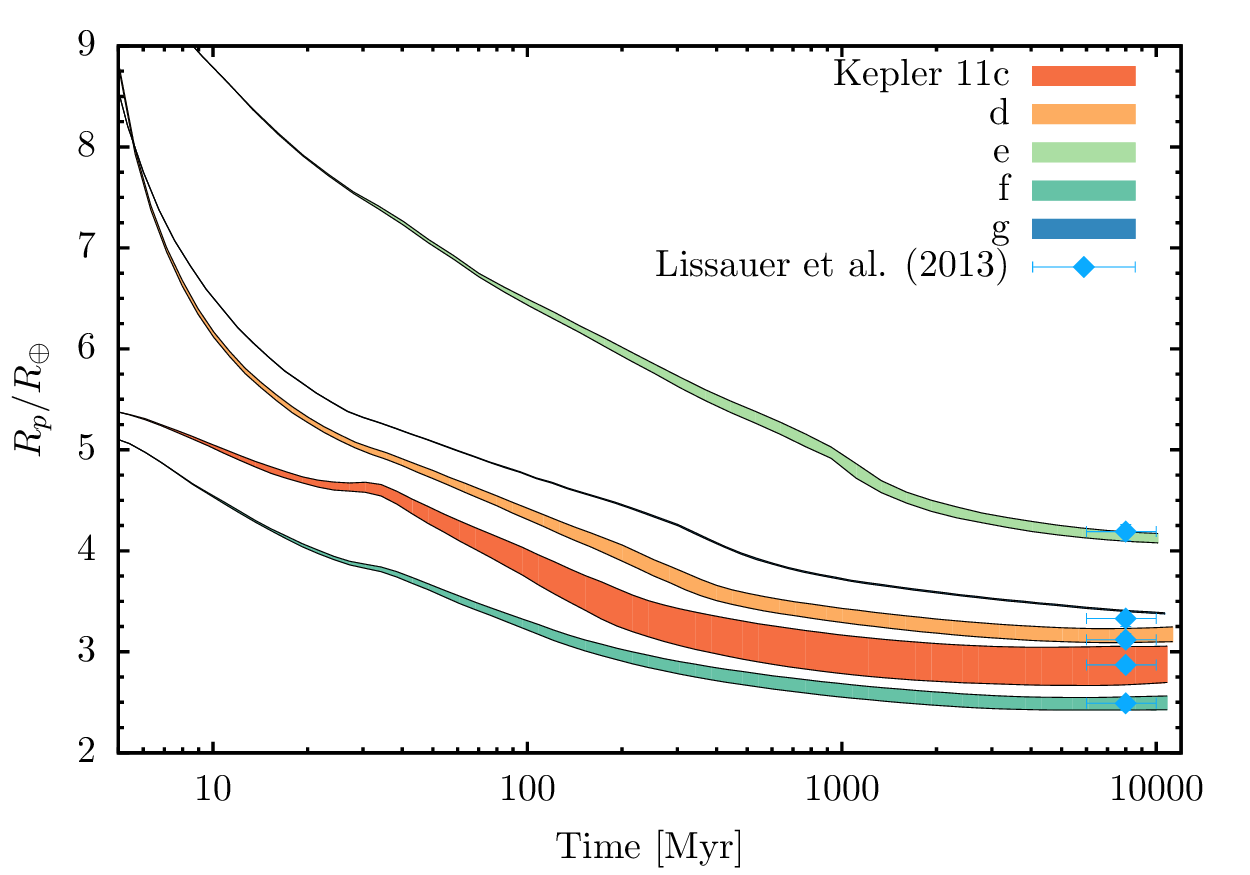}}
\caption{%
             Evolution of the radius of the simulated \kep\ planets during 
             the isolation phase for a range of evaporative mass-loss rates,
             $\dot{M}^{\mathrm{iso}}_{e}$. The efficiency parameter for the
             absorption of energetic stellar photons in \cieq{eq:mexuv} is,
             respectively, $\varepsilon=0.15$ (lower solid curves of pairs) and 
             $0.05$ (upper solid curves of pairs).
             }
\label{fig:rvse}
\end{figure}
The reference models use a standard value of $\varepsilon=0.1$
in the expression of $\dot{M}^{\mathrm{iso}}_{e}$,
to account for re-radiation of the EUV and X-ray stellar flux by
the envelope (see \cisec{sec:gl}).
Since this parameter is uncertain, the evolution in isolation 
was repeated by applying values of $\varepsilon=0.05$ and $0.15$. 
Results for the evolution of $R_{p}$ are plotted in \cifig{fig:rvse}. 
The case of \kep b is omitted for obvious reasons.
The solid curves of each pair (upper curve for the smaller efficiency) 
bracket the excursion of $R_{p}$ (colored regions).
Within the considered range of $\varepsilon$, the impact on the
radius at $8\,\Gyr$ is typically small.
For \kep g and e, the relative excursion of $M_{e}$ amounts to 
about $1$\% and $9$\%, respectively, and the percentile variations 
of $R_{p}$ are $0.4$\% and $2$\%.
The excursion of $M_{e}$ is larger for \kep d ($26$\%) and larger 
still for \kep f ($\approx 40$\%), owing to its small mass.
The relative change in $R_{p}$, though, is $\lesssim 5$\% for 
either. In all these cases, the simulated planets would still
provide reasonable matches to the observed planets, with $\Mp$ 
and $R_{p}$ lying within or proximate to measurement ranges.
Due to its vulnerability to mass loss, \kep c represents the most 
extreme case, with $M_{e}$ changing by a factor of three and 
$R_{p}$ by $\approx 14$\%, which would place the planet 
radius well outside of the measurement range.

The reference models are based on time-dependent stellar 
properties for a $1\,\Msun$ ([Fe/H]$=0.0$) star computed 
by \citet{siess2000}.
In order to determine the impact of the stellar evolution model, 
which enters the calculations through Equations~(\ref{eq:Teq}), 
(\ref{eq:mexuv}), and (\ref{eq:Tirr}), a simulation of \kep b was 
performed with the Yonsei-Yale stellar model \citep[][]{spada2013} 
in \cifig{fig:star} (see the figure's caption and \cisec{sec:esc} 
for further details), starting from the same disk's initial conditions
as in the reference model. 
The three equations depend on $L_{\star}$
($T^{4}_{\star}R^{2}_{\star}\propto L_{\star}$, although \cieq{eq:Tirr}
separately depends on $R_{\star}$ as well). 
The largest differences in stellar luminosity between the two stellar 
models occur for $t\lesssim 20\,\Myr$, i.e., during the disk-embedded 
phase and the early isolation phase of the planet. 
The Yonsei-Yale stellar model predicts lower luminosities, which tend 
to produce a somewhat cooler disk in regions where stellar irradiation 
is important in the energy budget of the gas (\cieq{eq:EEq}). 
The cooler temperature in turn reduces the disk thickness and hence 
affects the migration rate, since $|da/dt|\propto (a/H)^{2}$ (see \cieq{eq:dadt}).
The cooler temperatures during the nebular stage of the planet's
evolution also change the Bondi radius \Rbondi, and hence the 
planet accretion radius $\Rcapt$ in \cieq{eq:RA}.
By using the same initial orbital radius, $a_{i}$, as in the reference 
model (see Table~\ref{table:sumex}), the somewhat larger (in magnitude) 
migration speed requires a later start time, $\ts \approx 1.4\times 10^{5}\,\yr$ 
(the difference in luminosity is especially large for 
$t\lesssim 3\times 10^{5}\,\yr$, see \cifig{fig:star}). The resulting 
model provides an isolation time of $\tiso=4.1\,\Myr$ and values at 
$t=\tiso$ in very good agreement with those in Table~\ref{table:sumex_iso}.
During the isolation phase, the surface temperature of the planet 
differs by $\lesssim 20\,\K$ relative to that of the reference model. 
The \hhe\ envelope is entirely removed by $t\approx 30\,\Myr$,
as in \cifig{fig:ow1_sd}.
Because of the cooler gas temperatures and later start, the core contains
a slightly larger \ice\ fraction ($\approx 43$\% vs.\ $\approx 39$\% by mass).

Since the isolation phase of \kep b does not last long, 
a calculation with the Yonsei-Yale stellar model was also performed 
for the isolation phase of \kep c. The properties of the model
at $\tiso$ are those of the reference model listed in 
Table~\ref{table:sumex_iso}.
For most of the isolation phase ($t\gtrsim 50\,\Myr$),
the stellar luminosity of the Yonsei-Yale stellar model is 
around $9$\% lower, compared to $L_{\star}$ of the
\citeauthor{siess2000} model, and thus $T_{\mathrm{eq}}$ is only
marginally different. The calculation results in an envelope mass 
at $8\,\mathrm{Gyr}$ of $0.11\,\Mearth$ and a planet radius of 
$2.88\,\Rearth$. 
Both numbers are a little larger than the values in Table~\ref{table:sumex},
but within the errors of the measured mass and radius of \kep c. 
Clearly, the stellar evolution model does not affect the simulated 
planets significantly.

Dust grains, either entrained in the accreted gas and/or produced
by ablation of accreted solids, can pollute planetary envelopes
at temperature $T\lesssim 1500\,\K$ (assuming silicate grains).
In such cases, dust opacity 
regulates the envelope cooling rate and hence the planet's 
contraction timescale \citep[e.g.,][]{pollack1996,hubickyj2005}. 
The calculations presented here are
based on the opacity plotted in the top panel of \cifig{fig:opa}, 
which assumes a maximum grain radius of $1\,\mathrm{mm}$.  
To evaluate the effect of grain opacity in the envelope of the 
reference models, calculations were also performed by using 
the table plotted in the bottom panel of \cifig{fig:opa}, which 
assumes a maximum grain radius of $10\,\mathrm{mm}$.
The ratio of the grain-dominated, Rosseland mean opacities
in the two tables is about eight.
The lower opacity provided by the size distribution with larger
dust grains is expected to facilitate cooling and allow 
for higher gas accretion rates. The planet \kep b, which 
accumulates the least massive \hhe\ envelope, was simulated 
with the lower grain opacity.
By using the same initial conditions for both disk and planet
as for the reference model, the calculation provides values
of $\dot{M}_{e}$ that are initially a few times as large as those
of the reference model. The envelope mass achieves a maximum 
value of $M_{e}\approx 0.035\,\Mearth$ at a time $t\approx 2.6\,\Myr$,
after which the envelope loses mass to the disk as $R_{p}$
tends to exceed the accretion radius $\Rcapt$.
The planet attains isolation at a time $\tiso=4.1\,\Myr$. 
Despite the larger envelope mass during the planet's 
early accretion history, the core mass at \tiso\ is again 
$M_{c}=2.1\,\Mearth$ and $M_{e}=0.013\,\Mearth$
(about twice as large as that of the reference model), 
which is entirely removed by stellar radiation by an age 
of $\approx 33\,\Myr$. 

Although the opacity test indicates little impact on the simulated
\kep b planet, larger effects are to be expected for planets that
accumulate higher \hhe\ mass fractions. Indeed, the same test
repeated for the most massive planet, \kep e, results in an
entirely different outcome.
The planet reaches a crossover mass ($M_{e}=M_{c}$) of 
$\approx 7.2\,\Mearth$ at $t\approx 2\,\Myr$ and enters the transition
stage (see \cisec{sec:esc}) of fast -- and disk-limited -- gas accretion 
by $t\approx 2.2\,\Myr$. On its track to becoming a Hot Jupiter, 
the planet reaches a mass of $\approx 0.7\,\Mjup$ by $t\approx 2.5\,\Myr$.
The tendency to evolve into a giant planet, with the lower dust opacity
of \cifig{fig:opa} (bottom panel), is also obtained from in situ calculations.
Ex situ models matching the observed radius of \kep e with the lower 
opacity require a smaller core mass and, hence, a tighter initial orbit 
($a_{i}\approx 4.8\,\AU$). 
A calculation resulting in $R_{p}=4.18\,\Rearth$ at $8\,\Gyr$ provides
a mass $\Mp=5.55\,\Mearth$, below the one-standard-deviation lower 
limit of the measured value of $5.9\,\Mearth$. 
A calculation resulting in $\Mp=5.96\,\Mearth$ produces a radius 
$R_{p}\approx 4.5\,\Rearth$, well over the one-standard-deviation 
upper limit of the measured value of $4.26\,\Rearth$.
Similarly, in situ formation models indicate that $R_{p}$ can be 
matched with a mass $M_{p}$ somewhat smaller than 
the measured one-standard-deviation lower limit.

Both in situ and ex situ models assume formation in a disk
of $100\,\mathrm{km}$-radius planetesimals. A reduction in 
planetesimal size from $\approx 100$ to $\sim 1\,\mathrm{km}$ 
was studied by the authors in other contexts. First, the orbital
eccentricities and inclinations of these bodies decrease, 
and second, the cross section for planetesimal capture 
in the planetary envelope ($\mathcal{S}_{\mathrm{eff}}$ in 
\cieq{eq:dotmc}) increases \citep[see][Figure~6]{gennaro2014}. 
Both effects enhance the accretion rate of solids, $\dot{M}_{c}$. 
Whether the planetesimal size has an effect on the accreted gas 
mass depends on the ratio of the timescale for the buildup of the 
core to the disk lifetime. The core buildup timescale of the in situ 
models is already very short, thus the effect is negligible. 
In the ex situ models, the core accretion time would decrease
(considerably in the case of small planetesimals). The faster core
growth would increase the migration rate. The planet would be
driven more quickly toward the inner disk regions, where solids' 
densities are higher but the solids' mass available for accretion 
is lower. Therefore, the outcome is difficult to predict. We performed 
tests for \kep d with planetesimals of $10$ and $1\,\mathrm{km}$ 
in radius. Compared to the reference case discussed above, the 
final core mass increased by only about $10$\% and $15$\%, 
respectively.

\begin{figure}
\centering%
\resizebox{\linewidth}{!}{%
\includegraphics[clip]{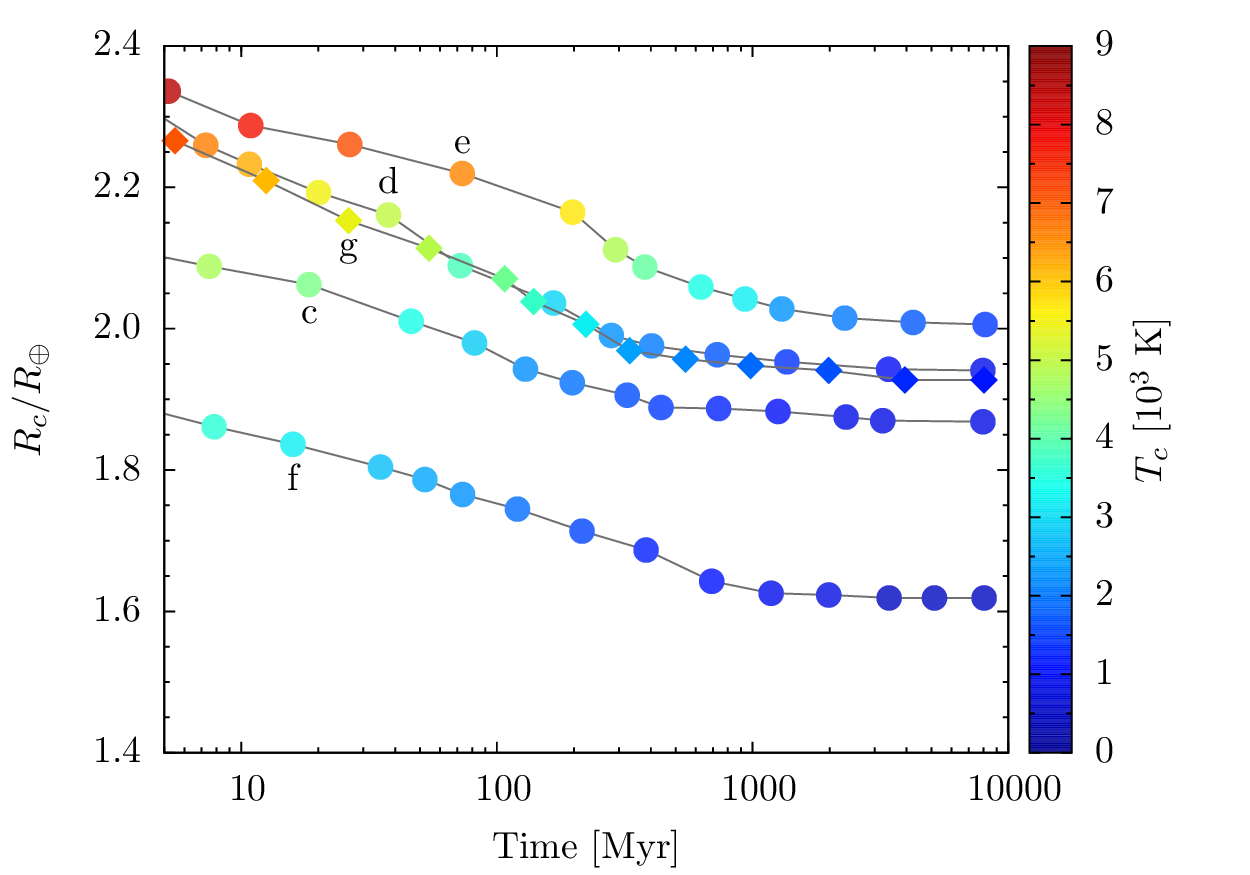}}
\caption{%
             Core radii of \kep\ planets, as indicated, calculated from 
             the thermal structures described in Appendix~\ref{sec:ics}. 
             The calculations apply pressure ($P_{c}$) and temperature 
             ($T_{c}$) at the core surface derived from ex situ models. 
             Data are color-coded by the temperature $T_{c}$.
             The core radius of \kep g is represented by diamonds instead 
             of circles. Nearly all the difference $\Delta R_{c}$ along each
             curve is caused by the contraction of the core's \ice\ shell.
             By $t\approx 0.5\,\Gyr$, $R_{c}$ is within $\lesssim 3$\% of
             its value at $8\,\Gyr$.
             }
\label{fig:rciso}
\end{figure}

The data listed in Tables~\ref{table:sumex} and \ref{table:sumex_iso} 
indicate that there can be marginal changes in core radii during the 
isolation phase. As mentioned in \cisec{sec:exsitu_res}, these
changes arise from differences in boundary pressures caused by 
contraction, resulting in core compression. Such small changes, 
however, may be offset by temperature effects through the core. 
In the calculations discussed above, core thermodynamics is ignored
and ex situ models include it only for phase transitions within the 
silicate mantle (see \cisec{sec:csc} and Appendix~\ref{sec:ics}).
The assumption of neglecting the core thermal stratification is
based on previous studies, which argued that thermal pressure 
should not significantly affect the core radius at these core masses
\citep[e.g.,][]{valencia2006,seager2007,sotin2007,sohl2014a}. 
Nonetheless, for completeness, improved structure models are 
presented and discussed in Appendix~\ref{sec:cot}.
These include temperature stratification, energy transfer, and 
temperature-dependent EoS in a self-consistent fashion; they
also include additional material phases.
Thermal effects (in \ice-rich cores) may be relatively important for 
the determination of $R_{c}$ during the early stages of the isolation 
phase ($t\lesssim 0.5\,\Gyr$, see \cifig{fig:rciso}). 
However, at an age of $\approx 8\,\Gyr$, the differences in $R_{c}$ 
between thermal and isothermal cores of the simulated \kep\ planets 
are small, typically $\lesssim 0.5$\% in most cases and somewhat 
less than $2$\% for \kep c.
Assessing the impact of thermodynamics for the case of \kep b 
is more difficult, because the expected steam envelope is not 
modeled. If the gas pressure and temperature at the bottom 
of the envelope were, respectively $\approx 1\,\mathrm{GPa}$ 
and $\approx 1000\,\K$, then $R_{c}$ would be $5$\% larger
than the value listed in Table~\ref{table:sumex} (see the discussion 
in Appendix~\ref{sec:cot}).

During isolation, planets gradually cool down. The pressure $P_{c}$ 
at the bottom of the envelope varies by factors of the order of unity, 
whereas temperature variations are larger. To make a simple 
assessment of the impact on $R_{c}$ of the varying conditions 
at the core-envelope boundary as planets cool, thermal structure 
calculations of the cores (see details in Appendix~\ref{sec:ics}) 
were performed by applying pressure and temperature at the core 
surface during the isolation phase.
\cifig{fig:rciso} indicates that there can be significant variations in
$R_{c}$, up to $\approx 0.34\,\Rearth$. The symbols in the figure 
are color-coded according to the core surface temperature, $T_{c}$, 
as obtained from ex situ calculations. However, in the figure, 
$92$--$94$\% of the total difference $\Delta R_{c}$ along each 
curve is due to the contraction of the cores' \ice\ shell. 
Additional details are given in Appendix~\ref{sec:cot}.
Assuming $\Delta R_{p}\approx \Delta R_{c}$, the inflated cores
may result in planet radii larger by $\approx 3$\% at around the 
isolation time, which may in turn enhance the evaporative mass-loss 
rate by $\approx 10$\% during early isolation times.
Results in \cifig{fig:rvse} suggest that the impact on $R_{p}$ at
$8\,\Gyr$ may be relatively small. 

\subsection{Effects of Changes in Initial Conditions}
\label{sec:sensi}

\begin{figure*}[]
\centering%
\resizebox{0.85\linewidth}{!}{%
\includegraphics[clip]{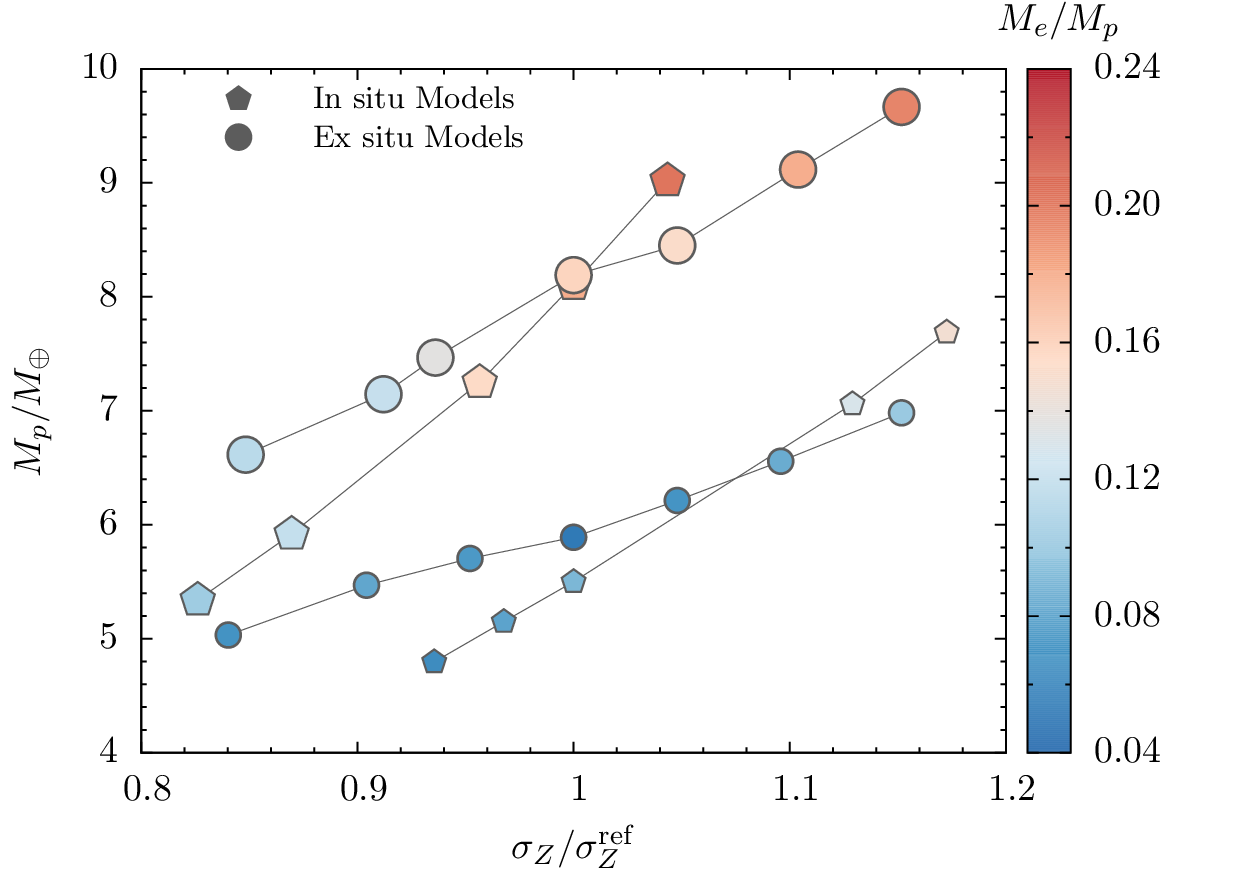}%
\includegraphics[clip]{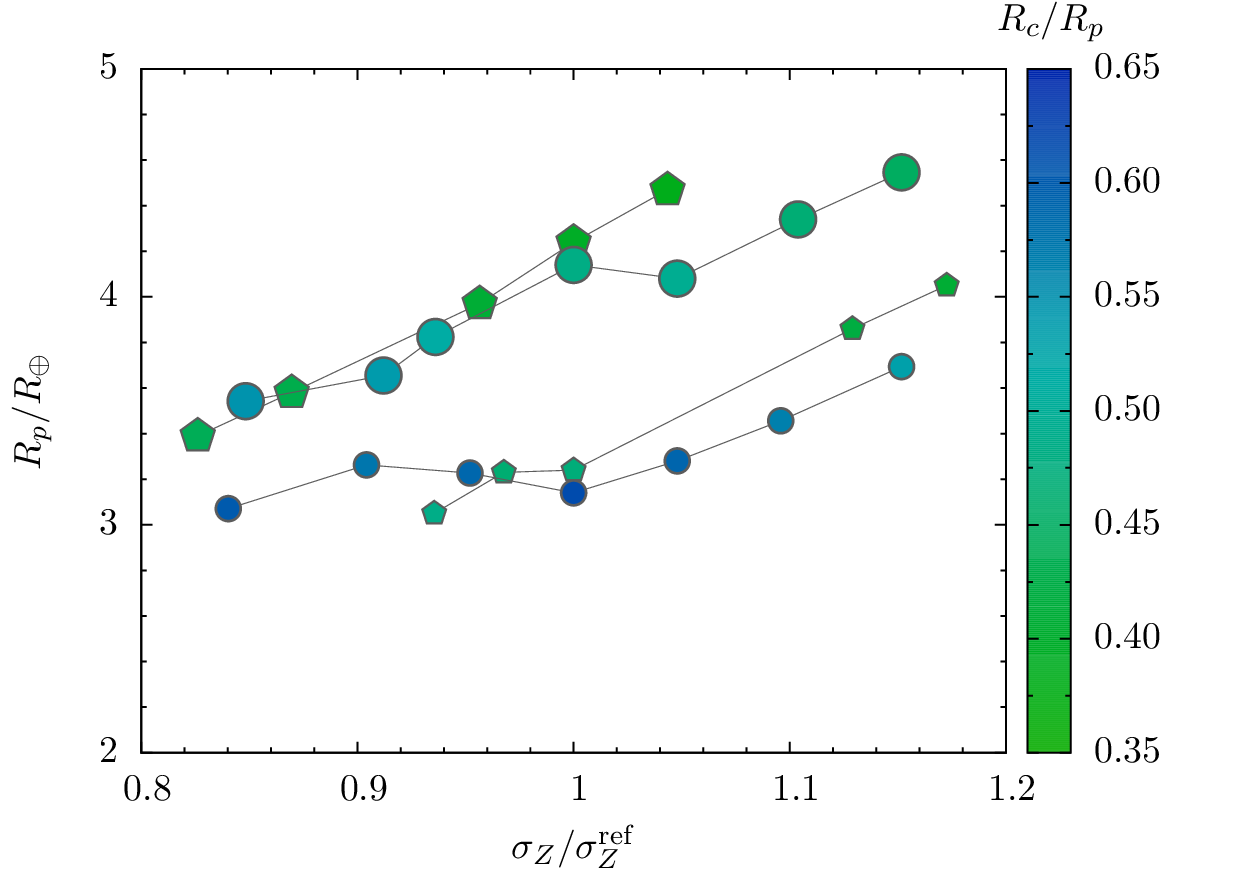}}
\resizebox{0.85\linewidth}{!}{%
\includegraphics[clip]{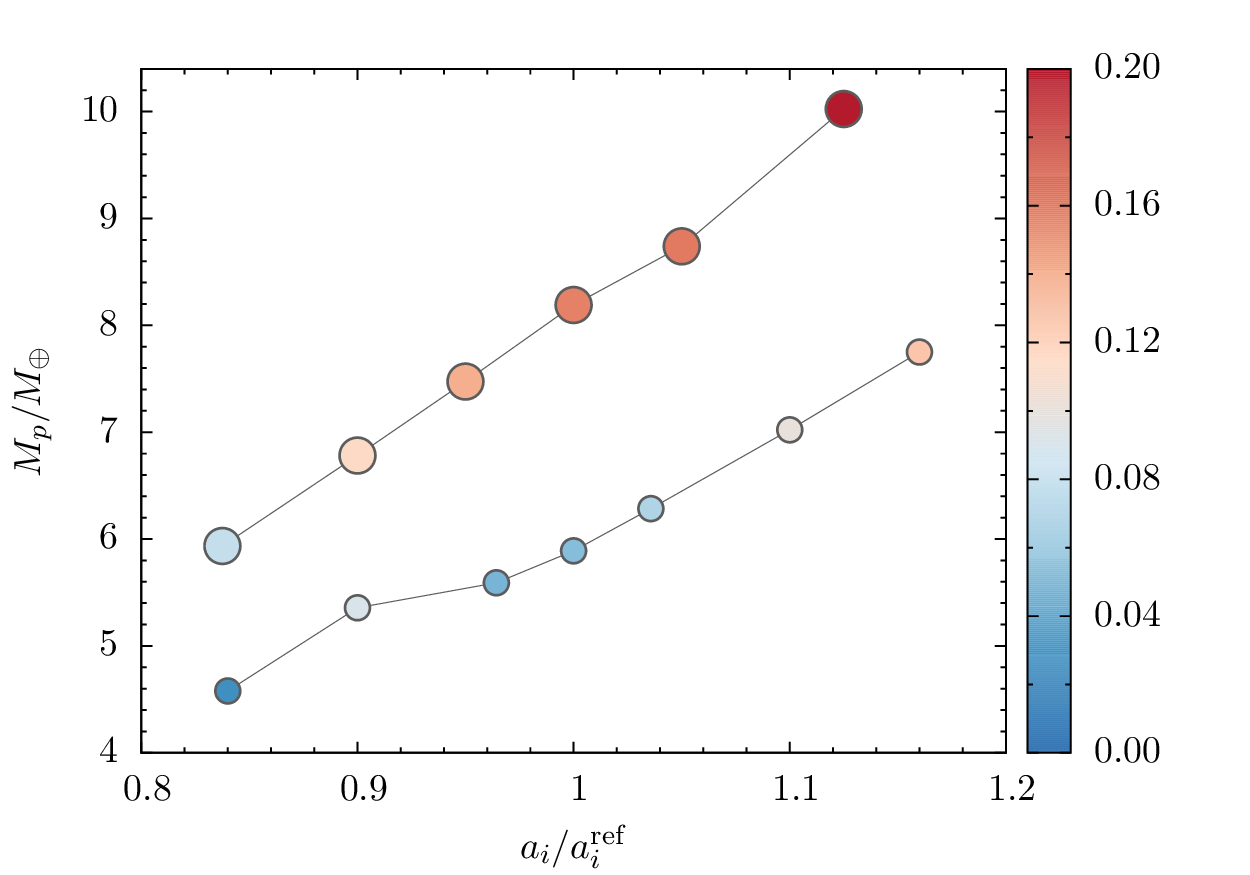}%
\includegraphics[clip]{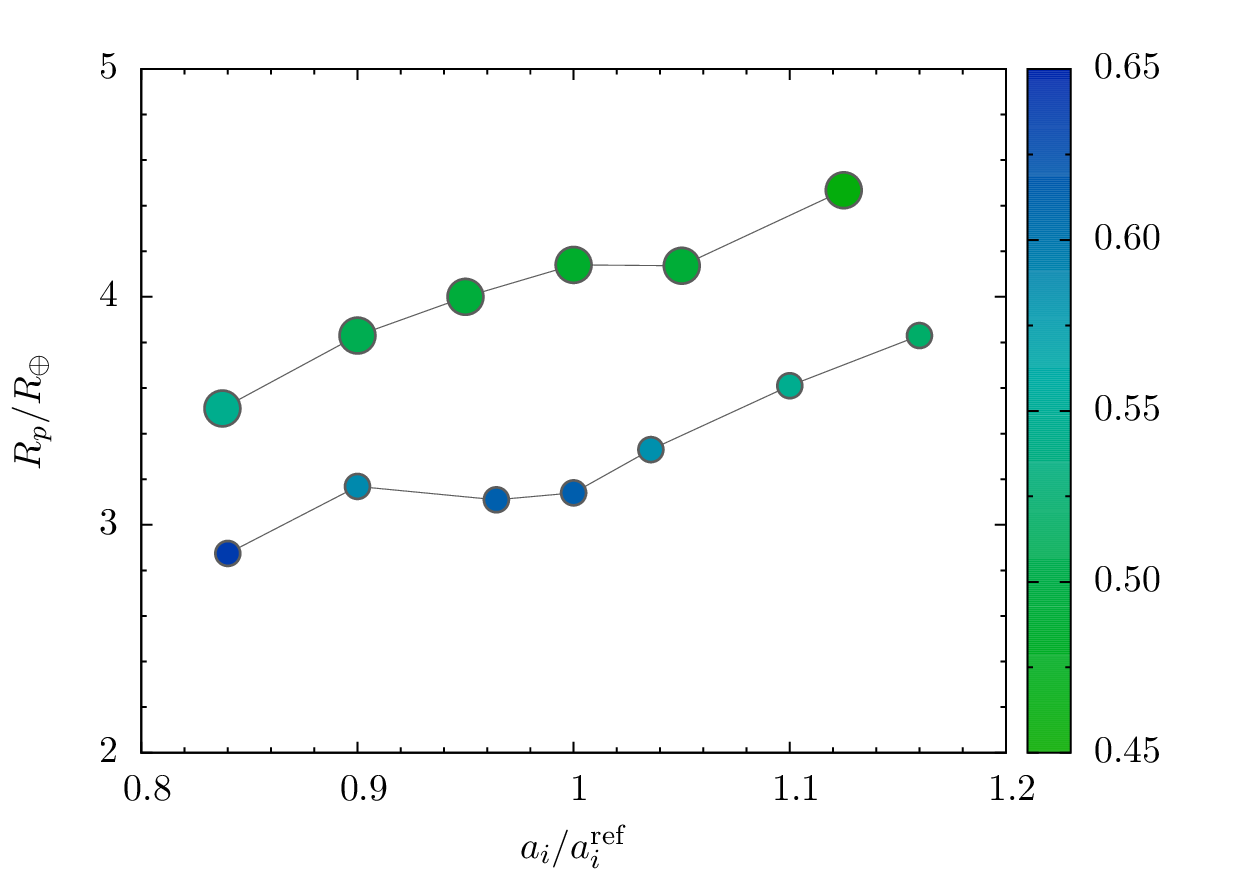}}
\caption{%
             Planet mass and radius at $8\,\Gyr$ versus the solids' surface density, 
             $\sigma_{Z}$ (top), and the orbit's initial radius, $a_{i}$ (bottom), for 
             \kep d and e (larger symbols). 
             Data are color-coded by the envelope-to-planet mass ratio (left) and 
             the core-to-planet radius ratio (right). The values of $\sigma_{Z}$ and 
             $a_{i}$ are scaled by those of the reference models listed in 
             Tables~\ref{table:sumin} and \ref{table:sumex}. In ex situ models, 
             $\sigma^{\mathrm{ref}}_{Z}$ is similar for the two planets, between 
             $3$ and $3.5\,\densu$.
             }
\label{fig:sensi}
\end{figure*}
\cifig{fig:sensi} illustrates how some properties of the models for \kep d and e
at $8\,\Gyr$ depend on the surface density of solids, $\sigma_{Z}$, and 
the initial orbital radius, $a_{i}$ (see the figure's caption for further details). 
Where relevant, in situ models are included as well.
In the figure, $\sigma^{\mathrm{ref}}_{Z}$ and $a^{\mathrm{ref}}_{i}$
are the reference values of the models discussed in Sections~\ref{sec:insitu} 
and \ref{sec:exsitu}.
In the calculations of the top panel, the initial gas density $\Sigma$ is
rescaled so to keep the ratio $\Sigma/\sigma_{Z}$ fixed. The planet mass 
increases monotonically as $\sigma_{Z}$ and $a_{i}$ increase. 
In either case, this is a consequence of the larger core mass that facilitates 
gas accretion, especially at early times. The final envelope mass of in situ 
models increases monotonically, as it generally (but not always) does 
also in ex situ models. In the latter calculations, because of the different
accretion history (and initial condition $a_{i}$), the final orbital radius 
$a_{f}$ varies, affecting the evolution of $M_{e}$ during isolation.
The radius $R_{p}$ tends to grow with the ratio $M_{e}/M_{c}$, which 
is not always a monotonic function of $\sigma_{Z}$ or $a_{i}$ in ex situ 
models (see the left panels of \cifig{fig:sensi}). Because of the small 
uncertainties on the observed radius, only initial conditions in the 
neighborhood of those adopted for the reference models can match 
observations. The final orbital distance varies by a factor of up to 
$\approx 1.8$ in the ex situ calculations of the top panels and 
$\approx 1.3$ in those of the bottom panels (relative to the reference 
values, $a_{f}=0.156$ and $0.194\,\AU$).

The in situ reference models of \cisec{sec:insitu} and those represented 
in \cifig{fig:sensi} assume an initial disk's gas-to-solid mass ratio of $200$. 
Experiments conducted on \kep e, by using $\sigma_{Z}=4600\,\densu$
(as in the reference case)
and varying the mass ratio from $50$ to $400$, resulted in mass and 
radius changes (at $8\,\Gyr$) of $\lesssim 10$\%. For a gas-to-solid 
mass ratio of $50$, $\Mp=7.7\,\Mearth$ and $R_{p}=3.98\,\Rearth$ 
whereas these values increase to $8.46\,\Mearth$ and $4.44\,\Rearth$, 
respectively, for a mass ratio of $400$. The planet mass is still within 
measurement errors, while the planet radius lies outside the observed 
range by at most a few percent.

\section{Discussion}
\label{sec:dac}

\subsection{Implications of In Situ and Ex Situ Formation}
\label{sec:ivse}

\begin{figure}
\centering%
\resizebox{\linewidth}{!}{%
\includegraphics[clip]{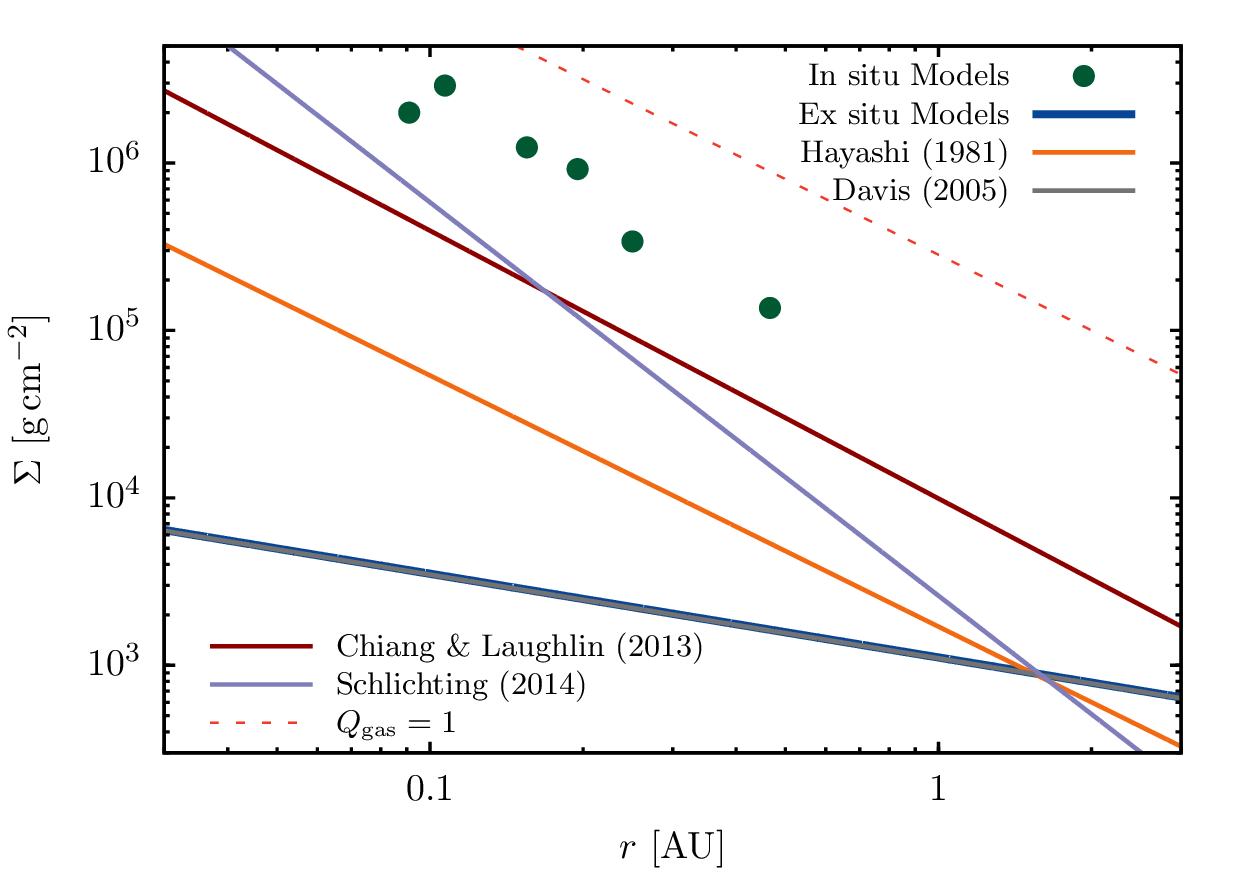}}
\caption{%
             Comparison of the gas surface density required for in situ formation
             (dots) and ex situ formation (thick blue line, see also \cifig{fig:devo_b}) 
             of \kep\ planets. Also shown in the plot are the minimum-mass solar 
             nebula density of 
             \citet[][]{hayashi1981} \citep[see also][]{weidenschilling1977a}, 
             and \citet{davis2005}, and the minimum-mass extrasolar nebula 
             density of \citet{chiang2013} and \citet{schlichting2014}.
             Fortuitously, the initial $\Sigma$ for ex situ models closely matches
             \citeauthor{davis2005}' density distribution.
             The dashed line represents the density threshold for gravitational
             instability of the initial gaseous disk, assuming $T=1000\,\K$.
             $Q_{\mathrm{gas}}$ is the Toomre stability parameter 
             \citep[e.g.,][pp.\ 607-622]{durisen2007}.
             }
\label{fig:sig_cmp}
\end{figure}
Both in situ and ex situ simulated planets result in radii, masses, and
orbital distances in agreement with measured values at the estimated
age of the system. Therefore, it is not possible to distinguish between
the two modes of formation from these final properties. The two 
formation scenarios do, however, provide entirely different perspectives
of the environment in which the planets grew, of their compositions,
and interior structures.

\cifig{fig:sig_cmp} shows the initial surface density of the gas applied
to the in situ (circles) and ex situ models (blue line). The figure also 
illustrates other reference surface densities 
\citep{weidenschilling1977a,hayashi1981,davis2005,chiang2013,schlichting2014}.
As noted above, $\Sigma$ at $t=0$ of the ex situ simulations 
fortuitously matches the surface density constructed by \citet{davis2005}.
It should be pointed out that, among these reference density distributions, 
only those of \citet[][]{chiang2013} and \citet{schlichting2014} were explicitly 
derived for close-in extrasolar planets. 
The other two are meant to apply to the solar system and are thus 
simple extrapolations at the short distances from the star of \kep\ planets.
The dashed line in the figure indicates the gas density above which
the disk would be gravitationally unstable to axisymmetric perturbations
according to the Toomre stability criterion \citep[pp.\ 607-622]{durisen2007}, using 
a constant gas temperature of $1000\,\K$. Even by applying 
$T=1000(0.1\,\AU/r)\,\K$, appropriate for $H/r$ nearly constant,
the initial $\Sigma$ inferred from in situ models would still be stable 
(although only marginally stable to non-axisymmetric perturbations).

In situ formation requires a very large $\sigma_{Z}$ within $\approx 0.5\,\AU$
of the star, as illustrated in \cifig{fig:sig_cmp} \citep[see also][]{bodenheimer2014}. 
The models discussed here suggest densities of solids between $700$ 
and $1.45\times 10^{4}\,\densu$ (see Table~\ref{table:sumin}).
If most of these solids had to form locally, the initial gaseous mass of 
the region had to be accordingly large. 
Given the short accretion timescales involved in the growth 
process, $\sim 10^{4}\,\yr$ (see Figures~\ref{fig:pw1} and \ref{fig:pw2}), 
most of the solids necessary for core assembly had to be available prior 
to the beginning of this process. 
Therefore, a gradual replenishment of the solids' reservoir from larger 
radii might not be a viable alternative to a large $\sigma_{Z}$. 
Moreover, this possibility would likely lead to a hierarchical system 
of planets with inwardly decreasing masses, which is inconsistent 
with the masses of \kep f and e (and probably of \kep g as well).

\begin{figure}
\centering%
\resizebox{\linewidth}{!}{%
\includegraphics[clip]{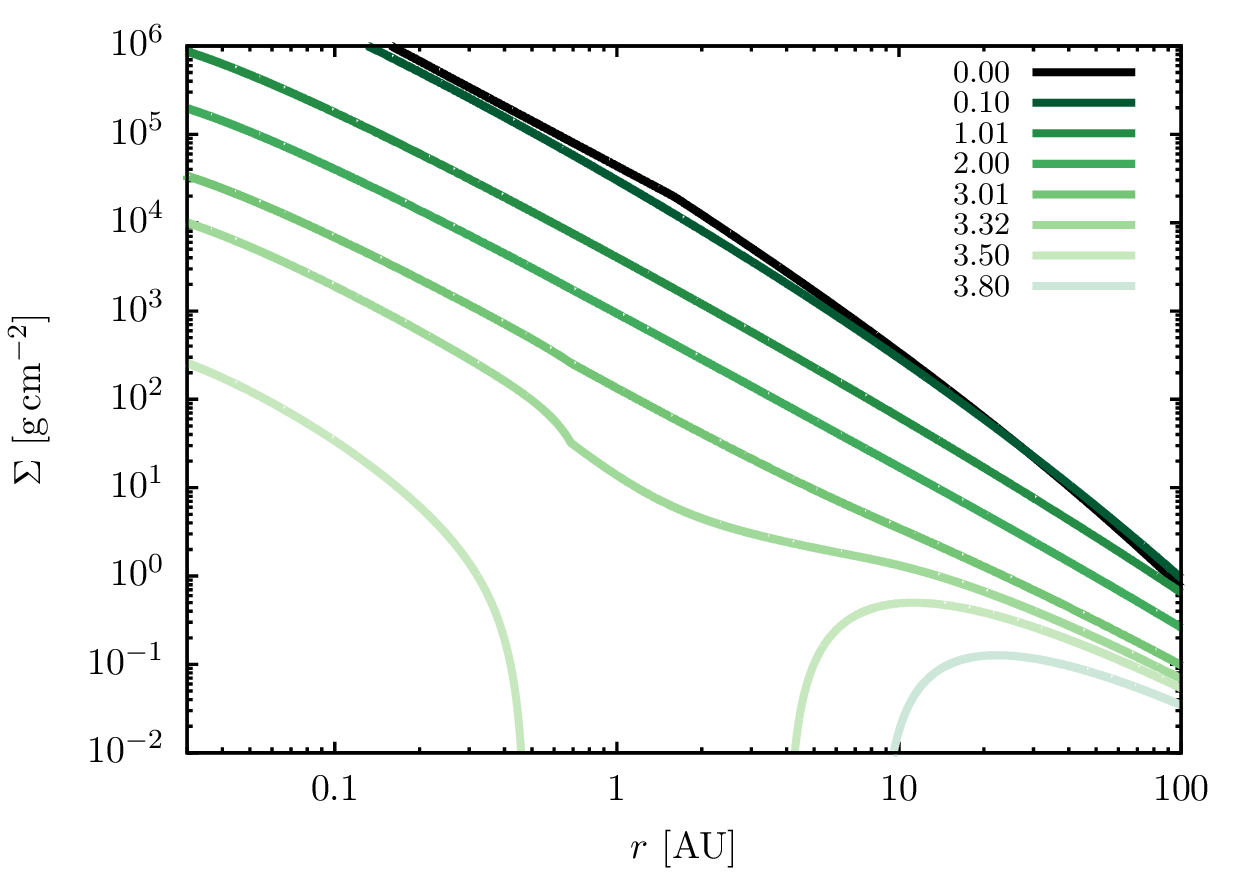}}
\caption{%
             Simulated evolution of the gas surface density applied to
             in situ models discussed in \cisec{sec:insitu}. 
             The initial $\Sigma$ was derived from the required $\sigma_{Z}$,
             augmented by the gas-to-solid mass ratio.
             See \cisec{sec:dac} for further details. 
             Times in the legend are in megayears.
             }
\label{fig:sig_insitu}
\end{figure}
A disk model aimed at mimicking the evolution of the initial in situ 
$\Sigma$ shown in \cifig{fig:sig_cmp} is presented in \cifig{fig:sig_insitu}. 
The disk has an initial mass of $\approx 0.18\,\Msun$ within 
$\approx 70\,\AU$. 
Both the kinematic viscosity of the gas and the emission rate of 
stellar EUV ionizing photons were chosen to induce photo-evaporation 
of the inner disk at around the age of $3.5\,\Myr$ (see \cisec{sec:insitu}). 
The region $r\lesssim 1\,\AU$ in the disk in \cifig{fig:sig_insitu}
is dispersed between $t\approx 3.5$ and $\approx 3.7\,\Myr$.
Gas temperatures are initially high, as can also be realized from 
simple arguments based on energy balance. At high densities,
the disk is optically thick in the vertical direction and the main 
source of energy is viscous heating. Hence \cieq{eq:EEq} reduces 
to $Q_{\nu}=Q_{\mathrm{cool}}$, which becomes
\begin{equation}
 \sigma_{\mathrm{SB}}\,T^{4}=\frac{27}{128}\left(\Sigma\kappa_{\mathrm{R}}+\frac{8}{3}\right)%
                                                 \nu\Sigma\Omega^{2},
 \label{eq:sbu}
\end{equation}
where $\Sigma\kappa_{\mathrm{R}}/2$ is the optical depth
of the disk's mid-plane. Since the initial accretion rate 
through the disk in this case is $3\pi\nu\Sigma\sim 10^{-7}\,\Msun\,\yr^{-1}$, 
\cieq{eq:sbu} and \cifig{fig:sig_insitu} imply that, at $r\approx 0.5\,\AU$, 
temperatures are initially in excess of $2500\,\K$ and drop below 
$1000\,\K$ at $t\gtrsim 1\,\Myr$. At $r\approx 0.1\,\AU$, temperatures 
become $\lesssim 1000\,\K$ at $t\gtrsim 2.5\,\Myr$.
These high gas temperatures would only allow for the presence in the 
disk of metals and highly refractory solid species, which would be 
reflected by the compositions of the cores.
The high temperatures also prevent the massive inner disk from 
becoming gravitationally unstable \citep[pp.\ 607-622]{durisen2007}.
Lowering the kinematic viscosity of the gas would somewhat reduce
the gas temperatures ($T\propto \nu^{1/4}$) but would also extend 
the disk lifetime (by halving $\nu$ it would take over $5\,\Myr$
to photo-evaporate the inner disk).
The effects of the initially high disk temperature and of the disk 
evolution on the in situ formation process require further calculations.

Ex situ formation can occur in low-mass disks, in which the initial 
$\sigma_{Z}$ increases from $\approx 5\,\densu$ at around $7\,\AU$ 
to $\approx 10\,\densu$ at $0.5\,\AU$, the disk region that provides 
the vast majority of the solids to assemble all the planetary cores. 
Initial gas densities at these radial distances are in the range from 
a few times $100\,\densu$ to $\approx 1.5\times 10^{3}\,\densu$ 
(see \cifig{fig:devo_b}). Cores are assembled over timescales 
of the order of $1\,\Myr$ (see Figures~\ref{fig:ow1_sd} and \ref{fig:ow2_sd}). 
Planets experience relatively low temperatures during their formation 
phase and spend a long enough time behind the ice condensation line 
to become enriched in \ice\ and other volatile substances (if available).

\begin{figure}
\centering%
\resizebox{\linewidth}{!}{%
\includegraphics[clip]{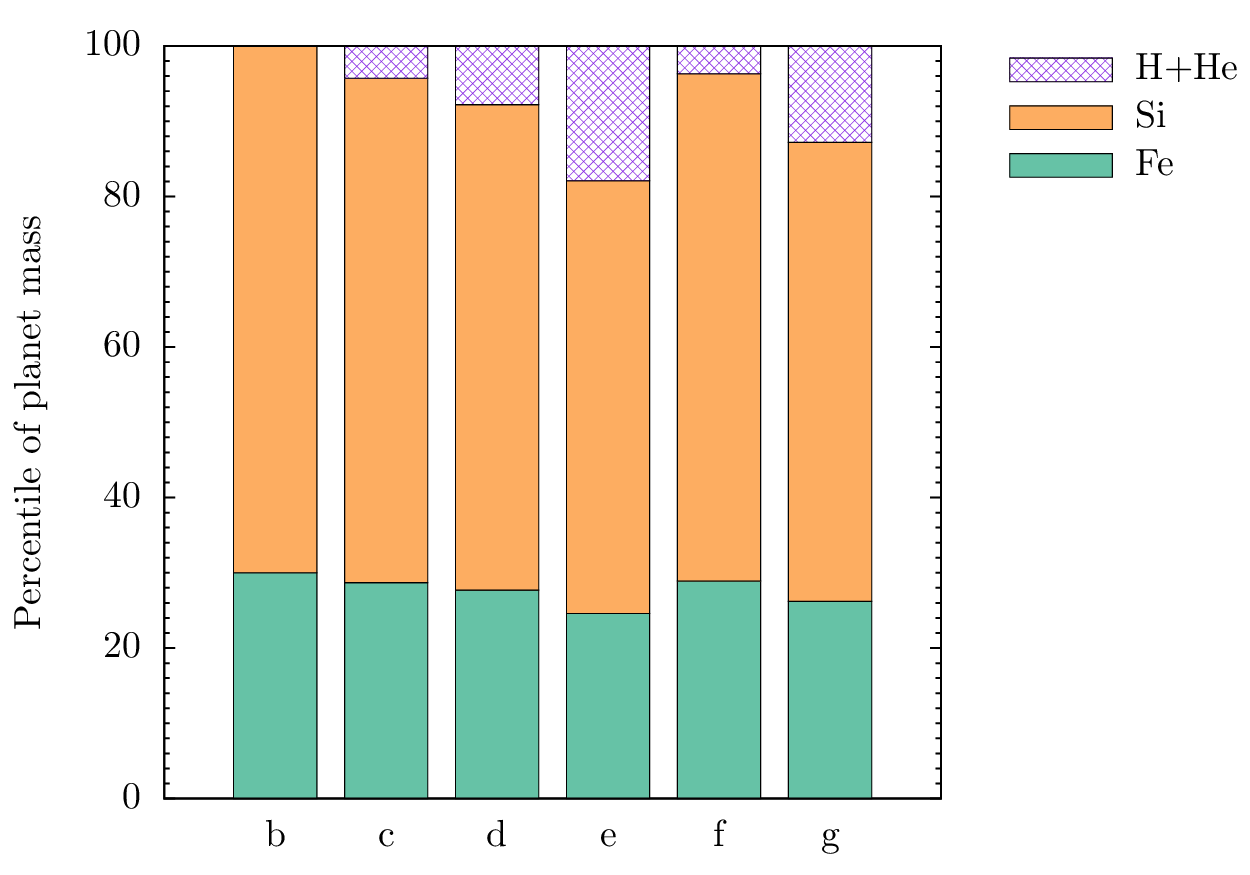}}\\
\resizebox{\linewidth}{!}{%
\includegraphics[clip]{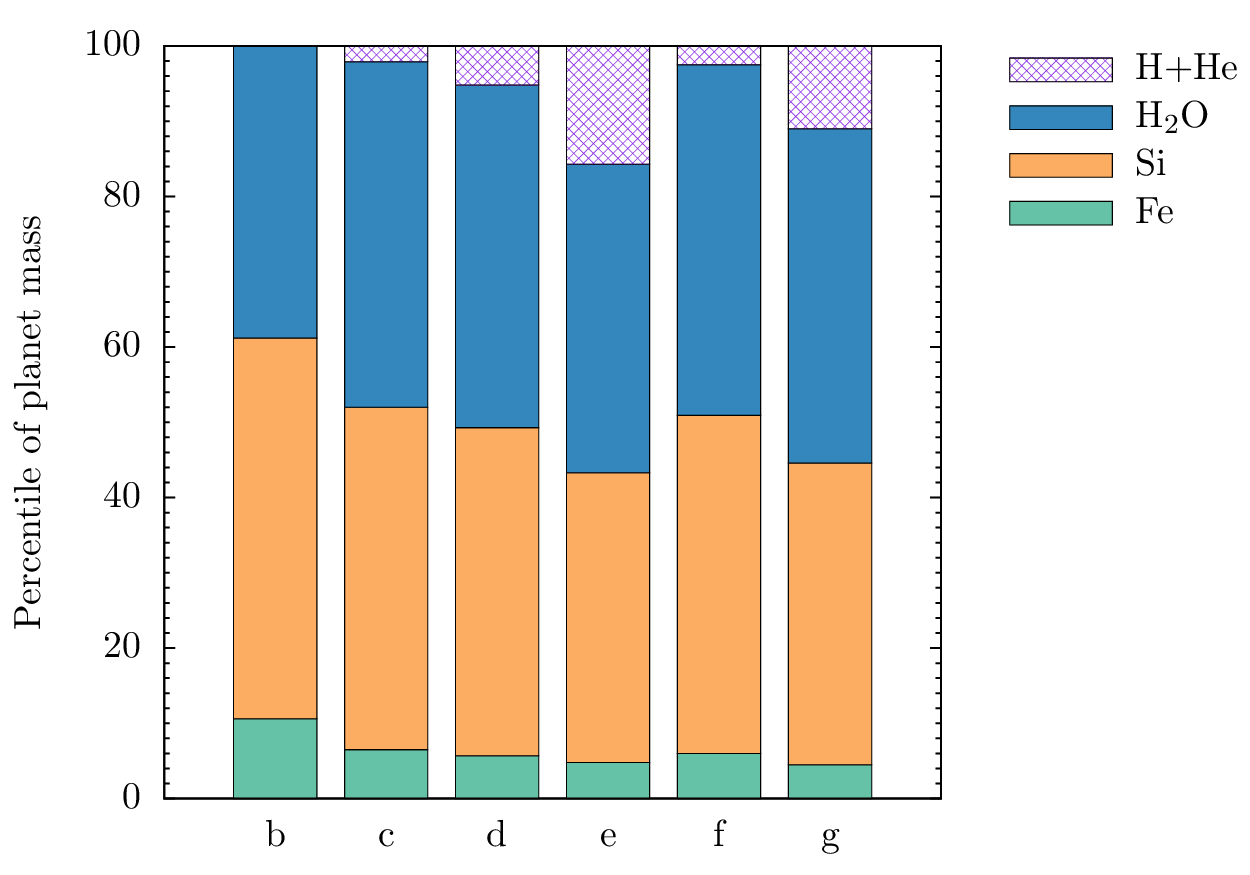}}
\caption{%
             Histogram of the fractional composition of simulated \kep\ planets.
             `Fe' indicates the iron nucleus of the core, `Si' the silicate
             mantle surrounding the nucleus, \ice\ the outer ice/water core shell, 
             and `\hhe' the planet's envelope gas. 
             In case of \kep b, the \ice\ mass fraction does not account for the possible 
             removal of atmospheric steam. The top/bottom panel displays results from
             in situ/ex situ models. In situ calculations assume fixed core mass
             fractions of Fe ($30$\% by mass) and silicates ($70$\% by mass).
             }
\label{fig:hist}
\end{figure}
The percentile composition of the six planets is represented in 
the histograms of \cifig{fig:hist}, for in situ as well as ex situ
models (see the figure's caption for details). Recall that in situ
simulations assume a fixed core composition of iron and silicates
of $30$\% and $70$\% by mass, respectively. In ex situ models,
\ice\ is assumed to be all condensed in the core, although
part of it should be in gaseous form mixed with \hhe\ gas, 
released by ablating solids passing through the envelope.
Both in situ and ex situ calculations predict that the percentile \hhe\ 
mass of \kep c is comparable to that of the smaller \kep f,
because most of the envelope is lost after formation. 
\kep b simulated in situ acquires a much larger envelope than 
does the ex situ simulated planet, because the core mass 
approaches its final value at very early times (compare the 
growth of $M_{e}$ in Figures~\ref{fig:pw1} and \ref{fig:ow1_sd}).

In order to reconcile observed and simulated radii of \kep b,
the in situ formation scenario requires that gas is sequestered 
in the core during formation and released afterwards. Outgassing
of at least $\approx 10^{-3}\,\Mearth$ of hydrogen would be sufficient 
to account for the observed radius. However, outgassing would still
need to compete against evaporative gas loss, which would 
operate at rates between $\approx 10^{-12}$ and 
$\approx 10^{-11}\,\Mearth\,\yr^{-1}$.
If outgassing began right after the removal of the primordial \hhe\ 
envelope (at $t\approx 40\,\Myr$) and continued for the age of the 
planet, the (average) outgassing rate would only need to be 
marginally higher than the (average) evaporative gas loss rate. 
This possibility, however, implies that the amount of gas 
sequestered in the core during formation ought to be significant 
in relative terms, between $\approx 0.01$ and $\approx 0.1\,\Mearth$.
The details regarding the processes of sequestration and outgassing 
remain to be investigated, but the result could have important 
implications concerning the in situ versus ex situ formation of \kep b.

The ex situ formation scenario predicts the presence of a steam
atmosphere (not modeled here). 
Based on previous assessments of the planet mass and radius
\citep[][]{lissauer2011a}, \citet{lopez2012} estimated that the planet 
radius can be matched by a water-world whose composition is about 
$40$\% \ice\ by mass, comparable to the \ice\ content of the simulated 
ex situ planet (see Table~\ref{table:sumex}). 
Assuming a hydrostatic and adiabatic  atmosphere with an ideal EOS 
for \ice, the mass $M_{e}$ necessary to match the observed radius is 
$\approx 7\times 10^{-3}\,\Mearth$, though non-ideal effects in the 
EOS may reduce this estimate. Because of the larger mean molecular 
weight, evaporative mass loss of \ice\ by stellar radiation should be less 
significant than is loss of lighter elements. Additionally, steam can be 
replenished by the condensed core.
Since hydrogen can mix with \ice\ at the pressures and temperatures 
of planetary interiors \citep{soubiran2015}, hydrogen may be present 
in the atmosphere of \kep b as well.

\subsection{On the Mass of \kep g}
\label{sec:oma}

\begin{figure}
\centering%
\resizebox{\linewidth}{!}{%
\includegraphics[clip]{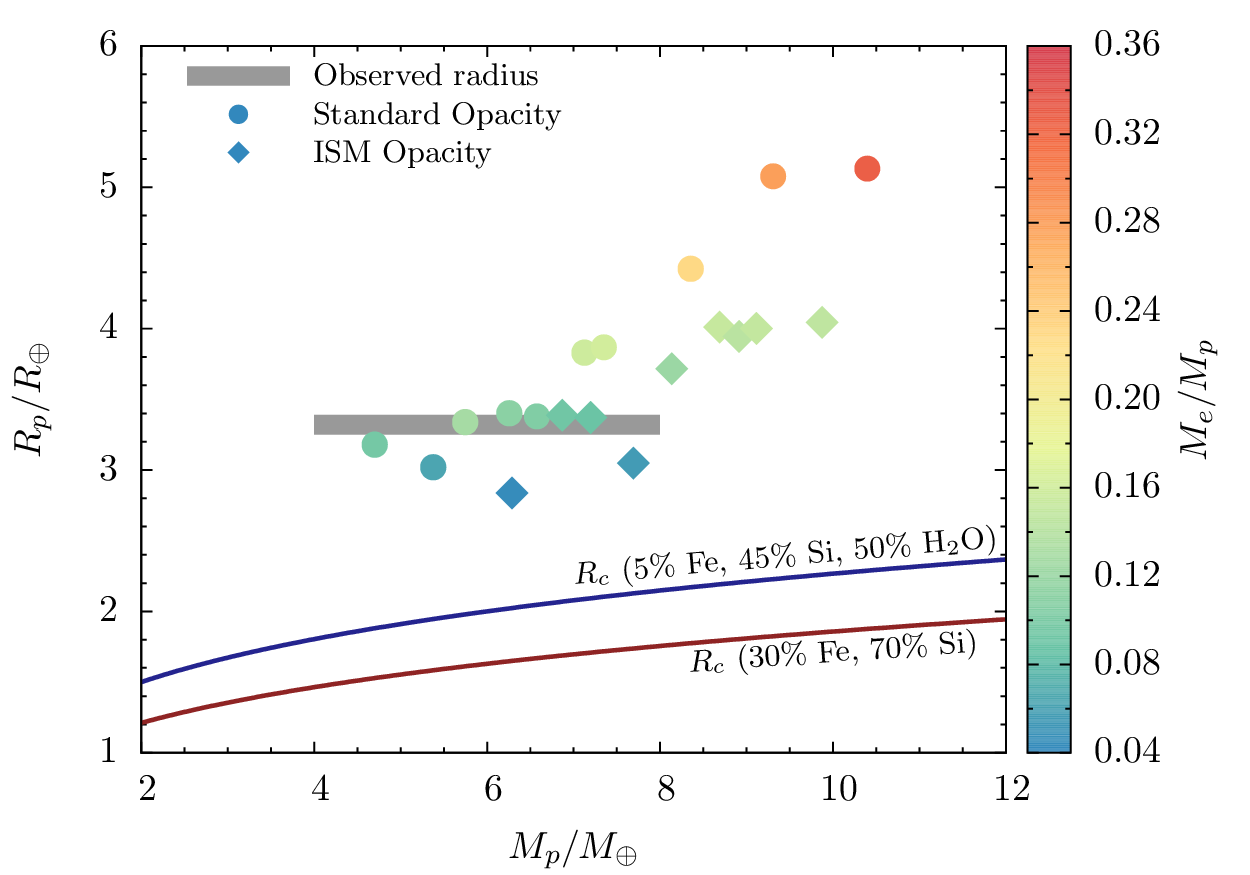}}
\caption{%
             Planet radius versus planet mass for various simulations
             of \kep g. Circles refer to models adopting the standard 
             opacity table (with dust grains up to $1\,\mathrm{mm}$ in
             radius, see \cifig{fig:opa}). 
             Diamonds refer to models adopting an interstellar
             medium-type distribution of grains (up to 
             $1\,\mathrm{\mu m}$ in radius), described in \citet{gennaro2013}. 
             The planet gas content is rendered by the color scale.
             Also plotted are the core radii for the interior compositions of 
             \kep g from in situ (lower solid curve) and ex situ simulations
             (upper solid curve).
             }
\label{fig:mg}
\end{figure}

Transit observations only place an upper limit of $\approx 25\,\Mearth$ 
on the mass of \kep g \citepalias{lissauer2013}. Assuming negligible 
amounts of gas in the envelope, an ex situ type composition of $5$\% iron, 
$45$\% silicates, and $50$\% \ice\ by mass, would result in a radius 
of $2.8\,\Rearth$, short of the observed value $R_{p}=3.33\,\Rearth$.
An in situ type composition would result in an even smaller radius
($2.3\,\Rearth$). Hence, the planet must bear a gaseous envelope.
 
Both in situ and ex situ models indicate that \kep g is somewhat less 
massive than \kep e and of comparable mass to \kep d. 
It is reasonable to imagine a scenario in which a lower core mass 
may attract a sufficiently large envelope to account for the observed 
radius. 
In fact, another simulation of \kep g (with different start conditions)
results in a radius only marginally smaller than that observed and 
$\Mp\approx 4.7\,\Mearth$. Even smaller masses might be attained, 
e.g., by lowering the dust opacity in the planet's envelope.
However, the opposite scenario of a much more massive planet and 
a smaller $M_{e}/\Mp$ ratio may be more difficult to realize.

\cifig{fig:mg} shows outcomes from additional ex situ simulations 
of \kep g.
In these models, the start conditions of the reference model
(\ts, $a_{i}$, and initial $\sigma_{Z}$) were changed with the 
aim of obtaining different planet masses and gas-to-condensible 
mass ratios. 
The final orbital radii are within roughly $10$\% of the observed 
orbital distance. Circles represent simulations that apply the
grain opacity of the top panel of \cifig{fig:opa}, as in reference models.
Data points are color-coded by their relative gas content, $M_{e}/\Mp$.
The solid lines indicate the core radius corresponding to the 
interior composition of in situ ($30$\% iron and $70$\% silicates)
and ex situ models (see Table~\ref{table:sumex}), assuming
a pressure at $R_{c}$ of $1\,\mathrm{GPa}$ and $T=T_{\mathrm{eq}}$. 
The observed radius of the planet, within measurement errors, 
is indicated by the gray-shaded area.
The data points include the in situ simulation of Table~\ref{table:sumin}
as well. This model has a comparatively large gas content 
for its radius, owing to the smaller value of $R_{c}$ 
(i.e., the difference $R_{p}-R_{c}$ is in fact comparably large).

From the data plotted in \cifig{fig:mg}, it appears difficult to reconcile 
observed and simulated radii for planets more massive than about 
$7\,\Mearth$. In fact, when $\Mp\gtrsim 7\,\Mearth$, the \hhe\ mass 
fraction $M_{e}/\Mp$ tends to exceed $\approx 16$\%, resulting in 
too large a radius.
In the calculations reported herein, the value of $M_{e}$ at the end 
of the formation phase is set by dispersal of the gaseous disk. 
The model with largest mass in \cifig{fig:mg} has 
$M_e\gtrsim 2.5\,\Mearth$ ($M_{e}/\Mp\gtrsim 0.28$)
and $R_{p}$ far larger than observed. 
If under different circumstances the accretion of gas was starved 
by disk dispersal at an earlier age and $R_{p}$ at $8\,\Gyr$ was 
similar to the observed value, the planet mass would not exceed 
$\approx 8\,\Mearth$ (an envelope mass somewhat smaller 
than $1\,\Mearth$ appears necessary to account for the observed 
radius).

According to \citetalias{lissauer2013}, an $\Mp\approx 15\,\Mearth$ 
planet with a $6.6$\% gas content by mass can match the radius of 
\kep g.
It may be possible to build a large core mass without attracting an 
excessively large envelope by augmenting the opacity in the envelope, 
which operates to slow contraction and inhibits gas accretion. 
This possibility was tested by using dust opacities calculated from 
an interstellar medium-type grain distribution, with dust grains 
of up to $1\,\mathrm{\mu m}$ in radius \citep[for details, see][]{gennaro2013}. 
Results from this second set of models are represented by diamonds 
in \cifig{fig:mg}. As expected, for a given $\Mp$, the ratio $M_{e}/\Mp$ 
is smaller than in the other set of simulations, yielding smaller planetary 
radii. 
Nonetheless, as $\Mp$ grows beyond $\approx 8\,\Mearth$, 
$M_{e}/\Mp\gtrsim 0.2$ and $R_{p}$ becomes too large to match
observations.
Clearly, other scenarios can be envisaged (e.g., most of $M_{c}$ is
accumulated toward the end of the disk lifetime, delaying envelope
growth), although they must be compatible with the formation of the 
inner planets. 
Within the assumptions adopted herein, both in situ and ex situ
models argue in favor of a mass smaller than or comparable to
\kep e's mass.

\section{Conclusions}
\label{sec:fine}

We constructed models of formation and long-term evolution of 
the six \kep\ planets. Both in situ and ex situ scenarios were
considered. The simulations presented here take into account 
many physical aspects of the formation and evolution processes
in detail (see \cisec{sec:np}). 
Approximations were nonetheless necessary in order
to render the problem tractable.
A major limitation of this study is the neglect of planet-planet 
interactions. Although it is not possible to speculate about the impact 
of this deficiency on the results, especially in regard to orbital stability
of planets and capture into mean-motion resonances, N-body simulations 
did show that compact planetary systems, and the \kep\ system in particular, 
may indeed originate in the presence of disk-driven orbital migration 
\citep{hands2014}.

Both in situ and ex situ models appear equally capable of generating 
planets whose radii, masses, and orbital distances (when relevant) 
at the estimated age of $\approx 8\,\Gyr$ agree with those measured 
via transit observations (see Sections~\ref{sec:insitu} and \ref{sec:exsitu}). 
Based exclusively on these final outcomes 
of the simulations, it seems difficult -- and certainly not obvious -- 
to argue in favor of one scenario over the other. 
The implications of the two formation scenarios are, however, 
significantly and profoundly different (see \cisec{sec:dac}). 

In situ formation may only work if a large amount of solids is available 
for accretion within $r\lesssim 0.5\,\AU$ of the star. The models built 
here predict a surface density of solids 
$7\times 10^{2}\le\sigma_{Z}\le1.45\times 10^{4}\,\densu$ 
(see Table~\ref{table:sumin}) and
a mass in solids of $\approx 50\,\Mearth$ inside $\approx 0.5\,\AU$.
Initial gas densities in the disk are expected to be accordingly high
($\gtrsim 10^{5}\,\densu$, see \cifig{fig:sig_insitu}).
Planets form at disk temperatures of over $1000\,\K$, hence
their cores may only contain metals and refractory materials. 
The large values of $\sigma_{Z}$ also result in extremely short 
assembly times of the cores (of the order of $10^{4}\,\yr$, see 
Figures~\ref{fig:pw1} and \ref{fig:pw2}). 
\kep b, which is completely stripped of its primordial \hhe\ envelope 
during the isolation phase (a fate that appears unavoidable!), must 
have a tenuous atmosphere continuously replenished over time,
or generated at a late age, by the release of gas sequestered in 
the core. More detailed modeling is required, taking 
into account the release of hydrogen as well as the evaporation of 
the atmosphere, to determine if this picture is reasonable. 

Ex situ formation may work in low-mass disks. The models built 
here are based on a disk whose distribution of solids at $t=0$ is 
$5\lesssim\sigma_{Z}\lesssim 10\,\densu$, proceeding inward from 
$\approx 7$ to $\approx 0.5\,\AU$. The initial gas density in this
region ranges from $\approx 100$ to $\approx 10^{3}\,\densu$
(see \cifig{fig:devo_b}). 
The disk's gas interior to $\approx 1\,\AU$ is dispersed in about 
$4\,\Myr$. Cores grow gradually, over timescales of the order of $10^{6}\,\yr$
(see Figures~\ref{fig:ow1_sd} and \ref{fig:ow2_sd}).
In all cases, most of the core assembly may be completed beyond 
the ice condensation line ($T\lesssim 150\,\K$). Consequently, 
ex situ formation predicts planetary cores to be rich in \ice, and 
possibly in other volatile materials, if hydrated planetesimals 
bear substantial amounts of ice (as assumed here).
Most of the \ice\ is at high pressures and temperatures 
(see Appendix~\ref{sec:cot}).
\kep b, which loses its entire envelope during isolation, is predicted 
to have a steam envelope originating from the release of \ice\ from 
the core's outer shell. 
The gaseous envelopes of the other planets may contain \ice\ 
mixed with hydrogen and helium as well.
In fact, since the critical temperature of ice is reached at
shallow depths ($T\approx 650\,\K$), passing solids should 
shed part of their mass in the outer envelope layers.

Simulations indicate that, from a formation standpoint, it may be 
difficult for \kep g to become more massive than $\approx 8\,\Mearth$,
collecting only a relatively light envelope (see \cisec{sec:dac}). 
Both in situ and ex situ models point at a planet mass not much 
greater than $\Mp\approx 7\,\Mearth$ and a percentile gas content 
between $\approx 10$ and $\approx 15$\%.

\acknowledgments

We thank Uma Gorti for numerous helpful discussions and for her precious 
guidance during the implementation of the disk photo-evaporation module.
We are grateful to an anonymous referee, whose insightful comments helped 
improve several parts of this paper.
G.D.\ thanks the Los Alamos National Laboratory for its hospitality.
G.D.\ acknowledges support from NASA Outer Planets Research Program
grant 202844.02.02.01.75 and from NASA Origins of Solar Systems Program 
grant NNX14AG92G.
Resources supporting this work were provided by the NASA High-End
Computing (HEC) Program through the NASA Advanced Supercomputing
(NAS) Division at Ames Research Center.

\appendix

\section{Core Structure Calculations}
\label{sec:ics}

\defcitealias{fei1993}{F93}
\defcitealias{fortes2005}{F05}
\defcitealias{frank2004}{F04}
\defcitealias{ichikawa2014}{I14}
\defcitealias{katsura2009a}{K09a}
\defcitealias{katsura2009b}{K09b}
\defcitealias{komabayashi2010}{K10}
\defcitealias{loubeyre1999}{L99}
\defcitealias{sohl2002}{S02}
\defcitealias{stamenkovic2011}{S11}
\defcitealias{stewart2005}{S05}
\defcitealias{stixrude1990}{S90}
\defcitealias{tanaka1998}{T98}
\defcitealias{wagner2002}{W02}
\defcitealias{wagner2011}{W11}
\begin{deluxetable}{cccccc}
\tablecolumns{6}
\tablewidth{0pc}
\tablecaption{Equations of state and related thermodynamics quantities\label{table:eos}}
\tablehead{
\colhead{Material}&\colhead{EoS type\tablenotemark{a}}&\colhead{EoS}&
\colhead{$\gamma$}&\colhead{$\theta_{\mathrm{D}}$}& \colhead{$\alpha$}
}
\startdata
$\epsilon$-Fe  & GR & \citetalias{wagner2011} &\citetalias{wagner2011}  & 
                               \citetalias{wagner2011} & Eq.~(\ref{eq:esp}) \B\\
$\gamma$-Fe      & BM & \citetalias{komabayashi2010}   & \citetalias{wagner2011} &
                               \citetalias{wagner2011} & Eq.~(\ref{eq:esp}) \B\\
$\alpha$-Fe          & BM& \citetalias{komabayashi2010}   &\citetalias{wagner2011} &
                               \citetalias{wagner2011} & Eq.~(\ref{eq:esp}) \B\\
Liquid~Fe              & BM& \citetalias{komabayashi2010}   &\citetalias{ichikawa2014} &
                                                                     & Eq.~(\ref{eq:esp}) \B\\
MgSiO$_{3}$ppv  & GR & \citetalias{wagner2011} &\citetalias{stamenkovic2011}  &
                               \citetalias{stamenkovic2011} & \citetalias{katsura2009b} \B\\
MgSiO$_{3}$pv     & GR & \citetalias{wagner2011} &\citetalias{stamenkovic2011}  & 
                               \citetalias{stamenkovic2011} & \citetalias{katsura2009b} \B\\
Mg$_{2}$SiO$_4$ & BM & \citetalias{wagner2011} &\citetalias{wagner2011}  & 
                               \citetalias{wagner2011}         & \citetalias{katsura2009a} \B\\
Ice X                       &  VN & \citetalias{loubeyre1999} & \citetalias{fei1993} &
                                \citetalias{fei1993} & \citetalias{frank2004} \B\\
Ice VII                     & BM & \citetalias{frank2004} & \citetalias{fei1993} &
                               \citetalias{fei1993}  & \citetalias{frank2004} \B\\
Ice VI                      & BM & \citetalias{sohl2002} & \citetalias{fei1993} &
                                \citetalias{fei1993} & \citetalias{stewart2005} \B\\
Ice V                       & BM & \citetalias{sohl2002} & \citetalias{fei1993} &
                                \citetalias{fei1993} & \citetalias{fortes2005} \B\\
Ice III                      & BM &\citetalias{sohl2002}  &  \citetalias{fei1993} &
                               \citetalias{fei1993} & \citetalias{fortes2005} \B\\
Ice Ih                       & BM & \citetalias{sohl2002} &  \citetalias{fei1993} &
                                \citetalias{fei1993} & \citetalias{tanaka1998} \B\\
Water                      & BM & \citetalias{stixrude1990} & \citetalias{stewart2005} &  &
                               \citetalias{stixrude1990} \B\\
Water                      & IAPWS & \citetalias{wagner2002} & \citetalias{wagner2002} &  &
                               Eq.~(\ref{eq:esp})
\enddata
\tablenotetext{a}{Generalized Rydberg (GR);  3rd-order Birch-Murnaghan (BM); Vinet (VN);
                           see \citet{stacey2008} for a review. The last entry is for IAPWS ordinary water.}
\tablerefs{
                \citet[\citetalias{fei1993}]{fei1993}; 
                \citet[\citetalias{fortes2005}]{fortes2005}; 
                \citet[\citetalias{frank2004}]{frank2004}; 
                \citet[\citetalias{ichikawa2014}]{ichikawa2014};
                \citet[\citetalias{katsura2009a}]{katsura2009a};
                \citet[\citetalias{katsura2009b}]{katsura2009b};
                \citet[\citetalias{komabayashi2010}]{komabayashi2010};
                \citet[\citetalias{loubeyre1999}]{loubeyre1999};
                \citet[\citetalias{sohl2002}]{sohl2002};
                \citet[\citetalias{stewart2005}]{stewart2005};
                \citet[\citetalias{stamenkovic2011}]{stamenkovic2011};
                \citet[\citetalias{stixrude1990}]{stixrude1990};
                \citet[\citetalias{tanaka1998}]{tanaka1998};
                \citet[\citetalias{wagner2002}]{wagner2002};
                \citet[\citetalias{wagner2011}]{wagner2011}.
                }
\end{deluxetable}
Here we describe our basic calculations of core structures, 
i.e., those of the condensible part of a planet,
along with improvements intended to check the validity of some of 
the assumptions and approximations applied in the models discussed 
above. 
An important cautionary note: the labels used in this appendix are 
unrelated to those used elsewhere in the paper.

As anticipated in \cisec{sec:csc}, the condensed interior of a planet 
consists of an iron nucleus (hereafter referred to as ``core'' for conformity
with geophysics terminology), 
surrounded by a silicate mantle, overlaid with an \ice\
shell. The core material can transition among the iron allotropes 
$\alpha$-Fe, $\gamma$-Fe , $\epsilon$-Fe\footnote{%
The crystal structure of these allotropes
is, respectively, body-centered cubic (bcc), face-centered cubic (fcc), 
and hexagonal close packed (hcp).}, and liquid iron,
according to the $P$-$T$ phase diagrams of \citet{kerley1993}
and \citet{anzellini2013}.
The silicate mantle can differentiate, with increasing pressure, into
olivine (Mg$_{2}$SiO$_{4}$), perovskite (MgSiO$_{3}$pv), and
post-perovskite (MgSiO$_{3}$ppv) layers. Here, the transition among 
these species is regulated by the phase diagrams of \citet{fei2004} and
\citet{tateno2009}. Post-post-perovskite phases 
\citep[e.g., see the discussion in][]{stamenkovic2011,wagner2012}
are neglected.
The \ice\ shell is composed of several types of ice (Ih, III, V, VI, VII, and X)
and water, according to the phase curves of \citet{loubeyre1999},
\citet{jlin2004}, \citet{jlin2005}, and \citet{choukroun2007}. 
The transition to the vapor phase is ignored.

Two fundamental and customary assumptions made in planetary structure 
calculations are those of spherical symmetry and hydrostatic 
equilibrium.
Indicating with $m$ the mass of the condensed matter interior to radius $R$, 
and with $P$, $Q$, and $T$, respectively, the pressure, the heat flux, 
and temperature at $R=R(m)$, the structure equations read
\begin{eqnarray}
 \frac{\partial R}{\partial m}& = &\frac{1}{4\pi \rho R^{2}}, \label{eq:cse_r}\\
 \frac{\partial P}{\partial m}& = &-\frac{G m}{4\pi R^{4}}, \label{eq:cse_p}\\
 \frac{\partial Q}{\partial m}& = &\left(\varepsilon\rho - \frac{2 Q}{R}\right)%
                                                    \frac{\partial R}{\partial m},\label{eq:cse_q}\\
 \frac{\partial T}{\partial m}& = &-\frac{Q}{\mathrm{N_{u}}k_{c}}%
                                                    \frac{\partial R}{\partial m}.\label{eq:cse_tc}\\
 \frac{\partial T}{\partial m}& = &\left(\frac{\partial T}{\partial m}\right)_{\!S}.\label{eq:cse_ts}
\end{eqnarray}
In \cieq{eq:cse_q}, $\varepsilon$ is the specific energy production 
rate (due to radiogenic heating produced by radioactive decay, 
to tidal heating, to accretion heating, to heating/cooling during 
phase changes, etc.).
Heat production in the \ice\ shell is set to zero, whereas it
accounts for radiogenic heating in the silicate mantle, where a constant 
value of $\varepsilon=7.38\times 10^{-12}\,\mathrm{W\,kg^{-1}}$ 
\citep[appropriate for the Earth,][]{turcotte2014} is applied.
In the iron core, $\varepsilon$ is chosen so that the heat
flux across the boundary with the silicate mantle (CMB) is
$Q_{\mathrm{CMB}}=-k_{c}(\partial T/\partial R)_{\mathrm{CMB}}$ 
\citep{valencia2006,wagner2011}. 
For the Earth test considered below, the power
through this boundary is $\approx 7\,\mathrm{TW}$, comparable to 
the lower limit estimated for the Earth \citep{tateno2009}.
Under the assumption of spherical symmetry, the gravitational 
acceleration $g=Gm/R^{2}$ is known from the solution $R=R(m)$.
In \cieq{eq:cse_tc}, $\mathrm{N_{u}}$ is the Nusselt number, which gives
the ratio of the total (conductive plus convective) to the conductive heat 
flux through the spherical surface of radius $R$, and $k_{c}$ is the 
thermal conductivity ($\mathrm{N_{u}}=1$ when the convective heat 
flux is zero).
Equation~(\ref{eq:cse_tc}) is applied to semi-convective layers 
whereas \cieq{eq:cse_ts}, which represents the adiabatic temperature 
gradient \citep[][]{anderson1989}
\begin{equation}
\left(%
 \frac{\partial T}{\partial m}\right)_{\!S} = \frac{\gamma T}{K_{S}}%
                                                                \frac{\partial P}{\partial m},
 \label{eq:cse_ta}
\end{equation}
is applied to vigorously convective layers. In the equation above, 
$\gamma$ is the Gr\"{u}neisen parameter and $K_{S}$ is the adiabatic 
bulk modulus \citep[e.g.,][]{anderson1989}.
Here the iron core is assumed to be fully convective (i.e., adiabatic) 
whereas the silicate mantle and the ice layers are semi-convective. 
``Fluid'' \ice\ layers are adiabatic \citep{fu2010}.

The adiabatic bulk modulus in \cieq{eq:cse_ta} can be written as
\begin{equation}
 K_{S}=\left(1+\alpha\gamma T\right) K_{T},
 \label{eq:KS}
\end{equation}
where $K_{T}=\rho\left(\partial P/\partial\rho\right)_{T}$ is the isothermal
bulk modulus and
\begin{equation}
 \alpha=\left(\frac{1}{K_{T}}\right)\left(\frac{\partial P}{\partial T}\right)_{\!\rho}
 \label{eq:esp}
\end{equation}
is the thermal expansivity \citep{anderson1989}. The expansivity is often 
approximated to a parameterized function $\alpha=\alpha(\rho,T)$
\citep[e.g.,][]{reynard1990,frank2004}.
These thermodynamical quantities relate to the specific heat at constant
pressure and volume, $C_{P}$ and $C_{V}$, according to 
$\gamma\rho C_{P}=\alpha K_{S}$ and $C_{P}K_{T}=C_{V}K_{S}$. 

If the heat flux carried via convection is written as 
$Q_{\mathrm{conv}}=%
-4\pi \rho R^{2} k_{v} \left[\partial T/\partial m - \left(\partial T/\partial m\right)_{S}\right]$ 
\citep[e.g.,][]{sasaki1986,abe1997}, then in \cieq{eq:cse_tc}
\begin{equation}
 \mathrm{N_{u}}k_{c}=\frac{(k_{c}+k_{v})Q}{Q-4\pi \rho R^{2} k_{v}\left(\partial T/\partial m\right)_{S}}.
 \label{eq:Nu}
\end{equation}
The thermal conductivity is, in general, a function of both pressure and temperature.
In the metallic core, $k_{c}$ is only used to determine $Q_{\mathrm{CMB}}$ 
(the heat flux at the core-mantle boundary) and is assumed to be dominated 
by the electronic contribution, according to the resistivity estimates of 
\citet{dekoker2012} and \citet{seagle2013}.
It is also assumed that electron-phonon scattering and electron-electron 
scattering equally contribute to the resistivity of iron \citep{zhang2015}.
In the mantle, $k_{c}$ combines contributions from lattice vibration (phonons), 
radiation, and electrons \citep{hofmeister1999,vandenberg2010}.
In the ice shell, $k_{c}$ only includes lattice vibration, according to the
formulation of \citet{hofmeister1999}.

The coefficient $k_{v}$ is zero if $|\partial T/\partial m|\le|(\partial T/\partial m)_{S}|$.
Otherwise, it is taken from the prescription for the eddy thermal diffusivity of 
\citet[pp.\ 215-230]{abe1995} and \citet{abe1997}
\begin{equation}
 k_{v}=\frac{4\pi}{18}%
           \left(\frac{\rho  g K_{S} R^{2}\alpha^{2} l^{4}_{\mathrm{mix}}}{\gamma\nu}\right)%
           \left[\left(\frac{\partial T}{\partial m}\right)_{S}-\frac{\partial T}{\partial m}\right],
 \label{eq:kv}
\end{equation}
which is based on a modified mixing length theory of thermal convection 
\citep{sasaki1986}. In \cieq{eq:kv}, $l_{\mathrm{mix}}$ plays the role of 
a mixing length \citep{tachinami2011}. Although not indicated above, the 
coefficient $k_{v}$ also includes a prescription for the limit of a vanishing 
viscosity \citep{abe1997}, when 
$\left(\partial T/\partial m\right)_{S} - \partial T/\partial m>%
81\nu^{2}/(16\pi \rho g \alpha R^{2} l^{4}_{\mathrm{mix}})$, in which
$k_{v}$ becomes independent of $\nu$ \citep[pp.\ 215-230]{abe1995}.
Estimates of $\nu=\nu(P,T)$,
when available, are affected by large uncertainties. \citet{stamenkovic2011}
presented a parametrization for the viscosity of perovskite in the diffusion 
creep regime, and argued that it also provides a reasonable approximation 
to the viscosity of the ppv phase. 
That parametrization is applied in the calculations to the entire silicate mantle.
For the Earth test discussed here, the dynamical viscosity ($\nu\rho$) of
the mantle ranges from $\sim 10^{20}$ to $\sim 10^{23}\,\mathrm{Pa\,s}$, 
in accord with the inferred values in the Earth's mantle 
\citep{stamenkovic2011,wagner2012}.
The complex rheology of ice, especially of the low-pressure phases,  
\citep[e.g.,][ and references therein]{barr2009},
is not meant to be fully accounted for in these calculations. For the purpose
of the semi-convection scheme, ice viscosity is assumed to be dominated 
either by diffusion or by dislocation creep, according to the parametrization 
of \citet{durham2001}.

\begin{deluxetable*}{ccclcccc}
\tablecolumns{8}
\tablewidth{0pc}
\tablecaption{Comparison of Measured and Computed Radii of 
Solar System Planets and Satellites\label{table:wgs_test}}
\tablehead{
\colhead{Planet/moon}&\colhead{$M/\Mearth$\tablenotemark{a}}&\colhead{$R/\Rearth$\tablenotemark{a}}&\colhead{$\mathrm{MoI}$\tablenotemark{b}}&\colhead{(Fe,Si,\ice)\%}&\colhead{$\Delta R/R$\tablenotemark{c}}&
\colhead{$\Delta R/R$\tablenotemark{d}}&\colhead{$\Delta\mathrm{MoI}/\mathrm{MoI}$\tablenotemark{d}}
}
\startdata
Mercury \B& $0.05527$ & $0.3829$ &$0.3359$\tablenotemark{i}&$(66.7,33.3,0)$\tablenotemark{i} & 
                                $-2\times 10^{-2}$& $-6\times 10^{-3}$&$-2\times 10^{-2}$\\
Venus    \B& $0.81500$ & $0.9499$ &$0.33$                               &$(23,77,0)$\tablenotemark{iii} & 
                                $\pp9\times 10^{-3}$& $\pp1\times 10^{-2}$&$\pp2\times 10^{-2}$\\
Earth     \B& $1.00000$ & $1.0000$ &$0.3308$\tablenotemark{ii}&$(32.7,67.3,0)$\tablenotemark{i} & 
                                $-8\times 10^{-3}$& $-3\times 10^{-3}$&$-2\times 10^{-2}$\\
Moon     \B& $0.01230$ & $0.2727$ &$0.3931$\tablenotemark{i}&$(2,98,0)$\tablenotemark{iv} & 
                                $\pp4\times 10^{-3}$ & $\pp6\times 10^{-3}$&$\pp3\times 10^{-3}$\\
Mars      \B& $0.10745$ & $0.5320$ &$0.3635$\tablenotemark{i}&$(22,78,0)$\tablenotemark{i} & 
                                $-3\times 10^{-3}$& $\pp2\times 10^{-3}$&$-3\times 10^{-2}$\\
Io           \B& $0.01496$ & $0.2859$ &$0.3768$\tablenotemark{v}&$(15,85,0)$\tablenotemark{v} &
                                $-7\times 10^{-3}$& $-5\times 10^{-3}$&$-2\times 10^{-2}$\\
Europa   \B& $0.00804$ & $0.2450$ &$0.346\pz$\tablenotemark{v}&$(10,80,10)$\tablenotemark{v} &
                                 $\pp3\times 10^{-2}$& $\pp3\times 10^{-2}$&$-5\times 10^{-2}$\\
Ganymede \B& $0.02481$ & $0.4130$ &$0.3105$\tablenotemark{v}&$(10,40,50)$\tablenotemark{v} & 
                                 $\pp5\times 10^{-3}$& $\pp6\times 10^{-3}$&$-1\times 10^{-2}$\\
Callisto   \B& $0.01802$ & $0.3783$ &$0.3549$\tablenotemark{v}&$(0,50,50)$\tablenotemark{v} & 
                                 $\pp7\times 10^{-3}$&$\pp3\times 10^{-2}$&$-1\times 10^{-1}$\\
Titan       \B& $0.02253$ & $0.4041$ &$0.3414$\tablenotemark{vi}&$(0,64,36)$\tablenotemark{vii} & 
                                 $-3\times 10^{-2}$& $-6\times 10^{-3}$&$-8\times 10^{-2}$\\
Triton        & $0.00358$ & $0.2124$ &$0.33\pz$                            &$(0,72,28)$\tablenotemark{viii} &
                                 $-5\times 10^{-2}$& $\pp2\times 10^{-2}$&$-5\times 10^{-2}$
\enddata
\tablenotetext{a}{From \href{http://ssd.jpl.nasa.gov/}{JPL Solar System Dynamics}.}
\tablenotetext{b}{Moment of inertia factor.}
\tablenotetext{c}{Results from simplified structure models.}
\tablenotetext{d}{Results from improved structure models.}
\tablerefs{\citet[i]{sohl2007}; \href{http://nssdc.gsfc.nasa.gov/planetary/}{Lunar \& Planetary Science} (ii);
                \citet[iii]{phillips1983}; \citet[iv]{righter2006};
                \citet[v]{sohl2002}; \citet[vi]{sohl2014b}; \citet[vii]{tobie2005}; \citet[viii]{sotin2004}.}
\end{deluxetable*}

The system of Equations~(\ref{eq:cse_r})-(\ref{eq:cse_ts}) is closed by 
equations of state (EoS) of the type $P=P(\rho,T)$, written as
\begin{equation}
 P(\rho,T)=P(\rho,300\,\K)+\Delta P_{\mathrm{th}}(\rho,T).
 \label{eq:PRT}
\end{equation}
The first term on the right-hand side is a room-temperature EoS, listed in 
Table~\ref{table:eos}, whereas the second term accounts for thermal 
corrections. 
The last entry in Table~\ref{table:eos} represents the revised release of 
the 1995 EoS for ordinary water from the International Association for 
the Properties of Water and Steam \citep[IAPWS,][]{wagner2002}, 
which already includes temperature dependence and therefore does 
not apply the second term on the right-hand side of \cieq{eq:PRT}.
The EoS in Table~\ref{table:eos} are valid up to pressures of at most
a few to several times $100\,\mathrm{GPa}$ 
\citep[e.g.,][]{seager2007,wagner2011}, which are easily exceeded 
in the deep interiors of super-Earths. (The inferred pressure at the center 
of the Earth is $\approx 360\,\mathrm{GPa}$, e.g., \citeauthor{prem1981}
\citeyear{prem1981}.)
Therefore, following \citet{seager2007}, each EoS is extended by 
extrapolation to larger pressures until they intersect the zero-temperature 
EoS of \citet{zapolsky1969}, which is based on the augmented formulation 
of the Thomas-Fermi-Dirac (TFD) theory of \citet{salpeter1967}.
In the high-pressure regime, where the TDF EoS is employed, 
finite-temperature corrections are not expected to be important at the
densities and temperatures encountered in these calculations 
\citep[e.g.,][]{cowan1957,decarvalho2014,boshkayev2016}.
It should be pointed out that the Generalized Rydberg EoS applied 
here to the high-pressure phases of iron and silicate is especially 
well suited to extrapolation at large pressures 
\citep[see, e.g.,][]{stacey2008,wagner2011}. 
All of the relevant EoS functions in Table~\ref{table:eos} intersect 
the corresponding TDF EoS.

In general, $\Delta P_{\mathrm{th}}$ in \cieq{eq:PRT} is a correction 
based on quasi-harmonic lattice vibration, according to 
the Mie-Gr\"{u}neisen-Debye theory 
\citep[e.g.,][]{anderson1989,jackson1996,stacey2008}
\begin{equation}
 \Delta P_{\mathrm{th}}=\gamma\rho\left[E_{\mathrm{th}}(\rho,T)%
                                     -E_{\mathrm{th}}(\rho,300\,\K)\right],
 \label{eq:DPth}
\end{equation}
where the specific internal energy is
\begin{equation}
 E_{\mathrm{th}}(\rho,T)=9n\left(\frac{k_{\mathrm{B}} T}{\mu m_{\mathrm{H}}}\right)%
   \left(\frac{T}{\theta_{\mathrm{D}}}\right)^{3}%
   \int_{0}^{\theta_{\mathrm{D}}/T} \frac{\xi^{3}d\xi}{e^{\xi}-1},
 \label{eq:Eth}
\end{equation}
in which the Debye temperature $\theta_{\mathrm{D}}$ is related
to the Gr\"{u}neisen parameter by 
\begin{equation}
 \gamma=\left(\frac{\partial \ln\theta_{\mathrm{D}}}{\partial \ln\rho}\right)_{\!T}.
 \label{eq:gruen}
\end{equation}
In the high-temperature limit, i.e., for liquids, $\theta_{\mathrm{D}}/T\ll 1$ 
\citep{stixrude1990} and \cieq{eq:Eth} becomes 
$E_{\mathrm{th}}(T)\approx 3 n k_{\mathrm{B}} T/(\mu m_{\mathrm{H}})$.
Typically $\theta_{\mathrm{D}}$ is derived via integration of \cieq{eq:gruen} 
once a suitable expression for $\gamma$ is known \citep[e.g.,][]{altshuler1987}. 
Here $\gamma$, and hence $\theta_{\mathrm{D}}$, is typically a function 
of $\rho$ only and its dependence on $T$ is neglected 
\citep[see][]{anderson1995}. 
However, for the IAPWS EoS, a fit to the specific heat at constant volume, 
$C_{V}=C_{V}(\rho,T)$, is also made available. In this case, the 
Gr\"{u}neisen parameter is calculated as 
$\gamma=\alpha K_{T}/(\rho C_{V})$, and hence depends on both
$\rho$ and $T$ ($\alpha$ and $K_{T}$ are directly calculated from the EoS).
Note that, in \cieq{eq:Eth}, the mean molecular weight of the substance, 
$\mu$, is expressed in units of the hydrogen mass, $m_{\mathrm{H}}$,
and that $\mu m_{\mathrm{H}}/n$ times the Avogadro's number is 
the mean atomic molar mass.
In the metallic core, $\Delta P_{\mathrm{th}}$ also accounts for
anharmonic vibration and electronic corrections \citep{dewaele2006,ichikawa2014}.

Table~\ref{table:eos} provides the sources for most of the the data 
needed in the structure calculations. At pressures of $P\gg 100\,\mathrm{GPa}$, 
the values of most thermodynamics functions are unknown. 
Uncertainty also affects their behavior at lower pressures but 
at  temperatures of $T\gg 1000\,\K$. Given the lack of information, 
the functions are simply extrapolated both in $P$ and $T$, as needed.

Since ex situ models of \kep\ planets contain large amounts of \ice,
whose high-pressure behavior has been studied via both
experiments and quantum molecular dynamics computations, 
the \ice\ EoS was compared to some available data.
(The IAPWS EoS, applied at $P\lesssim 1\,\mathrm{GPa}$, is 
a many-parameter fit to an extensive body of data,
\citeauthor{wagner2002} \citeyear{wagner2002}.)
In the pressure range between $\approx 50$ and 
$\approx 4\times 10^{3}\,\mathrm{GPa}$, where the transition 
to the TFD EoS occurs, the cold EoS of the high-pressure \ice\ 
phases differs by only a few percent or less from 
the density-functional theory data reported by \citet{seager2007}.
\cieq{eq:PRT} also reproduces within a few percent margin the 
liquid water, temperature-dependent EoS of \citet{abramson2004}.
In more extreme regimes, \citet{french2009} reported on 
computations of \ice\ EoS that account for temperature dependence 
at high pressures. 
The finite-temperature correction scheme applied here produces 
densities that typically agree within $\approx 5$\% of 
\citeauthor{french2009}'s tabled data for all pressures
(up to $10^{4}\,\mathrm{GPa}$) and  temperatures 
(up to $2.4\times 10^{4}\,\K$).

At high pressures ($P\gtrsim 50\,\mathrm{GPa}$) and temperatures 
($T\gtrsim 1500\,\K$) \ice\ should transition to a superionic phase, 
in which oxygen atoms remain on fixed sites in a lattice while hydrogen 
atoms can diffuse through the lattice, behaving fluid-like \citep{redmer2011}. 
Superionic phases separate fully liquid and fully solid \ice\ in the $P$-$T$ 
diagram \citep[e.g.,][]{wilson2013}.
The phase space of superionic \ice\ was identified according to 
the diagrams of \citet{wilson2013} and \citet{sun2015}. 
In superionic phases, it is simply assumed that the EoS of liquid \ice\ 
applies (which, as mentioned above, does reproduce available data 
at high pressure/temperature). 

\begin{figure*}[]
\centering%
\resizebox{0.85\linewidth}{!}{%
\includegraphics[clip]{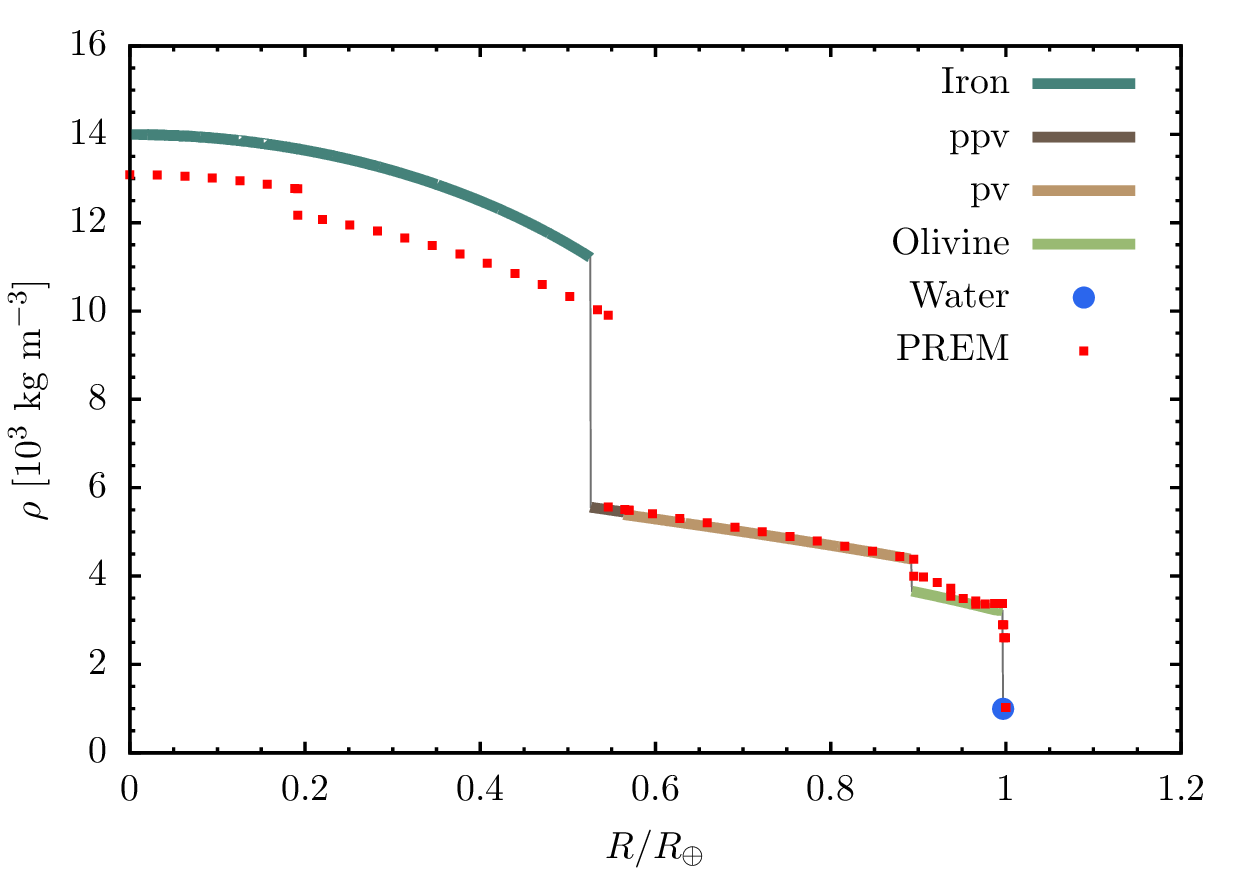}%
\includegraphics[clip]{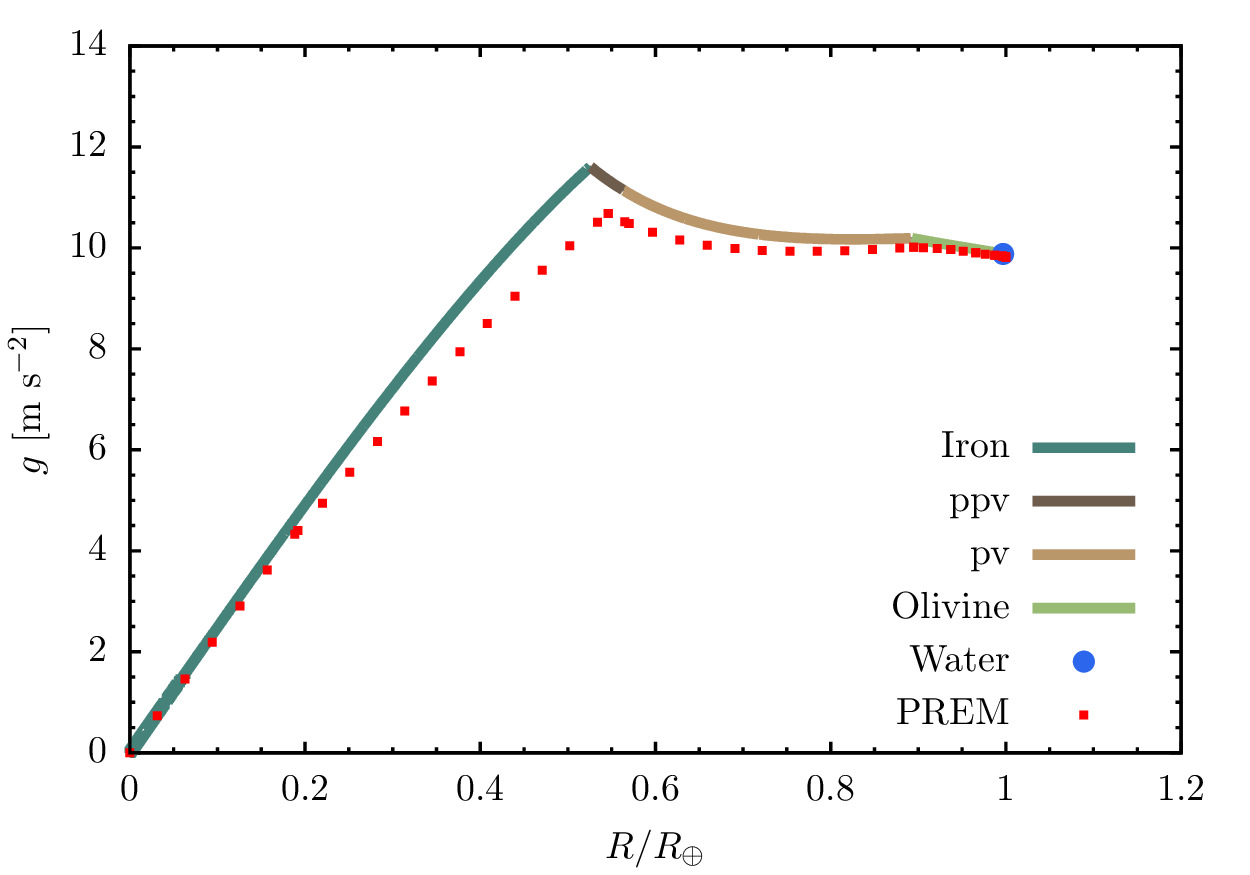}}
\resizebox{0.85\linewidth}{!}{%
\includegraphics[clip]{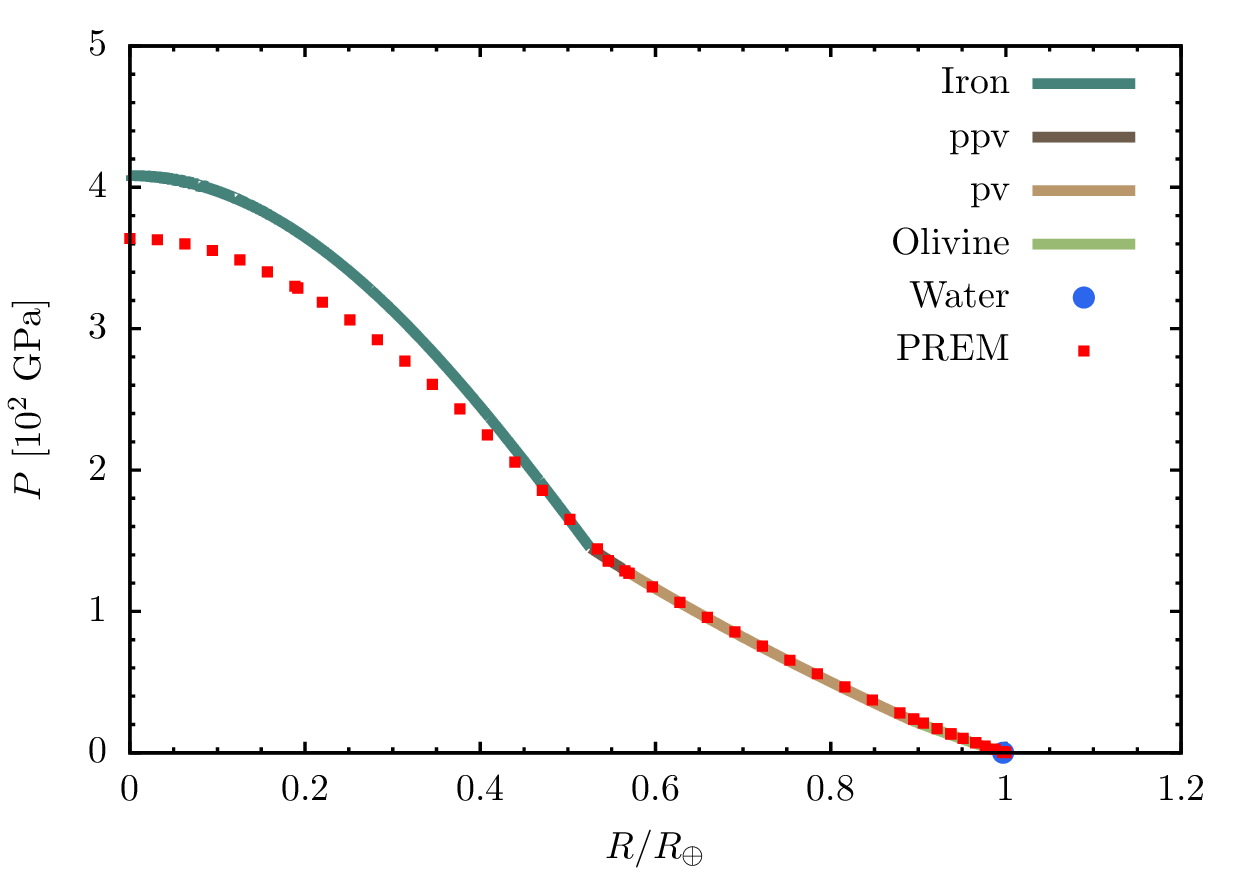}%
\includegraphics[clip]{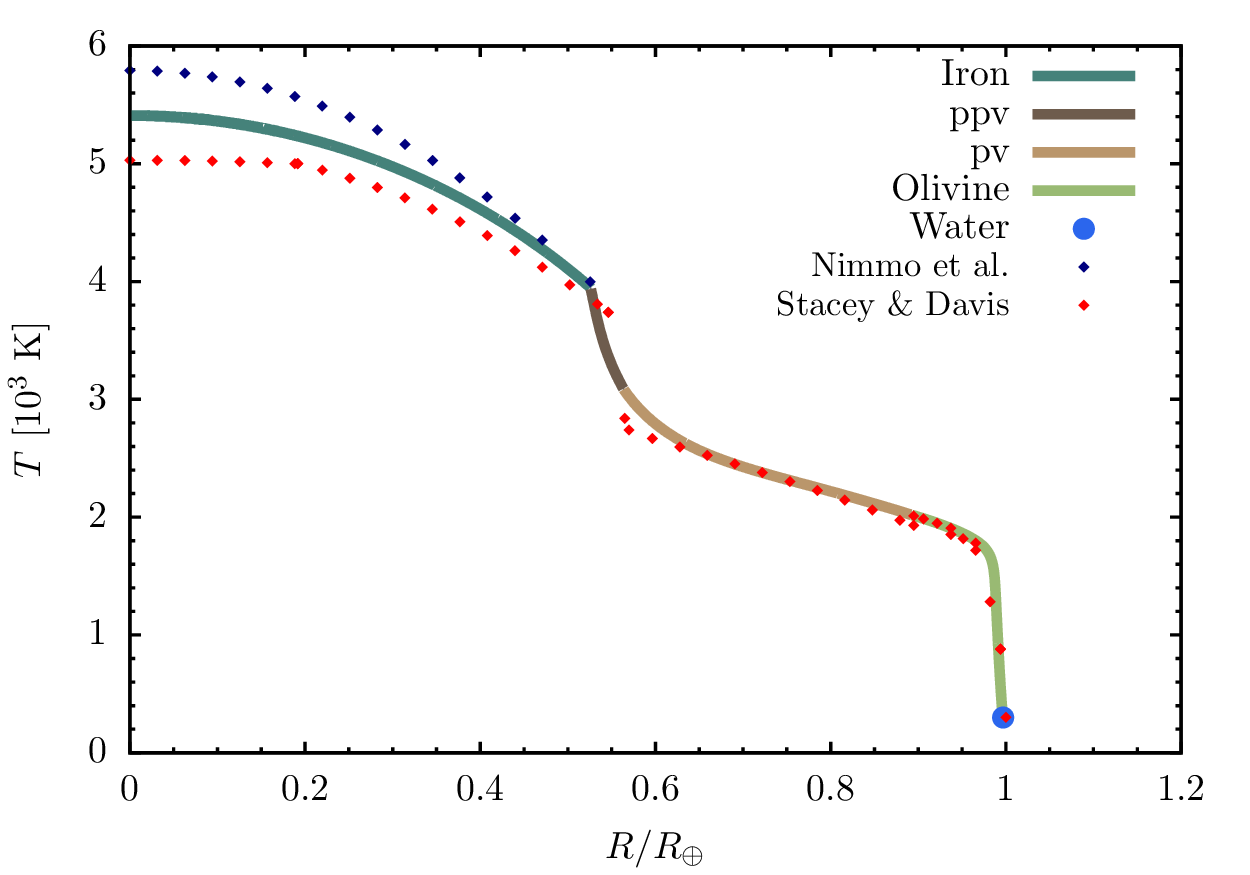}}
\caption{%
             Thermal structure calculation of the Earth's interior compared to the 
             density (top-left), gravitational acceleration (top-right), and pressure 
             (bottom-left) of the Preliminary Reference Earth Model 
             \citep[PREM,][]{prem1981}. The temperature stratification (bottom-right)
             is compared to the geotherm of \citet{stacey2008} and to the core
             adiabat of \citet{nimmo2004}. See the text for additional details.
             }
\label{fig:earth}
\end{figure*}

The structure Equations~(\ref{eq:cse_r})-(\ref{eq:cse_ts}), along with
\cieq{eq:PRT} for each layer, are integrated from $m=0$ to $m=M_{c}$ 
with boundary conditions $R(0)=0$, $P(0)=P_{0}$, $Q(0)=0$, $T(0)=T_{0}$,
$P(M_{c})=P_{c}$, and $T(M_{c})=T_{c}$. The quantities $P_{0}$ and
$T_{0}$ are the central pressure and temperature of the planet, 
whereas $P_{c}$ and $T_{c}$ are the pressure and temperature at 
the surface $R=R_{c}$ of the condensed part, i.e., at the bottom of the 
gaseous envelope, both provided by the envelope structure calculations.
The set of equations is numerically solved by means of a suite of 
implicit/explicit algorithms. In the case of a stiff problem, 
an algorithm based on backward differentiation formulas 
\citep[i.e., the Gear method,][]{gear1971} is employed. 
If the problem is non-stiff, the Adams-Moulton method \citep{hairer1993}
is used instead. If implicit methods encounter difficulties, a variable 
high-order extrapolation algorithm of variable step-size, based on 
the Gragg-Bulirsch-Stoer method \citep{hairer1993}, is used.
The marching step ($\Delta m$) is self-adaptive and is constrained to be 
smaller than the smallest of $|\partial \ln Y/\partial m|^{-1}$, computed locally
for $Y=(R,Q,P,T)$.
The integration is repeated, applying a search algorithm in order to adjust 
$P_{0}$ and $T_{0}$, until $P(M_{c})$ and $T(M_{c})$ are within 
a few percent ($\lesssim 10$\% for thermal calculations) of the boundary 
conditions $P_{c}$ and $T_{c}$.

\begin{deluxetable*}{cccccccc}
\tablecolumns{8}
\tablewidth{0pc}
\tablecaption{Structure Properties of the Condensed Part of \kep\ Planets\label{table:kep_comp}}
\tablehead{
\colhead{Planet}&\colhead{$M_{c}/\Mearth$}&\colhead{(Fe,Si,\ice)\%}&\colhead{$R_{c}/\Rearth$\tablenotemark{a}}&\colhead{$\mathrm{MoI}$\tablenotemark{a}}&\colhead{$R_{c}/\Rearth$\tablenotemark{b}}&
\colhead{$\mathrm{MoI}$\tablenotemark{b}}&\colhead{$Q_{c}/Q_{\oplus}$\tablenotemark{b}}
}
\startdata
b           & $2.10$ & $(10.6,50.6,38.8)$ & $1.47$ &$0.300$ & 
                                $1.55$& $0.281$&$0.50$\\
c           & $4.56$ & $(6.6,46.5,46.9)$ & $1.84$ &$0.306$ & 
                                $1.87$& $0.301$&$0.69$\\
d           & $5.58$ & $(6.0,46.0,48.0)$ & $1.93$ &$0.310$ & 
                                $1.94$& $0.308$&$0.77$\\
e           & $6.90$ & $(5.7,45.7,48.6)$ & $2.00$ &$0.316$ & 
                                $2.01$& $0.316$&$0.90$\\
f           & $2.74$ & $(6.1,46.1,47.8)$ & $1.62$ &$0.310$ & 
                                $1.62$& $0.310$&$0.53$\\
g           & $5.57$ & $(5.0,45.0,50.0)$ & $1.92$ &$0.315$ & 
                                $1.93$& $0.314$&$0.75$
\enddata
\tablenotetext{a}{Simplified structure models.}
\tablenotetext{b}{Improved structure models.}
\end{deluxetable*}

The core structure calculations used for the ex situ models, discussed 
in \cisec{sec:csc}, are simplified in that they use only mass continuity, 
\cieq{eq:cse_r}, and hydrostatic equilibrium, \cieq{eq:cse_p}, to compute 
$R_{c}$, imposing the conditions $\partial T/\partial m=\partial Q/\partial m=0$
($\Delta P_{\mathrm{th}}=0$). Moreover, they assume that the iron core 
is entirely made of $\epsilon$-Fe and the \ice\ shell of ice VII-X.
Here we intend to check how the improved structures (i.e., with 
thermodynamics and additional material phases) compare to those 
adopted to construct ex situ models.

Table~\ref{table:wgs_test} reports test results from the structures 
of planets and satellites of the solar system, both in the simplified 
and improved version.
Comparisons are carried out for the radius and the moment of
inertia factor (MoI).
The simplified models reproduce quite accurately the radii of these 
bodies. In the tests, as expected, the thermal structure of the 
improved models provides only minor (and sometimes minute) 
adjustments to the radius. Nonetheless, the results indicate that 
the thermal models produce reasonable interior structures.
Clearly, the adopted compositional partition into two/three layers
represents the single major factor determining the radius and the 
MoI of these bodies.
Of the simulated bodies, it should be noted that the iron mass 
fraction of the Moon is quite uncertain, yet both models provide 
a iron core radius of $\approx 350\,\mathrm{km}$, in agreement 
with detection from seismic analysis \citep{weber2011}. 
The MoI of Venus and Triton are undetermined, and the value
of $0.33$ in Table~\ref{table:wgs_test} is assumed as
representative of fully differentiated bodies \citep[e.g.,][]{chen2014}. 
The largest discrepancies are obtained for the MoI of Callisto 
and Titan, which is likely due to the adopted compositions and 
complete differentiation assumption. Callisto, for example, may 
have an intermediate silicate/ice mixed layer atop the silicate 
core \citep{sohl2002}.
None of the models in Table~\ref{table:wgs_test} predicts the
presence of the low-pressure $\alpha$-Fe phase, whereas 
$\gamma$-Fe is predicted in Mercury, Mars, and the satellites.
Therefore, only the high-pressure and liquid phases of iron are 
expected in the condensed part of the \kep\ planets discussed 
above.

\section{An Earth's Model}
\label{sec:EM}

\cifig{fig:earth} displays a more detailed comparison of the 
improved structure model of the Earth (solid line), compared to 
the Preliminary Reference Earth Model \citep[PREM,][]{prem1981} 
and to the geotherm of \citet{stacey2008}, represented as red 
squares and diamonds respectively (see the legend).
The structure calculation in the figure also includes an ``ocean'' 
layer (see the top-left panel), $0.0234$\% water by mass \citep{stacey2008}.
The agreement is generally good, with the largest differences 
confined to the metallic core.

Mantle semi-convection successfully reproduces the temperature
gradient in the Earth's outer layers \citep[see also][]{turcotte2014} 
and in the mantle in general.
The temperature at the CMB is $\approx 3900\,\K$, in accord with 
current estimates \citep{alfe2007,kamada2012}.
The discrepancies in pressure and density in the core are expected
\citep[see also][]{valencia2006,wagner2011} since only pure iron is 
considered and the presence of lighter elements, such as H, O, Si, 
and S \citep[e.g.,][pp.\ 117-147]{alfe2007,vocadlo2015}, is 
neglected\footnote{It is also estimated that the Earth's core may 
contain up to about $10$\% Ni by mass \citep[pp.\ 117-147]{vocadlo2015}.}.
These impurities are believed to reduce the core density by roughly 
$7$\% relative to that of pure iron, at the same conditions of pressure
and temperature \citep{alfe2007}. The inner part of the metallic core 
is believed to be less polluted than is the outer (liquid) part.
The calculated core radius is $3.7$\% 
smaller than predicted by the PREM, $R=3480\,\mathrm{km}$. 
The absence of iron alloyed with lighter elements also prevents 
the formation of an outer molten core. In fact, the melting temperature 
of Fe at the CMB pressure of $\approx 135\,\mathrm{GPa}$ is around 
$4200\,\K$ \citep{anzellini2013}, whereas the melting temperature
at that pressure of, e.g., iron sulfide (FeS) is around $3200\,\K$ 
\citep{anderson1996}.
A calculation conducted with the Vinet EoS for $\epsilon$-Fe
of \citet{anderson2001} yields core pressures, temperature, and 
densities that differ by $\lesssim 1$\% from those in \cifig{fig:earth}.

Core temperatures are  quite uncertain, with suggested maximum 
values ranging up to $\approx 6000\,\K$, and maybe above 
\citep{sola2009,anzellini2013}.
As a reference, the bottom-right panel also shows the core adiabat of 
\citet[][]{nimmo2004}, anchored at the CMB temperature.
Again, the general agreement indicates that the thermal model performs
reasonably well.

\section{Thermal Structures of the Solid Interiors of Kepler 11 Planets}
\label{sec:cot}

\begin{figure*}[]
\centering%
\resizebox{\linewidth}{!}{%
\includegraphics[clip]{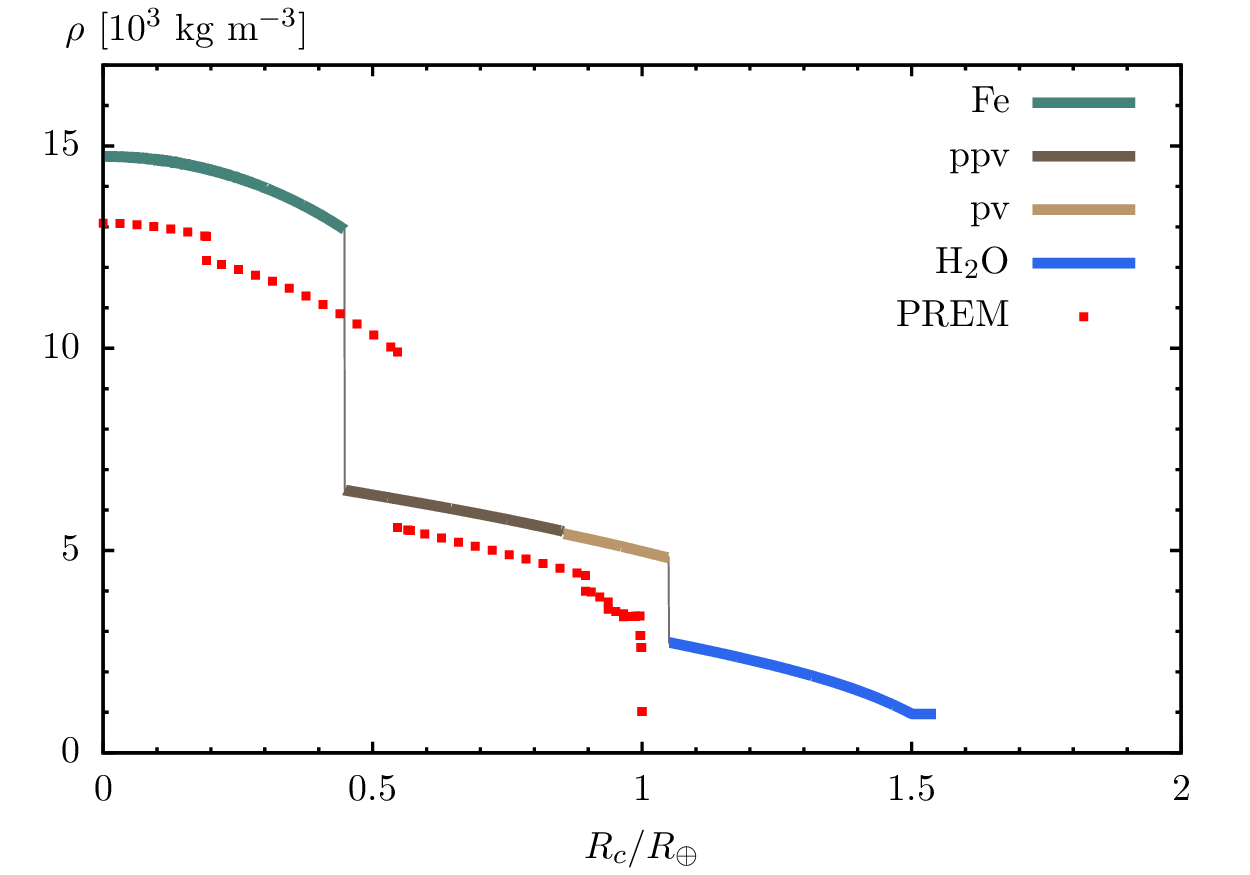}%
\includegraphics[clip]{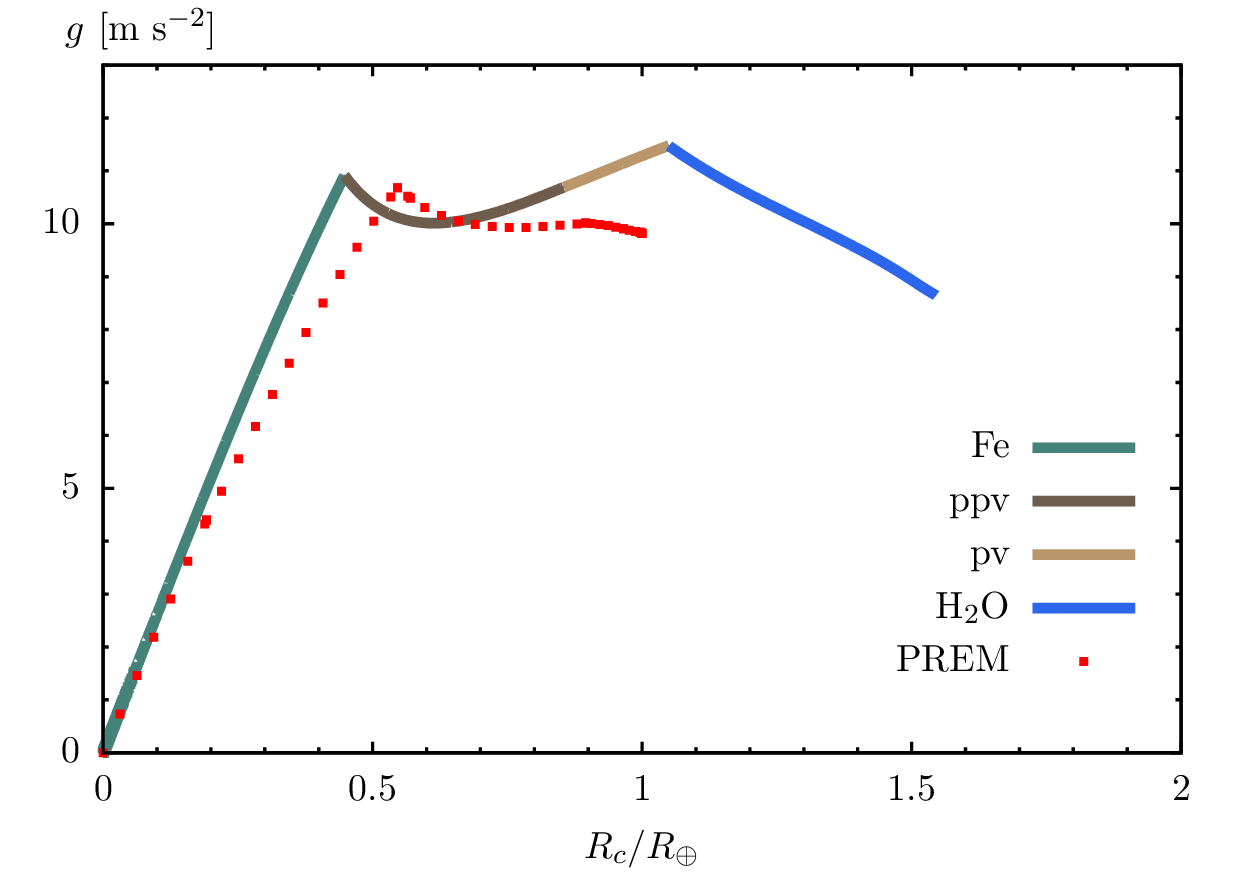}%
\includegraphics[clip]{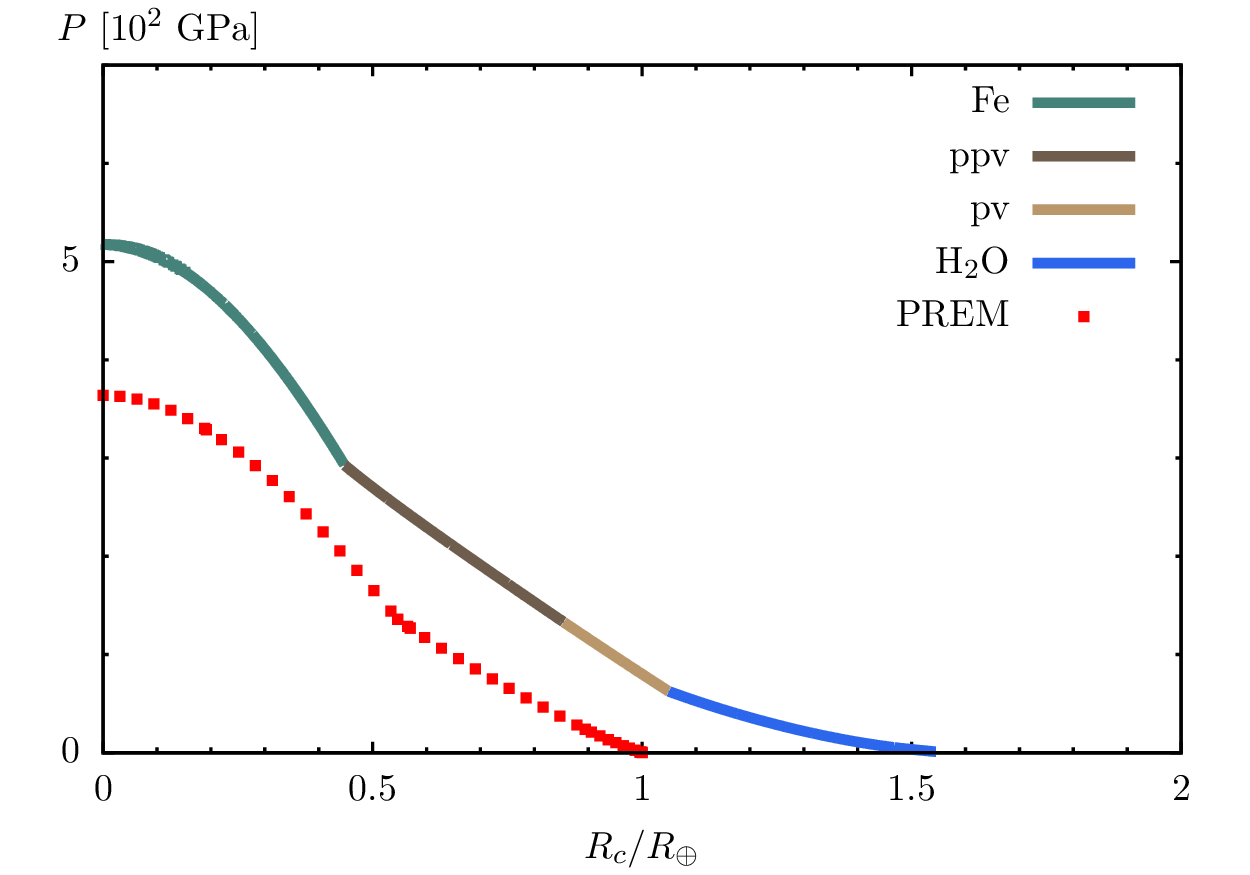}%
\includegraphics[clip]{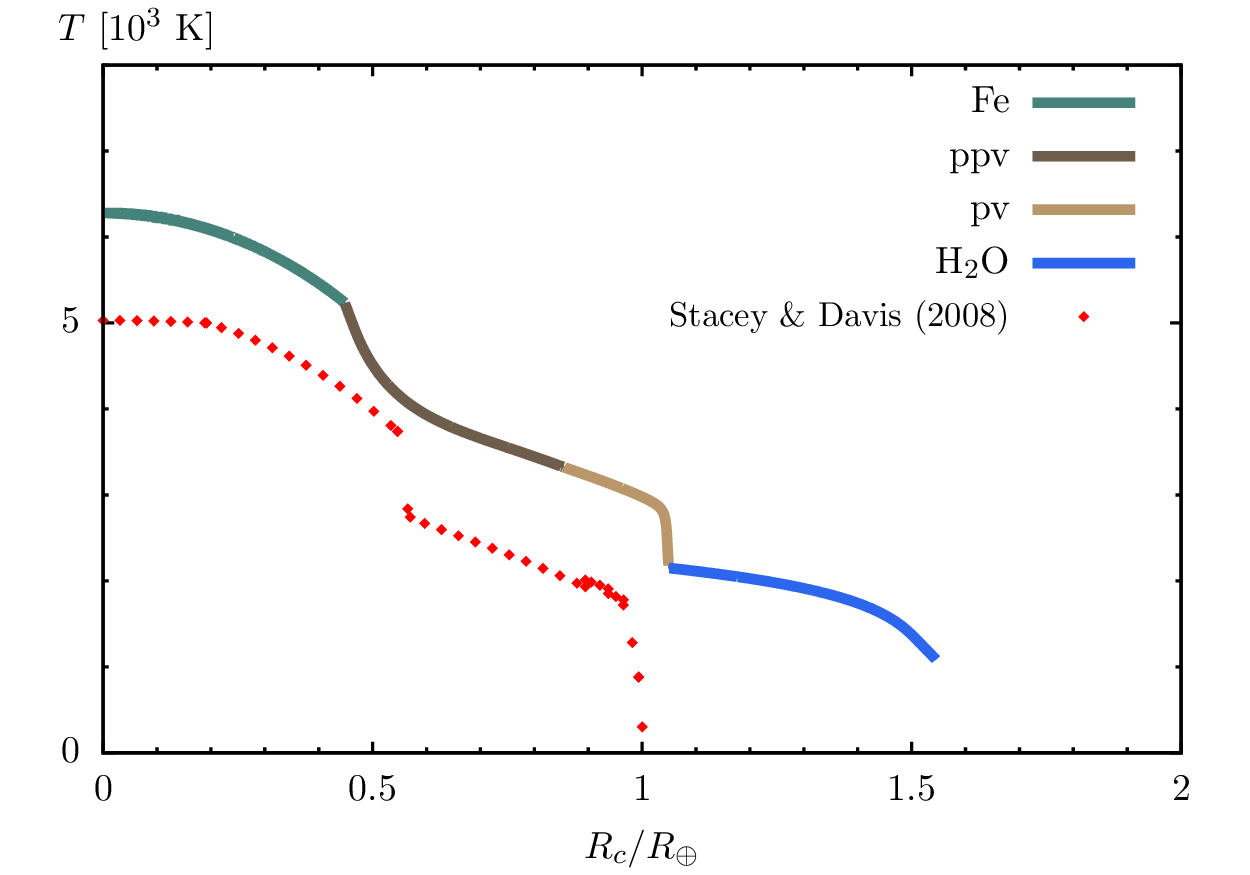}}
\resizebox{\linewidth}{!}{%
\includegraphics[clip]{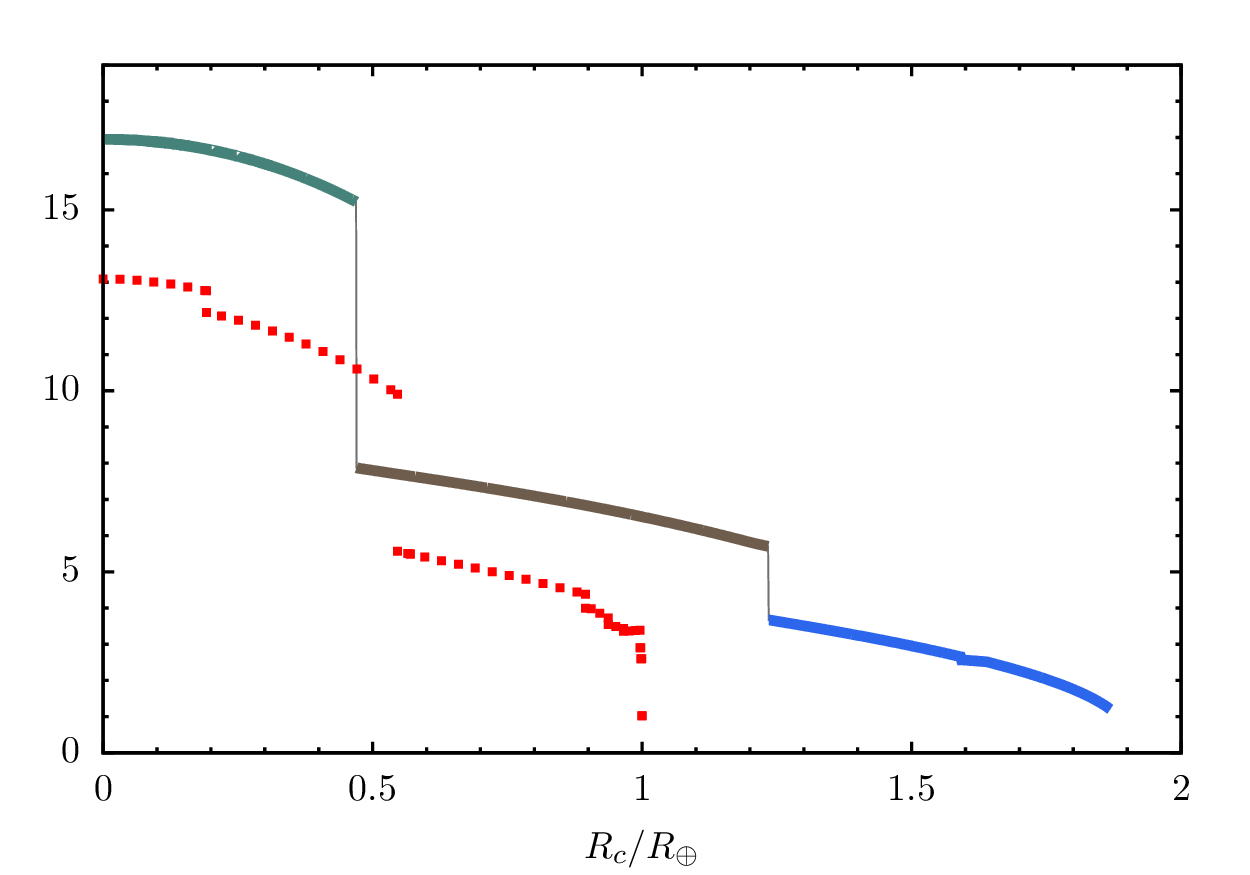}%
\includegraphics[clip]{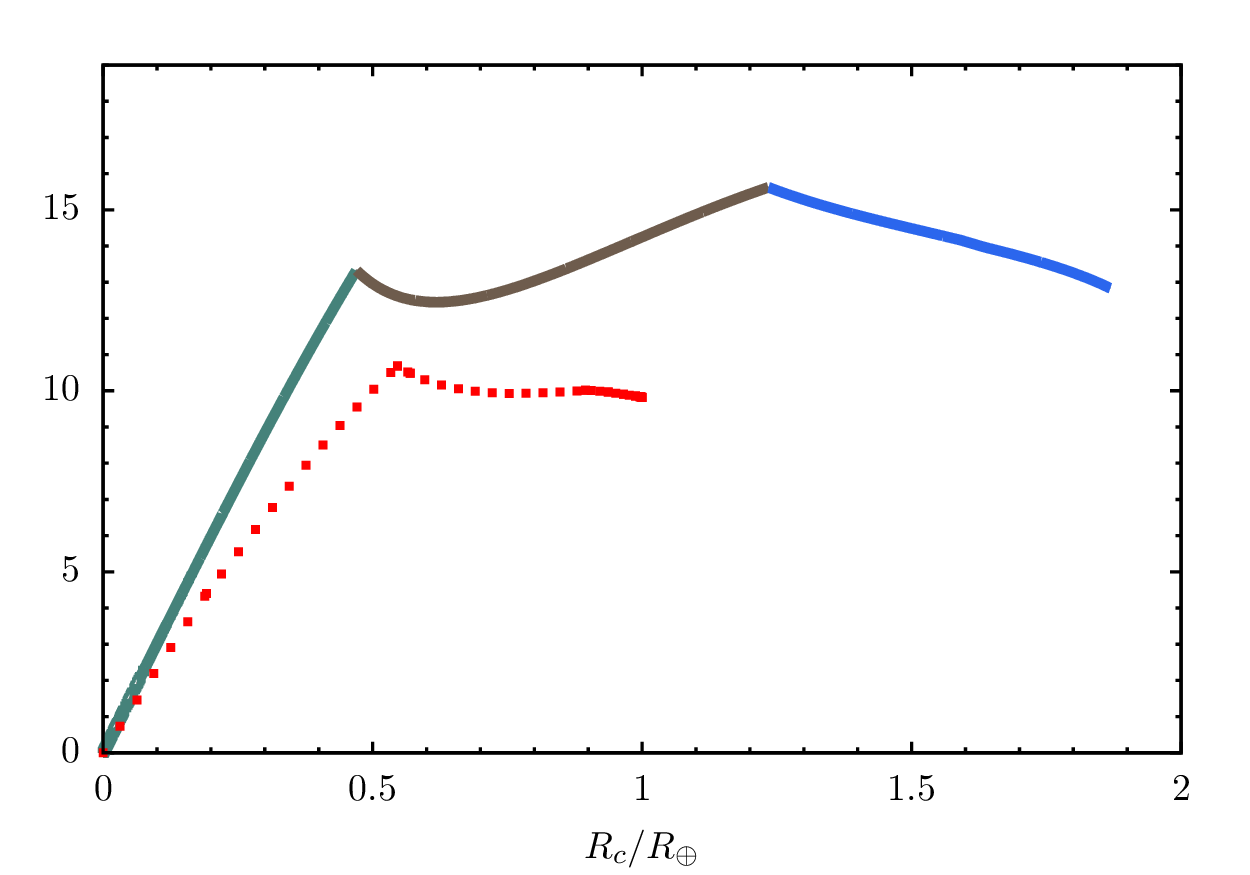}%
\includegraphics[clip]{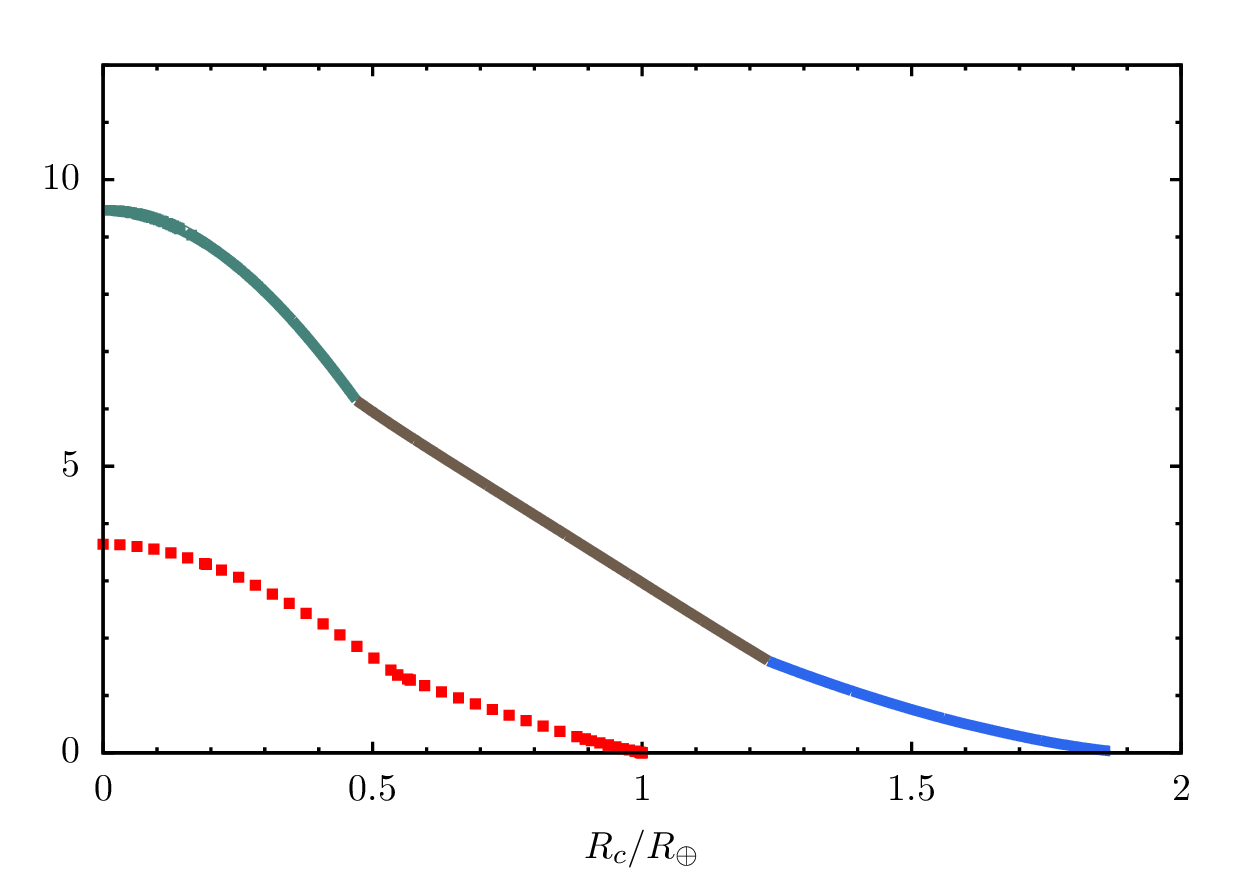}%
\includegraphics[clip]{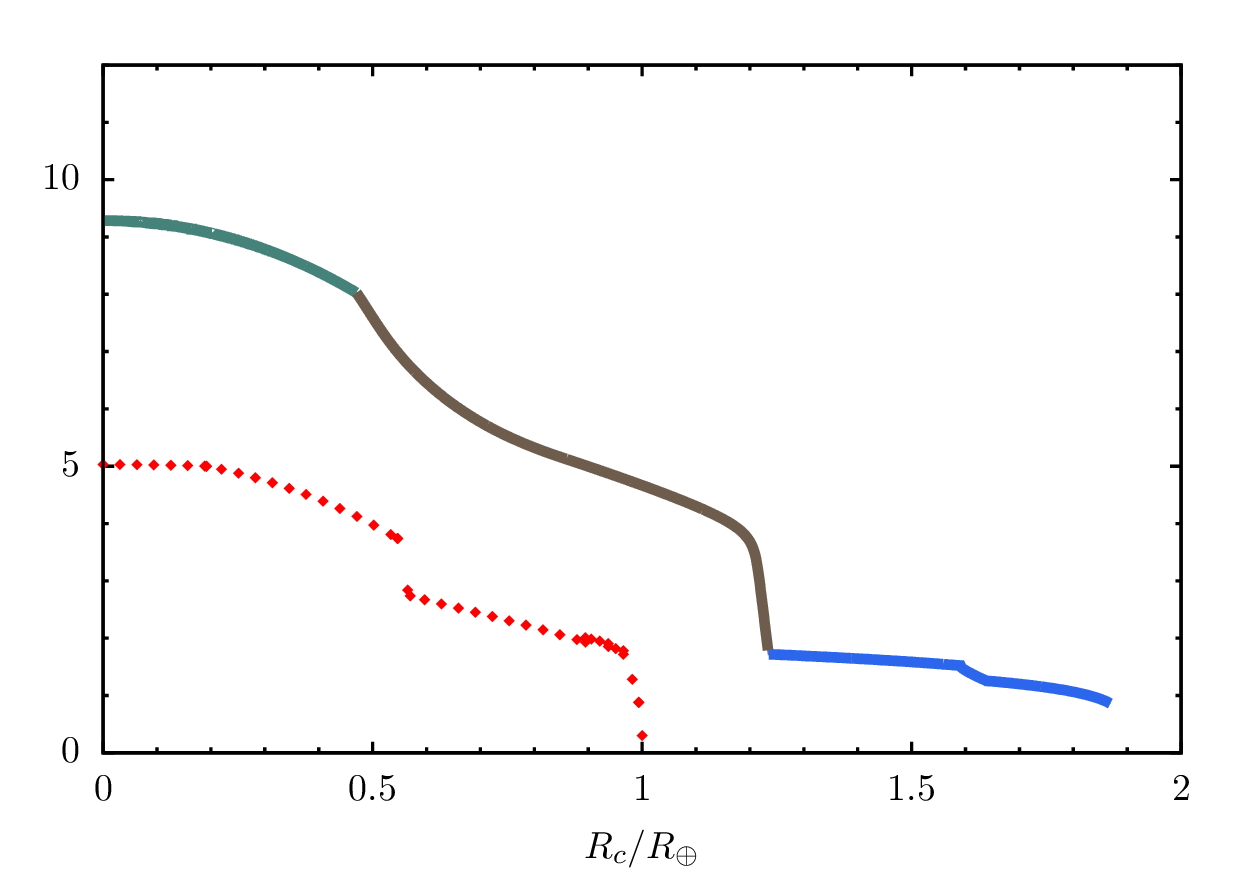}}
\resizebox{\linewidth}{!}{%
\includegraphics[clip]{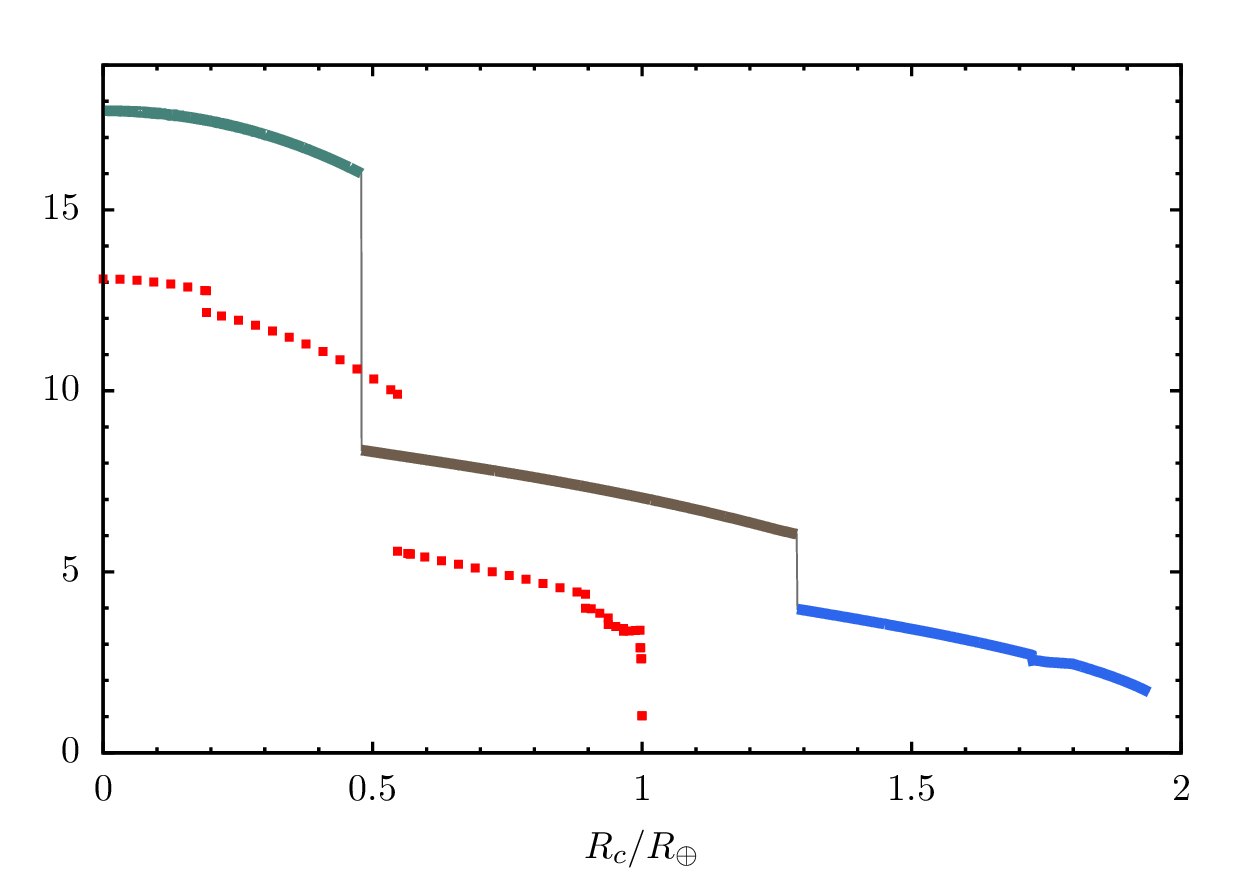}%
\includegraphics[clip]{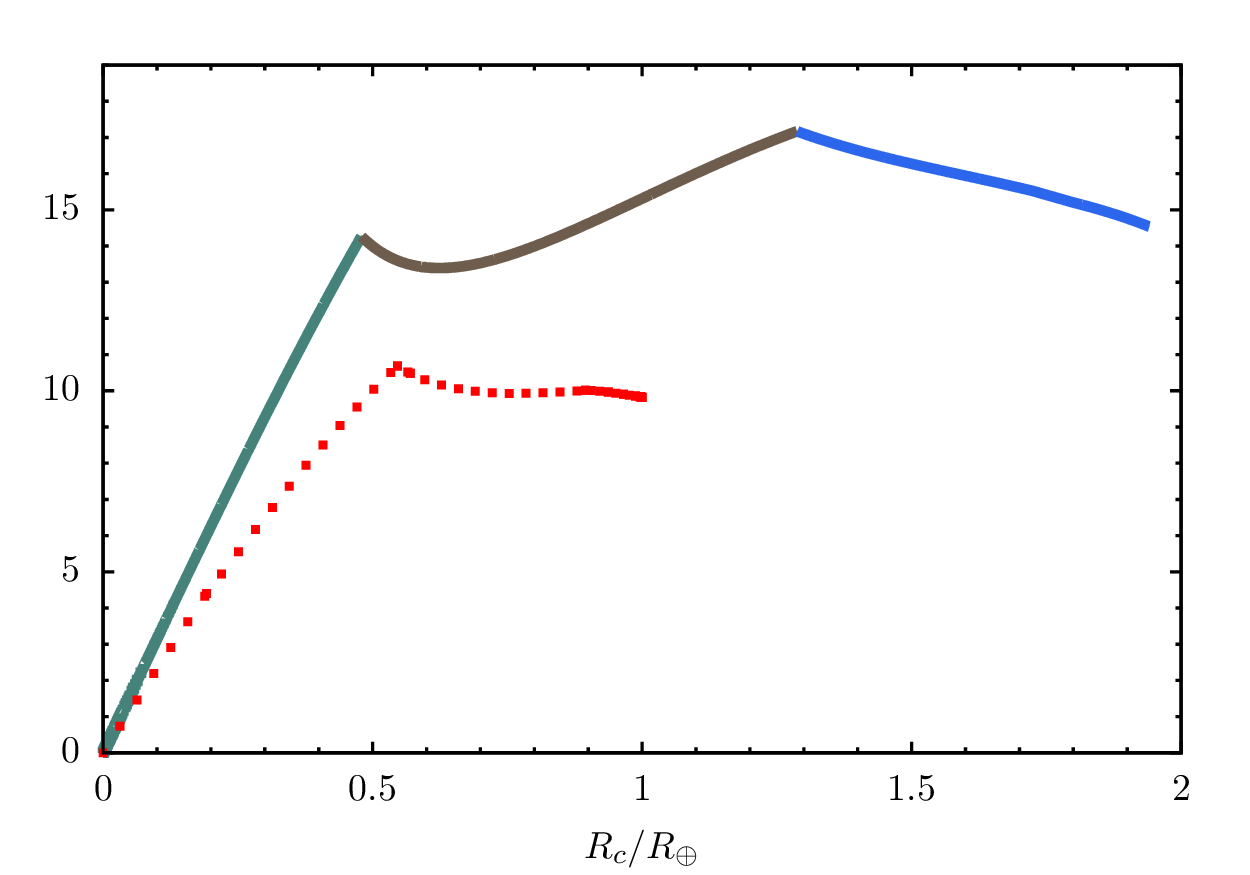}%
\includegraphics[clip]{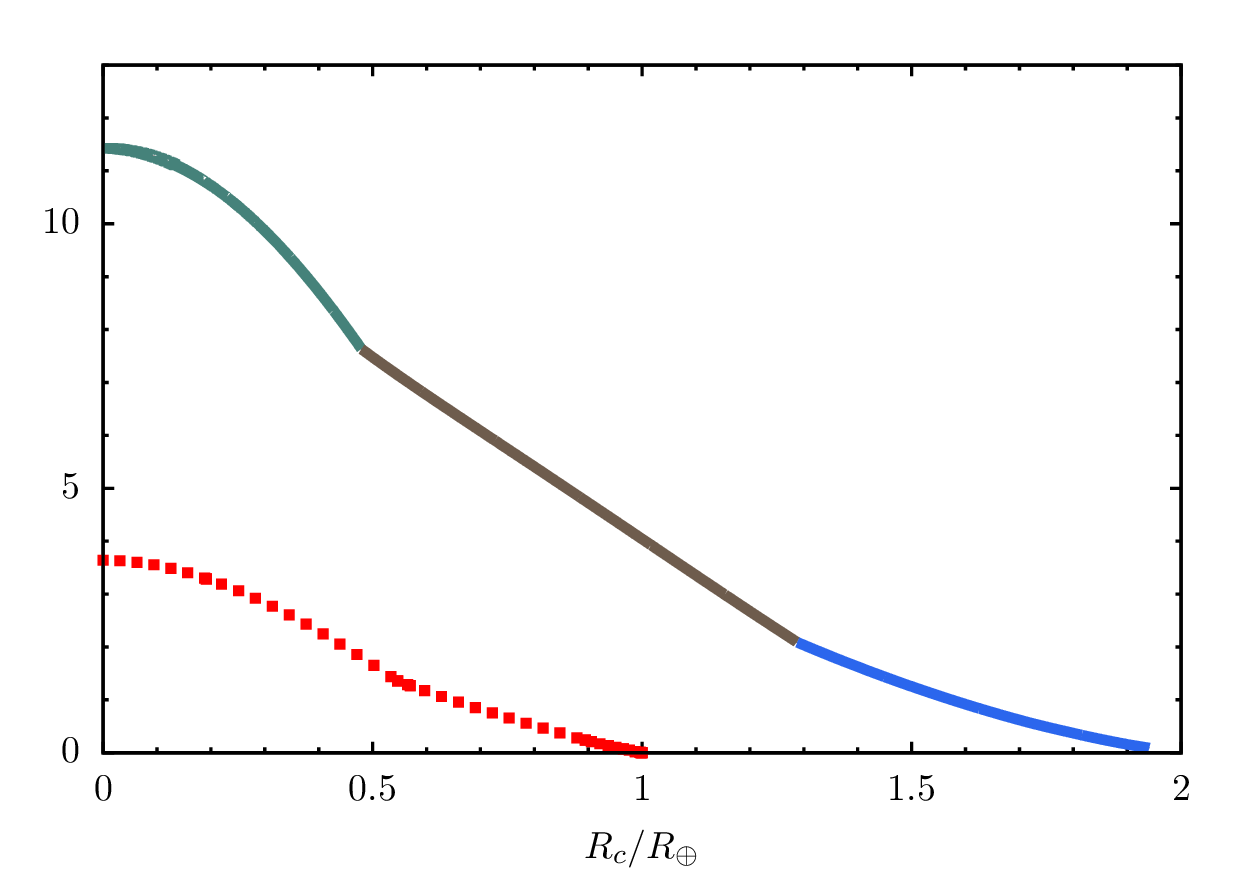}%
\includegraphics[clip]{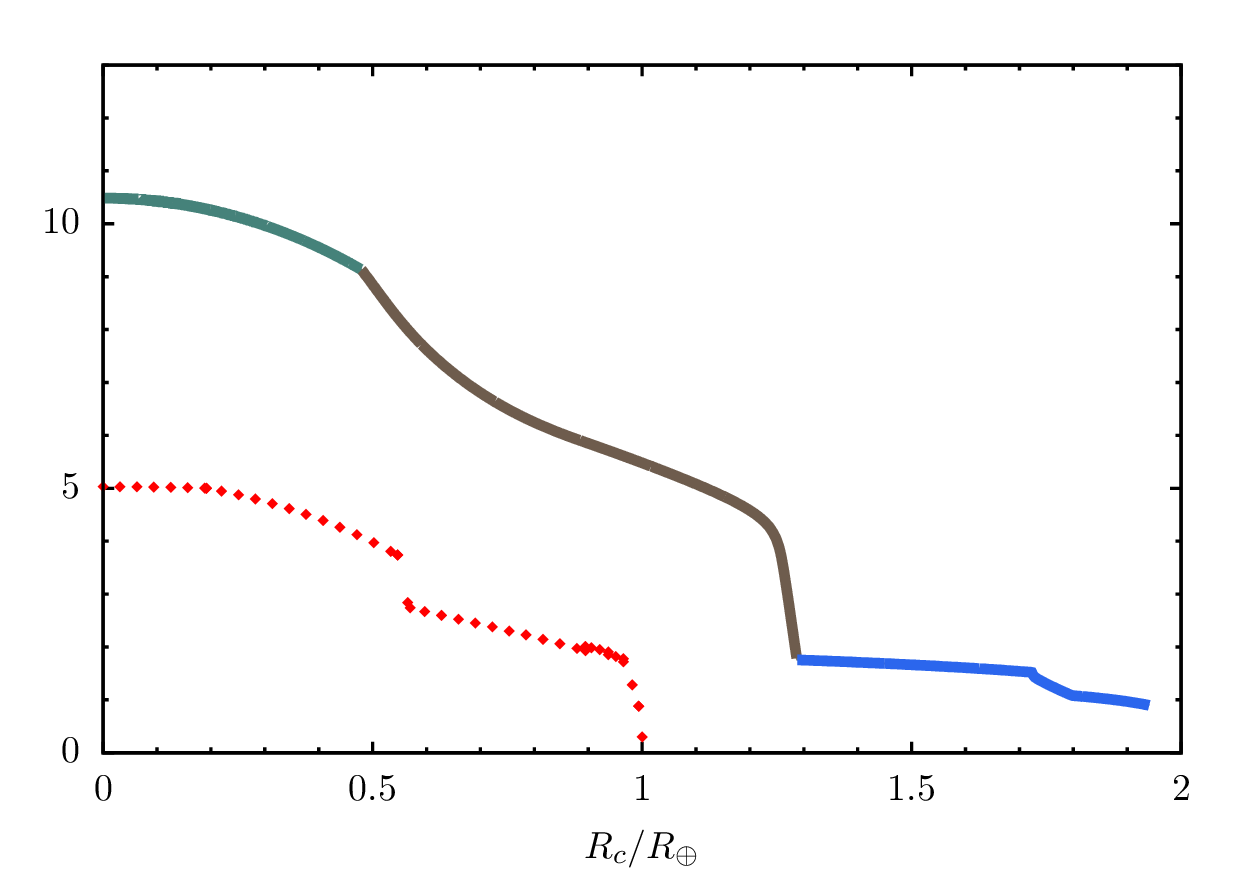}}
\resizebox{\linewidth}{!}{%
\includegraphics[clip]{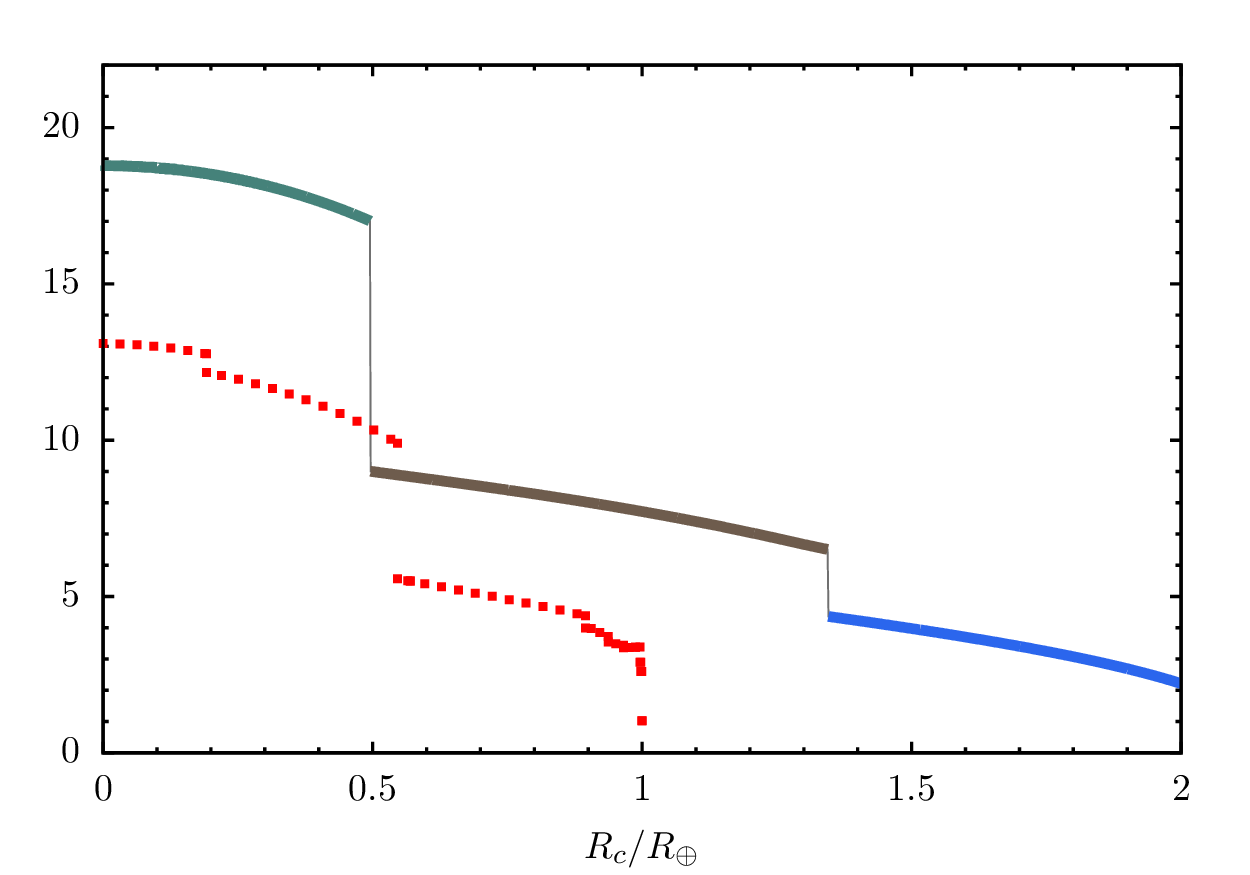}%
\includegraphics[clip]{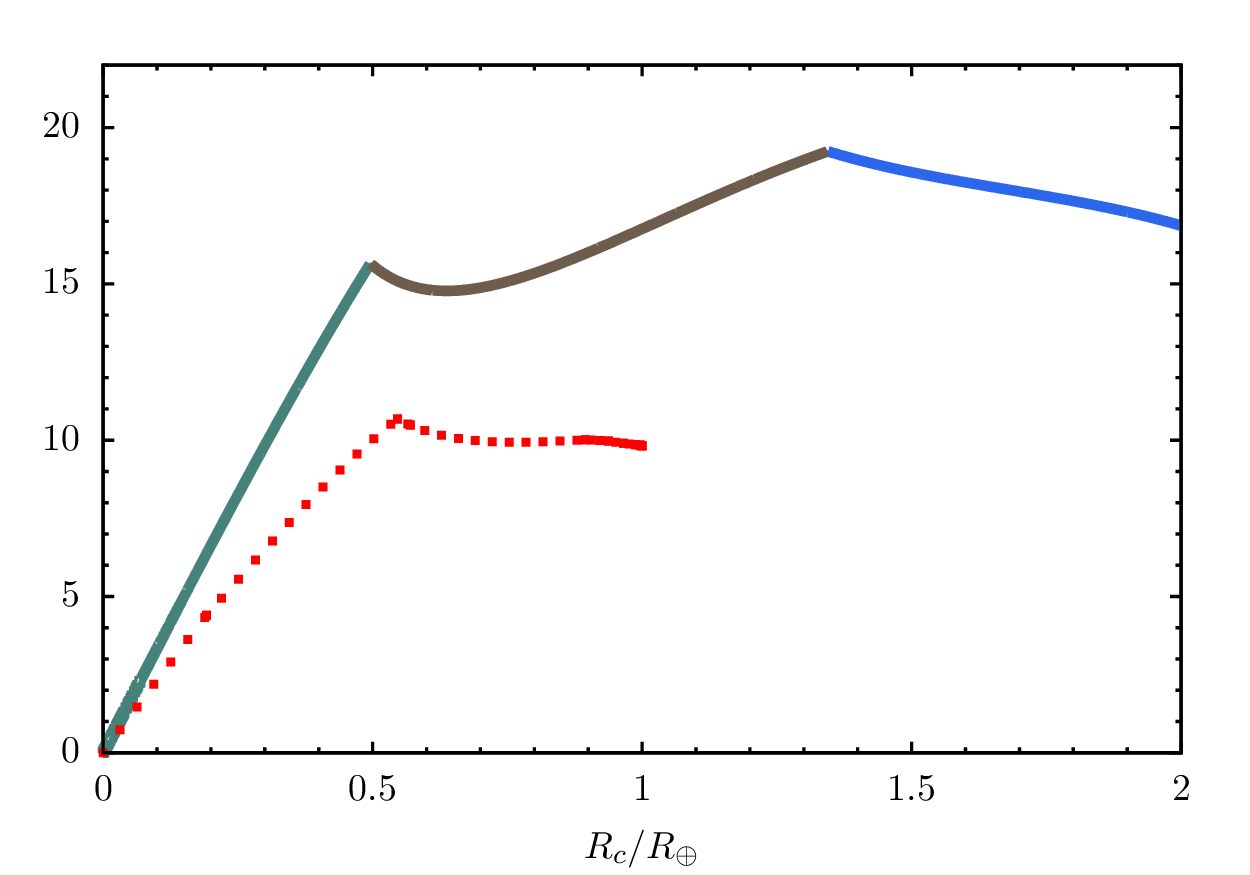}%
\includegraphics[clip]{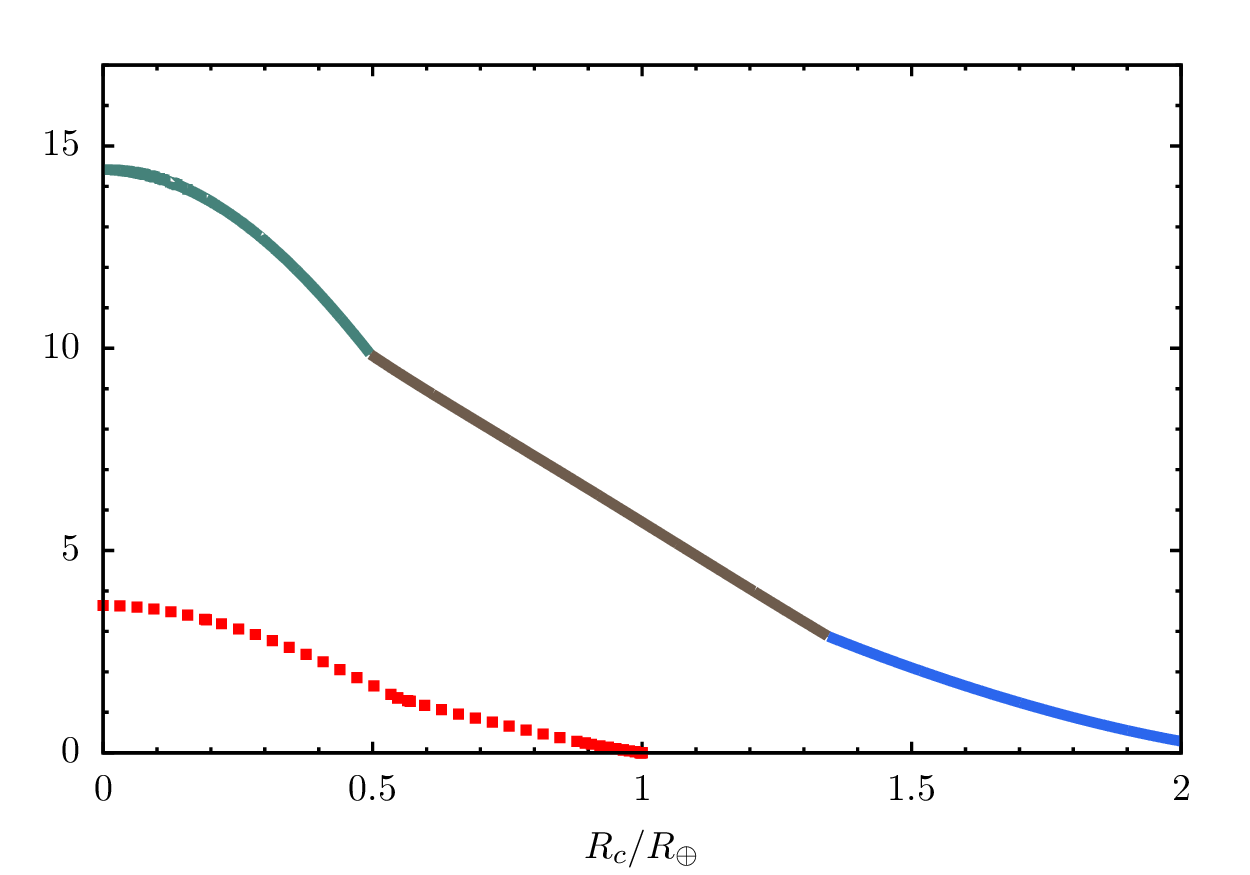}%
\includegraphics[clip]{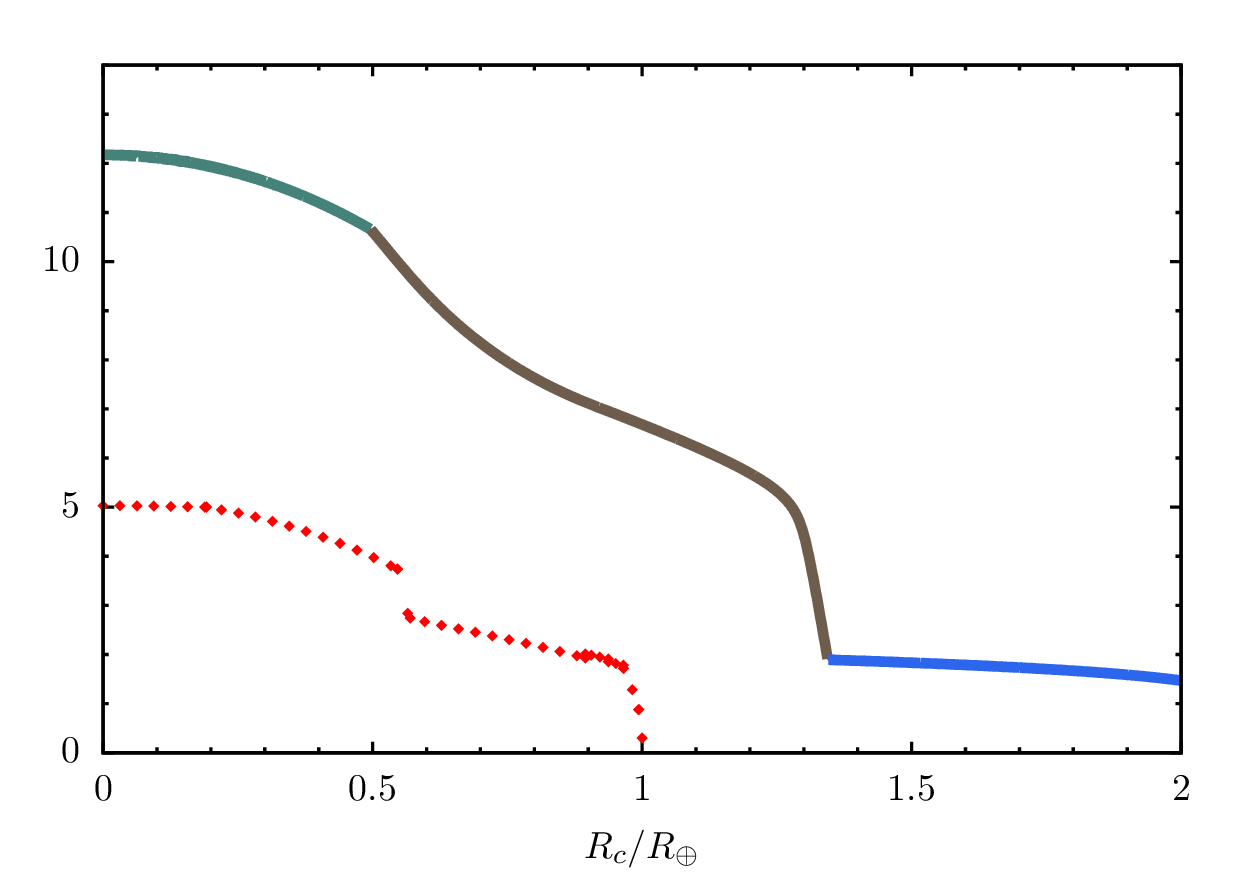}}
\resizebox{\linewidth}{!}{%
\includegraphics[clip]{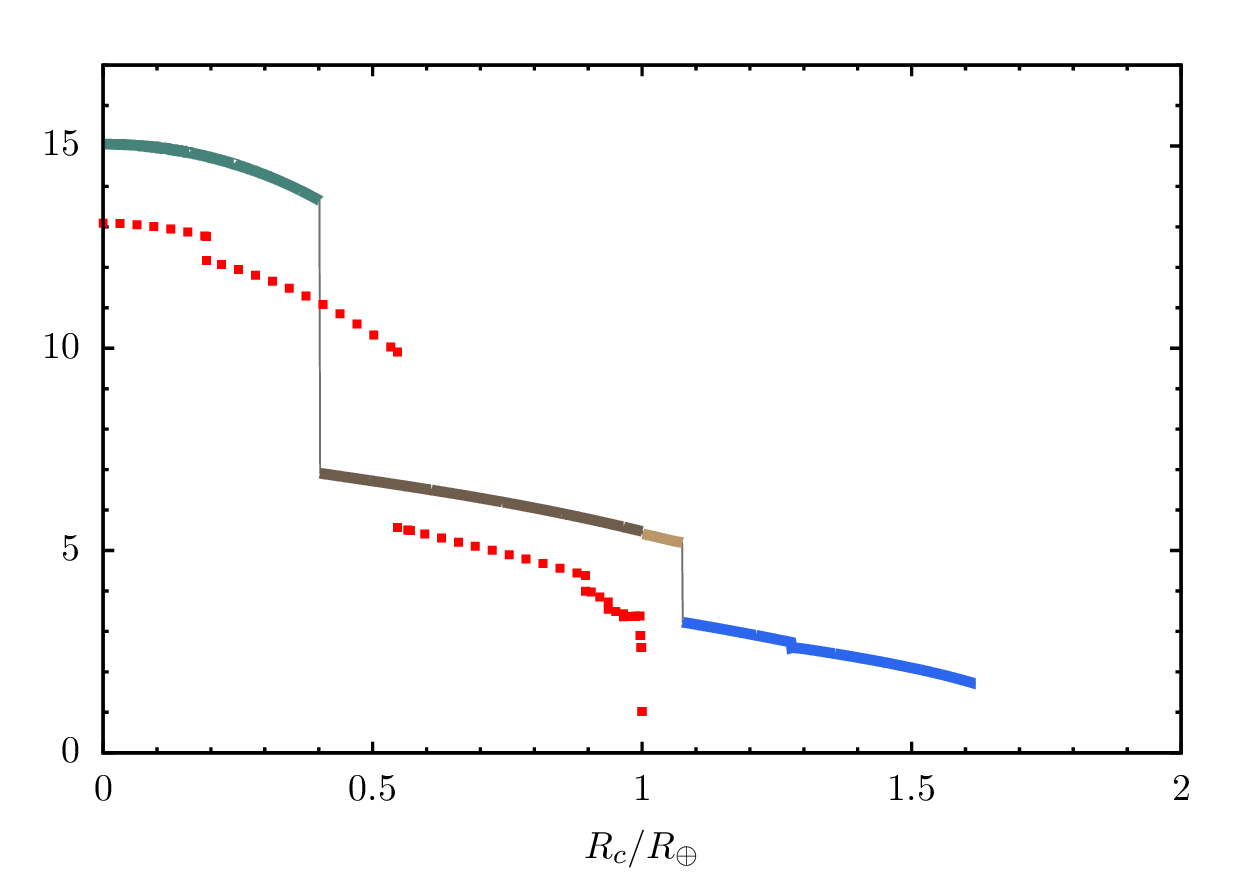}%
\includegraphics[clip]{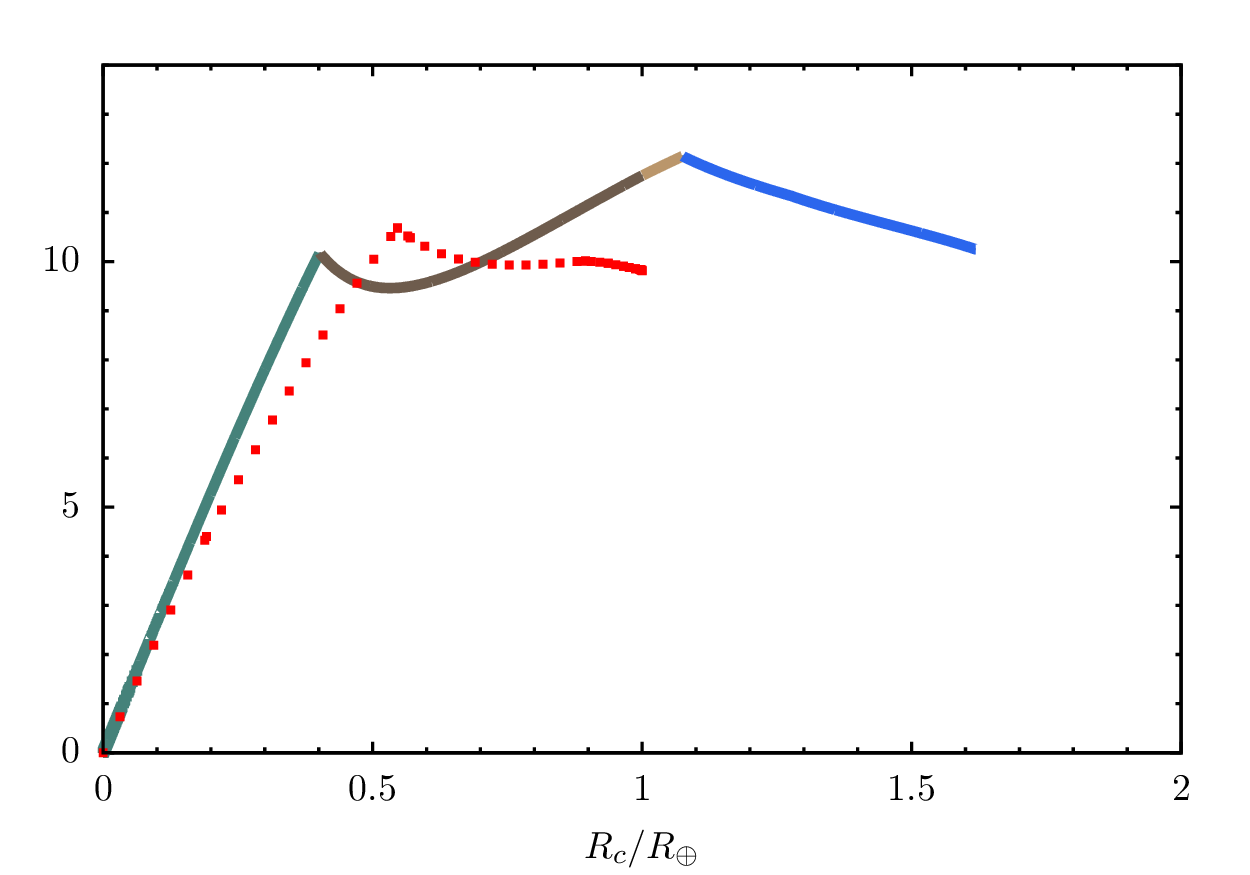}%
\includegraphics[clip]{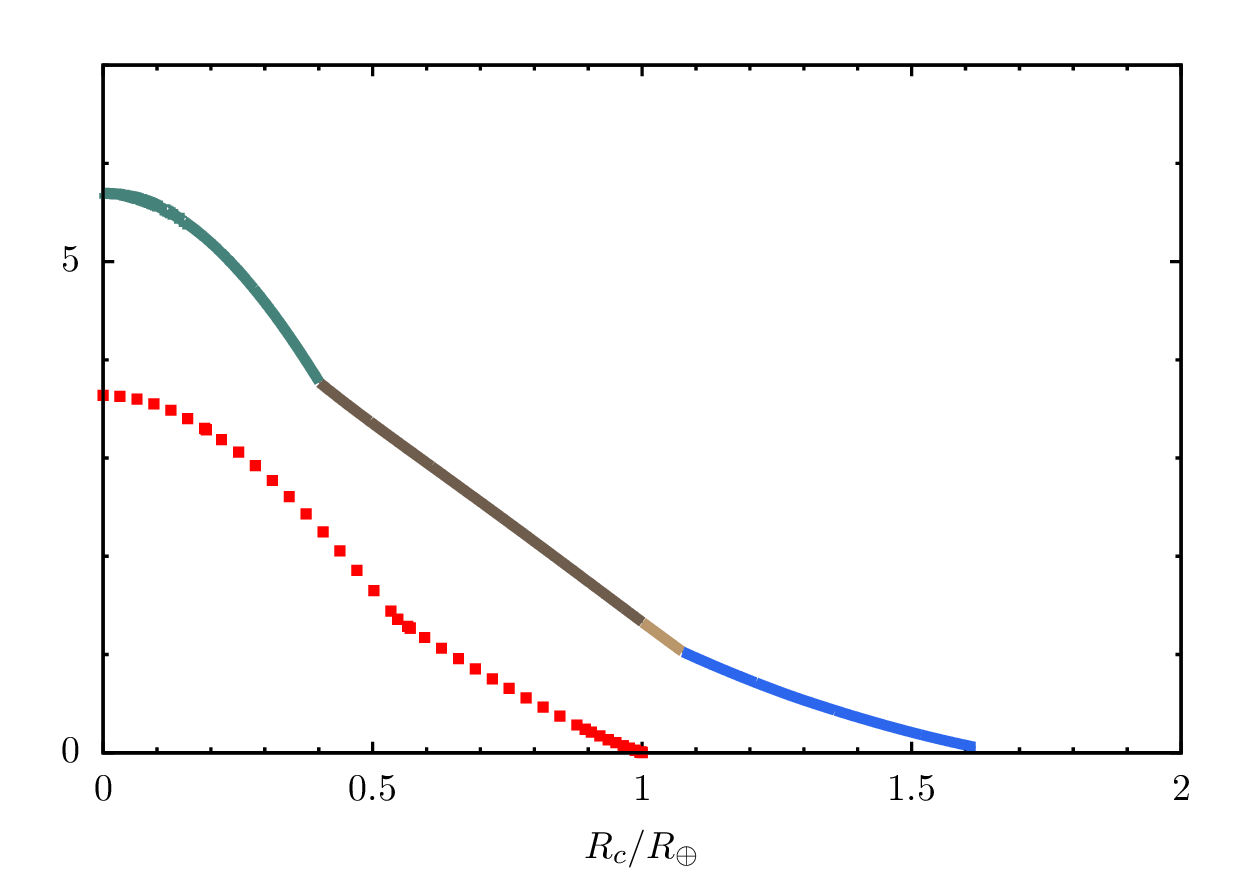}%
\includegraphics[clip]{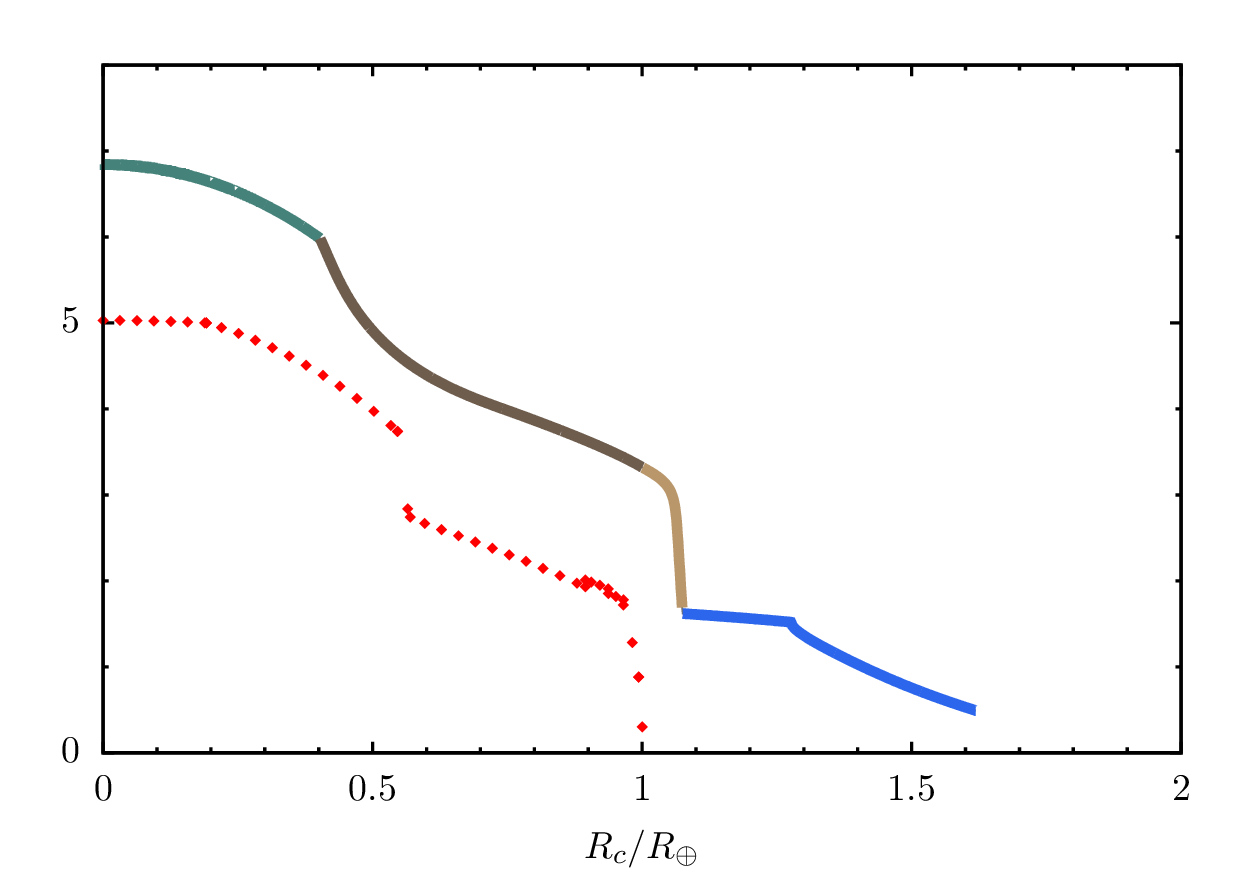}}
\resizebox{\linewidth}{!}{%
\includegraphics[clip]{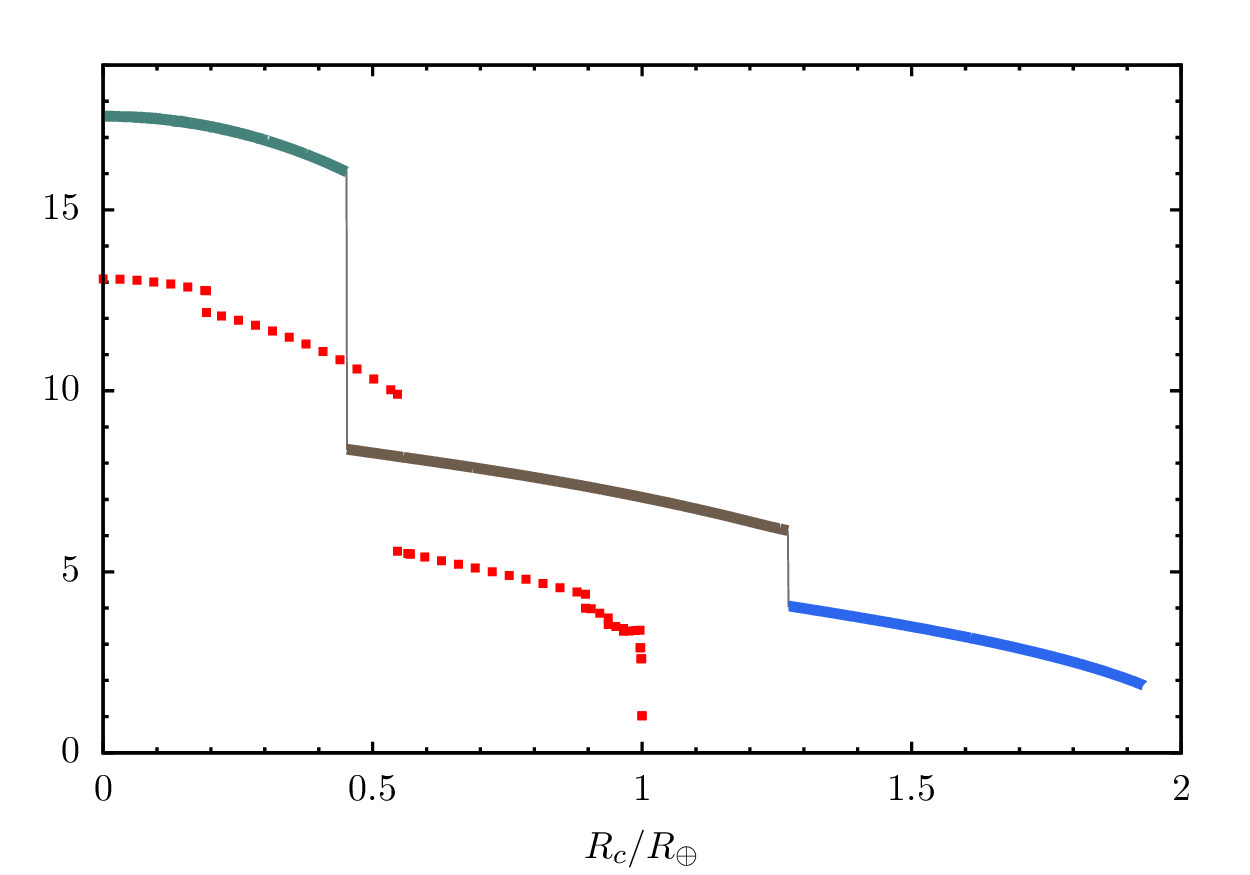}%
\includegraphics[clip]{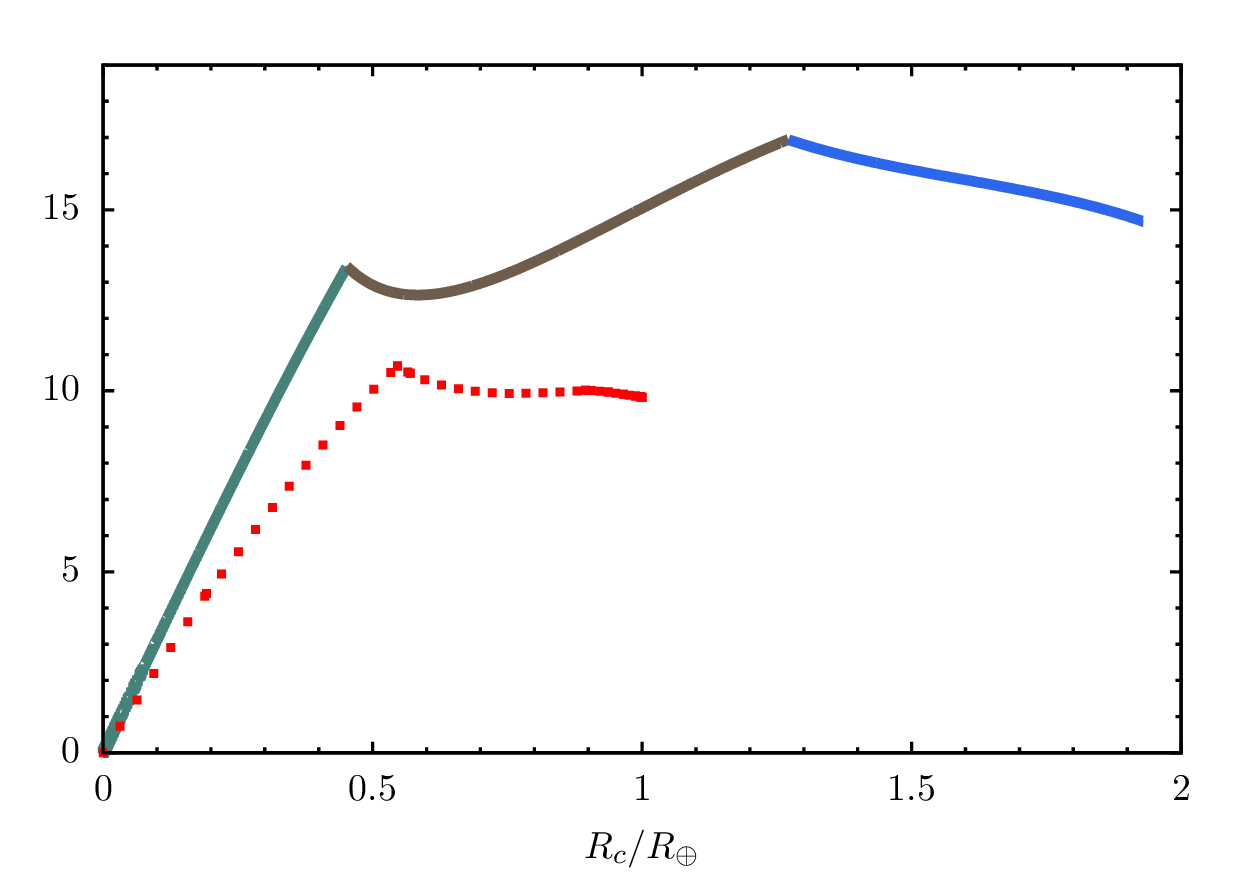}%
\includegraphics[clip]{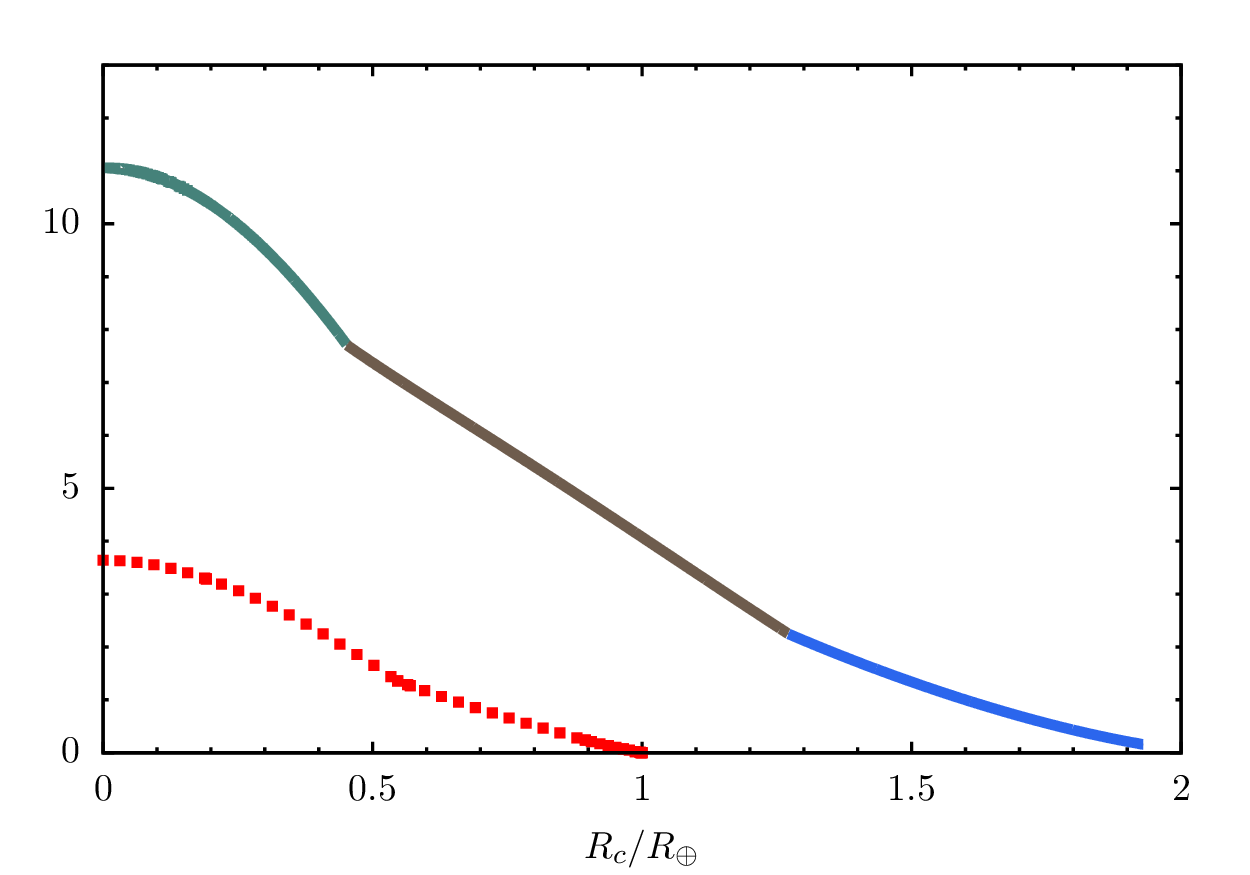}%
\includegraphics[clip]{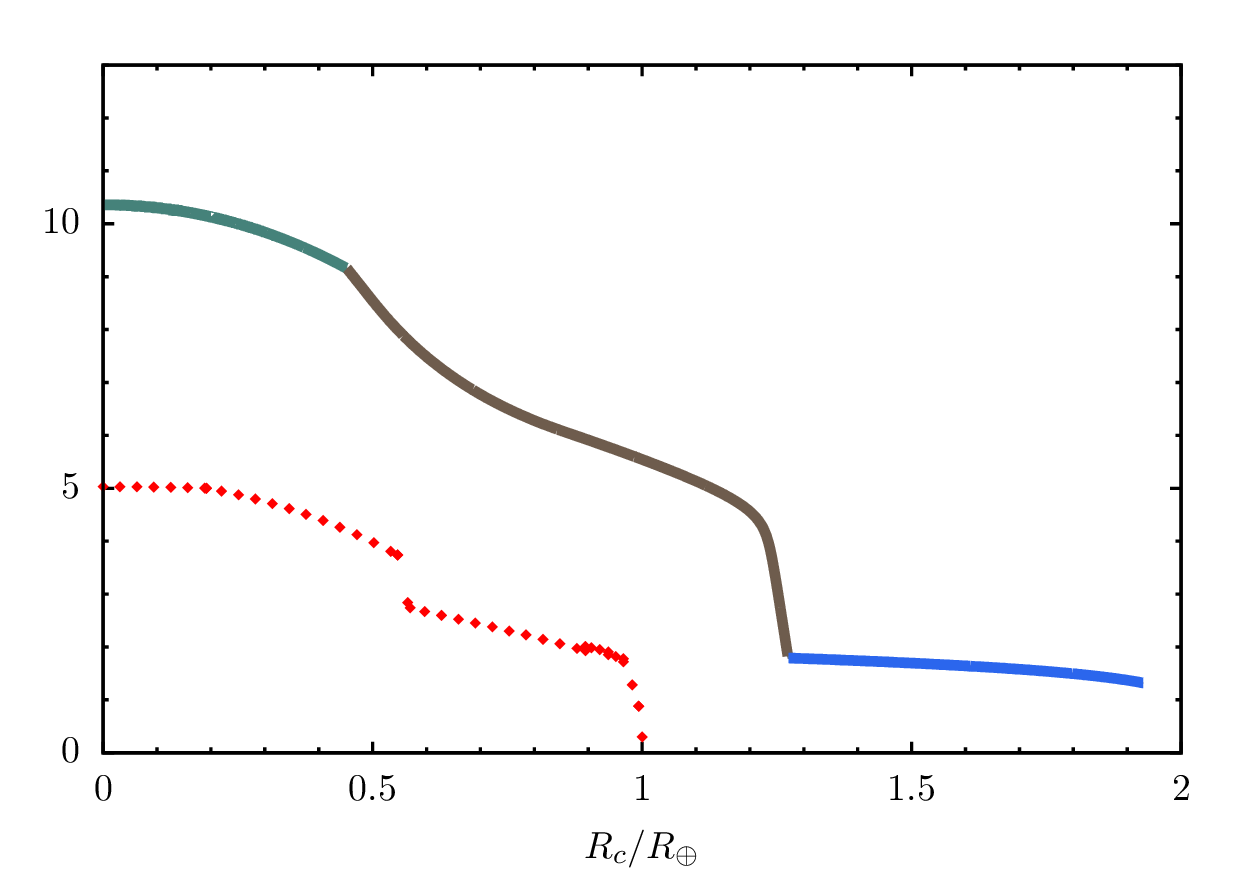}}
\caption{%
             Thermal structure calculations of the condensed cores of \kep\ 
             planets at an age of $8\,\Gyr$. See the text for further details. 
             Each row of panels illustrates the stratification of density (left), 
             gravitational acceleration (center-left), pressure (center-right), and 
             temperature (right). 
             From top to bottom, the panels refer to the interiors of \kep b through g.
             }
\label{fig:thc}
\end{figure*}

As mentioned above, at least $\approx 80$\% of \kep\ planets' total 
mass consists of condensed material, hence the importance of the
condensed structure of the planets in determining their radii. 
In this appendix, the radii and moment of inertia factors derived 
from isothermal models of the condensed interiors, applied in 
the main text, are compared to those derived from the improved 
structure models described in Appendix~\ref{sec:ics}. 
It should be pointed out that not all condensed layers of the simulated 
planets are in a solid phase, as liquid iron and liquid/superionic \ice\ 
may exist as well.
Results from this comparison are listed in Tables~\ref{table:kep_comp}.

In order to asses the impact of interior thermodynamics, 
the structure of the cores of the simulated planets was 
re-calculated applying core-envelope boundary pressures 
and temperatures ($P_{c}$ and $T_{c}$) obtained from the 
ex situ calculations discussed in \cisec{sec:exsitu}, at the age 
of $8\,\Gyr$ (see \cifig{fig:thc}).
The pressure $P_{c}$ ranges from $\approx 3$ to 
$\approx 30\,\mathrm{GPa}$, whereas the temperature
$T_{c}$ is between a little over $500$ and $\approx 1500\,\K$.
Clearly, the values are unconstrained in the case of \kep b, 
since no light-element envelope is present at $8\,\Gyr$. 
However, assuming the presence of a relatively thick steam 
atmosphere, the values $P_{c}=1\,\mathrm{GPa}$ 
and $T_{c}=1100\,\K$ are applied.
The differences between thermal and isothermal core radii are 
small, less than $0.5$\% in most cases and somewhat less than 
$2\%$ for \kep c. 
The addition of thermal pressure, the second term on 
the right-hand side of \cieq{eq:PRT}, tends to reduce density 
and increase $R_{c}$. However, there are additional effects
to consider when comparing the two types of models.
For example, \ice\ in isothermal cores is all in condensed form,
whereas the liquid and superionic phases are, in fact, predominant 
in thermal models.

The case of \kep b is more difficult to evaluate. The choice of
$P_{c}$ and $T_{c}$ (stated above) leads to a $5$\% larger $R_{c}$, 
compared to its isothermal counterpart. Upward and downward 
variations of a factor of two in boundary pressure produce a total 
change in core radius $\Delta R_{c}\approx 0.05\,\Rearth$.
If a thinner steam atmosphere could account for the observed radius, 
assuming a near-isothermal structure at the equilibrium temperature,
$T_{c}\approx T_{\mathrm{eq}}\approx 820\,\K$, a boundary pressure 
$P_{c}\approx 0.1\,\mathrm{GPa}$ would still provide a core radius only 
about $5$\% larger than that in Table~\ref{table:kep_comp} (and 
$\approx 10$\%, $\approx 0.15\,\Rearth$, larger than the isothermal 
radius).
Significantly more inflated cores, and \ice\ shells in particular, would 
require much thinner atmospheres \citep[e.g., $R_{c}\approx 1.76\,\Rearth$ 
for $P_{c}\approx 25\,\mathrm{MPa}$, see also][]{thomas2016}, 
although it is not clear whether this possibility is relevant to \kep b.

The results of Table~\ref{table:kep_comp} also indicate that the
moment of inertia at $8\,\Gyr$ is well approximated by isothermal 
models. 
The last column lists the energy flux at the core surface, $Q_{c}$,
normalized to the average surface flux of the Earth, 
$Q_{\oplus}=8.7\times 10^{-2}\,\mathrm{W\,m^{-2}}$ \citep{turcotte2014}.
As explained above, these values are based on radiogenic heating 
rates in the planets' silicate mantles estimated for the Earth
(at its present age).
The total energy output of the cores, $4\pi R^{2}_{c}Q_{c}$, is 
negligible compared to the luminosities reported in Table~\ref{table:sumex}.
Experiments conducted on \kep b and e with heating rates reduced 
by a factor of two ($\varepsilon=3.7\times 10^{-12}\,\mathrm{W\,kg^{-1}}$) 
resulted in very similar interiors ($R_{c}$ and the MoI factor differing
by $\lesssim 0.3$\%, $P$ and $\rho$ at $R=0$ differing
by $\lesssim 1$\%, and $T$ at $R=0$ differing by $1$--$2$\%).
The surface heat flux, $Q_{c}$, was instead reduced by a factor
of $\approx 1.8$.

\cifig{fig:thc} shows some structural properties of \kep\ condensed 
interiors at $8\,\Gyr$. The density, gravitational acceleration, pressure, 
and temperature are plotted along with the PREM and the reference 
geotherm of \citet{stacey2008}, as indicated in the legends of the top 
panels.
As mentioned above, the core temperatures of the Earth may actually 
be higher. The temperature at the inner core boundary of the Earth 
\citep[$P\approx 330\,\mathrm{GPa}$, $R=1221\,\mathrm{km}$,][]{prem1981} 
was estimated to be $6230\pm 500\,\K$ \citep{anzellini2013}, based
on an experimental determination of the melting curve of Fe
\citep[though uncertainties persist, e.g., ][pp.\ 117-147]{alfe2009,aquilanti2015,vocadlo2015}.

The iron nucleus of all simulated planets is in the solid phase.
The interior models of \cifig{fig:rciso}, however, have totally 
or partially molten nuclei at early isolation times, 
$t\lesssim 50$--$100\,\Myr$.
Moreover, if iron in the nucleus was alloyed with lighter elements 
(such as sulfur or oxygen), a significant volume fraction would 
likely be in the liquid phase, due to lower melting temperatures 
\citep[e.g.,][]{anderson1996,sotin2007,morard2014}. 
It is also likely that the presence of iron alloys would cause a 
(partially) molten nucleus in (at least some of) the planets at 
$8\,\Gyr$. For example, the CMB pressures of \kep b and f are 
$\approx 300$ and $\approx 370\,\mathrm{GPa}$, respectively. 
The melting temperature of FeS at these pressures is between 
$\approx 4000$ and $\approx 4300\,\K$ \citep{anderson1996}, 
lower than the calculated CMB temperatures of those planets.
It should be stressed, though, that sulfur and oxygen are among 
the impurities that produce the largest depression of the melting 
curve compared to that of pure iron \citep{terasaki2016}.

The silicate mantles contain only the high-pressure phases 
perovskite (pv) and post-perovskite (ppv). In fact, the transition between
olivine and perovskite happens at pressures around $23\,\mathrm{GPa}$
\citep[for $T\approx 2000\,\K$,][]{fei2004}, whereas pressures at the 
bottom of the \ice\ shell are in excess of $60\,\mathrm{GPa}$.
For the largest planets \kep c, d, e, and g, the pressure at the base of
the \ice\ shell is $\gtrsim 160\,\mathrm{GPa}$, so that ppv is the only 
phase present in their silicate mantles. Post-post-perovskite phases
are predicted for $P\gtrsim 900\,\mathrm{GPa}$ \citep{wagner2012},
a value reached only at the bottom of the mantle of \kep e
\citep[although the stability field of post-ppv phases is also 
determined by temperature,][]{stamenkovic2011}.
The \ice\ shell is mostly in fluid/superionic phase, except for
the presence of ice VII layers in the outer parts of \kep c, d and f.

The interior structures in \cifig{fig:thc} can be qualitatively compared to 
those presented by \citet[, see their Figure~1]{wagner2012}, since the 
structural and thermal model applied here share a number of similarities 
to theirs. However, a detailed comparison is not possible. 
They considered super-Earths with an Earth-like composition and Earth-like 
conditions at the surface. Masses are also different from those of \kep\ 
planets. Because of the assumed composition, their planets have radii 
smaller than those obtained here, for similar core masses. Consequently, 
their central pressures are higher, $\propto M^{2}_{c}/R^{4}_{c}$.
Mantle temperatures are different, which could depend on the boundary
conditions at the surface, on the presence of the \ice\ shell, and on the 
behavior of semi-convection in the outer layers of the silicate mantles. 
In fact, these layers behave as a stagnant lid, in which heat is mainly 
transported via conduction. The temperature gain across the lid typically 
accounts for $20$--$30$\% of the temperature gain across the entire 
mantle. 
The details of the temperature gradient in the lid depend on the thermal 
conductivity $k_{c}$, which is still poorly constrained in the high-pressure
and temperature ranges of pv and ppv phases 
\citep[e.g., see the discussion in][]{vandenberg2010,stamenkovic2011,hunt2012,wagner2012}. 
Heat transport becomes more efficient as convection grows more and 
more vigorous with depth, underneath the stagnant lid. However, details of 
mantle convection depend on the assumed viscosity parametrization, 
which is different between the two studies.

The temperatures deep down in the mantles may exceed the melting 
temperatures of silicates so that, instead of \cieq{eq:cse_tc}, 
the adiabatic temperature gradient in \cieq{eq:cse_ts} would apply to 
those layers. The $P$--$T$ curves of the mantles in \cifig{fig:thc} 
were compared to the high-pressure melting curve of silica (SiO$_{2}$) 
recently determined by \citet{millot2015} via shock-compression experiments. 
It was found that in none of the planets' mantles the temperature rises 
above the melting point. The situation may be different at earlier epochs, 
though, before the planets cool down. In the case of \kep e, it was found 
that over most of the evolution in isolation, $t \gtrsim 0.2\,\Gyr$, mantle 
temperatures lie below the melting curve.  At earlier times, however, the 
mantle is entirely molten \citep[according to the experiments of][]{millot2015} 
and, hence, it may be approximated as adiabatic. For the less massive 
planets, \kep c and f, mantle temperatures drop below the melting
curve at earlier times, $t \approx 20$--$30\,\Myr$.

\cifig{fig:rciso} shows that the condensed cores of \kep\ planets formed 
ex situ (i.e., with significant \ice\ content) can undergo substantial 
contraction during the early stages of evolution in isolation. 
By $t\approx 0.5\,\Gyr$, however, the cores are within a few percent 
of their current radii in Table~\ref{table:kep_comp}.
As mentioned in \cisec{sec:effects}, almost all of the contraction takes 
place in the outer \ice\ shells, and only $6$--$8$\% of the total 
difference $\Delta R_{c}$ shown in \cifig{fig:rciso} is caused by 
the contraction of nucleus and mantle. 
In fact, assuming that the condensed core of \kep g (which undergoes 
a contraction corresponding to $\Delta R_{c}\approx 0.34\,\Rearth$) 
was $30$\% Fe and $70$\% Si by mass, the difference in radii between 
$t\approx 5\,\Myr$ and $\approx 8\,\Gyr$ would be only 
$\Delta R_{c}=0.03\,\Rearth$!
The average gravitational energy per unit time released by the contraction 
of the cores in \cifig{fig:rciso}, during the first $1\,\Gyr$ of evolution, is
$\approx 10^{-11}\,L_{\odot}$, becoming much smaller during the following 
$7\,\Gyr$.



\end{document}